\pdfoutput=1
\documentclass[a4paper,11pt]{book}
\usepackage[latin1]{inputenc}
\usepackage[ngerman,USenglish]{babel}
\usepackage{amsmath}
\usepackage{amsfonts}
\usepackage{amssymb}
\usepackage{myslashed}
\usepackage{cancel}
\usepackage{graphicx}
\usepackage{slashed}
\usepackage{picins}
\usepackage{eso-pic} 
\usepackage{overpic}
\usepackage{wick}
\usepackage{simplewick}
\usepackage{xspace}
\usepackage{rotating}
\usepackage[toc]{appendix}
\usepackage{setspace}

\usepackage{floatflt}
\usepackage[justification=justified]{caption,subfig}
\usepackage{nonfloat}

\usepackage{ifthen}

\usepackage{listings}

\usepackage{pdfpages}

\DeclareMathAlphabet{\mathscr}{OT1}{pzc}{m}{it}

\usepackage{dsfont}
\newcommand{\Eins}{\mathds{1}}
\usepackage{xcolor}

\usepackage[nottoc]{tocbibind}

\usepackage{fancyhdr}
\usepackage[inner=35mm,outer=25mm,nofoot,bmargin=3cm,tmargin=35mm,headheight=1cm,headsep=10mm,footnotesep=10mm]{geometry}

\usepackage[plainpages=false,pdfpagelabels]{hyperref}

\newcounter{fnnumber}

\newcommand\topalignbox[1]{\leavevmode\vtop{\vskip0ex\hbox{#1}}}

\newcommand{\crossout}[1]{\setbox0\hbox{#1} \hbox to \wd0{\rlap{\rule[.6ex]{\wd0}{.5pt}}\box0}} 

\newcommand{\terminol}[1]{#1}
\newcommand{\toolkit}[1]{{\tt #1}}

\newcommand{\units}[1]{\ensuremath{\,\mathrm{#1}}}
\newcommand{\bra}[1]{\left\langle #1 \right|}
\newcommand{\ket}[1]{\left| #1 \right\rangle}

\newcommand{\braKet}[3]{#3\langle#1#3|#2#3\rangle}
\newcommand{\leftind}[1]{{\vphantom{|}}_{#1}\!}
\newcommand{\Tr}{\mathrm{tr}}
\newcommand{\BTr}{\mathrm{Tr}}
\newcommand{\elll}{\ell}
\newcommand{\latcfn}{\ensuremath{C}}
\newcommand{\TMD}{TMD PDF\xspace}
\newcommand{\TMDs}{TMD PDFs\xspace}

\newcommand{\MI}[1]{{\lfloor\hspace{-0.3ex}#1\hspace{-0.3ex}\rfloor}}
\newcommand{\prp}{\perp}
\newcommand{\GammaOp}{\Gamma^{\text{op}}}
\newcommand{\GammaDist}{\ensuremath{\Gamma^\text{d}}}
\newcommand{\GammaTwo}{\ensuremath{\Gamma^\text{2pt}}}
\newcommand{\GammaThr}{\ensuremath{\Gamma^\text{3pt}}}
\newcommand{\GammaDiq}{\ensuremath{\Gamma^\text{diq}}}
\newcommand{\lat}{{\text{lat}}}
\newcommand{\ren}{{\text{ren}}}
\newcommand{\unren}{{\text{unren}}}
\newcommand{\norm}{{\text{norm}}}
\newcommand{\Eu}[1]{{\bar{#1}}}
\newcommand{\muE}{\Eu{\mu}}
\newcommand{\nuE}{\Eu{\nu}}
\newcommand{\alphaE}{\Eu{\alpha}}
\newcommand{\betaE}{\Eu{\beta}}
\makeatother
\newcommand{\slashedE}[1]{\myslashed{{\rlap{/}{\hskip 0.15em}/}}{#1}}
\newcommand{\cdotE}{{\bar\cdot}}
\makeatletter
\newcommand{\tcdot}{{\cdot}}

\newcommand{\myRe}{\ensuremath{\mathrm{Re}}}
\newcommand{\myIm}{\ensuremath{\mathrm{Im}}}

\newcommand{\dlangle}{{\rlap{\big\langle}\hskip 0.1em\big\langle}}
\newcommand{\drangle}{{\rlap{\big\rangle}\hskip 0.1em\big\rangle}}

\newcommand{\Wline}[1]{\ensuremath{{\mathcal{U}}[#1]}}
\newcommand{\WlineC}[1]{\ensuremath{{\mathcal{U}}{[#1]}}}
\newcommand{\WlineClat}[1]{\ensuremath{{\mathcal{U}}^\text{lat}{[#1]}}}
\newcommand{\WlineI}[2]{\ensuremath{{\mathcal{U}}#1{[#2]}}}
\newcommand{\Wlineren}[1]{\ensuremath{{\mathcal{U}}^\text{ren}[#1]}}

\newcommand{\mquad}{\hspace{2mm}}

\newcommand{\quark}{q}
\newcommand{\Quark}{Q}
\newcommand{\Afield}{A}

\newcommand{\MDM}{\mathcal{M}}
\newcommand{\tMDM}{\tilde{\mathcal{M}}}
\newcommand{\tAmp}{\tilde{A}}
\newcommand{\TFM}{{\hat{\mathbb{T}}}}
\newcommand{\renZ}{{\mathscr{Z}}}

\newcommand{\mean}[1]{\ensuremath{\left\langle{#1}\right\rangle}}
\newcommand{\vect}[1]{\ensuremath{\boldsymbol{#1}}}
\newcommand{\vprp}[1]{\vect{#1}_\prp}

\newcommand{\mmin}{{\text{min}}}

\newcommand{\plateau}[1]{\mathop{\mathrm{\text{plateau}}}\left[\,{#1}\,\right]}

\newcommand{\toddmark}[1]{\left[#1\right]}

\newcommand{\doctitle}{%
	Transverse Momentum Distributions\\inside the Nucleon\\from Lattice QCD
	}
\newcommand{\docshorttitle}{%
	Transverse Momentum Distributions from Lattice QCD
	}
\newcommand{\docauthor}{Bernhard Musch}

\definecolor{MyDarkBlue}{rgb}{0,0.16,0.5} 
\definecolor{MyDarkGreen}{rgb}{0,0.50,0.16} 
\definecolor{MyBrown}{rgb}{0.2,0.45,0} 
\hypersetup{%
	pdfstartview={Fit},%
	pdftitle={\docshorttitle},%
	pdfauthor={\docauthor},%
	pdfkeywords={particle physics, transverse momentum dependent parton distribution functions, nucleon, quarks, lattice, QCD},%
	colorlinks={true},%
	linkcolor={MyDarkBlue},%
	citecolor={MyDarkGreen},%
	filecolor={MyBrown},%
	urlcolor={blue}%
}

\def\clap#1{\hbox to 0pt{\hss#1\hss}}

\def\mathclap{\mathpalette\mathclapinternal}

\def\mathclapinternal#1#2{%
\clap{$\mathsurround=0pt#1{#2}$}}

\begin{document}

\setcounter{tocdepth}{3}
\setcounter{secnumdepth}{3}

\pagenumbering{alph}
\pagestyle{empty}

\includepdf[fitpaper=true,pages=-]{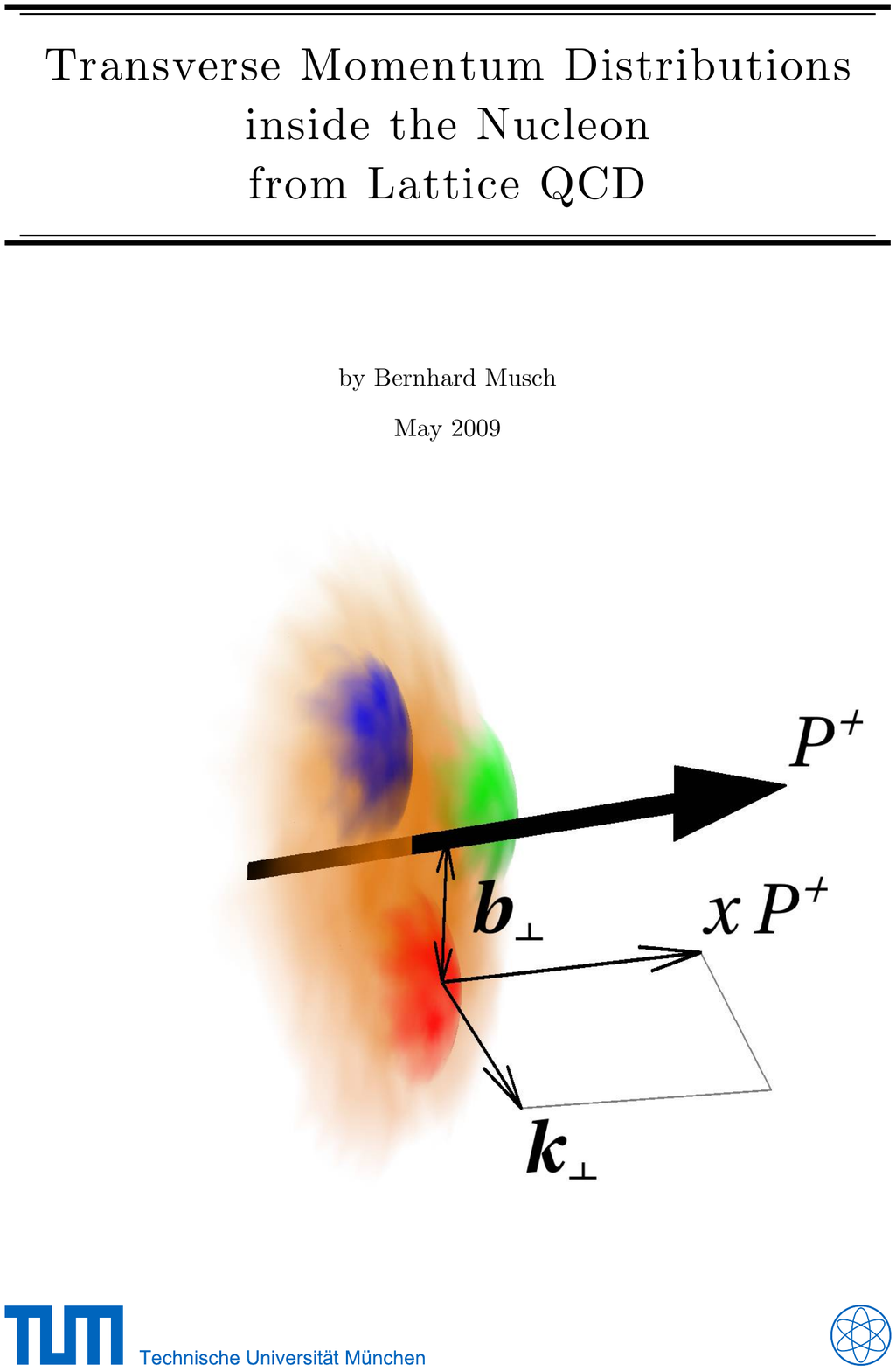}
\newcommand{\mskt}{\ensuremath{\langle \vec k_T^2 \rangle}}

\pagestyle{empty}

\selectlanguage{ngerman}

\vbox to \textheight{
\includegraphics[height=10mm]{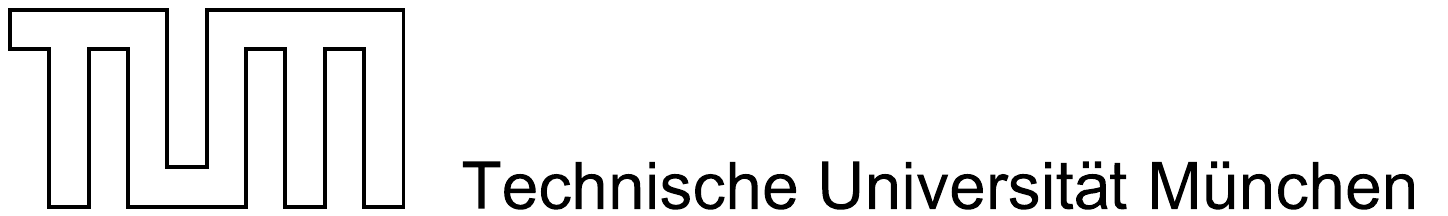}\hfill
\includegraphics[height=10mm]{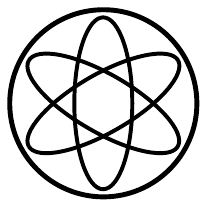}\par
\vspace{1cm}
{\centering%
Physik Department\\
Institut f\"ur Theoretische Physik T39\\
Univ.-Prof. Dr. Wolfram Weise\par
}%
\vspace{2cm}
{\centering%
{\bfseries\huge\doctitle\par}%
\vspace{1cm}
}
{\centering%
Dipl.-Phys. (Univ.) Bernhard Ulrich Musch\par
}%
\vfil
{
Vollst\"andiger Abdruck der von der Fa\-kul\-t\"at f\"ur Phy\-sik der Tech\-ni\-schen Uni\-ver\-si\-t\"at M\"un\-chen zur Er\-lang\-ung des aka\-demi\-schen Grades eines \par
\begin{centering}
	\emph{Doktors der Naturwissenschaften (Dr. rer. nat.)}\par
\end{centering}
genehmigten Dissertation.\par
}
\vspace{1cm}
{%
\doublespacing
{\centering\begin{tabular}{lcrl}
Vorsitzender: & \hspace*{1em}& & Univ.-Prof. Dr. Reiner Kr\"ucken\\
Pr\"ufer der Dissertation: & & 1. & TUM Junior Fellow Dr. Philipp H\"agler\\
 & & 2. & Univ.-Prof. Dr. Thorsten Feldmann\\
\end{tabular}\par}
}
\vspace{1cm}
{%
Die Dissertation wurde am 12.05.2009 bei der Technischen Universit\"at M\"unchen eingereicht und durch die Fakult\"at f\"ur Physik am 29.05.2009 angenommen.\par
}
\singlespacing
}
\newpage

\selectlanguage{USenglish}

\pagestyle{empty}


\vspace*{3cm}

\selectlanguage{USenglish}

{\Large\bfseries%
Summary
}

Nucleons, i.e., protons and neutrons, are composed of quarks and gluons, whose interactions are described by the theory of quantum chromodynamics (QCD), part of the standard model of particle physics. This work applies lattice QCD to compute quark momentum distributions in the nucleon. The calculations make use of lattice data generated on supercomputers that has already been successfully employed in lattice studies of spatial quark distributions ("nucleon tomography"). In order to be able to analyze transverse momentum dependent parton distribution functions, this thesis explores a novel approach based on non-local operators. One interesting observation is that the transverse momentum dependent density of polarized quarks in a polarized nucleon is visibly deformed. A more elaborate operator geometry is required to enable a quantitative comparison to high energy scattering experiments. First steps in this direction are encouraging.

\vspace{3cm}

\selectlanguage{ngerman}
{\Large\bfseries%
Zusammenfassung
}

Nukleonen, also Protonen und Neutronen, bestehen aus Quarks und Gluonen, deren Wechselwirkung durch die Quantenchromodynamik (QCD) innerhalb des Standardmodells der Teilchenphysik beschrieben wird. Diese Arbeit nutzt Gitter-QCD zur Berechnung von Quark-Impulsverteilungen im Nukleon. Dabei wird auf Gitterdaten zurückgegriffen, die auf Hochleistungsrechnern erstellt und bereits erfolgreich für die Analyse der räumlichen Quarkverteilung ("Nukleontomographie") eingesetzt wurden. Für die Untersuchung von Transversalimpuls-abhängigen Partonverteilungsfunktionen stellt diese Arbeit ein neuartiges Verfahren, basierend auf nicht-lokalen Operatoren, vor. Die Ergebnisse zeigen u.a. eine sichtbare Verformung der Transversalimpuls-abhängigen Dichte polarisierter Quarks im polarisierten Nukleon. Ein quantitativer Vergleich mit Streuexperimenten erfordert kompliziertere Operatorgeometrien. Erste Schritte in diese Richtung sind vielversprechend.

\selectlanguage{USenglish}
\newpage


\pagestyle{fancyplain}

\renewcommand{\chaptermark}[1]{\markboth{#1}{}}
\renewcommand{\sectionmark}[1]{\markright{\thesection\ #1}{}}
\lhead[\fancyplain{}{\thepage}]
	{\fancyplain{}{\rightmark}}
\chead{}
\rhead[\fancyplain{}{\leftmark}]
	{\fancyplain{}{\thepage}}
\lfoot{}
\cfoot{}
\rfoot{}

\pagenumbering{arabic}
\setcounter{page}{1}

\tableofcontents
\newpage


\chapter{Introduction}
\label{chap-intro}
What does a proton look like inside? Today, it is well established that \terminol{nucleons}, i.e., protons and neutrons, are composed of \terminol{quarks} and \terminol{gluons}, whose interactions are described by the theory of \terminol{quantum chromodynamics (QCD)}, part of the \terminol{standard model of particle physics}. Quarks carry a charge called ``color'', and are kept confined in color-neutral bound states by the strong interaction mediated by the gluons. Consequently, we can never observe isolated quarks in a detector.
The proton has a radius of the order of $1 \units{fm} = 1 \times 10^{-15}\units{m}$, inside of which the quarks are confined. According to Heisenberg's uncertainty principle, this means that the quarks cannot be permanently at rest in the nucleon. There is an intrinsic motion of quarks inside the nucleon, or, more precisely, the quark momenta follow distributions of non-zero width. In this work, we investigate this momentum distribution with theoretical means, using lattice QCD.

\captionsetup[figure]{format=default,justification=centering}
\begin{floatingfigure}{4cm}
	\includegraphics*[width=4cm,trim=150 0 70 350,clip=true]{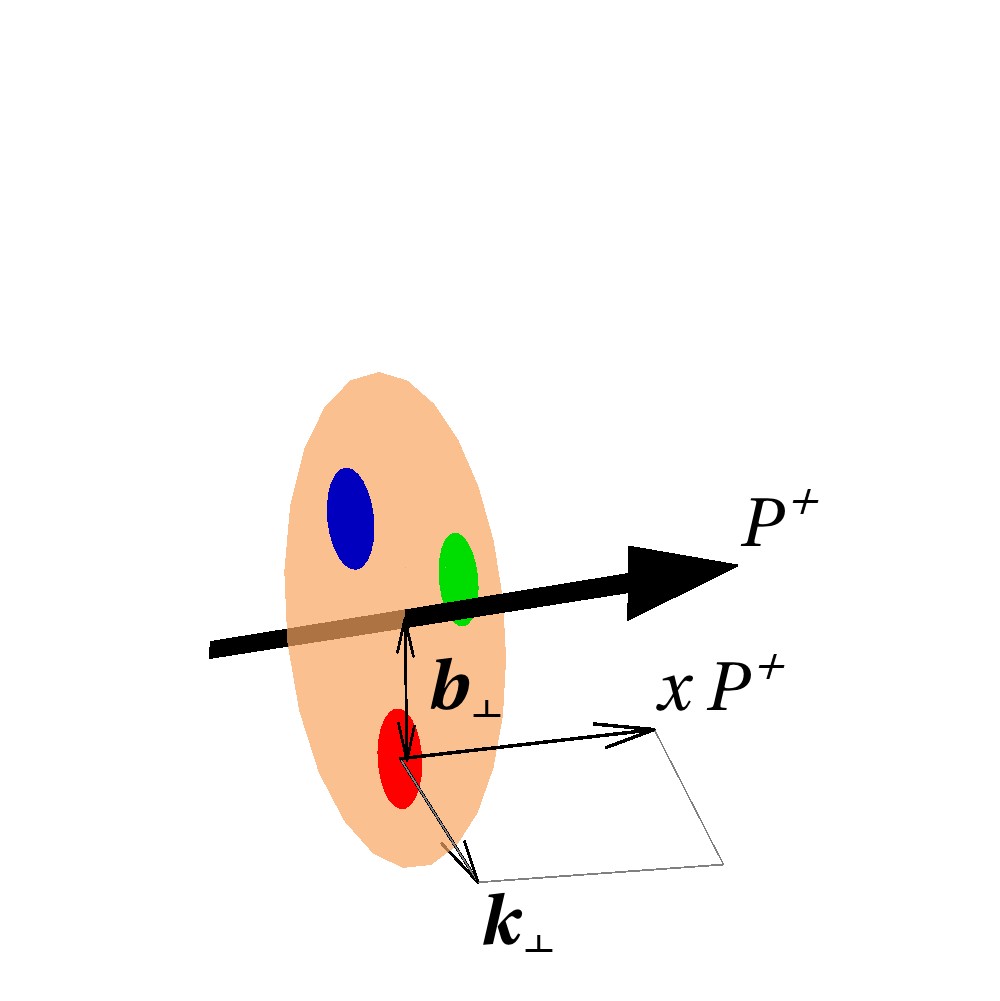}\par
	\caption{Illustration of a nucleon with large momentum $P$.}
	\label{fig-partonpicture}
	\end{floatingfigure}
\captionsetup[figure]{justification=justified}
Experimental information about the internal structure of the nucleon is obtained in \terminol{high energy scattering experiments}, in which we can probe the nucleon at large velocities close to the speed of light. Due to the Lorentz contraction, such a fast nucleon appears flat like a disk, as illustrated in Fig.~\ref{fig-steplike}. During the scattering experiment, a hard interaction takes place with one of the quarks inside the nucleon. At the instant of the interaction, the quark carries a \terminol{momentum fraction} $x$ of the nucleon momentum $P$, and has an \terminol{intrinsic transverse momentum} $\vprp{k}$ perpendicular to the direction of flight of the nucleon. The spatial location of the quark in the transverse plane is given by a vector $\vprp{b}$, the so-called \terminol{impact parameter}. The distribution of quarks with respect to the longitudinal momentum fraction $x$ is experimentally most easily accessible, e.g., in \terminol{fully inclusive} \terminol{deeply inelastic scattering} (\terminol{DIS}) experiments. It is parametrized in terms of conventional \terminol{parton distribution functions} (\terminol{PDF}s), such as $f_1(x)$. In general, PDFs tell us how likely it is to find a quark that carries a given momentum fraction $x$ of the nucleon. The concept of PDFs can be extended to \terminol{transverse momentum dependent parton distribution functions} (\TMDs) such as $f_1(x,\vprp{k}^2)$, which describe quark distributions with respect to both longitudinal momentum $x$ and transverse momentum $\vprp{k}$. Effects of the intrinsic transverse quark momentum are observable, e.g., in the angular distribution of the measured final state particle in semi-inclusive deeply inelastic scattering (SIDIS) experiments. Another extension of PDFs are the generalized parton distribution functions (GPDs). Those enable us, amongst other things, to visualize the nucleon in terms of $\vprp{b}$-dependent quark densities \cite{Burkardt:2000za,Ralston:2001xs,Burkardt:2002hr,Diehl:2005jf} (``nucleon tomography''). 

In the study at hand we attempt to calculate transverse momentum dependent parton distributions from first principles, i.e., from the theory of quantum chromodynamics. 
Although the laws of QCD are given by a few elegant mathematical expressions, it is very difficult to calculate properties of nucleons or other \terminol{hadrons}, due to the strong interaction. In recent years, great progress has been made within \terminol{lattice QCD}, a method that allows us to perform quantum field theoretical ``simulations'', albeit at an extreme computational cost. In particular, it has become possible to calculate GPDs \cite{Gockeler:2003jfa,Hagler:2003is,Schroers:2003mf,Renner:2004ck,Gockeler:2005cj,Diehl:2005ev,Gockeler:2006zu,Hagler:2007xi}, which yield ``tomographic'' pictures of the nucleon, showing an interesting deformation in the spin sensitive channels, see, e.g., Ref. \cite{Gockeler:2006zu}. 


The goal of this work is to devise methods within lattice QCD that enable us to obtain similar results for \TMDs. The $\vprp{k}$-dependent distributions provide information complementary to the $\vprp{b}$-dependence encoded in GPDs\footnote{\TMDs and GPDs are complementary in the sense that there is no simple transformation among them \cite{Ji:2004gf}. However, there exist certain non-trivial relations between GPDs and \TMDs \cite{Burkardt:2003uw,Burkardt:2003je,Meissner:2007rx}. In principle, it is conceivable to study $\vprp{b}$- and $\vprp{k}$-dependence simultaneously in the context of Wigner distributions \cite{Wigner:1932eb,Ji:2003ak,Belitsky:2003nz,Ji:2004gf}.}. 
The ability to visualize the $\vprp{k}$-dependence of quark distributions inside the nucleon has been an important incentive for our study of \TMDs on the lattice. In contrast to experiment, it is particularly easy on the lattice to study spin dependent effects. Analyzing polarized quarks in a polarized nucleon, we shall find an analogous deformation as seen, previously, with nucleon ``tomography'' in the $\vprp{b}$-plane. 

The motivation to study \TMDs is not solely based on our interest in the nucleon itself. 
In many scattering experiments, including those exploring physics beyond the standard model at the LHC, it is neccesary to give an accurate description of hadronic subprocesses. Computer programs modeling these subprocesses, such as the Monte Carlo event generator \toolkit{Herwig++} \cite{Bahr:2008pv}, rely on assumptions about the intrinsic motion of quarks inside hadrons, see, e.g., Ref. \cite{Gieseke:2007ad}. On the lattice, we can test these assumptions, e.g., how well the factorization $\text{``}f_1(x,\vprp{k}^2) \propto f_1(x) f_1^{(1)}(\vprp{k}^2)\text{''}$ is fulfilled or whether a Gaussian function is an adequate description of the $\vprp{k}$-dependence.

Note, however, that most of the results we present have been obtained with a simplified operator, which is particularly easy to realize on the lattice. There is still much debate about the precise operator needed to define \TMDs that are suitable in the description of scattering processes. For a quantitative comparison of results from the lattice and from scattering experiments, we will have to go beyond the simplified operator. We shall briefly show first steps in that direction towards the end of this work. 

Numerical lattice calculations are carried out in different stages. The production stage requires months on a supercomputer. These expensive computations are justified by the wide range of questions that can be addressed with the resulting data. In our case, we profit from the fact that the essential building blocks needed to calculate GPDs can be reused to analyze \TMDs, at a comparatively small additional computational cost.

This research project is based on \terminol{configurations} generated by the \terminol{MILC} collaboration \cite{Ber01} and \terminol{propagators} generated by the \terminol{Lattice Hadron Physics Collaboration (LHPC)} \cite{Hagler:2007xi}. The production of these large files has required an immense computational effort. To store the files locally, we have equipped one of our computers with 3.5 terabytes of space on hard drives. For the primary steps of our analysis of \TMDs, we have developed programs in C++ using the \toolkit{Chroma} and \toolkit{QDP++} libraries \cite{Edwards:2004sx} for parallelized lattice computations. To run them, we have combined personal computers of our theory department to small clusters with the \toolkit{MPICH2} implementation of the \toolkit{Message Passing Interface}. For the final analysis we have used \toolkit{Mathematica}.

\section{Outline}

The next chapter will familiarize the reader with the concept of \TMDs in the context of scattering experiments and factorization theorems. It will also become clear that the correct definition and regularization of \TMDs is still under debate.

The third chapter will start with a brief introduction to lattice QCD, before we will specialize to the calculation of nucleon structure. We will give some specifications of the ensembles and propagators that were kindly provided to us by the MILC and LHPC collaborations, and will explain our techniques to estimate statistical uncertainties.

In the main part of this thesis, chapter \ref{chap-straightlinks}, we will work with a simplified definition of \TMDs, suitable for first explorations on the lattice. The \TMDs we thus obtain are not strictly identical to those used in the literature and for the description of, e.g., semi-inclusive deeply inelastic scattering.
Nevertheless, they do provide qualitative insights into, e.g., the spin structure of the nucleon. A large part of this chapter will be devoted to \terminol{renormalization}, a necessary step in order to be able to present results which are independent of the details of our lattice calculation. 

The fifth chapter explores future concepts. In particular, we will discuss the results of a test calculation which goes beyond our simplified definition of \TMDs.

We summarize in chapter \ref{chap-conclusion}. Notes on conventions and some details are provided in the appendix. It is planned to create a web page with documentation of the file formats and software tools developed for this research project \cite{MuschTmddoc}.

\section{Basics of QCD}
\label{sec-basicQCD}

In general, electroweak interactions have little influence on the structure of hadrons. Thus the theory we need for the description of the structure of the nucleon is QCD, which describes just quark and gluon fields. For details on QCD, we refer to textbooks such as Ref.~\cite{Muta87}. The QCD Lagrangian reads
\begin{equation}
	\mathcal{L}_\text{QCD}[\bar \quark,\quark,\Afield](x) = \sum_{\quark = u,d,s,\ldots} \bar \quark(x) \left(i \slashed{D} - m_\quark  \right) \quark(x)  
	- \frac{1}{4}\, F_{\mu \nu\, a}(x)\, F^{\mu \nu}_a(x)\ ,
\end{equation}
where $\mu,\nu$ are Lorentz indices and where the index $a=1..(N_c^2-1)$ refers to the adjoint representation of the color group $\text{SU}(N_c)$, $N_c=3$. The quark fields $\bar u(x), \bar d(x), \ldots$ and $u(x), d(x), \ldots$ implicitly carry a color index $i=1..N_c$ and a Dirac index $\alpha=1..4$. In functionals and functional integrals, we will use the symbols $\bar \quark$ and  $\quark$ to refer collectively to all quark degrees of freedom. 
The quark masses $m_u$, $m_d$, $m_s$, $\ldots$ are fundamental constants. The covariant derivative operator
\begin{equation}
	D_\mu = \partial_\mu - i\, g\, \Afield_\mu
\end{equation}
introduces an interaction of quarks and gluons. Here the gluon field $\Afield_\mu(x)$ is a $3\times 3$ color matrix, which can be expressed in terms of real fields $\Afield_{\mu\,a}(x)$ according to $\Afield_\mu(x) = \Afield_{\mu\,a}(x)\,T_a$, where the $T_a$  are the 8 generators  of SU(3).\footnote{In terms of the Gell-Mann matrices $\lambda_a$, the generators are defined as $T_a = \lambda_a /2 $.} The coupling strength is given by the constant $g$. The field strength tensor $F_{\mu\nu\,a}(x)$ is defined in terms of the gluon field as
\begin{equation}
	F_{\mu\nu\,a} = \partial_\mu \Afield_{\nu\,a} - \partial_\nu \Afield_{\mu\,a} + g\ f_{a b c}\, \Afield_{\mu\, b}\, \Afield_{\nu\, c} \ ,
\end{equation}
where $f_{a b c}$ are the structure constants of SU(3). 
The Lagrangian $\mathcal{L}_\text{QCD}$ is invariant under local gauge transformations of the form
\begin{align}
	\quark(x) & \rightarrow \quark'(x) = W(x)\, \quark(x) \label{eq-gaugetrans-q} \ ,\\
	\bar \quark(x) & \rightarrow \bar \quark'(x) = \bar \quark(x)\, W^\dagger(x) \label{eq-gaugetrans-qbar} \ ,\\
	\Afield_\mu(x) & \rightarrow \Afield'_\mu(x) = W(x) \left( \Afield_\mu(x) - \frac{i}{g}\, W^{-1}(x)\left(\partial_\mu W(x)\right)\, \right) W^{-1}(x) \ , \label{eq-gaugetrans-A}
\end{align}
where the unitary $3\times3$ matrix $W(x)$ is an element of SU(3). 
As we will see in section \ref{sec-gaugetrafolat}, local gauge transformations of the gluonic degrees of freedom will look much simpler in the lattice formulation.

\chapter{Nucleon Structure from Deeply Inelastic Scattering Experiments}
\label{chap-dis}
This chapter briefly reviews the role \TMDs play in our understanding of scattering processes. The correlators introduced here will be the starting point for our lattice calculations described in the rest of this work. 
However, as we will explain, specifying the correlators precisely is a difficult task and still a matter of ongoing research. 

\section{Experimental Setups}

\begin{figure}[btp]
	\centering%
	\hfill%
	\subfloat[][]{%
		\label{fig-DrellYan-process}%
		\includegraphics[width=0.35\textwidth,trim=0 0 10 0,clip=true]{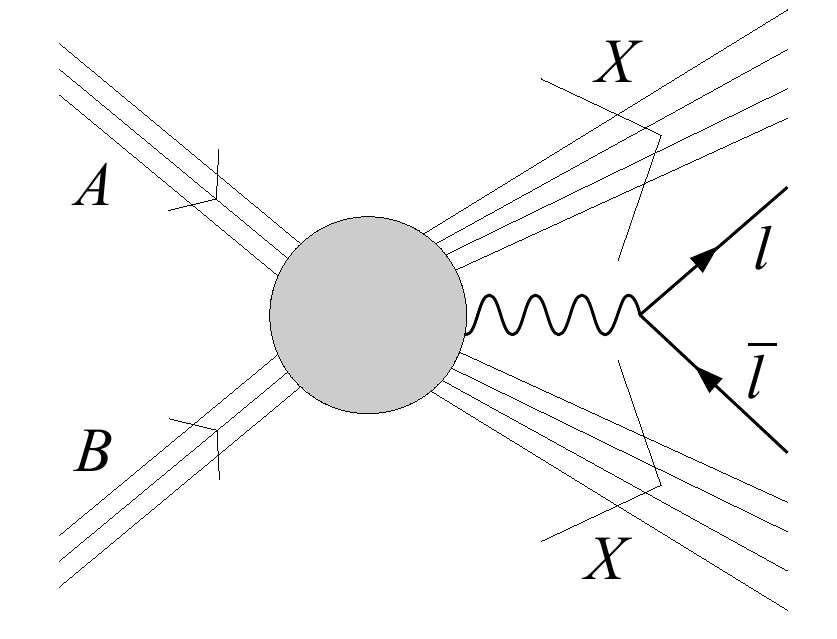}%
		}%
	\hfill%
	\subfloat[][]{%
		\label{fig-SIDIS-process}%
		\includegraphics[width=0.35\textwidth,trim=0 0 10 0,clip=true]{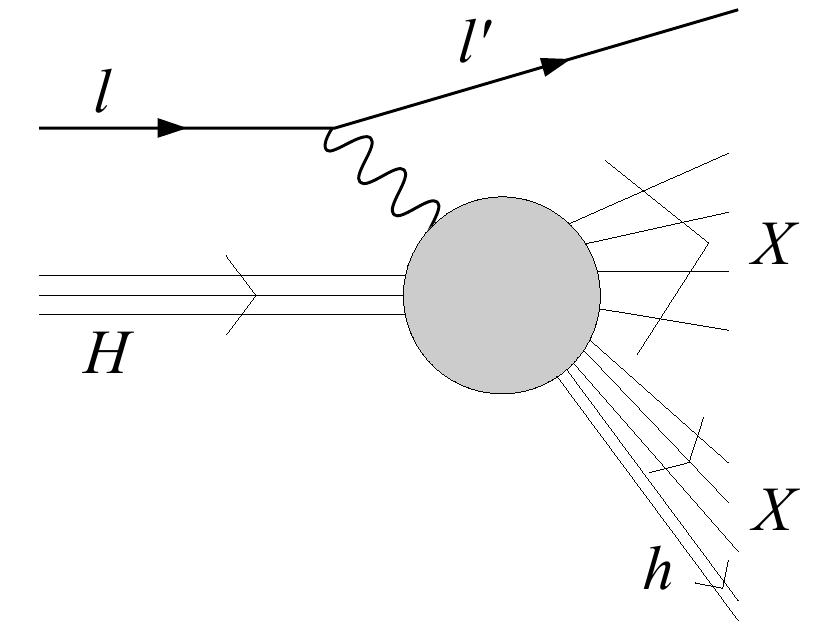}%
		}%
	\hspace*\fill\par%
	\caption[unpolarized]{%
		 Two examples of processes sensitive to \TMDs. We draw the leading contributions, in which a single electroweak gauge boson (wiggled lines) is exchanged.\par
		\subref{fig-DrellYan-process}\mquad%
			The Drell-Yan process. Two hadrons $A$ and $B$ collide to form a lepton pair $l$, $\bar l$ and a number of other particles subsumed in $X$.\par			
		\subref{fig-SIDIS-process}\mquad%
			The SIDIS process. A lepton $l$ scatters off a hadron $H$ (typically a proton), leading to the production of a hadron $h$ as part of a jet. Apart from $h$, all particles in this jet and in the debris of $H$ are collected in $X$.
			\par%
		\label{fig-TMD-processes}
		}
	\end{figure}
To determine the structure of the nucleon experimentally, it is advantageous to probe the nucleon with interactions that can be described accurately in perturbation theory. Reactions of the nucleon with leptons are therefore of primary interest. The interaction between the lepton and the parton is mediated by an electroweak particle, i.e. a photon, a $W$ or a $Z$ boson. The following reactions are sensitive to \TMDs:\par
\begin{tabular}{cll}
	$\bullet$ & $A+B\rightarrow l+\bar l + X$, & the \terminol{Drell-Yan process}, and \\
	$\bullet$ & $l+H\rightarrow l'+h+X$, & called \terminol{1-particle inclusive} or \\
		& &\terminol {semi-inclusive deeply inelastic scattering (SIDIS)}.\\
\end{tabular}\par
The two processes given above are illustrated and explained in Fig.~\ref{fig-TMD-processes}. Note that we restrict ourselves to the leading contributions, where only a single electroweak gauge boson is exchanged.

In the following, let us have a closer look at SIDIS. For reasons of clarity, we restrict ourselves to the special case that the leptons are electrons $e^-$, the incoming hadron is a nucleon, and the exchanged virtual gauge boson is a photon $\gamma^*$. 

\begin{figure}[btp]
	\centering%
	\subfloat[][]{%
		\label{fig-SIDISkinematics}%
		\begin{minipage}{0.4\textwidth}
		\vbox{
		\vspace*{1cm}
		\includegraphics[width=\textwidth]{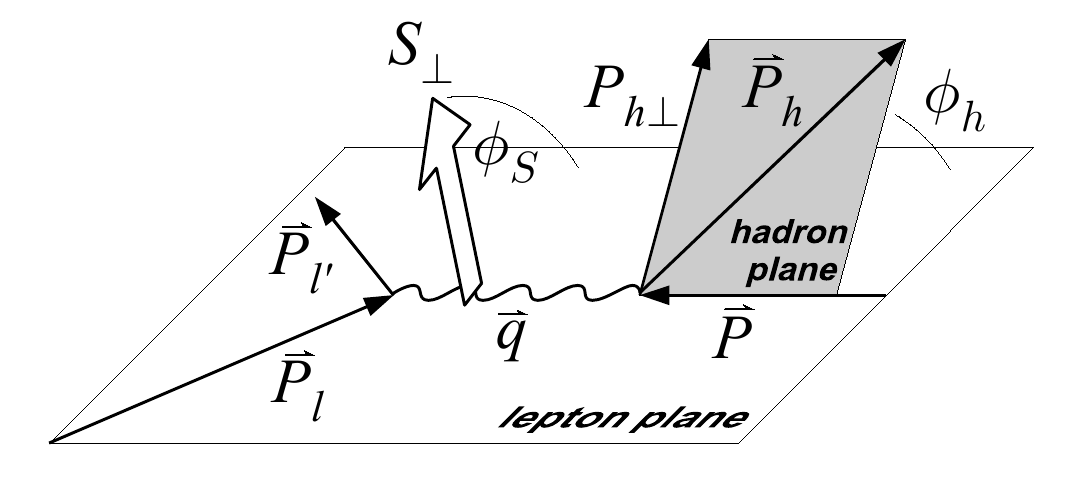}\par
		}\par
		\end{minipage}%
		}
	\hfill%
	\subfloat[][]{%
		\label{fig-SIDISdiagram}
		\includegraphics[width=0.57\textwidth]{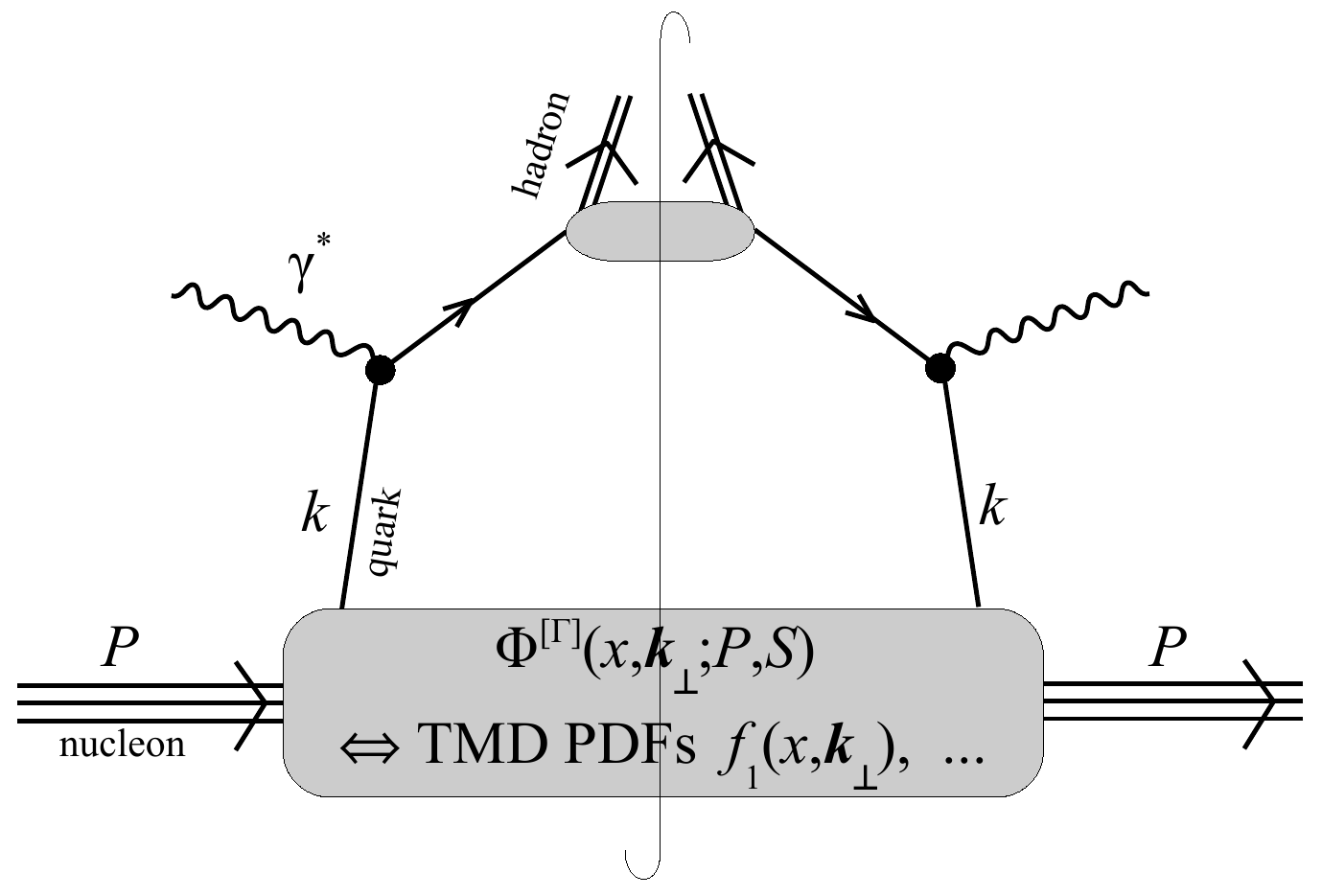}\par
		}%
	\caption[SIDIS diagram]{%
		\subref{fig-SIDISkinematics}\mquad%
			Kinematics of SIDIS following the Trento conventions \cite{Bacchetta:2004jz}.\par%
		\subref{fig-SIDISdiagram}\mquad%
			Simplified factorized tree level diagram of the hadron tensor in SIDIS.\par			
		\label{fig-SIDISdetails}%
		}
\end{figure}

The kinematics of the reaction is depicted in Fig.~\ref{fig-SIDISkinematics}, see Ref. \cite{Bacchetta:2004jz}. The momenta $P_l$ and $P_{l'}$ of the in- and outgoing electrons together with the nucleon momentum $P$ define the lepton plane. We choose to align the nucleon momentum $P$ and the photon momentum $q=P_l-P_{l'}$ with the $z$-axis\footnote{A common choice of this type is the Breit frame, with the only non-vanishing component of the photon four-momentum $q^3 = -Q$.}. Then the momentum of the produced hadron $P_h$ can have components $P_{h\prp}$ in the $xy$-plane, transverse to the nucleon momentum. Thus we can define an azimuthal angle $\phi_h$ between $P_{h\prp}$ and the lepton plane. For a polarized nucleon target, the transverse spin components of the nucleon form an angle $\phi_S$ with the lepton plane.

In the SIDIS cross section
\begin{equation}
	\frac{d\sigma}{d^3 P_h\ d^3 P_{l'}} \propto L_{\mu\nu}\ W^{\mu\nu}\ ,
\end{equation}
the lepton tensor $L_{\mu\nu}$ is calculable in perturbation theory. All the non-perturbative information related to hadron structure is encoded in the hadron tensor
\begin{align}
	W^{\mu\nu}(P,q,P_h)  = &\  
	\delta^{(4)}(q+P-P_h-P_X)\ \sum_X\  \bra{N(P,S)} J^\mu(0) \ket{X\,h(P_h,S_h)}\nonumber \\
	\times&\ \bra{X\,h(P_h,S_h)} J^\nu(0) \ket{N(P,S)} \nonumber \\
	= &\ \int \frac{d^4 \elll}{(2 \pi)^4}\, e^{i q \cdot \elll}\  \sum_X\ \bra{N(P,S)} J^\mu(\elll) \ket{X\,h(P_h,S_h)}\nonumber \\ \times&\ \bra{X\,h(P_h,S_h)} J^\nu(0) \ket{N(P,S)}\ ,
	\label{eq-hadrontensor}
\end{align}
where $\ket{N(P,S)}$ is a nucleon state with momentum $P$ and spin $S$, $h(P_h,S_h)$ is a hadron with momentum $P_h$ and spin $S_h$, and where $J^\mu(\elll)$ is the electromagnetic current of the quarks at position $\elll$. Any one of the two matrix elements on the right hand side of eq.~(\ref{eq-hadrontensor}) corresponds to the gray blob in Fig.~\ref{fig-SIDIS-process}.
Note that the process is called \emph{semi}-inclusive because we sum over $X$, which is only part of the final state $\ket{X\,h(P_h,S_h)}$.

\section{A Parton Model inspired Factorization Ansatz}

How can we extract information about the structure of the nucleon from the hadron tensor? 
We need to decompose (\terminol{``factorize''}) the reaction further into perturbative (\terminol{hard}) and non-perturbative (\terminol{soft}) subprocesses. A simplified sketch of a factorized hadron tensor is given in Fig.~\ref{fig-SIDISdiagram}. The left and right halves of the diagram are mirror images. Each half is related to one of the matrix elements on the right hand side of eq.~(\ref{eq-hadrontensor}). Let us concentrate on the left half of the diagram. The central assumption we want to make is that the virtual photon $\gamma^*$ couples with the electromagnetic current of a single quark of the nucleon, which initially carries momentum $k$. We want to look at cases where the virtuality $Q^2\equiv -q^2$ of the photon is large, $Q^2 \gg m_N^2$. In this case the quark-photon-interaction is a \terminol{hard} process, i.e., it can be treated perturbatively and it is drawn outside the gray blobs. The struck quark cannot appear as a real particle in the detector because it carries a color charge. According to the confining property of QCD, color charges cannot be isolated from each other over distances more than about a femtometer. Thus the struck quark undergoes a process called \terminol{hadronization}, or \terminol{fragmentation}, depicted in the upper gray blob of the diagram and parametrized in terms of \terminol{fragmentation functions} (also called \terminol{decay functions}). Here quark-antiquark pairs emerge from the vacuum and guarantee that the particles appearing in the detector are color neutral objects like hadrons. A jet of particles forms, of which only one hadron is detected individually. The rest of the jet (upper blob) and the remnants of the incoming nucleon (lower blob) are subsumed in $X$ and are typically ignored in the analysis of the experiment.

The proposed decomposition of the hadron tensor in Fig.~\ref{fig-SIDISdiagram} corresponds to an early factorization ansatz of the cross section suggested by Collins \cite{Collins:1992kk}\footnote{To keep the discussion simple, we quote the unpolarized cross section.}:
\begin{equation}
	\frac{d \sigma}{d^3 P_{l'}\ d^3 P_h} = 
	\frac{4 x_B}{P_{l'}^0\, P_h^0\, Q^2} \sum_{\quark=u,d,s,\ldots} \int d^2 \vprp{k}\ f_{1,\quark}(x_B,\vprp{k};Q^2)\ \frac{d\hat \sigma}{d \Omega}\ D_{h,\quark}(z,\vprp{k}+\vprp{q};Q^2)\ .
	\label{eq-earlycollinsfac}
\end{equation}
Here we have used the following kinematic variables: 
\begin{align}
	Q^2 &\equiv -q^2\ , & 
	x_B & \equiv \frac{Q^2}{2 P \cdot q}\ , & 
	z & \equiv \frac{P \cdot P_h}{P \cdot q} \ ,& 
	q_\prp & \equiv q - \frac{P_h \cdot q}{P_h \cdot P} P - \frac{P \cdot q}{P \cdot P_h} P_h\ .
\end{align}
In eq.~(\ref{eq-earlycollinsfac}), the cross section $\hat \sigma $ collects the hard pieces of the reaction, i.e., the short-distance part of electron-quark scattering. $D_{h,\quark}(z,\vprp{k}+\vprp{q},Q^2)$ is a fragmentation function for quark type $\quark$ and corresponds to the upper blob in Fig.~\ref{fig-SIDISdiagram}. Finally, $f_{1,\quark}(x_B,\vprp{k};Q^2)$ describes the density of quarks $\quark$ in the nucleon. It corresponds to the lower blob in Fig.~\ref{fig-SIDISdiagram} and is the prototype of a \TMD. 

Why do we parametrize the quark distribution $f_{1,\quark}$ in terms of $x_B$ and $\vprp{k}$? Let us understand this in the parton model, where we assume that the moment of interaction with the lepton is so short that the nucleon can be described as a collection of free particles, the ``partons'' (quarks or gluons). Consider, for example, the rest frame of the incident lepton, $\vect{P_l} = 0$. In this frame, the momentum of the nucleon will be very large, which we can write as $P^+ \gg m_N$ using the light cone coordinates specified in section~\ref{sec-lightconecoord}. The entire momentum transfer $q$ is passed on to a single parton, a quark with initial momentum $k$. Because we work in a large momentum frame of the nucleon, the relative magnitudes of the components of $k$ will behave as $k^+ : \vprp{k} : k^- \sim P^+/m_N : 1 : m_N/P^+$. 
We neglect the suppressed momentum component $k^-$ in the kinematics of the process.
To specify $k^+$, we introduce the dimensionless variable $x \equiv k^+ / P^+$.
%
Now, let us require that the parton be not too far off-shell, i.e., $k^2 \approx 0$ and $(k+q)^2 \approx 0$. 
This means $0 \approx k^2 + 2 k \tcdot q + q^2 \approx 2 x P \tcdot q - Q^2$, from which follows $x \approx x_B$. Thus the so-called \terminol{Bjorken scaling variable} $x_B$ can be approximately identified with the \terminol{longitudinal momentum fraction} $x$ of the parton in the nucleon. 

We have already remarked that $\vprp{k}$ is suppressed by a factor $m_N/P^+$ as compared to the longitudinal quark momentum. In many cases, the average intrinsic transverse momenta of quarks in the nucleon and the final state are negligible in the kinematic description of a scattering process. In such cases, one can work with the conventional, ``\terminol{integrated}'' PDFs and fragmentation functions like $f_{1,\quark}(x;Q^2)$ and $D_{h,\quark}(z;Q^2)$, compare section \ref{sec-relationPDFs} below. 
However, when we consider SIDIS for a small transverse jet momentum ($|\vprp{q}| \ll Q$), we can no longer ignore intrinsic transverse momenta, and we need ``\terminol{unintegrated}'', transverse momentum dependent distributions. Just like an integrated parton distribution function $f_{1,\quark}(x_B;Q^2)$, the transverse momentum dependent distribution $f_{1,\quark}(x_B,\vprp{k};Q^2)$ follows an evolution equation in $Q^2$, which has been studied in Ref.~\cite{Ceccopieri:2005zz}. 

To promote the factorization ansatz to a factorization theorem, one can stick to the following strategy \cite{Collins:1989gx}: First, one evaluates the cross section in perturbation theory. To make this possible, incoming and outgoing hadrons are replaced by quarks. Then, one shows that the factorization ansatz is indeed valid for the perturbative calculation. This involves proving that the leading contributions come from Feynman graphs with momenta that are either far off-shell, or inside non-overlapping ``leading regions'', each of which is attributed to one of the non-perturbative contributions $f_\quark$ and $D_{h,\quark}$. One can now  identify the contributions from $f_\quark$ and $D_{h,\quark}$ in the cross section and divide them out. For the remaining short-distance contribution $\hat \sigma$ it is irrelevant whether we are using quarks or hadrons as external particles. Thus finally factorization is established and a perturbative expression for the short-distance part $\hat \sigma$ is available.

However, due to the important role of gluons, it turns out that Fig.~\ref{fig-SIDISdiagram} and eq.~(\ref{eq-earlycollinsfac}) need to undergo modifications before they can be used as an ansatz for a factorization theorem. We shall discuss these difficult issues later. 

\section{The Fundamental Quark-Quark Correlator}
\label{sec-corr}


\subsection{Definition}

Let us stick to the na\"ive picture of the previous section for a moment and look once more at the lower blob in Fig.~\ref{fig-SIDISdiagram}. It is a quark-quark correlation function of the nucleon. In its general form, we write it as
\begin{align}
	\Phi^{[\GammaOp]}_\quark(k,P,S) & = \int \frac{d^4 \elll}{(2\pi)^4} \ 
	e^{-ik \cdot \elll}\ 
	\underbrace{ \frac{1}{2} \bra{N(P,S)}\ \bar \quark(\elll)\, \GammaOp\ \WlineC{\mathcal{C}_\elll}\ \quark(0)\ \ket{N(P,S)} }_{\displaystyle \tilde \Phi^{[\GammaOp]}_\quark(\elll,P,S) }\ ,
	\label{eq-corr}
\end{align}
where the $\GammaOp$ is a Dirac matrix. The \terminol{Wilson line}  $\WlineC{\mathcal{C}_{\elll}}$ as defined in eq.~(\ref{eq-WLineAlongPath}) runs along a path $\mathcal{C}_\elll$ from $\elll$ to the origin. It is necessary to connect the quark operators via a Wilson line to ensure gauge invariance. The meaning of the Wilson line, also called \terminol{gauge link}, and the choice of the contour $\mathcal{C}_\elll$ will be discussed in the sections to follow. 

We have already mentioned that $k^-$ is negligible in the large momentum frame of the nucleon. Therefore, we are going to average over it in the correlator. We can also understand this as follows: Imagine the reaction takes place in the $xy$-plane. The time it takes for the nucleon to traverse this plane is proportional to $m_N/P^+$. Boosting to the nucleon rest frame, we find that the region of spacetime locations $\elll$ where the reaction takes place has a negligible extent in the ``$+$''-direction, suppressed by $(m_N/P^+)^2$. 
Therefore, the important information for the description of the high momentum collision is given by the correlator $\tilde \Phi^{[\GammaOp]}_\quark(\elll,P,S)$ at $\elll^+ = 0$.
After the Fourier transform, this is equivalent to an integration over $k^-$. Thus we define
\begin{equation}
	\Phi^{[\GammaOp]}_\quark(x,\vprp{k};P,S) \equiv \int dk^- \Phi^{[\GammaOp]}_\quark(k,P,S)\ ,
	\label{eq-TMDcorr}
\end{equation}
with the longitudinal momentum fraction $x \equiv k^+/P^+$ as introduced before. The calculation of the transverse momentum dependent correlators $\Phi^{[\GammaOp]}_\quark(x,\vprp{k};P,S)$ will be the focus of the study at hand. In order to state the results in a form independent of an explicit choice of a frame of reference (in particular, independent of $P^+$), these correlators are parametrized in terms of \TMDs, see section~\ref{sec-param}. \TMDs are profile functions that kinematically only depend on $x$ and $\vprp{k}^2$. One \TMD we have already met is $f_{1,\quark}(x,\vprp{k})$. It appears in 
\begin{align}
	\Phi^{[\gamma^+]}_\quark(x,\vprp{k};P,S) & = f_{1,\quark}(x,\vprp{k}^2) + \langle\text{spin dependent terms}\rangle \label{eq-f1intro} 
\end{align}
and is commonly interpreted as the density of quarks of flavor $\quark$ in the nucleon, see section~\ref{sec-probInt}. It is easy to understand why the correlator with $\GammaOp=\gamma^+$ is so important:
As in the case of the quark momentum $k$, the ``$+$''-component of the current $\bra{N(P,S)}\bar \quark \gamma^\mu \quark \ket{N(P,S)}$ is enhanced by a factor $P^+/m_N$.
Note that $\Phi^{[\GammaOp]}_\quark$ and $f_{1,\quark}$ are also functions of renormalization and factorization scales $\mu, Q^2, \ldots$, as we will see later. For the moment, we omit these additional variables in our notation.

\subsection{Relation to PDFs}
\label{sec-relationPDFs}

If we integrate not only over $k^-$, but also over $\vprp{k}$ in eq.~(\ref{eq-corr}), the components $l^+$ and $\vprp{\elll}$ become zero:
\begin{align}
	\Phi^{[\GammaOp]}_\quark(x;P,S) & \equiv \int dk^- \int d^2\vprp{k}\ \Phi^{[\GammaOp]}_\quark(k,P,S) \nonumber \\
	& = \int \frac{d \elll^-}{2\pi} \ 
	e^{-ixP^+\elll^-}\ 
	\frac{1}{2} \bra{N(P,S)}\ \bar \quark(\elll^-\hat n_-)\, \GammaOp\ \WlineC{\mathcal{C}_{\elll^-\hat n_-}}\ \quark(0)\ \ket{N(P,S)}\ .
	\label{eq-PDFcorr}
\end{align}
This correlator is used to introduce conventional, $\vprp{k}$-integrated PDFs such as $f_{1,\quark}(x) \equiv \Phi^{[\gamma^+]}_\quark(x;P,S)$. Thus, na\"ively, one would think that relations between PDFs and \TMDs such as
\begin{equation}
	\text{``}\quad f_{1,\quark}(x) = \int d^2\vprp{k}\ f_{1,\quark}(x,\vprp{k}^2)\quad \text{''}
	\label{eq-intpdfrelation}
\end{equation}
hold. However, the integral above is undefined. This is due to the behavior of $f_{1,\quark}(x,\vprp{k}^2)$ at large $\vprp{k}$, where perturbation theory is applicable. As explained in, e.g., in Ref. \cite{Bacchetta:2008xw,Diehl:2008sk}, one finds:
\begin{equation}
	f_{1,\quark}(x,\vprp{k}) \sim \frac{1}{\vprp{k}^2} f_{1,\quark}(x)
	\qquad \text{for }\vprp{k}^2 \rightarrow \infty\ .
\end{equation}
The interesting, non-perturbative information is encoded in \TMDs at low $\vprp{k}$.
Introducing a cutoff in eq.\,(\ref{eq-intpdfrelation}),
\begin{equation}
	\quad f^\text{cut}_{1,\quark}(x;Q) \equiv \int_{|\vprp{k}|<Q} d^2\vprp{k}\ f_{1,\quark}(x,\vprp{k}^2)
	\label{eq-intpdfrelationm}
\end{equation}
regularizes the integral and leads to a $Q$-dependence that corresponds to the DGLAP evolution  \cite{Gribov:1972ri,Parisi:1974sq,Dokshitzer:1977sg,Altarelli:1977zs} of conventional, integrated PDFs. However,
for $f^\text{cut}_{1,\quark}(x;Q)$ it is no longer evident that the correlator simplifies as in eq.~(\ref{eq-PDFcorr}), see also the discussion in \cite{Collins:2003fm}.

The approaches mentioned in sections~\ref{sec-Hautmann}, \ref{sec-Chay}, and \ref{sec-Chered} promise to elucidate the relation between PDFs and \TMDs. For the calculation of integrated PDFs in practice, the divergence in $\vprp{k}$ does not pose a problem, because the correlator eq.~(\ref{eq-PDFcorr}) is renormalized and evaluated directly, without the detour via \TMDs. 

\subsection{Mellin Moments}
\label{sec-defmellin}

Just as for PDFs, it can be useful to analyze the $x$-dependence of the correlator in terms of so-called \terminol{Mellin moments}. The $n^\text{th}$ Mellin moment is defined as
\begin{multline}
	\Phi^{[\GammaOp](n)}_\quark(\vprp{k};P,S)  \equiv \int_{-1}^{1} dx\ x^{n-1}\ \Phi^{[\GammaOp]}_\quark(x,\vprp{k};P,S) \\
	 = \int_{0}^{1} dx\ x^{n-1}\ \left( \Phi^{[\GammaOp]}_\quark(x,\vprp{k};P,S) + (-1)^{n-1}\, \Phi^{[\GammaOp]}_{\quark}(-x,\vprp{k};P,S) \right)\ .
	\label{eq-Mellincorr}
\end{multline}
Given that $\Phi^{[\GammaOp]}_\quark(x,\vprp{k};P,S)$ vanishes for $|x| > 1$, the integration limits in the definition above are inessential. As will be explained in section \ref{sec-lfquant}, we associate negative values of $x$ to antiquarks. Thus the second line of the above equation indicates that the Mellin moments introduced this way are a combination of quark and antiquark distributions.
Of particular interest to us is the \terminol{first Mellin moment}, where we simply integrate over $x=k^+/P^+$: 
\begin{align}
	\Phi^{[\GammaOp](1)}_\quark(\vprp{k};P,S) & = \int \frac{dk^+}{P^+} \int dk^-\ \Phi^{[\GammaOp]}_\quark(k,P,S) \nonumber \\
	& = \frac{1}{P^+} \int \frac{d^2\vprp{\elll}}{(2\pi)^2}\ e^{i \vprp{k}\cdot \vprp{\elll} } \ \tilde \Phi^{[\GammaOp]}_\quark(\elll,P,S) \big\vert_{\elll^+ = \elll^- = 0}\ .
	\label{eq-deffirstmellin}
\end{align}
The quark separations $\elll$ appearing in the correlator now lie in the transverse plane and are purely spatial -- a good premise for lattice calculations. The correlator describes the distribution of quarks (and antiquarks) in transverse momentum space, irrespective of their longitudinal momentum. For example, using the relation $f_{1,\bar \quark}(x,\vprp{k}^2)=-f_{1,\quark}(-x,\vprp{k}^2)$ \cite{Tangerman:1994eh}, we have in the unpolarized case
\begin{align}
	\Phi^{[\gamma^+](1)}_\quark(\vprp{k};P,S) = f_{1,\quark}^{(1)}(\vprp{k}^2) + \langle \text{spin dependent terms} \rangle\ ,\nonumber \\
	f_{1,\quark}^{(1)}(\vprp{k}^2) = \int_{0}^{1} dx\ f_{1,\quark}(x,\vprp{k}^2) - \int_{0}^{1} dx\ f_{1,\bar \quark}(x,\vprp{k}^2)\ .
\end{align}
Thus we interpret $f_{1,\quark}^{(1)}(\vprp{k}^2)$ as the difference of two $\vprp{k}$-dependent densities, namely the quark density $\int_{0}^{1} dx\ f_{1,\quark}(x,\vprp{k}^2)$ and the antiquark density $\int_{0}^{1} dx\ f_{1,\bar \quark}(x,\vprp{k}^2)$.
We present lattice results for the first Mellin moment in section \ref{sec-mellin}.

\section{Improved Definitions of \TMDs}
\label{sec-tmddef}

\subsection{A Starting Point: The Straight Wilson Line}

An na\"ive guess for the Wilson line $\WlineC{\mathcal{C}_\elll}$ in eq.~(\ref{eq-corr}) is a straight line from $\elll$ to $0$, i.e. $\WlineC{\mathcal{C}_\elll} = \Wline{\elll,\ 0}$ in the notation of eq.~(\ref{eq-WLineStraight}).
The resulting correlator is gauge invariant, and serves us for first exploratory studies on the lattice, see chapter~\ref{chap-results}. However, definitions of \TMDs designed for the description of real scattering experiments require more complicated gauge link structures, see below.

\subsection{The Physical Role of the Wilson Line}
\label{sec-linkrole}

\begin{figure}[btp]
	\centering%
	\subfloat[][]{%
		\label{fig-SIDISdiagram-glue}%
		\includegraphics[width=0.4\textwidth]{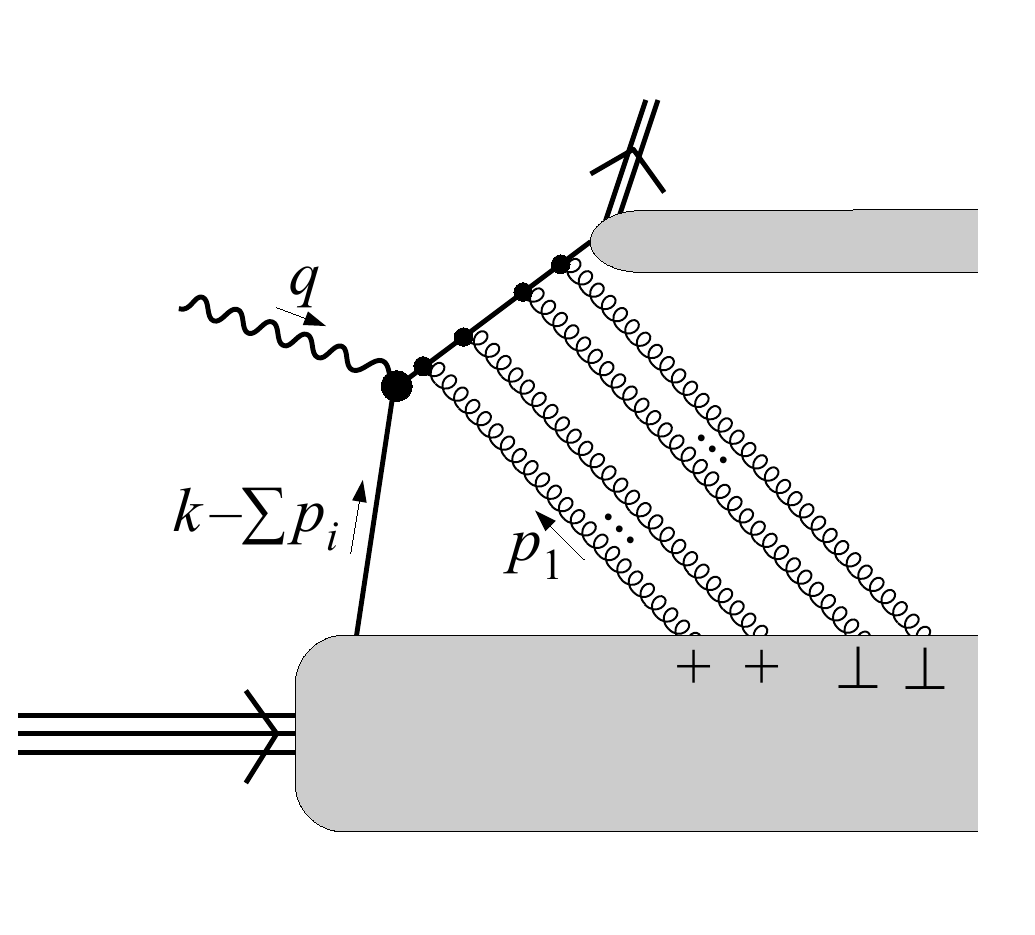}\par
		}
	\hfill%
	\subfloat[][]{%
		\label{fig-SIDISdiagram-eikonal}
		\includegraphics[width=0.57\textwidth]{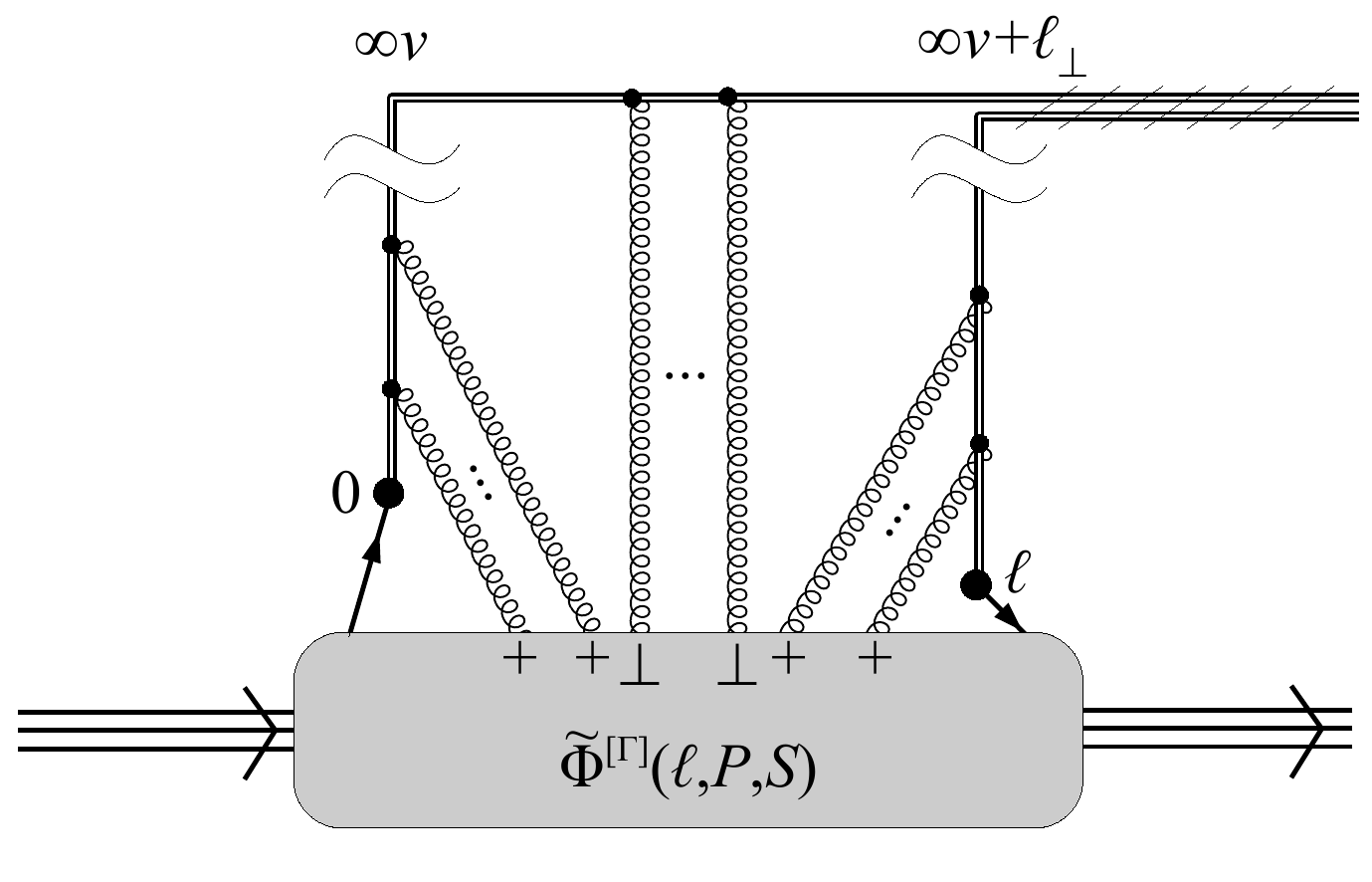}\par
		}%
	\vspace{-10pt}
	\caption[SIDIS diagram with glue]{%
		\subref{fig-SIDISdiagram-glue}\mquad%
			Cut SIDIS diagram with gluons at leading order in $M/Q$. Longitudinally polarized gluons are indicated by ``+'', transversely polarized gluons by ``$\prp$''. There can be any number of longitudinally and transversally polarized gluons. Note that the total momentum carried by the quark and the gluons leaving the blob is $k$.\par%
		\subref{fig-SIDISdiagram-eikonal}\mquad%
			Diagram of the quark-quark correlator with gauge links and gluons. In SIDIS, the link direction $v$ is either identical or close to $\hat n_-$, depending on the regularization prescription. Partial cancellation of the sections of the gauge link which run to transverse infinity has been disputed in Refs. \cite{Cherednikov:2007tw,Cherednikov:2008ua}.\par
		\label{fig-SIDISgluons}%
		}
	\vspace{-10pt}
\end{figure}

The Wilson line $\WlineC{\mathcal{C}_\elll}$ in the quark-quark correlator arises naturally from diagrams involving gluon loops that contribute at leading order in $m_N/Q$. In the following, we briefly motivate this statement in the context of SIDIS. Details can be found, e.g., in Ref.~\cite{Pijlman:2006tq}.
We will find that the Wilson line represents ``final state interactions'', namely soft gluons exchanged between the remnants of the nucleon and the hadronizing parton (compare, e.g., Refs.  \cite{Brodsky:2002cx,Brodsky:2002ue}).

In the attempt of proving a factorization theorem for SIDIS, diagrams of the type shown in Fig.~\ref{fig-SIDISdiagram-glue} turn out to be important. Consider $j$ gluons with momenta $p_1$, $p_2$, $\ldots$, $p_j$. From the quark propagator before the $i^\text{th}$ gluon vertex, we get a denominator of the form 
\begin{equation}
\frac{1}{(k+q-p_i-p_{i+1}-\ldots-p_j)^2-m_\quark^2 +i \epsilon} \approx 
\frac{1}{-2 q^-}\ \frac{1}{\hat n_- \cdot (p_i+\ldots+p_j)  - i \epsilon}\ .
\label{eq-eikonalqarkdenom}
\end{equation}
The approximation above is called \terminol{eikonal approximation} (see Ref.~\cite{Collins:1981uk} for details) and is applicable for soft gluons (the $p_i$ are small) and quarks almost on-shell ($(k+q)^2-m_\quark^2 \approx 0$). Now let us look at a straight Wilson line running from the origin to infinity along a four-vector $v$:
\begin{align}
	\Wline{\infty v,0} & = \mathcal{P}\ \exp \left( -i\,g\int_{\infty}^{0} d\lambda\ v \tcdot \Afield(\lambda v) \right) = \overline{\mathcal{P}}\ \exp \left( i\,g\int_{0}^{\infty} d\lambda\ v \tcdot \Afield(\lambda v) \right) \nonumber \\
	& = 1 + \sum_{j=1}^\infty (ig)^j \int_0^\infty d\lambda_1 \cdots \int_{\lambda_{j-1}}^\infty d\lambda_j \ 
	v\tcdot \Afield(v\lambda_j)\, \cdots\, v\tcdot \Afield(v\lambda_1) \nonumber \\
	& = 1 + \sum_{j=1}^\infty \int \frac{d^4 p_1}{(2\pi)^4} \cdots \int \frac{d^4 p_j}{(2\pi)^4} \ \frac{g\, v\tcdot \tilde \Afield(p_j)}{v \tcdot p_j - i \epsilon}
	 \cdots \frac{g\, v\tcdot \tilde \Afield(p_1)}{v \cdot (p_1+\ldots+p_j) - i \epsilon}\ .
	\label{eq-Wline-expanded}
\end{align}
Here $\overline{\mathcal{P}}$ denotes reverse path ordering and $\tilde \Afield(p) \equiv \int d^4 x\, e^{ipx} \Afield(x)$. The last line has an interpretation in terms of Feynman rules \cite{Collins:1989gx,Collins:1981uw,Curci:1980uw,Baulieu:1979mr}: The denominators $(v \tcdot (p_i+\ldots+p_j) - i \epsilon)^{-1}$ are displayed as double lines and are called \terminol{eikonal lines}. The gluon vertex is proportional to $v^\mu$, i.e., only gluons polarized along $v$ can couple to the eikonal lines. The denominators in the last line of the equation above remind us of eq.~(\ref{eq-eikonalqarkdenom}) if we set $v=\hat n_-$. Indeed, we are able to encode the exchange of the longitudinally polarized gluons (coupling with $\Afield^+$ and marked with a ``+'' in Fig.~\ref{fig-SIDISdiagram-glue}) in our \TMDs by including a Wilson line $\Wline{\infty v,0}$ in the definition of our quark-quark correlator $\tilde \Phi^{[\GammaOp]}(\elll,P,S)$.  

There is a trick frequently used in the literature to hide the Wilson line. We can define quark fields with a ``gauge history'' 
\begin{equation}
	\psi_{\quark,v}(\elll) \equiv \Wline{\infty v + \elll,\elll} \quark(\elll)
	\label{eq-quarkgaugehist}
\end{equation}
and write our quark-quark correlator in eq.~(\ref{eq-corr}) as
\begin{equation}
	\tilde \Phi_\quark^{[\GammaOp]}(\elll,P,S) \equiv \frac{1}{2} \bra{N(P,S)}\ \bar \psi_{\quark,v}(\elll)\, \GammaOp\ \psi_{\quark,v}(0)\ \ket{N(P,S)}\ .
	\label{eq-corrimplicitWline}
\end{equation}
However, for $v=\hat n_-$ this trick only works in gauges where the transverse gauge fields $\Afield_\prp(\infty \hat n_- + \elll_\prp)$ vanish at light cone infinity \cite{Ji:2004wu}. In other gauges (including light-cone gauge $\Afield^+=0$), it has been noticed that there are also leading contributions from transversely polarized gluons with negligible momentum in ``+''-direction \cite{Belitsky:2002sm,Boer:2003cm,Pijlman:2006tq}. We can absorb these effects into the soft correlator using the transverse Wilson line $\WlineC{\infty \hat n_- + \infty \elll_\prp,\infty \hat n_-}$. Figure~\ref{fig-SIDISgluons} shows the resulting quark-quark nucleon correlator for SIDIS. Together with the other side of the cut diagram, the complete Wilson line in eq.~(\ref{eq-corr}) reads
\begin{equation}
	\WlineC{\mathcal{C}_\elll} = \Wline{\elll,\ \infty v + \elll,\ \infty v +  \infty \elll}\ \Wline{\infty v + \infty \elll_\prp,\ \infty v,\ 0} \ ,
	\label{eq-StdSIDISWilsonLine}
\end{equation}
where $v=\hat n_-$ and where we have used the notation introduced in  eqns.~(\ref{eq-WLineStraight}) and (\ref{eq-WLineConcat}). In some works, it has been assumed that the transverse sections of the Wilson line cancel partially, so that
\begin{equation}
	\WlineC{\mathcal{C}_\elll} = \Wline{\elll,\ \infty v + \elll_\prp,\ \infty v,\ 0} \ .
	\label{eq-StdSIDISWilsonLineC}
\end{equation}
The shape of this Wilson line is a staple extending out to infinity. Note, however, that Refs.~\cite{Cherednikov:2007tw,Cherednikov:2008ua} claim that the cancellation of transverse sections is incorrect due to additional divergences produced by a cusp in the Wilson line at transverse infinity (see section~\ref{sec-WilsonLineRen} for the discussion of renormalization properties of Wilson lines). In any case, the Wilson line now forms a continuous connection between the two quark fields, so that we end up with a gauge invariant definition of \TMDs. 

In our considerations above we have focused on the SIDIS process, where we obtain a Wilson line that corresponds to final state interactions. It should be remarked that for the Drell-Yan process, the lightlike part of the Wilson line runs in the other direction as compared to SIDIS, i.e., $v=-\hat n_-$ \cite{Collins:2002kn}.
In this case, the Wilson line represents ``initial state interactions'' (compare, e.g., Ref. \cite{Brodsky:2002rv}).
In general, an incoming quark turns into a Wilson line coming in from $-\infty \hat n_-$, while an outgoing quark creates a Wilson line out to $\infty \hat n_-$ \cite{Bomhof:2006dp,Pijlman:2006tq,Chay:2004zn}. 

\subsection{Rapidity Divergences in the Lightlike Wilson Line}
\label{sec-rapidity}

In the eikonal approximation eq.~(\ref{eq-eikonalqarkdenom}), the ejected quark is treated like a massless particle moving along the ``$-$''-direction. It was realized \cite{Collins:1981uw} that this procedure removes a physical, process dependent cutoff, and thus leads to a severe divergence, sometimes termed \terminol{rapidity divergence}, see, e.g., Ref.~\cite{Collins:2007ph}. The cause of the divergence can be traced back to gluons with unphysically large momentum in the ``$-$''-direction \cite{Collins:1981uw}. The divergence cannot be regularized by the introduction of a gluon mass or with dimensional regularization. Two types of strategies have been developed to handle this divergence:
\begin{enumerate}
\item Following a suggestion by Collins, Soper and Sterman \cite{Collins:1981uk,Collins:1981uw} in the context of fragmentation functions, the longitudinal gauge link can be placed slightly off the light cone, i.e., with $v$ not exactly equal to $\hat n_-$. The distributions defined in this way depend on $\zeta \equiv (P \cdot v)^2/|v^2|\approx (P^+)^2 | v^- / 2 v^+ |$, which acts as a cutoff parameter and is chosen large but finite. An evolution equation can be derived to describe the dependence of the distributions on $\zeta$, see e.g. eqns. (6.4) and (6.6) in Ref.  \cite{Collins:1981uk}.
\item The divergence can be cancelled by a subtraction factor, introduced as vacuum expectation values of Wilson lines. In general, in such an approach the parameter controlling the rapidity cutoff will appear in the subtraction factor. 
\end{enumerate}
A review of these approaches can be found in Ref. \cite{Collins:2003fm}.

\subsection{Definitions of \TMDs Proposed in the Literature}
\label{sec-proposeddefs}

Let us look at some concrete proposals in the literature how the Wilson lines can be arranged to obtain well-behaved definitions of \TMDs. 


\subsubsection{The Factorization Formula by Ji, Ma and Yuan}
\label{sec-JiMaYuan}

\begin{figure}[btp]
	\centering
	\includegraphics{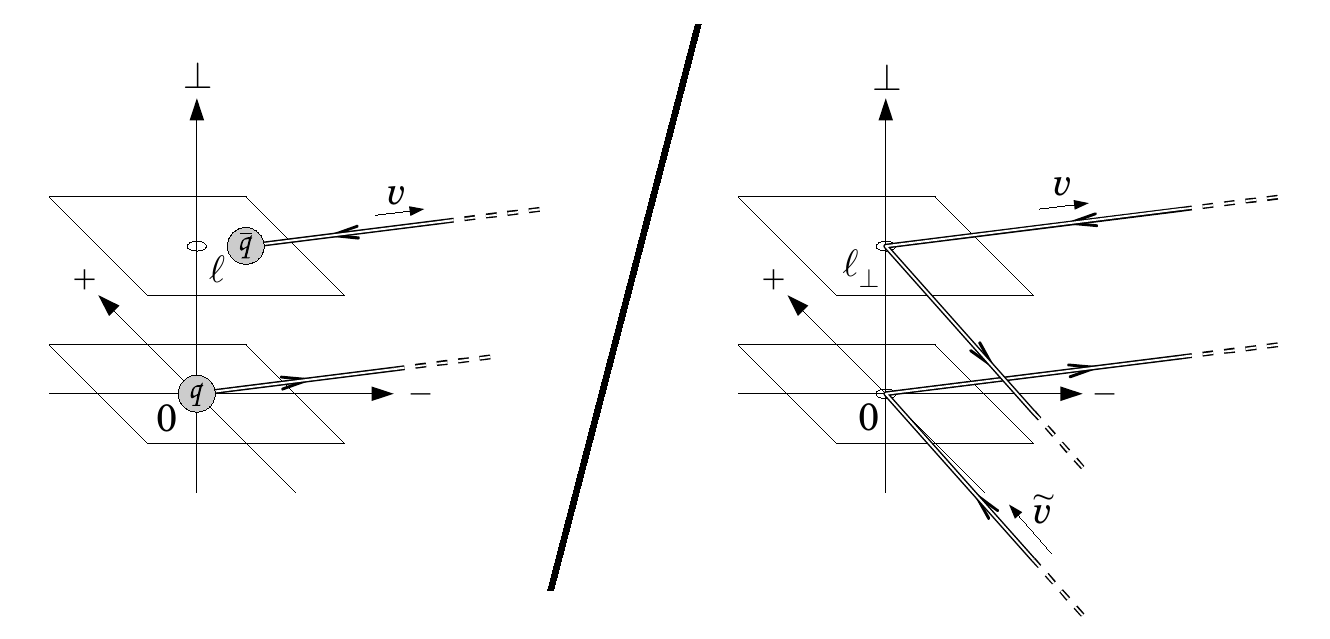}\par
	\caption{Arrangement of Wilson lines in the definition of \TMDs by Ji, Ma and Yuan \cite{Ji:2004wu,Ji:2004xq}.
	\label{fig-LinkPathsJi}
	}
\end{figure}

\begin{figure}[btp]
	\centering
	\includegraphics[width=0.5\textwidth]{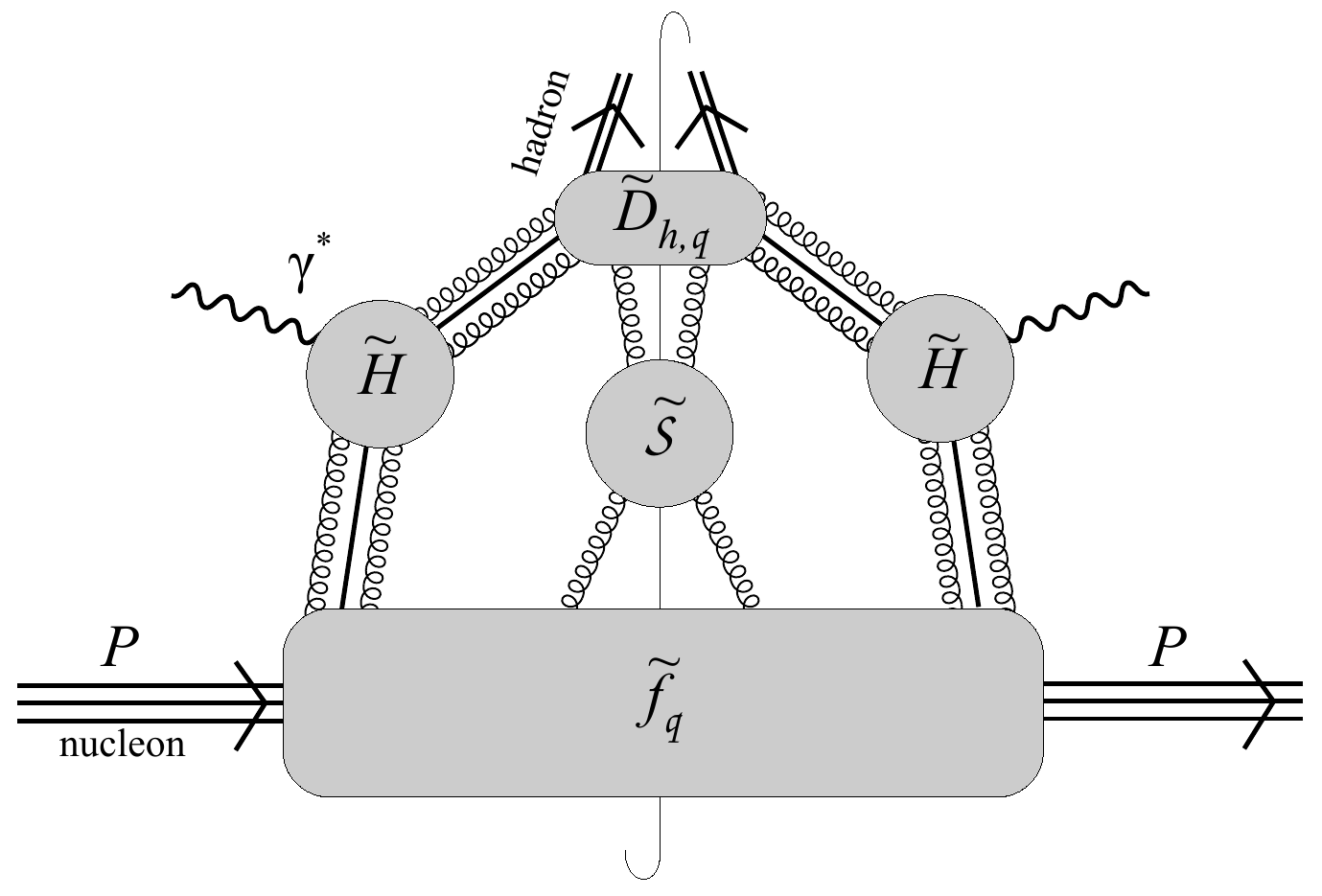}\par
	\caption{The leading regions for SIDIS after soft and collinear factorizations. 
	\label{fig-JiLeading}
	}
\end{figure}

Ji, Ma and Yuan, \cite{Ji:2004wu,Ji:2004xq} have proposed a factorization prescription where both Wilson lines slightly off the light cone and subtraction factors appear. Let us study their results for SIDIS. 
In their definition of \TMDs, which we illustrate in Fig.~\ref{fig-LinkPathsJi}, they introduce two slightly timelike link directions $v=v^+ \hat n_+ + v^- \hat n_- \approx \hat n_-$ and $\tilde v= \tilde v^+ \hat n_+ + \tilde v^- \hat n_- \approx \hat n_+$. 
To factorize the hadron tensor, Ji, Ma and Yuan take the most general setup of leading regions into account, as depicted in Fig.~\ref{fig-JiLeading}. 
As an example, consider the unpolarized leading structure function $F$ in the hadron tensor $W^{\mu\nu} = - \frac{1}{2} g_\prp^{\mu\nu} F(x_B,z,\vect{P}_{h\prp},Q^2) + \ldots$. In factorized form, the Fourier transform of $F$ with respect to $\vect{P}_{h\prp}$ reads
\begin{align}
	\tilde F(x_B,z,\vprp{\elll},Q^2) & = \sum_{\quark = u,d,s,\ldots} e^2_\quark\  
	\tilde f_{1,\quark}(x_B,z\,\vprp{\elll};\mu^2,Q^2 \rho,\rho)\ \tilde D_{h,\quark}(z,\vprp{\elll};\mu^2,Q^2\rho,\rho)\ \nonumber \\ & \times
	\tilde{\mathscr{S}}(\vprp{\elll};\mu^2,\rho)\ \tilde H(Q^2,\mu^2,\rho)\ .
	\label{eq-JiFactorization}
\end{align}
Here $e_\quark$ is the electric charge of quark $\quark$, and $\mu$ is a renormalization and factorization scale. The dependence on $\zeta$, which pa\-ra\-me\-tri\-zes the direction of $v$, is hidden in the dependence on $Q^2$ and the parameter $\rho\equiv\sqrt{v^- \tilde v^+/v^+ \tilde v^-}$ via a special choice of coordinates. 
The soft factor $\tilde{\mathscr{S}}$ appears not only in the equation above, but also as subtraction factor in the definition of the fragmentation function $\tilde D_{h,\quark}$ and in the definition of the \TMD:
\begin{multline}
	\tilde f_{1,\quark}(x,\vprp{\elll};\mu^2,Q^2\rho,\rho)  \equiv \frac{1}{\tilde{\mathscr{S}}(\vprp{\elll},\mu^2,\rho)} \int \frac{d\elll^-}{(2\pi)}\ e^{-ix P^+\elll^-} \\  \times
	\frac{1}{2} \bra{N(P,S)}\ \bar \quark(\elll) \Wline{\elll,\infty v+\elll}\ \gamma^+\ \Wline{\infty v,0} \quark(0)\ \ket{N(P,S)} \Big\vert_{\elll^+=0} \ .
\end{multline}
Since Ji, Ma and Yuan work in a gauge where gauge fields vanish at infinity, the transverse sections of Wilson lines at infinity need not be specified. Thus up to the soft factor, this definition is equivalent to the one in eqns.~(\ref{eq-corr}),\,(\ref{eq-f1intro}). 
The soft factor reads
\begin{equation}
	\tilde{\mathscr{S}}(\vprp{\elll};\mu^2,\rho) = \frac{1}{N_c} \bra{0} \Tr\ \Wline{-\infty \tilde v+\elll_\prp,\ \elll_\prp,\ \infty v + \elll_\prp}\ \Wline{\infty v,\ 0,\ -\infty \tilde v}\ \ket{0}\ .
\end{equation}
Ji, Ma and Yuan provide evolution equations for the $\rho$- and $\mu$-dependence (see also Ref. \cite{Idilbi:2004vb}), and give arguments that their formula is valid to all orders in perturbation theory. 

\subsubsection{Subtracted \TMDs by Hautmann, Collins and Metz}
\label{sec-Hautmann}

\begin{figure}[btp]
	\centering
	\includegraphics{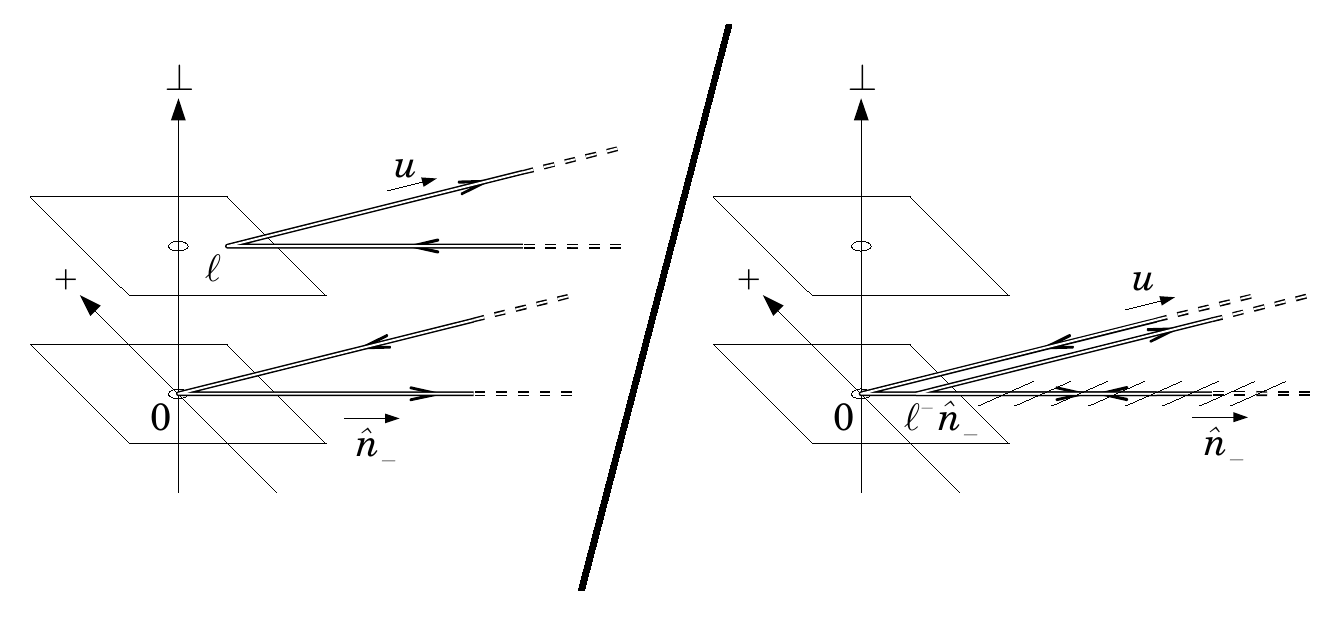}\par
	\vspace{-10pt}
	\caption{Arrangement of Wilson lines in the subtraction factor appearing in the definition of \TMDs by Hautmann \cite{Hautmann:2007uw}.
	\label{fig-LinkPathsHautmann}
	}
\end{figure}

Hautmann \cite{Hautmann:2007uw} follows suggestions put forward in Refs.~\cite{Collins:1999dz,Collins:2000gd,Collins:2004nx} and proposes a quark distribution of the form
\begin{align}
	\tilde f_{1,\quark}(x,\vprp{\elll};\zeta_u) & = \int \frac{d\elll^-}{(2\pi)} e^{-ix P^+\elll^-} \nonumber \\ & \times
	\frac{1}{2} \bra{N(P,S)}\ \bar \quark(\elll)\, \Wline{\elll,\infty \hat n_-+\elll}\ \gamma^+\ \Wline{\infty \hat n_-,0}\, \quark(0)\ \ket{N(P,S)}
	\nonumber \\ & \times \frac{\bra{0} \Tr\ \Wline{\infty u + \hat n_- \elll^-,\ \hat n_- \elll^-,\ \infty \hat n_-}\ \Wline{\infty \hat n_-,\ 0,\ \infty u} \ket{0}}{\bra{0} \Tr\ \Wline{\infty u+\elll,\ \elll,\ \infty \hat n_- + \elll}\ \Wline{\infty \hat n_-,\ 0,\ \infty u} \ket{0}} \Big\vert_{\elll^+=0}\ .
	\label{eq-HautmannTMD}
\end{align}
Again, a gauge with vanishing gauge fields at infinity is employed.
The subtraction factor in the last line of the above equation is illustrated in fig.~\ref{fig-LinkPathsHautmann}.
Its purpose is to cancel ``endpoint singularities'' occurring in \TMDs for $x\rightarrow 1$.
In contrast to the approach by Ji, Ma and Yuan, the vector $v$ employed here is exactly $\hat n_-$. The auxiliary non-lightlike direction $u = u^+ \hat n_+ + u^- \hat n_-$ appearing in the subtraction factor introduces a dependence on the regularization parameter $\zeta_u \equiv (P^+)^2 u^- / 2 u^+$. However, the subtraction factor cancels and the dependence on $\zeta_u$ disappears for the $\vprp{k}$-integrated parton distribution function, which one obtains by setting $\elll_\prp=0$ in eq.~(\ref{eq-HautmannTMD}). Thus the definition by Hautmann promises to provide a relation between \TMDs and regular, $\vprp{k}$-integrated PDFs.  

\subsubsection{\TMDs in Soft-Collinear Effective Theory by Chay}
\label{sec-Chay}

\begin{figure}[btp]
	\centering
	\includegraphics[trim=0 15 0 0,clip=true]{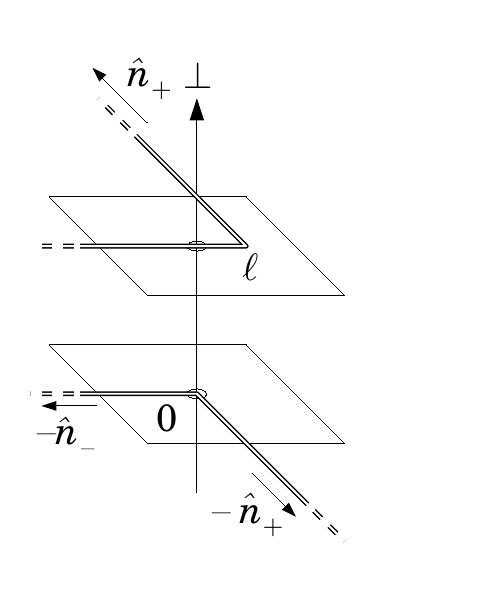}\par
	\caption{Arrangement of Wilson lines in the subtraction factor of Chay's definition of \TMDs in SCET \cite{Chay:2007ty}.
	\label{fig-LinkPathsChay}
	}
\end{figure}

\vspace{-5pt}
Another proposal \cite{Chay:2007ty} promising a relation to regular PDFs employs \terminol{soft-collinear effective theory (SCET)} to regularize the parton distributions, using exclusively light-like Wilson lines. The definition of the quark distribution contains the subtraction factor depicted in fig.~\ref{fig-LinkPathsChay} and 
reads
\begin{align}
	f_{1,\quark}(x,\vprp{k},\kappa) & = \int d\omega \int \frac{d\elll^-}{(2\pi)^2} \frac{d^2\vprp{\elll}}{(2\pi)^2} e^{i(\omega/2-xP^+ +\kappa)\elll^-/2+i \vprp{k} \cdot \vprp{\elll}} \nonumber \\
	& \times \bra{N(P,S)} \bar \chi_{\hat n_+}(\elll_\prp) \delta(\omega - \mathscr{P}_+) \frac{\gamma^+}{2} \chi_{\hat n_+}(0) \ket{N(P,S)} \nonumber \\
	& \times \frac{1}{N_c} \bra{0} \Tr\, \Wline{\infty \hat n_+ + \elll,\ \elll,\ -\infty \hat n_-+\elll}\ \Wline{-\infty \hat n_-,\ 0,\ -\infty \hat n_+} \ket{0} \Big\vert_{\elll^+=0}\ .
\end{align}
Here $\chi_{\hat n_+}$ are quark fields within SCET. Their definition includes a ``gauge history'' analogous to eq.~(\ref{eq-quarkgaugehist})%
. The appearance of $\omega$ and the large momentum label operator  $\mathscr{P}_+$ is a SCET specific feature. The additional scale $\kappa$ in the above equation is related to soft gluon emission. Note that the Wilson line in the soft factor starts at $-\infty \hat n_+$ but ends at $\infty \hat n_+ + \elll$. Thus there is no obvious way to close the Wilson line to a manifestly gauge invariant loop. 

\subsubsection{Definition of \TMDs by Cherednikov and Stefanis}
\label{sec-Chered}

\begin{figure}[btp]
	\centering
	\includegraphics[trim=0 20 0 0,clip=true]{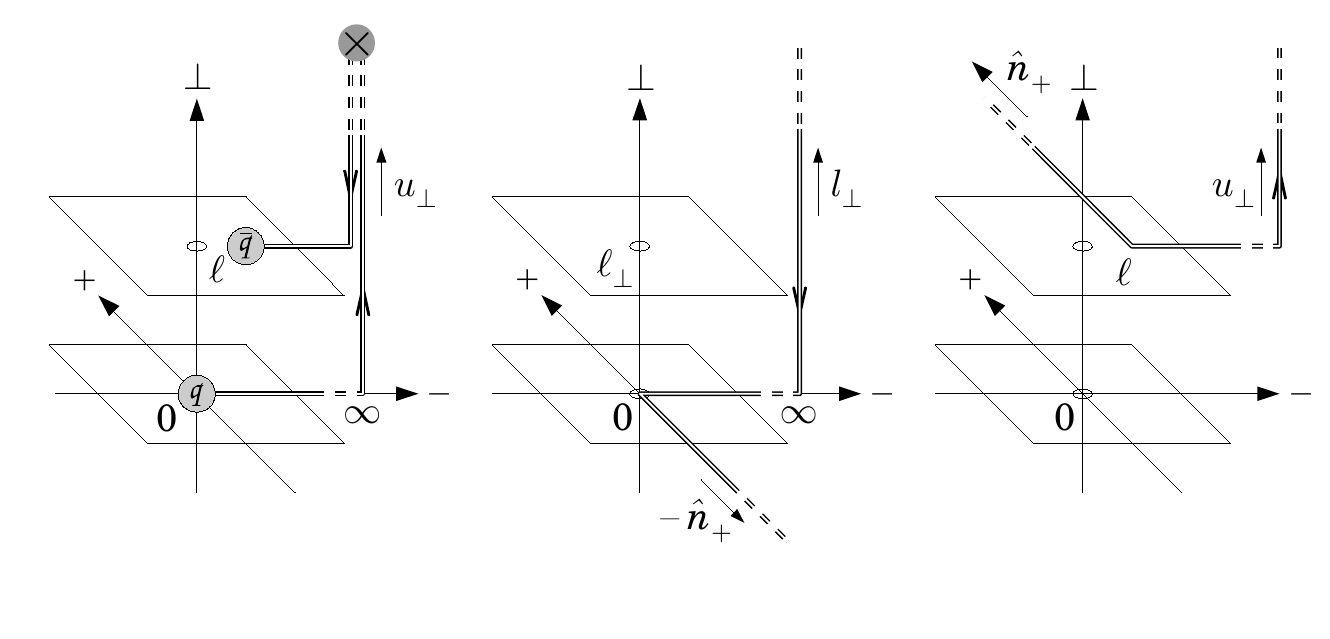}\par
	\vspace{-5pt}
	\caption{Arrangement of Wilson lines in the definition of \TMDs by Cherednikov and Stefanis \cite{Cherednikov:2007tw,Cherednikov:2008ua}. The cross at the end of the transverse link indicates a non-smooth connection of the line at infinity.
	\label{fig-LinkPathsCherednikov}
	}
\end{figure}

\vspace{-5pt}
As already mentioned, the authors of Refs.~\cite{Cherednikov:2007tw,Cherednikov:2008ua} deduce from an analysis of anomalous dimensions that the Wilson lines in eqns.~(\ref{eq-StdSIDISWilsonLine}) and (\ref{eq-StdSIDISWilsonLineC}) are not equivalent. They suggest a definition of parton densities with lightlike Wilson lines ($v=\hat n_-$) and include soft subtraction factors with cusps at $0$ and $\elll$:
\begin{align}
	\tilde f_{1,\quark}(x,\vprp{\elll};\mu,\eta) & = \int \frac{d\elll^-}{(2\pi)} e^{-ix P^+\elll^-} \nonumber \\
	& \times \frac{1}{2} \bra{N(P,S)}\ \bar \quark(\elll)\, \Wline{\elll,\ \infty \hat n_-+\elll_\prp,\ \infty \hat n_-+\infty u_\prp+\elll_\prp}\ \gamma^+\ \nonumber \\ 
	& \times \Wline{\infty \hat n_- + \infty u_\prp,\ \infty \hat n_-,\ 0}\, \quark(0)\ \ket{N(P,S)} \nonumber \\ & \times \frac{1}{N_c}\bra{0} \Tr\ \Wline{-\infty \hat n_+,\ 0,\ \infty \hat n_-,\ \infty \hat n_- + \infty \elll_\prp} \ket{0} \nonumber \\
	& \times \frac{1}{N_c} \bra{0} \Tr\ \Wline{\infty \hat n_- + \infty u_\prp+\elll,\ \infty \hat n_- + \elll,\ \elll,\ \infty \hat n_+ + \elll} \ket{0}\ .
\end{align}
Here the mass scale $\eta$ is hidden in the regularization of rapidity divergences with a pole prescription and plays a role similar as $\zeta$ in section~\ref{sec-rapidity}. The renormalization scale $\mu$ is needed for dimensional regularization. The choice of the transverse direction $u_\prp$ is completely arbitrary. An illustration of the Wilson lines is given in fig.~\ref{fig-LinkPathsCherednikov}.
As in Chay's approach, there is no obvious way to complete the Wilson lines such that the soft factors become manifestly gauge invariant expressions. Cherednikov and Stefanis show that their \TMDs fulfill -- at least formally -- the simple relation to integrated distributions eq.~(\ref{eq-intpdfrelation}).

\subsection{Wilson Line Self-Energy}
\label{sec-selfenergy}

As will be discussed in section \ref{sec-WilsonLineRen}, the Wilson line exhibits a divergence dependent on its length $L$ due to self-energy graphs. It can be removed by a factor $\exp( - \delta m\, L )$, where $\delta m$ is a renormalization constant which vanishes for dimensional regularization, but not for cutoff schemes as in lattice QCD. The divergence may be eliminated by a subtraction factor included in the definition of the \TMD.
In accordance with the suggestion in Ref. \cite{Collins:2008ht}, consider the definition
\begin{align}
	f_{1,\quark}(x,\vprp{k};y_n) & = \int \frac{d\elll^-}{(2\pi)} \int \frac{d^2\vprp{\elll}}{(2\pi)^2} e^{-ix P^+\elll^- + i\vprp{k} \cdot \vprp{\elll}} \lim_{\eta\rightarrow\infty} \nonumber \\ & \times 
	\frac{1}{2} \frac{\bra{N(P,S)}\ \bar \quark(\elll)\, \gamma^+ \Wline{\elll,\ \eta v+\elll,\ \eta v,\ 0}\ \quark(0)\ \ket{N(P,S)}}{ \sqrt{\mathscr{loop}(\eta ,\elll)} } \Big\vert_{\elll^+=0}\ .
	\label{eq-NewCollins}
\end{align}
The numerator in the second line is our $\tilde \Phi^{[\gamma^+]}$, however the Wilson line $\WlineC{\mathcal{C}_\elll}$ does not run all the way out to infinity. 
The rapidity parameter $y_n$ symbolically reminds us that we must take care of rapidity divergences, e.g., by choosing a non-lightlike $v$. 
The Wilson loop $\mathscr{loop}(L,\elll)$ is given by:
\begin{align}
	\mathscr{loop}(\eta,\elll) \equiv \frac{1}{N_c} \bra{0} \Tr\ \Wline{\eta v,\ -\eta v,\ -\eta v+\elll,\ \eta v+\elll,\ \eta v} \ket{0}\ .
\end{align}
This loop is twice as long as the link path in the numerator of the second line of eq.~(\ref{eq-NewCollins}). Thus the square root in the denominator cancels the self-energy divergence.
Note that the loop expectation value becomes $\eta$-independent for $v=\hat n_-$. 

We will try out the idea presented above on the lattice in section~\ref{sec-stapleresults}, with a setup of link paths as illustrated in Fig. \ref{fig-staple}. 

\begin{figure}[btp]
	\centering
	\includegraphics{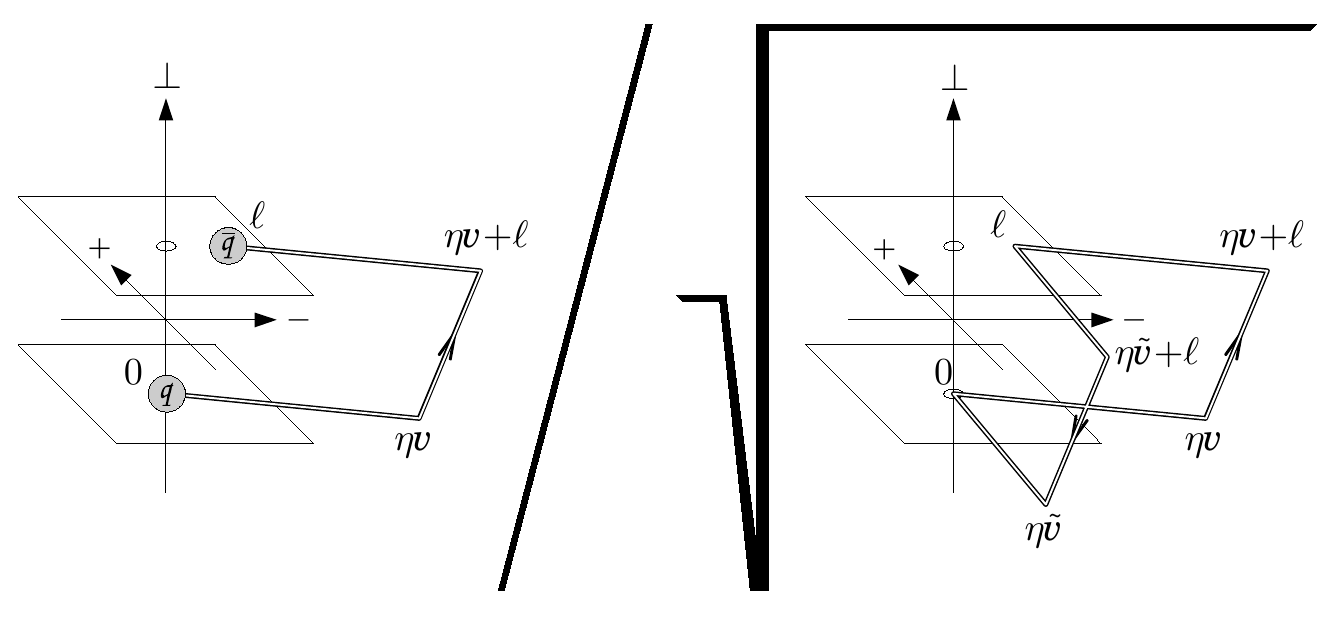}\par
	\caption{Cancellation of self-energy contributions in the definition of \TMDs, as suggested in Ref. \cite{Collins:2008ht} but generalized to two different directions $v$ and $\tilde v$ in the Wilson loop subtraction factor.
	\label{fig-LinkPathsCollinsNewMod}
	}
\end{figure}

It might be interesting to generalize the idea of the loop subtraction factor to cusped loops of the form
\begin{align}
	\mathscr{loop}(\eta,\elll) \equiv \frac{1}{N_c} \bra{0} \Tr\ \Wline{0,\,\eta \tilde v,\,\eta \tilde v + \elll,\, \elll,\,\eta v + \elll,\,\eta v,\,0 } \ket{0}\ ,
\end{align}
where we have introduced an auxiliary direction $\tilde v$. The loop has now approximately the shape of a folded rectangle, see Fig.~\ref{fig-LinkPathsCollinsNewMod}. With an appropriate choice of $\tilde v$, this loop structure cancels the self-energy divergence and, taking into account that transverse pieces can be left out in the limit $\eta\rightarrow\infty$ in appropriate gauges, shows similarity to the subtraction factors introduced by Ji, Ma, and Yuan (cf. section~\ref{sec-JiMaYuan}) as well as Hautmann, Collins and Metz (cf. section~\ref{sec-Hautmann}). However, these authors do not put their subtraction factors under a square root. 

As a side remark, note that the middle section $\Wline{\eta v+\elll,\ \eta v}$ of the gauge link in eq.\,(\ref{eq-NewCollins}) is not exactly transverse if $\elll^- \neq 0$. We could have maintained a transverse middle section by taking the gauge link $\Wline{\elll,\ \eta v+\elll_\prp,\ \eta v,\ 0}$ instead. 
In the limit $\eta \rightarrow \infty$, the two versions are formally equivalent, because $\infty v + \elll =  \infty v^+ \hat n_+ + (\infty v^- + \elll^-) \hat n_- + \elll_\prp = \infty v + \elll_\prp$.
If we were to insist on the gauge link $\Wline{\elll,\ \eta v+\elll_\prp,\ \eta v,\ 0}$, an appropriate Wilson loop subtraction factor $\mathscr{loop}(\eta,\elll)$ would look slightly more complicated, and the simple transformation properties of the gauge link under discrete symmetries in eq.\,(\ref{eq-stapletrafoprops}) below would not hold exactly any more.

\subsection{Remarks}

Quoting from a recent proceedings article by Collins \cite{Collins:2008ht} -- ``to allow non-perturbative methods in QCD to be used to estimate parton densities, operator definitions of parton densities are needed that can be taken literally''. As we have seen, there is now a variety of proposals available for transverse momentum dependent parton correlators. We hope that from these ideas a generally accepted definition of \TMDs will emerge that remains finite-valued and well-defined beyond perturbation theory.



\section{Probability Interpretation of \TMDs}
\label{sec-probInt}

\subsection{The Parton Picture using Light-Front Quantization}
\label{sec-lfquant}


We usually think of \TMDs as probability distributions of quarks inside the nucleon. 
This notion has its origin in light-front quantization \cite{Kogut:1969xa,Dirac:1949cp,Brodsky:1997de,Burkardt:1995ct}:
As mentioned in section~\ref{sec-corr}, the reaction at high momentum is sensitive to the nucleon structure close to the plane $\elll^+=0$. The Dirac spinor $\quark(\elll)$ on that plane has only two independent dynamical components. These ``\terminol{good components}'' are projected out according to $\quark_{(+)}(\elll) \equiv \frac{1}{2} \gamma^- \gamma^+\ \quark(\elll)$. The helicity projected Fourier transforms
\begin{equation}
	\tilde \quark_{(+),\lambda}(k^+,\vprp{k})\ \equiv\ \frac{1}{2}(1+\lambda \gamma^5)\ \int \frac{d\elll^-}{\sqrt{2\pi}} \int \frac{d^2\vprp{\elll}}{2\pi}\ e^{i k^+ \elll^- - i \vprp{k} \cdot \vprp{\elll}}\ \quark_{(+)}(\elll^- \hat n_- + \elll_\prp)
\end{equation}
are written in terms of quark annihilation operators $b_{\lambda,\quark}(k^+,\vprp{k})$ and antiquark creation operators $d^\dagger_{\lambda,\quark}(k^+,\vprp{k})$ of helicity $\lambda$ and quark flavor $\quark$ according to
\begin{align}
	\sqrt{2 |k^+|}\ \tilde \quark_{(+),\lambda}(k^+,\vprp{k})\ \equiv \
	\left\{ \begin{array}{ll} U_{(+),\lambda}(k^+,\vprp{k})\, b_{\lambda,\quark}(k^+,\vprp{k})\ & : \ k^+ > 0 \\ V_{(+),\lambda}(-k^+,-\vprp{k})\, d^\dagger_{\lambda,\quark}(-k^+,-\vprp{k})\ & : \ k^+ < 0 \end{array} \right.\ ,
	\label{eq-lightfrontquant}
\end{align}
where $U_{(+),\lambda}(k^+,\vprp{k})$ and $V_{(+),\lambda}(k^+,\vprp{k})$ are appropriate Dirac spinors, and where we attribute anticommutation relations to the creation and annihilation operators as in Ref. \cite{Burkardt:1995ct}. In light-front quantization, the distinction between matter and antimatter is made according to the sign of $k^+$ (or $x = k^+/P^+$). Note that the sign of $k^+$ is equal to the sign of $k^0$ if  the dispersion relation $k^2 = m_\quark^2$ is satisfied.  It is now easy to show that for $x > 0$
\begin{equation}
	\sum_\lambda \frac{\bra{N(P,S)}\, b^\dagger_{\lambda,\quark}(xP^+,\vprp{k})\, b_{\lambda,\quark}(xP^+,\vprp{k})\, \ket{N(P,S)}}{\braKet{N(P,S)}{N(P,S)}{}}\ = \frac{(2\pi)^3}{P^+}\  \Phi^{[\gamma^+]}_{\mathcal{U}=\Eins}(x,\vprp{k};P,S) \ ,
	\label{eq-LFcorr}
\end{equation}
i.e., we obtain our usual quark correlator $\Phi^{[\gamma^+]}(x,\vprp{k};P,S)$, though we have to set the gauge link $\WlineC{\mathcal{C}_\elll}$ to unity. The operator $b^\dagger_{\lambda,\quark}(xP^+,\vprp{k})\, b_{\lambda,\quark}(xP^+,\vprp{k})$ simply gives the number density of quarks with helicity $\lambda$, momentum fraction $x$ and transverse momentum $\vprp{k}$. Thus the corresponding \TMD $f_{1,\quark}(x,\vprp{k})$ has an interpretation as a density of quarks%
\footnote{The prefactor $(2\pi)^3/P^+$ is a consequence of the convention to operate on our distributions with integrals $\int dx$ and $\int d^2\vprp{k}$ instead of the usual momentum integrals $\int dk^+/(2\pi)$ and $\int d^2 \vprp{k}/(2\pi)^2$.}%
. 


In order to ensure that the gauge link on the right hand side of eq.~(\ref{eq-LFcorr}) is unity, we have to fix an appropriate gauge. The most natural gauge to use together with light-front quantization is the light cone gauge, $\Afield^+=0$, which, however, is affected by the same divergences as lightlike Wilson lines, see section~\ref{sec-rapidity} and Ref.~\cite{Collins:2003fm}. 
We caution the reader that there are conceptual difficulties related to the application of light-front quantization to QCD, see, e.g., Ref.~\cite{Burkardt:1997bd,Burkardt:1995ct}.


\subsection{Positivity of the Quark-Quark Correlator}

Ref.~\cite{Bacchetta:1999kz} gives very useful bounds on \TMDs. In particular, it is shown that $f_{1,q}(x,\vprp{k})$ is positive. Here we reexamine the positivity argument, paying special attention to the role of the Wilson line. 

\subsubsection{Concept of the Proof}
\label{sec-probproofmech}

Consider Dirac matrices $\GammaOp$ for which a matrix $\GammaDist$ exists such that
\begin{equation}
	\frac{1}{2} \gamma^0\, \GammaOp = (\GammaDist)^\dagger\, \GammaDist\ .
	\label{eq-proofreq}
\end{equation}
For example, to analyze $f_{1,q}(x,\vprp{x})$, we select $\GammaOp=\gamma^+$, then $\GammaDist=2^{-1/4}(\frac{1}{2}\gamma^-\gamma^+)$. 
In the following, whenever we work with a specific gauge fixing condition $G$, we will add an index $G$ to the state vectors.
In all our considerations, we restrict ourselves to the plane $\elll^+=0$ and to $x>0$.
The Wilson line running from $\elll$ to the origin in the quark-quark correlator eq.~(\ref{eq-corr}) will have to be split into two pieces according to $\Wline{\mathcal{C}_\elll} = \Wline{\mathcal{C}_{1,\elll}}\, \Wline{\mathcal{C}_{2,\elll}}$ with suitable definitions of the paths $\mathcal{C}_{1,\elll}$ and $\mathcal{C}_{2,\elll}$ discussed in the following sections. Let $\{\ket{n}\}$ be a complete set of momentum eigenstates with momenta $p_n$, normalized according to $\braKet{n}{n'}{}=\delta_{nn'}$. The quark-quark correlator can now be written as:
\begin{multline}
	\tilde \Phi^{[\GammaOp]}(\elll) = \bra{N(P,S)} \quark^\dagger(\elll)\,\WlineC{\mathcal{C}_{1,\elll}}\,(\GammaDist)^\dagger\ \GammaDist\,\WlineC{\mathcal{C}_{2,\elll}}\,\quark(0) \ket{N(P,S)} = \\
	\leftind{G}\bra{N(P,S)} \quark^\dagger(\elll)\,\WlineC{\mathcal{C}_{1,\elll}}\,(\GammaDist)^\dagger\ \left( \sum_n \ket{n}_G\ \leftind{G}\bra{n} \right)\ \GammaDist\,\WlineC{\mathcal{C}_{2,\elll}}\,\quark(0) \ket{N(P,S)}_G = \\
	\sum_n  \Big( \leftind{G}\bra{n} \GammaDist\,\Wline{\mathcal{C}_{1,\elll}}^\dagger\,\quark(\elll) \ket{N(P,S)}_G\Big)^*\
	\Big( \leftind{G}\bra{n} \GammaDist\,\WlineC{\mathcal{C}_{2,\elll}}\,\quark(0) \ket{N(P,S)}_G \Big) = \\
	\sum_n e^{i (P-p_n)\cdot \elll} \Big( \leftind{G}\bra{n} \GammaDist\ \WlineC{\mathcal{C}_{1,\elll}-\elll}^\dagger \quark(0) \ket{N(P,S)}_G\Big)^*\
	\Big( \leftind{G}\bra{n} \GammaDist\,\WlineC{\mathcal{C}_{2,\elll}}\, \quark(0) \ket{N(P,S)}_G \Big)\ .
	\label{eq-probproof3}
\end{multline}
In the last step, it is required that the gauge fixing condition $G$ is translation invariant. Otherwise $G$ would change into a different condition $G'$ in the matrix element on the left. Suppose that our prescription to split the Wilson line fulfills 
\begin{equation}
	\WlineC{\mathcal{C}_{1,\elll}-\elll}^\dagger = \WlineC{\mathcal{C}_{2,\elll}} \equiv \WlineC{\mathcal{C}_3}\ ,
	\label{eq-WLineSplitting}
\end{equation}
where the path $\mathcal{C}_3$ is $\elll$-independent. Then
\begin{equation}
 	\tilde \Phi^{[\GammaOp]}(\elll) = \sum_n  e^{i (P-p_n)\cdot \elll}\ \sum_{\alpha,i} \Big\vert\ \leftind{G}\bra{n}\ \GammaDist_{\alpha\beta}\,\WlineC{\mathcal{C}_3}_{ij}\, \quark_{\beta j}(0)\ \ket{N(P,S)}_G\ \Big\vert^2\ .
\end{equation}
For clarity, we have made color indices $i,j$ and Dirac indices $\alpha$, $\beta$ explicit in the last line.
If the condition $G$ is independent of $\elll$, we can now Fourier-transform to momentum space
\begin{multline}
	\Phi^{[\GammaOp]}(x,\vprp{k}) = \int \frac{d\elll^-}{2\pi} \int \frac{d^2 \vprp{\elll}}{(2\pi)^2}\ e^{-i\, k\cdot \elll}\ \tilde \Phi^{[\GammaOp]}(\elll)\ \Big\vert_{\elll^+=0} =  \\
	\sum_n\ \int \frac{d\elll^-}{2\pi} \int \frac{d^2 \vprp{\elll}}{(2\pi)^2}\ e^{i\, (P-p_n-k)\cdot \elll}\ 
	\Big\vert\ \leftind{G}\bra{n} \GammaDist\,\WlineC{\mathcal{C}_3}\, \quark(0) \ket{N(P,S)}_G\ \Big\vert^2 \ \Big\vert_{\elll^+=0} = \\
	\sum_n\  
	\Big\vert\ \leftind{G}\bra{n}\ \GammaDist\,\WlineC{\mathcal{C}_3}\, \quark(0)\ \ket{N(P,S)}_G \Big\vert^2 \ \delta(p_n^+ - (1-x)P^+)\ \delta^{(2)}(\vect{p}_{n\prp} + \vprp{k}) \ \geq 0\ .
	\label{eq-probproof4}
\end{multline}
Apart from proving positivity, the last line gives another inspiration for an interpretation as a parton distribution. The delta functions restrict the momenta of the 
states $\ket{n}$. Obviously these states carry the longitudinal momentum of the nucleon, except for a momentum fraction $x$, which is missing, and a transverse momentum $-\vprp{k}$. This indicates that the quark field operator $\GammaDist \Wline{\mathcal{C}_3}\quark(0)$ annihilates a ``parton'' with longitudinal momentum $x P^+$ and transverse momentum $\vprp{k}$. 


Ref.~\cite{Bacchetta:1999kz} uses the argument above for various spin combinations to obtain a whole set of bounds for various \TMDs. 


\subsubsection{Wilson Line out to Infinity}

We can accommodate for a Wilson line of the form given in eq.~(\ref{eq-StdSIDISWilsonLine}) in the proof above by choosing a gauge $G$ where the gauge fields vanish at infinity, e.g., the Feynman gauge. Then the transverse pieces are unity, and it is sufficient to define $\WlineC{\mathcal{C}_{1,\elll}} = \Wline{\elll,\ \infty v + \elll}$ and $\WlineC{\mathcal{C}_{2,\elll}} = \Wline{\infty v,\ 0}$. Obviously this choice satisfies the requirements above, with $\WlineC{\mathcal{C}_{3,\elll}} = \Wline{\infty v,\ 0}$, provided $v$ has no functional dependence on $\elll$. Thus positivity of the respective quark-quark correlator is shown. 


The argument above has been presented without the inclusion of any subtraction factors that appear in the improved correlator definitions of section \ref{sec-proposeddefs}. It is clear, however, that a subtraction factor as part of the quark-quark correlator can destroy the positivity argument if it introduces an additional $\elll$-dependence.

\subsubsection{Straight Wilson Line}
\label{sec-probproofstraight}

For the quark-quark correlator with a straight Wilson line $\Wline{C_\elll} = \Wline{\elll,0}$, it is not possible to find a prescription to split the Wilson lines in such a way that eq.~(\ref{eq-WLineSplitting}) is satisfied, so we need to take a different route. Splitting the line ``at infinity'', we choose $\WlineC{\mathcal{C}_{1,\elll}} = \Wline{\elll,\ \infty\elll}$, $\WlineC{\mathcal{C}_{2,\elll}} = \Wline{\infty\elll,0}$. Let us introduce again quark fields with a ``gauge history''
\begin{align}
	\Psi_{\quark,\elll}(\xi) \equiv \Wline{\infty\elll + \xi,\ \xi} \quark(\xi)\ .
\end{align}
The gauge invariance of the $\Psi_{\quark,\elll}(\xi)$ is guaranteed if we can assume that the gauge fields vanish at $\infty\elll $ \cite{PhysRev.139.B476}.\footnote{To maintain this, valid gauge rotations $W(x)$ of the gauge transformations eqns.~(\ref{eq-gaugetrans-q})--(\ref{eq-gaugetrans-A}) must become constant at $\elll \infty$.} This assumption is problematic for lightlike $\elll$.\footnote{Keep in mind that for $\elll^2 = 0$, the physical, Lorentz invariant distance $(\infty \elll)^2 = \infty \tcdot 0$ is undefined.} Therefore, let us assume that $\elll$ is spacelike for the moment. The idea of gauge invariant fermion fields dates back to Dirac \cite{Dirac:1955uv}. Gauge invariant fermion fields with Wilson lines out to infinity have been discussed within QED, e.g., in Refs. \cite{Mandelstam:1962mi,PhysRev.139.B476}, see also Ref. \cite{Ji:2004wu}. 

We can now rewrite the first steps of our proof eq.~(\ref{eq-probproof3}) as 
\begin{multline}
	\tilde \Phi^{[\GammaOp]}(\elll) = \bra{N(P,S)} \Psi^\dagger_{\quark,\elll}(\elll)\,(\GammaDist)^\dagger\ \GammaDist\,\Psi_{\quark,\elll}(0) \ket{N(P,S)} = \\
	\sum_n  \Big( \bra{n} \GammaDist\,\Psi_{\quark,\elll}(\elll) \ket{N(P,S)}\Big)^*\
	\Big( \bra{n} \GammaDist\, \Psi_{\quark,\elll}(0) \ket{N(P,S)} \Big) = \\
	\sum_n e^{i (P-p_n)\cdot \elll} \Big|\bra{n} \GammaDist\ \Psi_{\quark,\elll}(0) \ket{N(P,S)}\Big|^2 \ .
	\label{eq-probproof3str}
\end{multline}
Owing to the gauge invariant quark fields, we have been able to perform these steps without fixing a gauge up to now. However, in order to carry out the Fourier transform of $\tilde \Phi^{[\GammaOp]}(\elll)$ with respect to $\elll$, we need to show that the $\elll$-dependence of the squared matrix element in the equation above is superficial.

To this end, we now fix the gauge to the \terminol{radial gauge} $G$ (also called \terminol{Fock-Schwinger-gauge}) \cite{Fock:1937dy,Schwinger:1951nm,Cronstrom:1980hj}. To obtain this gauge, we set $W(\xi) = \Wline{0,\xi}$ in eqns.~(\ref{eq-gaugetrans-q})--(\ref{eq-gaugetrans-A}). This procedure fixes the gauge completely up to a global color rotation, see also Refs. \cite{Cronstrom:1980hj,Leupold:1996hx}. The gauge field $\Afield'(\xi)$ in the radial gauge satisfies $\xi^\mu \Afield'_\mu(\xi) = 0$ for any $\xi$, from which follows immediately that radial Wilson lines $\WlineI{'}{\elll,0}$ become unity for any $\elll$. Thus we obtain 
\begin{equation}
	\sum_n e^{i (P-p_n)\cdot \elll} \Big(\ \leftind{G}\bra{n} \GammaDist\, \quark'(0) \ket{N(P,S)}_{G}\ \Big)^2 \ .
\end{equation}
We would now like to continue the proof as in eq.~(\ref{eq-probproof4}). However, $\elll$ can become lightlike in the integral of the Fourier transform. If the integrand is regular for lightlike $\elll$, this does not cause any difficulties, because we can exclude null sets. 
If we make this assumption, eq.~(\ref{eq-probproof4}) is applicable, where now $\WlineC{\mathcal{C}_3}=\Eins$.
In this last step, translation invariance of the gauge condition $G$ is not required. 

We have to acknowledge that the use of gauge invariant quark fields relies on non-trivial assumptions, which deserve to be checked again carefully in the future.



\subsection{Difficulties regarding the Interpretation of \TMDs}
\label{sec-probprob}

Let us distinguish three different questions regarding the interpretation of \TMDs:
\begin{enumerate}
	\item Are we able to isolate features of the nucleon with \TMDs?
	\item Are \TMDs universal, i.e., is the same set of distributions applicable for the description of a multitude of scattering processes?
	\item Do \TMDs have a mathematical interpretation in terms of probability distributions?
\end{enumerate}

Concerns that we might have to answer the first question in the negative were raised when it became known that final (or initial) state interactions play an important role in parton distribution functions even at leading twist \cite{Brodsky:2002ue}. It was argued that, due to the influence of the final state, parton distribution functions do not solely contain information encoded in the wave function of the nucleon. However, according to our present understanding, the relevant final state interactions are encoded in the gauge link, see, e.g. Ref. \cite{Collins:2002kn}. Therefore, it is not necessary to know the wave function of the final state in order to calculate parton distributions. Thus we may still think of \TMDs as properties specific to the nucleon. 

Regarding the question about universality, it has been argued that the same \TMDs describe both the Drell-Yan and the SIDIS process, apart from known sign changes due to the reflection of the gauge link \cite{Collins:2004nx}. The situation appears to be more complicated for other processes, see Ref. \cite{Bomhof:2007xt}.


Let us now discuss the third question. The parton densities given by \TMDs must be positive and normalizable if we want to interpret them as probability distributions in a mathematical sense. In the previous section, we have discussed an approach to establish positivity of parton densities. Concerning normalizability, we encounter the same difficulties as already mentioned in section \ref{sec-relationPDFs}: In $\int dx \int d^2 \vprp{k}\ \Phi^{[\gamma^+]}_\quark(x,\vprp{k};P,S)$, the $\vprp{k}$-integral is undefined. The integral becomes finite if we restrict the integration range to $|\vprp{k}|<Q$. Whether there is a more advantageous way to introduce a regularization scale is still a matter of ongoing research. With the Gaussian ansatz we will use in section \ref{sec-mellin}, the $\vprp{k}$-integral is finite without an explicit cutoff. Here the scale dependence is hidden in the ansatz, which we know to be inapplicable at large $\vprp{k}$.

In general, parton densities can only be interpreted as probability distributions within the context of an appropriate regularization and renormalization scheme and with respect to the corresponding scales.

For all practical purposes in this work, we will give our results an interpretation in terms of quark densities intrinsic to the nucleon in the sense of probability distributions. The discussion above should make the reader aware of the more subtle issues regarding this point of view.

\section{More \TMDs: Parametrization of \texorpdfstring{$\Phi^{[\GammaOp]}$}{Phi[GammaOp]}}
\label{sec-param}

In the previous sections, $f_{1}(x,\vprp{k})$ served as an important example of a \TMD. Seven more of these profile functions are needed to describe spin dependent effects at leading twist. 

\subsection{Lorentz-Invariant Amplitudes}
\label{sec-linvamp}

In order to find out which \TMDs exist, the first step is a parametrization of the correlator $\Phi^{[\GammaOp]}$ in terms of Lorentz-invariant amplitudes. Here we will briefly review this procedure, using a formulation that will be convenient for our lattice calculations and paying special attention to the role of the Wilson line. Let us look at a somewhat generalized quark-quark correlator that also allows spin-transitions of the nucleon:
\begin{equation}
	\Phi^{[\GammaOp]}(k,P,S,S')  = \int \frac{d^4 \elll}{(2\pi)^4} \ 
	e^{-ik \cdot \elll}\ 
	\frac{1}{2} \bra{N(P,S')}\ \bar \quark(\elll)\, \GammaOp\ \WlineC{\mathcal{C}_\elll}\ \quark(0)\ \ket{N(P,S)}\ .
	\label{eq-corrssprime}
\end{equation}
We can parametrize this object in terms of Lorentz-invariant amplitudes that are independent of the nucleon spin vectors $S$ and $S'$. This can be seen by rewriting the correlator in the form 
\begin{equation}
	\Phi^{[\GammaOp]}(k,P,S,S')  = \frac{1}{2}\ \bar U(P,S')\, \MDM_{\GammaOp}(k,P;\mathcal{C}_\elll)\, U(P,S)\ .
	\label{eq-MDMdef}
\end{equation}
Here $U(P,S)$ are the Dirac spinors introduced in section~\ref{sec-diracspinorsmink} and $\MDM_{\GammaOp}(k,P;\mathcal{C}_\elll)$ is a Dirac matrix.
Notice that $\MDM_{\GammaOp}(k,P;\mathcal{C}_\elll)$ is not $\elll$-dependent; rather, the appearance of $\mathcal{C}_\elll$ in the argument of $\MDM$ reminds us that there is a prescription how to connect the quark fields for any given quark separation $\elll$. 
The symmetry transformation properties of $\MDM_{\GammaOp}(k,P;\mathcal{C}_\elll)$ under Hermitian conjugation~$(\dagger)$, parity~$(\mathscr{P})$ and time reversal~$(\mathscr{T})$ are
\begin{align}
	&(\dagger): & \left[ \MDM_\Gamma(k,P;\mathcal{C}_\elll) \right]^\dagger 
	& = \gamma^0\ \MDM_{\gamma^0\,\Gamma^\dagger\,\gamma^0} (k,P;{\tilde{\mathcal{C}}}_{-\elll}+\elll)\ \gamma^0\ , \label{eq-Mhermit} \\
	&(\mathscr{P}): & \MDM_\Gamma(k,P;\mathcal{C}_\elll) 
	&= \gamma^0\ \MDM_{\gamma^0\,\Gamma\,\gamma^0} (\overline{k}, \overline{P};\overline{\mathcal{C}_{\bar \elll}})\ \gamma^0\ , \\
	&(\mathscr{T}): &\left[ \MDM_\Gamma(k,P;\mathcal{C}_\elll) \right]^* 
	&= \gamma^5 C\ \MDM_{C^\dagger \gamma^5\,\Gamma^*\,\gamma^5 C} (\overline{k},\overline{P};-\overline{\mathcal{C}_{-\bar \elll}})\ C^\dagger \gamma^5\ . 
	\label{eq-Mtimerev}
\end{align}
Here the bar indicates sign change of the spatial components of a four-vector, e.g., $k=(k^0,\vect{k})$ $\Rightarrow$ $\bar k \equiv (k^0,-\vect{k})$. The path $\tilde{\mathcal{C}}_\elll$ is meant to be the reverse path of $\mathcal{C}_\elll$, i.e., $\tilde{\mathcal{C}}_\elll(\lambda) \equiv \mathcal{C}_\elll(1-\lambda)$, and $C$ is the charge conjugation matrix $\gamma^2 \gamma^0$.

\subsubsection{Straight Wilson Lines}


For the straight Wilson line $\WlineC{\mathcal{C}_\elll} = \Wline{\elll,0}$, the Dirac matrix $\MDM_{\GammaOp}(k,P;\mathcal{C}_\elll)$ only depends on the four-vectors $k$ and $P$. For a given $\GammaOp$, we now express $\MDM_{\GammaOp}(k,P;\mathcal{C}_\elll)$ as a linear combination of all Lorentz-covariant structures that can be formed from $k$ and $P$, weighted with invariant amplitudes $A_i(k^2,k\tcdot P)$. 
We eliminate all structures that are incompatible with the transformation properties $(\dagger)$, $(\mathscr{P})$ and $(\mathscr{T})$ given in eqns.~(\ref{eq-Mhermit})-(\ref{eq-Mtimerev}). Note that the straight Wilson line fulfills $\WlineC{\mathcal{C}_\elll} = \WlineC{{\tilde{\mathcal{C}}}_{-\elll}+\elll} = \WlineC{\,\overline{\mathcal{C}_{\bar \elll}}\,} = \Wline{-\overline{\mathcal{C}_{-\bar \elll}}\,}$, so all three equations (\ref{eq-Mhermit})-(\ref{eq-Mtimerev}) provide useful constraints. Inserting the structures into eq.~(\ref{eq-MDMdef}) and using the Gordon identities in eqns.~(\ref{eq-gordonidfirst})-(\ref{eq-gordonidlast}), we find that some of the structures are redundant for the parametrization of $\Phi^{[\GammaOp]}(k,P,S,S')$, and we leave them out. We thus arrive at the parametrization given in eq.\,(\ref{eq-Mstructs}) in the appendix. In terms of $\Phi^{[\GammaOp]}(k,P,S)$, the structures thus obtained read\footnote{Note that no additional amplitudes appear in the correlator $\Phi^{[\GammaOp]}(k,P,S,S')$ as compared to the ``spin diagonal'' correlator $\Phi^{[\GammaOp]}(k,P,S)$.}
\begin{align}
	\Phi^{[\Eins]}(k,P,S) & = 
		2\, m_N\ A_1 \ ,\nonumber\\
	\Phi^{[\gamma^\mu]}(k,P,S) & =
		2\ A_2\ P^\mu
		+ 2\ A_3\ k^\mu 
		+ \left[ \frac{2}{m_N}\ A_{12}\ \epsilon^{\mu \nu \alpha \beta} S_\nu P_\alpha k_\beta \right] \ ,\nonumber\\
	\Phi^{[\sigma^{\mu \nu}]}(k,P,S) & =
		  \left[ \frac{2}{m_N}\ A_4\ (P^\mu k^\nu - P^\nu k^\mu) \right]
		+ 2\ A_9\ \epsilon^{\mu \nu \alpha \beta} S_\alpha P_\beta \nonumber\\ &
		+ 2\ A_{10}\ \epsilon^{\mu \nu \alpha \beta} S_\alpha k_\beta  
		+ \frac{2}{{m_N}^2}\ A_{11}\ \epsilon^{\mu \nu \alpha \beta} k_\alpha P_\beta (k \cdot S)  \ ,\nonumber\\
	\Phi^{[\gamma^\mu \gamma^5]}(k,P,S) & = 
		- 2\, m_N\ A_6\ S^\mu
		- \frac{2}{m_N}\ A_7\ P^\mu (k \cdot S)
		- \frac{2}{m_N}\ A_8\ k^\mu (k \cdot S) \ ,\nonumber\\
	\Phi^{[\gamma^5]}(k,P,S) & = \left[ 2i\ A_5 \ (k\cdot S) \right]\ .
	\label{eq-phitraces}
	\end{align}
The structures above correspond to the parametrization worked out in Refs. \cite{Ralston:1979ys,Tangerman:1994eh,Muld95}. Powers of $m_N$ have been inserted to make the $A_i$ dimensionless. All amplitudes $A_i$ above are real valued because of the requirement $A_i^* = A_i$ following from the hermiticity constraint $(\dagger)$. The structures proportional to $A_4$, $A_5$ and $A_{12}$ (highlighted by square brackets) change sign under time reversal $(\mathscr{T})$, and are thus incompatible with eq.~(\ref{eq-Mtimerev}). An experimental measurement giving a non-zero value for such a \terminol{$\mathscr{T}$-odd amplitude} can be interpreted as a clear sign that a non-trivial Wilson line is needed in the definition of the quark-quark correlator, as we will see in the next section.

\subsubsection{Wilson Lines out to Infinity}
\label{sec-paramextended}

Let us consider a Wilson line of the shape $\WlineC{\mathcal{C}_{\elll}} \equiv \WlineC{\mathcal{C}_{\elll,v}} \equiv \Wline{\elll,\eta v+\elll, \eta v, 0}$ for $\eta\rightarrow \infty$ as in eq.\,(\ref{eq-NewCollins}) and, for $\eta = \infty$, in eq.\,(\ref{eq-StdSIDISWilsonLineC}). 
This contour introduces an additional dependence on $v$, so the Lorentz-invariant amplitudes are now functions $A_i(k^2,k\tcdot P,v\tcdot k,v^2,v\tcdot P)$. Only the directional information contained in $v$ is relevant, so we may rescale $v$ by $|v\tcdot P|^{-1}$, and the amplitudes take the form
\begin{equation}
	A_i\left(k^2,k\tcdot P,\frac{v\tcdot k}{|v\tcdot P|},\frac{v^2}{|v\tcdot P|^2},\frac{v\tcdot P}{|v\tcdot P|}\right) =
	A_i\left(k^2,k\tcdot P,\frac{v\tcdot k}{|v\tcdot P|},\zeta^{-1},\mathrm{sgn}(v \tcdot P)\right) \ .
\end{equation}
Note that for $v\approx\hat n_-$, we have $v\tcdot k/v\tcdot P\approx x$ and $\zeta^{-1}\approx 0$.
The Wilson line fulfills 
\begin{gather}
	(\dagger):\quad \WlineC{{\tilde{\mathcal{C}}}_{-\elll,v}+\elll} = \Wline{\elll,\ \eta v+\elll,\ \eta v,\ 0}\ = \WlineC{\mathcal{C}_{\elll,v}} \ ,\nonumber \\
	(\mathscr{P}):\quad \WlineC{\,\overline{\mathcal{C}_{\bar \elll,v}}\,} = \WlineC{\mathcal{C}_{\elll,\bar{v}}} \ , \qquad
	(\mathscr{T}):\quad \WlineC{-\overline{\mathcal{C}_{-\bar \elll,v}}\,} = \WlineC{\mathcal{C}_{\elll,-\bar{v}}} \ .
	\label{eq-stapletrafoprops}
\end{gather}
from which we conclude that $v$ transforms as $(\dagger)$: $v \rightarrow v$, $(\mathscr{P})$: $v \rightarrow \bar v$, $(\mathscr{T})$: $v \rightarrow -\bar v$ in eqns.~(\ref{eq-Mhermit})-(\ref{eq-Mtimerev}). We recognize that time reversal converts a future-pointing Wilson line into a past-pointing Wilson line \cite{Collins:2002kn}. The arguments of the amplitudes $A_i$ remain invariant under application of $(\dagger)$ and $(\mathscr{P})$, but the third and the last argument change sign under $(\mathscr{T})$, because $ v \tcdot k \rightarrow -\bar v \tcdot \bar k = - v \tcdot k$ and $ v \tcdot P \rightarrow -\bar v \tcdot \bar P = - v \tcdot P$. Obviously, time reversal $(\mathscr{T})$ provides a relation between two subtypes of amplitudes, $A_i(\ldots,+1)$ and $A_i(\ldots,-1)$, rather than a relation that limits the number of allowed structures. This means that amplitudes such as the ones in square brackets in eq.~(\ref{eq-phitraces}) become possible \cite{Collins:2002kn}. 

A complete parametrization of $\Phi^{\GammaOp}$, which also takes the $v$-dependence of the Lorentz-covariant structures into account, has been published in Ref.~\cite{Goeke:2005hb} and involves 32 amplitudes in total. In a lattice calculation with non-straight Wilson lines, the existence of all these amplitudes must be taken into account.
Initial studies with a gauge link $\Wline{\elll,\eta v+\elll, \eta v, 0}$ on the lattice will be presented in section~\ref{sec-staples}.

\subsection{From Amplitudes to \TMDs}
\label{sec-contampstotmds}

The next step is to rewrite the $k^-$ integral eq.~(\ref{eq-TMDcorr}) for the different structures in (\ref{eq-phitraces}) in Lorentz invariant form to isolate the profile functions. The procedure will be shown in more detail in sec.~\ref{sec-paramconnect}, when we address the Fourier transformed amplitudes. 
At leading twist, one obtains the following \TMDs \cite{Ralston:1979ys,Tangerman:1994eh,Muld95,Goeke:2005hb}:
\begin{align}
	\Phi^{[\gamma^+]}(x,\vprp{k};P,S) & = f_1(x,\vprp{k}^2) - \toddmark{\frac{\vect{\epsilon}_{\prp ij}\, \vect{k}_{\prp i}\, \vect{S}_{\prp j}}{m_N}\ f_{1T}^\prp(x,\vprp{k}^2)} \label{eq-phigammaplus}\ , \\
	\Phi^{[\gamma^+\gamma^5]}(x,\vprp{k};P,S) & = \Lambda\, g_{1L}(x,\vprp{k}^2) + \frac{\vprp{k} \cdot \vprp{S}}{m_N}\ g_{1T}(x,\vprp{k}^2) \label{eq-phigammaplusgfive} \ , \\
	\Phi^{[\sigma^{i+}]}(x,\vprp{k};P,S) & = \epsilon_\prp^{ij} S_j\ h_{1T}(x,\vprp{k}^2) + \frac{\epsilon_\prp^{ij} k_j}{m_N} \left( \Lambda\, h_{1L}^\prp(x,\vprp{k}^2) + \frac{\vprp{k} \cdot \vprp{S}}{m_N}\, h_{1T}^\prp(x,\vprp{k}^2) \right) \nonumber \\ & + \toddmark{\frac{k^i}{m_N} h_1^\prp(x,\vprp{k}^2)} \ ,\label{eq-phisigmaplusi} 
\end{align}
where $i,j=1,2$ are indices denoting transverse directions. The nucleon spin has been decomposed as in eq.~(\ref{eq-spinparam}) in the appendix.
Again, the structures in square brackets are $\mathscr{T}$-odd and therefore not present for correlators implemented with a straight gauge link.
Note that the integrated PDF corresponding to $g_{1L}$ is called $g_1$, and that $h_1$ is the integrated PDF corresponding to $h_{1T} + (\vprp{k}^2/2m_N^2) h_{1T}^\prp$.
The list of \TMDs at \emph{leading twist} given above is complete, and does not change when the $v$-dependence is taken fully into account \cite{Goeke:2005hb}.

\section{Azimuthal Asymmetries from Transverse Momentum Dependence}

Experimentally, we have access to the transverse momentum dependence of parton distributions through certain azimuthal asymmetries. Which asymmetries can be exploited to extract specific \TMDs is summarized in Ref.~\cite{Boer:1997nt}. Below, we list some asymmetries of particular interest.

\subsection{The Cahn Effect}
\label{sec-cahn}

Even in unpolarized measurements, we see evidence of the intrinsic transverse motion of quarks \cite{Cahn:1978se}. The dependence of the SIDIS cross section on $\phi_h$ is governed by the $\vprp{k}$-dependence of $\Phi^{[\gamma^+]}_{\text{unpol}}(x,\vprp{k};P,S) = f_1(x,k_\prp^2)$. If the quarks carried no intrinsic transverse momentum, i.e., if $\Phi^{[\gamma^+]}_{\text{unpol}}(x,\vprp{k};P,S) = f_1(x) \delta^{(2)}(\vprp{k})$, then the dependence on $\phi_h$ would be trivial, c.f. eq.~(43) in Ref.~\cite{Kotzinian:1994dv}. 

Making the Ansatz 
\begin{equation}
	f_{1,\quark}(x,\vprp{k}^2) = f_{1,\quark}(x) \frac{1}{\pi \mean{k_\prp^2}} \exp\left( -\frac{k_\prp^2}{\mean{k_\prp^2}} \right)
\end{equation}
motivated by the Fermi motion of partons in hadrons \cite{Cahn:1978se,Ceccopieri:2007ek}, the authors of Ref.~\cite{Anselmino:2005nn} find that a value
\begin{equation}
	\mean{k_\prp^2}^{-1/2} = 0.5 \units{GeV}
\end{equation}
for the \terminol{root mean square (RMS)} transverse momentum agrees best with data from EMC \cite{Aubert:1983cz,Arneodo:1986cf,Ashman:1991cj} and FNAL~E665 \cite{Adams:1993hs}.
Based on the same ansatz, Ref.~\cite{Ceccopieri:2007ek} addresses the question how $\mean{k_\prp^2}$ changes under evolution of $Q^2$.

Lattice results related to the Cahn effect will be presented in section \ref{sec-gaussresults}.

\subsection{\texorpdfstring{$\mathscr{T}$}{T}-Odd Effects}
\label{sec-todd}

The bracketed term in eq.~(\ref{eq-phigammaplus}) involves the so called \terminol{Sivers function} $f_{1T}^\prp(x,\vprp{k}^2)$, introduced in Refs.~\cite{Sivers:1989cc,Sivers:1990fh}. This leading twist $\mathscr{T}$-odd \TMD is proportional to $\vprp{k} \times \vprp{S}$ and thus describes the correlation of the intrinsic transverse momentum of unpolarized quarks with the transverse spin component of the nucleon. In SIDIS, the Sivers effect is accessible from the transverse spin asymmetry $\propto \sin(\phi_h - \phi_S)$. Therefore, a polarized nucleon target is required. Experimental information about the Sivers function has been obtained from Belle, HERMES and COMPASS data, see e.g., \cite{D'Alesio:2008jz}. Further experimental knowledge may also come from the PAX, PHENIX, RHIC and STAR experiments \cite{Collins:2005ie,Collins:2005rq}. 

Another leading twist $\mathscr{T}$-odd \TMD, the \terminol{Boer-Mulders function} $h_1^\prp(x,\vprp{k})$, is accessible from unpolarized experiments by measuring a $\cos(2\phi_h)$ asymmetry \cite{Boer:1997nt}. It can be interpreted as a correlation proportional to $\vprp{k}\times\vprp{s}$ in the quark density, where $\vprp{s}$ represents the spin of a transversely polarized quark. Experimental results come from NA10 at CERN as well as E165 and E866/NuSea at FNAL, see,  e.g., Ref. \cite{Zhang:2008nu,D'Alesio:2008jz}.

The existence of non-vanishing $\mathscr{T}$-odd \TMDs, like the Sivers function, was ruled out at first \cite{Collins:1992kk}. Ideas that $\mathscr{T}$-odd \TMDs may exist none the less due to final state interactions \cite{Boer:1997nt,Brodsky:2002cx} could be cast into mathematical form, once the directional change of the Wilson line under time reversal had been discovered \cite{Collins:2002kn}. As mentioned before, the direction $v$ of the gauge link switches from future-pointing to past-pointing when comparing SIDIS with the Drell-Yan process. Therefore the time reversal transformation $(\mathscr{T})$ transforms SIDIS \TMDs into Drell-Yan \TMDs. The $\mathscr{T}$-even distributions are equal for the two processes, while the $\mathscr{T}$-odd distributions change sign \cite{Collins:2002kn}.

Obviously, $\mathscr{T}$-odd \TMDs cannot be explored on the lattice with the straight Wilson line we use in chapter \ref{chap-results}. With a staple-shaped Wilson line, we will get a first glimpse into the realm of $\mathscr{T}$-odd contributions from the lattice in section~\ref{sec-oddratio}.

\subsection{The Quark Density in the Polarized Nucleon from \texorpdfstring{$g_{1T}(x,\vprp{k}^2)$}{g1T(x,kprp)}}
\label{sec-poldens}

In the pursuit of obtaining a picture of the quark density with respect to transverse momentum, let us write down the operator
\begin{multline}
	\left( \frac{\Eins+\lambda\gamma^5}{2}\, \quark_{(+)}(\elll) \right)^\dagger\, \WlineC{\mathcal{C}_\elll}\,
	\left( \frac{\Eins+\lambda\gamma^5}{2}\, \quark_{(+)}(0) \right) = \\
	\quark^\dagger_{(+)}(\elll)\, \WlineC{\mathcal{C}_\elll}\, \frac{\Eins + \lambda \gamma^5}{2}\, \quark_{(+)}(0)  = 
	\frac{1}{\sqrt{2}}\ \bar \quark(\elll)\, \WlineC{\mathcal{C}_\elll}\, \frac{\gamma^+ + \lambda \gamma^+ \gamma^5}{2}\, \quark(0)\ .
\end{multline}
This tells us that the density of longitudinally polarized quarks with helicity $\lambda$ is obtained from
\begin{equation}
	\Phi^{[(\gamma^+ + \lambda \gamma^+ \gamma^5)/2]}(x,\vprp{k};P,S) = \frac{1}{2}\,\Phi^{[\gamma^+]}(x,\vprp{k};P,S) + \frac{\lambda}{2}\,\Phi^{[\gamma^+\gamma^5]}(x,\vprp{k};P,S)\ .
	\label{eq-polarizedquarkdens0}
\end{equation}
If the nucleon is polarized in transverse direction ($\Lambda=0$), this becomes
\begin{equation}
	\rho_{TL}(x,\vprp{k};\vprp{S},\lambda) \equiv \frac{1}{2} f_{1}(x,\vprp{k}^2) + \frac{\lambda}{2} \frac{\vprp{k} \cdot \vprp{S}}{m_N}\ g_{1T}(x,\vprp{k}^2) - \toddmark{\frac{1}{2}\,\frac{\vect{\epsilon}_{\prp ij}\, \vect{k}_{\prp i}\, \vect{S}_{\prp j}}{m_N}\ f_{1T}^\prp(x,\vprp{k}^2)} \ ,
	\label{eq-polarizedquarkdens}
\end{equation}
where the bracketed term is $\mathscr{T}$-odd (and vanishes for straight gauge links). We now recognize that $g_{1T}(x,\vprp{k}^2)$ parametrizes a correlation between quark helicity, transverse nucleon spin and transverse momentum, which introduces an axial asymmetry in the density. The structure with $g_{1T}(x,\vprp{k}^2)$ vanishes if we average $\rho_{TL}(x,\vprp{k};\vprp{S},\lambda)$ over $\vprp{k}$. Accordingly, an integrated PDF corresponding to $g_{1T}(x,\vprp{k}^2)$ does not exist. Reference~\cite{Bacchetta:1999kz} gives the following bounds on $g_{1T}(x,\vprp{k}^2)$:
\begin{equation}
	\frac{\vprp{k}^2}{m_N^2} \left( (g_{1T})^2 + (f_{1T}^\prp)^2 \right) \leq (f_1 + g_{1L})(f_1 - g_{1L}) \leq (f_1)^2 \ ,
	\label{eq-g1Tbounds}
\end{equation}
from which follows
\begin{equation}
	\left| g_{1T} \frac{\vprp{k}}{m_N} \right| \leq | f_1 | \ .
\end{equation}
It is clear that this inequality must hold in order to guarantee positivity of eq.~(\ref{eq-polarizedquarkdens}).

Experimentally, $g_{1T}(x,\vprp{k}^2)$ is accessible from the azimuthal asymmetry $\cos(\phi_h - \phi_S)$ of the cross section. Its measurement requires longitudinally polarized leptons and a transversely polarized target \cite{Boer:1997nt}. 

In section~(\ref{sec-gaussresults}), we will compute the first $x$-moment of $g_{1T}(x,\vprp{k}^2)$ on the lattice using straight Wilson lines and will find that eq.\,(\ref{eq-g1Tbounds}) numerically also holds for the first Mellin moment. 



\section{Model Predictions}
\label{sec-models}

\begin{figure}[btp]
	\centering%
	\subfloat[][]{%
		\label{fig-diquark-f1}%
		\includegraphics[clip=true]{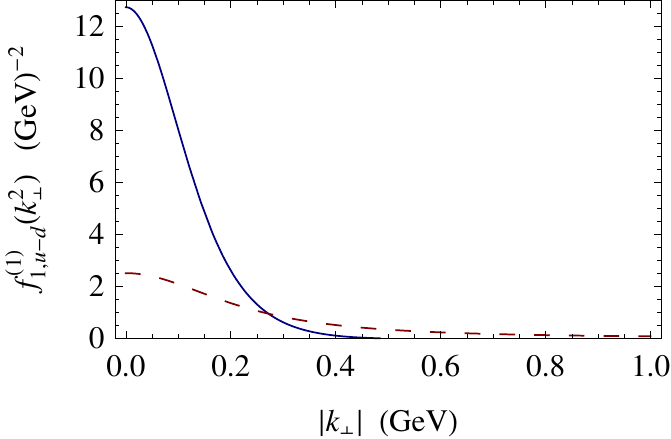}%
		}\hfill%
	\subfloat[][]{%
		\label{fig-diquark-g1T}%
		\includegraphics[clip=true]{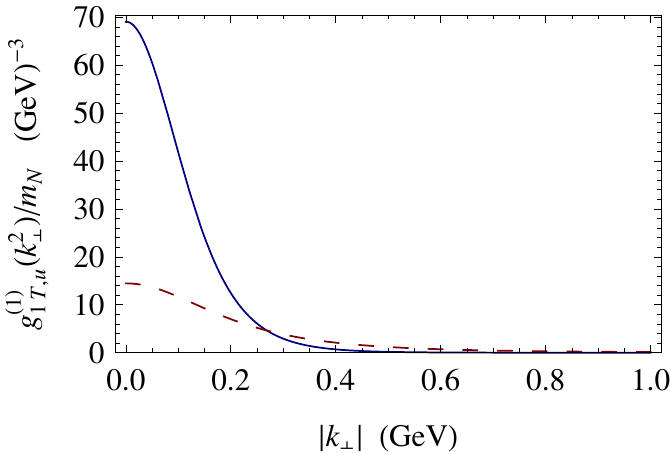}%
		}\par%
	\caption[Scalar Diquark Model]{%
		Examples for predictions for the first Mellin moment from the scalar diquark model \cite{Jakob:1997wg} using the expressions and parameters specified in the text. The solid graphs were evaluated with $\alpha=2$ and are normalized to satisfy quark counting eq.~(\ref{eq-diqmodelnorm}). The dashed curves correspond to $\alpha=1$ and normalization with the cutoff $1\units{GeV}$. \par
		\label{fig-diquark}
		}
\end{figure}

\TMDs have also been addressed within models, see, e.g., the appendices of Ref. \cite{Meissner:2007rx}. To give an example, the authors of Ref. \cite{Jakob:1997wg} calculate the correlator eq.~(\ref{eq-corr}) within a spectator model (termed ``scalar diquark model'' in Ref. \cite{Meissner:2007rx}), and find
\begin{align}
	f_{1,R}(x,\vprp{k}^2) & = N_R^2 \frac{(1-x)^{2\alpha -1}}{16\pi^3} \frac{(x m_N + m)^2 + \vprp{k}^2}{\left(\vprp{k}^2 + \lambda_R^2(x)\right)^{2\alpha}} \ ,\\
	g_{1T,R}(x,\vprp{k}^2) & = \alpha_R m_N N_R^2 \frac{(1-x)^{2\alpha -1}}{8\pi^3} \frac{x m_N + m}{\left(\vprp{k}^2 + \lambda_R^2(x)\right)^{2\alpha}} \ ,
\end{align}
where $\lambda_R(x) \equiv \Lambda^2 (1-x) + x M_R^2 - x (1-x) m_N^2$ and where the index $R=a$ or $R=s$ denotes the type of spectator. The \TMDs for individual quark flavors are obtained from $f_{1,u} = (3/2) f_{1,s} + (1/2) f_{1,a}$, $f_{1,d} = f_{1,a}$ and analogously for $g_{1T}$. At the given order, the antiquark \TMDs (ascribed to values $x<0$) are zero. The authors of Ref. \cite{Jakob:1997wg} fix the constants $N_R$ from the condition
\begin{equation}
	\int_0^1 dx\ \int d^2 \vprp{k}\  f_{1,R}(x,\vprp{k}^2) = 1 \ ,
	\label{eq-diqmodelnorm}
\end{equation}
which ensures that the total number of up quarks is $\int dx \int d^2 \vprp{k}\,f_{1,u}(x,\vprp{k}) = 2$ and the total number of down quarks is $\int dx \int d^2 \vprp{k}\,f_{1,d}(x,\vprp{k}) = 1$.
Following this strategy and taking a typical set of values $\Lambda = 0.5 \units{GeV}$, $m = 0$, $m_N = 0.938 \units{GeV}$, $M_s = 0.6 \units{GeV}$, $a_s = 1$, $M_a = 0.8\units{GeV}$, $a_a = - 1/3$ and $\alpha = 2$ from Ref.~\cite{Jakob:1997wg}, we evaluate the first Mellin moment (see section \ref{sec-defmellin}) of the distributions numerically and plot them in Fig.~\ref{fig-diquark}. The rederived results in Ref.~\cite{Meissner:2007rx} have $\alpha$ set to $1$ and thus reproduce the perturbative tail $f_{1}(x,\vprp{k}^2) \sim 1/\vprp{k}^2$ at large $\vprp{k}$ as mentioned in section~\ref{sec-relationPDFs}.
To allow for a qualitative comparison, we plot them in Fig.~\ref{fig-diquark}, leaving all parameters except for $\alpha$ unchanged. To be able to normalize them, we restrict the integration range to $|\vprp{k}|<1\units{GeV}$ in eq.~(\ref{eq-diqmodelnorm}).

\chapter{Nucleon Structure from Lattice QCD}
\label{chap-lattice}
\section{Basics of Lattice QCD}

\subsection{The Path Integral formulation}
\label{sec-pathintform}

How do we calculate correlation functions within QCD? The \terminol{path integral} formulation of quantum field theory provides a very elegant prescription, and serves here as a starting point for the development of numerical methods.\par

Suppose $O$ is some expression in terms of quark fields $\bar \quark$, $\quark$ and gluon fields $\Afield$, for example $O := \bar u(0)\, d(0)\, \bar d(y)\, u(y)$. According to the path integral formulation, we can calculate the vacuum expectation value $\bra{\Omega}\mathcal{T}\hat O\ket{\Omega}$ of the corresponding time ordered operator $\mathcal{T} \hat O$ from
\begin{equation}
	\dlangle O \drangle \equiv \frac{ \int \mathcal{D}\Afield\ \int \mathcal{D}\bar \quark\ \int \mathcal{D}\quark\ O[\bar \quark,\quark,\Afield]\ \exp(i\, S_\text{QCD}[\bar \quark,\quark,\Afield]) }{\int \mathcal{D}\Afield\  \int \mathcal{D}\bar \quark\ \int \mathcal{D}\quark\  \exp(i\,S_\text{QCD}[\bar \quark,\quark,\Afield])} = \bra{\Omega}\mathcal{T}\hat O\ket{\Omega}\ .
	\label{eq-cont-pathint}
\end{equation}
In the path integral above, $\bar \quark$, $\quark$ and $\Afield$ denote functions of space-time (the \terminol{field configurations}), and the expression $O$ is a functional of them. The dynamics of the system arises from constructive interference in the integral over all field configurations; here $\mathcal{D}\bar \quark$, $\mathcal{D}\quark$ and $\mathcal{D}\Afield$ represent appropriate integration measures. The \terminol{action} $S_\text{QCD}$ is also a functional of $\bar \quark$, $\quark$ and $\Afield$, and defined as 
\begin{equation}
	S_\text{QCD}[\bar \quark, \quark,\Afield] \equiv \lim_{T \rightarrow \infty \exp(-i \theta)}\ \int_{-T}^{T} dx^0\ \int_{-\infty}^{\infty} d^3 x\ \mathcal{L}_\text{QCD}[\bar \quark, \quark,\Afield](x)\ .
	\label{eq-cont-action}
\end{equation}
The infinitesimal constant $\theta>0$ implements a tiny Wick rotation which makes sure that the state $\ket{\Omega}$ in eq.\,(\ref{eq-cont-pathint}) corresponds to the vacuum, see e.g. Ref. \cite{Peskin:1995ev}, chapter 9.2.

%


\subsection{Path Integral in Euclidean Space}

For our purposes here, lattice quantum field theory serves as a tool to calculate matrix elements in QCD numerically.
One of the problems of a numerical evaluation of eq.\,(\ref{eq-cont-pathint}) is the oscillatory term $\exp(i\, S_\text{QCD}[\bar \quark,\quark,\Afield])$. We therefore set $\theta=\pi$ in eq.\,(\ref{eq-cont-action}), or equivalently, we substitute
\begin{equation}
	x_\Eu{4} \equiv i x^0 = i x_0\,,\hspace{10pt}
	x_\Eu{1} \equiv x^1 = - x_1\,,\hspace{10pt}
	x_\Eu{2} \equiv x^2 = - x_2\,,\hspace{10pt}
	x_\Eu{3} \equiv x^3 = - x_3\,.
\end{equation}
The new coordinates $x_\muE$ are Euclidean: $x_\mu x^\mu = - x_\muE x_\muE = -(x_\Eu{1} x_\Eu{1} + x_\Eu{2} x_\Eu{2} + x_\Eu{3} x_\Eu{3} + x_\Eu{4} x_\Eu{4})$. 
It is customary to define the Euclidean action $S^\text{E}_\text{QCD}$ as
\begin{align}
	S^\text{E}_\text{QCD}[\bar \quark,\quark,\Afield] & = -i\, S_\text{QCD}[\bar \quark,\quark,\Afield]\big\vert_{\theta=\pi} \nonumber\\ 
	& = \int d^4 x \left( \sum_{\quark = u,d,s,\ldots} \bar \quark(x) \left( \slashedE{D} + m_\quark  \right) \quark(x)  
	+ \frac{1}{4}\, F_{\muE \nuE\, a}(x)\, F_{\muE \nuE\,a}(x) \right) \ ,
	\label{eq-cont-eu-act}
\end{align}
where, in the last line, we have switched to Euclidean notation as explained in appendix~\ref{sec-euklidean}. Equation (\ref{eq-cont-pathint}) becomes
\begin{equation}
	\dlangle O \drangle = \frac{\int \mathcal{D}\Afield\  \int \mathcal{D}\bar \quark\ \int \mathcal{D}\quark\  O[\bar \quark,\quark,\Afield]\ \exp(- S^\text{E}_\text{QCD}[\bar \quark,\quark,\Afield]) }{\int \mathcal{D}\Afield\  \int \mathcal{D}\bar \quark\ \int \mathcal{D}\quark\  \exp(-S^\text{E}_\text{QCD}[\bar \quark,\quark,\Afield])}\ .
	\label{eq-cont-eu-pathint}	
\end{equation}
The exponential term is now well-behaved, since $S^\text{E}_\text{QCD}[\bar \quark,\quark,\Afield]$ is non-negative and real. Note that the $90^\circ$ Wick rotation we have performed requires the analytic continuation of the fields $\bar \quark$, $\quark$ and $A$ to Euclidean space. For the evaluation of $O[\bar \quark,\quark,\Afield]$, the fields $\bar \quark(x)$, $\quark(x)$ and $A(x)$ are in practice only available at Euclidean coordinates $x$. For example, the two-point correlation function obtained by setting $O:=\bar u(0)\, d(0)\, \bar d(x)\, u(x)$ can only be calculated for Euclidean separations $x$, since $x_\Eu{4}$ is real.

%
%

\subsection{Discretization of Free Fermions}
By discretizing the action, we are able to replace the functional integrals in eq.\,(\ref{eq-cont-eu-pathint}) by a large number of ordinary integrals.
Let us introduce a lattice $\lbrace x_{(0)},x_{(1)},x_{(2)},\ldots \rbrace \equiv \mathbb{L} \equiv a\,\mathbb{Z}^4$ of points in four-dimensional Euclidean space with uniform lattice spacing $a$, and abbreviate vectors of length $a$ along the Euclidean axes with $\hat{\mu} \equiv a\, \hat{e}_\muE$. Consider the action of free quarks of a single flavor
\begin{equation}
	S_\text{free}[\bar \quark,\quark] = \int d^\Eu{4} x\ \bar \quark(x) \left( \gamma_\muE \partial_\muE + m_\quark  \right) \quark(x) \ .
	\label{eq-freefermcont}
\end{equation}
A naively discretized version of eq.~(\ref{eq-freefermcont}) is
\begin{equation}
	S_\text{free}^\lat[\bar \quark,\quark] = a^4 \sum_{x\in\mathbb{L}} \left( \sum_{\mu=1}^4 \bar  \quark(x)\,  \gamma_\muE\, \frac{\quark(x+\hat{\mu}) - \quark(x-\hat{\mu})}{2a}  + \bar  \quark(x)\, m_q\, \quark(x) \right) \ ,
	\label{eq-naivefreeactunscal}
\end{equation}
where $\partial_\muE$ has been replaced by a central difference. 
$S_\text{free}^\lat$ depends on a countable number of field variables, namely $\bar \quark(x_{(0)}),$ $\quark(x_{(0)}),$ $\bar \quark(x_{(1)})$, $\quark(x_{(1)}),\ldots$. It is convenient to work with dimensionless variables on the lattice. Therefore, we will follow the convention that the quark fields on the lattice are rescaled according to $\quark(x) \rightarrow a^{-3/2} \quark(x)$, and the masses are replaced by $\hat{m}_\quark \equiv a m_\quark$. The lattice action then reads
\begin{equation}
	S_\text{free}^\lat[\bar \quark,\quark] = \sum_{x\in\mathbb{L}} \left( \sum_{\mu=1}^4 \bar  \quark(x)\,  \gamma_\muE\, \frac{\quark(x+\hat{\mu}) - \quark(x-\hat{\mu})}{2}  + \bar  \quark(x)\, \hat{m}_\quark\, \quark(x) \right) \ .
	\label{eq-naivefreeact}
\end{equation}
Note that this action is bilinear in $\bar \quark$ and $\quark$, and can thus be written in the form
\begin{equation}
	S_\text{free}^\lat[\bar \quark,\quark] = \sum_{x,y \in \mathbb{L}}\ \sum_{\alpha,\beta}\ \bar \quark_\alpha(x)\ K_{(\alpha,x),(\beta,y)}\ \quark_\beta(y) \ ,
	\label{eq-freeactbil}
\end{equation}
where we have made Dirac indices $\alpha$ and $\beta$ explicit, and where the $K_{(\alpha,x),(\beta,y)}$ form a large matrix of coefficients.

For given $\bar \quark$ and $\quark$, the discretized action $S_\text{free}^\lat$ converges to the continuum action $S_\text{free}$ in the limit $a \rightarrow 0$. 
Even though, this naive action $S_\text{free}^\lat$ does not provide the desired continuum limit in the path integral. Reasons for this will become clear in the following.


\subsection{The Fermion Doubling problem}
\label{sec-fermdbl}

The dispersion relation that belongs to the lattice action $S_\text{free}^\lat$ in the previous section has an unwanted feature: It has 15 additional energy minima at nonzero three-momentum. The problem is a consequence of the use of the central difference operation in eq.~(\ref{eq-naivefreeact}). 
The shortest wavelength that can be realized in one lattice direction  is $2a$. The central difference $\quark(x+\hat{\mu}) - \quark(x-\hat{\mu})$ vanishes for such a field configuration, because the central difference extends over $2$ lattice spacings rather than just one. In other words, the central difference is blind to certain modes on the Brillouin zone. 
If we were to carry out a lattice calculation with the na\"ive fermion action $S_\text{free}^\lat$, all 16 low energy modes present in the action would be populated and would contribute to our measurements. We would have 16 fermion species instead of one. A number of techniques has been developed to deal with this problem. Here we list just three of them (For details, see, e.g., Ref. \cite{Roth,DeGrand:2006zz}.) 


\begin{itemize}
	\item \terminol{Wilson fermions}: We can add a term 
	\begin{equation}
	 -a^4 \sum_{x\in\mathbb{L}}\sum_{\muE=1}^4  a \hat r\,\bar \quark(x) \frac{\quark(x-\hat \mu) - 2 \quark(x) + \quark(x+\hat \mu) }{2a^2} 
	\label{eq-wilsonterm}
	\end{equation} to the naive fermion action eq.~(\ref{eq-naivefreeactunscal}). Here $\hat r$ is a dimensionless parameter. The expression above is a discretized version of $-(a \hat r/2)\bar \quark(x) \partial_\muE \partial_\muE \quark(x)$. This term vanishes in the limit $a\rightarrow0$. At finite lattice spacing it alters the dispersion relation, raising the energy of the spurious modes. Thus only the physical fermion species survives. The drawback of the Wilson action is that it does not exhibit \terminol{chiral symmetry}: When we set the quark masses $m_\quark$ to zero, QCD is invariant under the global transformation $\quark \rightarrow \exp(i\epsilon\gamma_\Eu{5})\quark$. At nonzero lattice spacing, the Wilson term destroys this invariance explicitly. 
	\item \terminol{Staggered fermions} \cite{Kogut:1974ag}: Consider again the naive fermion action eq.~(\ref{eq-naivefreeact}). It is invariant under translations of step size $a$ along the lattice axes. It turns out that the fermion matrix $K_{(\alpha,x),(\beta,y)}$ defined in eq.~(\ref{eq-freeactbil}) can be block diagonalized into four blocks. This means that there are four independent groups of fermion species, without any mutual interactions between those groups. Therefore, we can simply remove three of those groups. The resulting action is invariant under translations of step size $2a$, i.e., the unit cell of the action is of size $(2a)^4$. This ``staggered'' action still exhibits $16/4=4$ fermion species, called \terminol{tastes}. Quark correlation functions calculated with the staggered action exhibit \terminol{taste splittings}, i.e., depending on the taste degrees of freedom that have been used to set up the correlator, the expectation value will be a little higher or lower. The taste splittings should vanish in the continuum limit $a \rightarrow 0$. The staggered action is invariant under a certain modified chiral transformation. A serious problem of staggered fermions is multiplicity: In fermionic loops, all four tastes give a contribution. To avoid this, one commonly makes use of the \terminol{fourth root trick}: In the Monte Carlo sampling step (see sections \ref{sec-fermionintegration} and \ref{sec-MCsim} below), one takes the fourth root of the fermion determinant. Whether it is guaranteed that this procedure reproduces QCD in the continuum limit is a matter of ongoing debate \cite{Creutz:2008nk,Golterman:2008gt}. At present, we use staggered actions in spite of this open issue, because they are computationally exceptionally cheap, and thus permit us to explore parameter ranges and lattice sizes that would otherwise be inaccessible with present computing resources.
	\item \terminol{Domain wall fermions} \cite{Shamir:1993zy}: 
	This action (just as the overlap action) is able to establish a modified version of chiral symmetry on the lattice and is at the same time doubler free. The modified chiral symmetry transformation is $\quark \rightarrow \exp[i\epsilon\gamma_\Eu{5}(1 - (a/2 \hat r_0) D )]\quark$, where the operator $D$ fulfills the Ginsparg-Wilson relation \cite{Ginsparg:1981bj}, namely $\{D, \gamma_\Eu{5}\}= (a/ \hat r_0) D \gamma_\Eu{5} D$. Chiral symmetry on the lattice is desirable because it reduces operator mixing (see section \ref{sec-opmix} below), simplifies renormalization and is valuable for chiral extrapolation. The idea behind domain wall fermions is to separate left-handed and right-handed quarks spatially in an auxiliary fifth lattice dimension. The additional dimension comes at a considerable computational cost. The larger the lattice size $\hat L_5$ in this auxiliary dimension is chosen, the more accurately chiral symmetry is fulfilled. Before correlation functions are evaluated, the lattice is projected onto the usual four dimensions.
\end{itemize}

We note that the fermion actions above are still of the bilinear form of eq.\,(\ref{eq-freeactbil}). The calculations in this work will be carried out using a hybrid action, with domain wall valence quarks on top of a staggered sea, as explained in section \ref{sec-lattices} below.

\subsection{The Gauge Principle on the Lattice}
\label{sec-gaugetrafolat}

Upon discretization, some symmetries of the continuum theory are lost, but they are restored in the limit $a \rightarrow 0$. Concerning local gauge symmetry, it turns out that we can construct a discretized action that retains gauge invariance even at finite lattice spacing. Consider the free fermion action $S_\text{free}^\lat$ in eq.\,(\ref{eq-freeactbil}). First, let us focus on quark bilinears involving neighboring sites, such as $\bar \quark_\alpha(x)\, K_{(\alpha,x),(\beta,x+\hat{\mu})}\, \quark_\beta(x+\hat{\mu})$. Under the gauge transformation eq.\,(\ref{eq-gaugetrans-q}) and (\ref{eq-gaugetrans-qbar}), this term becomes $\bar \quark_\alpha(x)\, W^\dagger(x)\,K_{(\alpha,x),(\beta,x+\hat{\mu})}\, W(x+\hat{\mu})\,\quark_\beta(x+\hat{\mu})$. To make it gauge invariant, we introduce a new set of fields $U_\muE(x)$ of $\mathrm{SU}(3)$ color matrices, which transform according to
\begin{equation}
	U_\muE(x) \rightarrow U'_\muE(x) = W(x)\,U_\muE(x)\, W^\dagger(x+\hat{\mu}) \ .
	\label{eq-gaugetransf-u}
\end{equation}
With appropriate insertions of the $U$ fields, we can modify the quark bilinear terms in such a way that they become gauge invariant. For example, the terms
\begin{equation*}
	\bar \quark_\alpha(x)\, K_{(\alpha,x),(\beta,x+\hat{\mu})}\, U_\muE(x)\, \quark_\beta(x+\hat{\mu})
	\qquad\text{and}\qquad
	\bar \quark_\alpha(x+\hat{\mu})\, K_{(\alpha,x+\hat{\mu}),(\beta,x)}\, U^\dagger_\muE(x)\, \quark_\beta(x)
\end{equation*}
are gauge invariant. The $U_\muE(x)$ are called \terminol{link variables}, and are depicted as lines connecting neighboring lattice sites, see Fig.~\ref{fig-thelattice}. For later convenience, we introduce the notation
\begin{equation}
	U_\muE(x) \equiv U(x,x+\hat{\mu})\ , \qquad
	U^\dagger_\muE(x) \equiv U(x+\hat{\mu},x) \ .
\end{equation}
Quark bilinears of non-neighboring sites require several insertions of adjacent link variables. For example, $\bar \quark_\alpha(x) K_{(\alpha,x),(\beta,x+2\hat{\mu})} U(x,x+\hat{\mu}) U(x+\hat{\mu},x+2\hat{\mu}) \quark_\beta(x+2\hat{\mu})$ is gauge invariant. Note that the prescription of rendering bilinears invariant is not unique; any connected path between the quark fields could be chosen. Finally, the resulting fermion action can be written in the form
\begin{equation}
 	S_F^\lat[\bar \quark,\quark,U] = \sum_{x,y \in \mathbb{L}}\ \sum_{\alpha,\beta}\ \sum_{i,j}\ \bar \quark_{\alpha,i}(x)\ K_{(\alpha,i,x),(\beta,j,y)}[U]\ \quark_{\beta,j}(y) \equiv
	\sum_{\MI{1},\MI{2}} \bar \quark_\MI{1}\, K_{\MI{1}\MI{2}}[U]\, \quark_\MI{2} \ ,
	\label{eq-fermactbil}
\end{equation}
where we have made color indices $i$ and $j$ explicit. The products of link variables needed to maintain gauge invariance of the individual quark bilinears have been combined with our former coefficient matrix to form an object $K_{(\alpha,i,x),(\beta,j,y)}[U]$. On the right hand side, the indices $\MI{1}$ and $\MI{2}$ each abbreviate a combination of a Dirac index, a color index and a lattice site. 

On the lattice, we get a rather intuitive understanding of the gauge principle. The quark field variable $\quark_\alpha(x)$ is a color vector with respect to a frame of reference of color coordinates. Local gauge transformations allow us to rotate this frame of reference independently at each lattice site. The link variable $U_\muE(x)$ tells us how the color frames of two neighboring lattice sites are rotated relative to each other. Thus they enable us to form expressions involving quark fields at different sites that remain unaffected by local color coordinate transformations, i.e., local gauge transformations.

\subsection{The Gauge Action on the Lattice}

Now we still need an action $S_G^\lat[U]$ that determines dynamics of the link variables. 
We can give the link variables an interpretation in terms of the gluon fields $\Afield_\mu(x)$ of the continuum theory by writing
\begin{equation}
	U(x,x+\mu) \hat = \exp \left( i g a \Afield_\muE(x) \right) \ .
\end{equation}
Making this identification, we request that $S_G^\lat[U]$ corresponds to the continuum gluonic part of the continuum action eq.\,(\ref{eq-cont-eu-act}) when formally taking the limit $a\rightarrow 0$:
\begin{equation}
	\lim_{a \rightarrow 0}\  S_G^\lat[U]\ \hat =\ \int d^4 x\ \frac{1}{4} F_{\muE\nuE\,a}(x)\, F_{\muE\nuE\,a}(x)\ .
\end{equation}
A simple action fullfilling this requirement is
\begin{equation}
	S_G^\lat[U] = \frac{\beta}{6} \sum_{x\in\mathbb{L}} \sum_{\muE,\nuE} \Tr_c \left( \Eins - U_{\muE\nuE}(x) \right) \ .
	\label{eq-latgaugeact}
\end{equation}
Here $\beta = 6/g^2$ is the coupling constant, $\Tr_c$ the trace over color indices, and the \terminol{plaquette} $U_{\muE\nuE}(x)$ a rectangular loop of four link variables:
\begin{equation}
	U_{\muE\nuE}(x) \equiv 
		U(x,x+\hat{\mu})\,
		U(x+\hat{\mu},x+\hat{\mu}+\hat{\nu})\,
		U(x+\hat{\mu}+\hat{\nu},x+\hat{\nu})\,
		U(x+\hat{\nu},x) \ .
\end{equation}
Since $U_{\muE\nuE}(x) = U_{\nuE\muE}^\dagger(x)$, the terms in the gauge action add up to a real, non-negative value.\footnote{As a side remark, consider the limit $\beta\rightarrow \infty$. In this limit, all plaquettes are forced to unity. Consequently, any gauge link made up of link variables connecting two given lattice sites gives the same $\mathrm{SU}(3)_c$-matrix. In that sense, color space is ``flat'', and we recover the free fermion action. }

Together with the fermion action, we have now a lattice action $S^\lat[\bar \quark,\quark,U] \equiv S_F^\lat[\bar \quark,\quark,U] + S_G^\lat[U]$.
Using this discrete action in the path integral eq.\,(\ref{eq-cont-eu-pathint}), we can replace the functional integrals by ordinary integrals:
\begin{equation*}
	\int \mathcal{D}\Afield \longrightarrow \mathop{\prod_{x \in \mathbb{L}}}_{\mu=1..4} \left( \int d U_\muE(x) \right) \equiv \int dU,
	\qquad
	\int \mathcal{D}\quark \longrightarrow \prod_n \left( \int d \quark_n \right) \equiv \int d\quark 
\end{equation*}
and analogously for $\bar \quark$. The path integral then reads
\begin{equation}
	\dlangle O\drangle = \frac{ \int dU\  \int d\bar \quark\ \int d\quark\ O[\bar \quark,\quark,U]\ \exp(- S^\lat[\bar \quark,\quark,U]) }{  \int dU\ \int d\bar \quark\ \int d\quark\ \exp(-S^\lat[\bar \quark,\quark,U])} \ .
	\label{eq-lat-pathint}	
\end{equation}

\subsection{Finite Volume}

\begin{figure}[tbp]
	\centering
	\includegraphics[scale=1.375,trim= 5 10 10 5,clip=true]{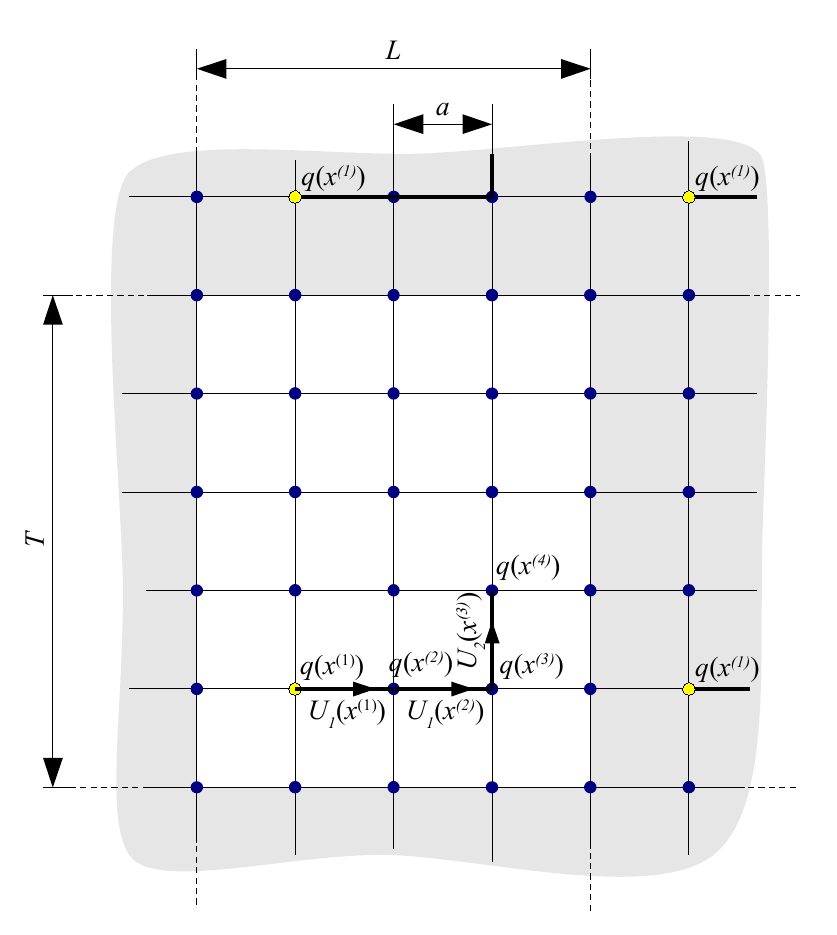}
	\caption{Illustration of the periodic lattice and its degrees of freedom.\label{fig-thelattice}}
\end{figure}
We cannot simulate an infinite number of field variables. Therefore, we restrict ourselves to a finite lattice volume. We achieve this by introducing periodic (or antiperiodic) boundary conditions on the borders of a four-dimensional box. Typically, a lattice volume has three equal spatial side lengths of $\hat L$ lattice units and a temporal extent of $\hat T$ lattice units. A two-dimensional illustration of the periodic lattice is shown in Fig.~\ref{fig-thelattice}. The number of integration variables in the path integral is now finite. 

In order to analyze nucleon structure, the nucleon should fit inside the lattice volume. From the point of view of \terminol{chiral effective field theory}, the nucleon is a core dressed by a cloud of pions. If the lattice box size is too small, our observables will be affected by pion mediated interactions of the nucleon with its mirror images on the periodically continued lattice. Therefore, the pion compton wavelength should be small compared to the lattice box size: $2 \pi / m_\pi \lesssim L$. The lattices considered in this work feature $m_\pi L \gtrsim 6$, a size at which volume effects are usually smaller than statistical errors. In the following, we will make no effort to estimate systematic uncertainties from finite volume effects.

\subsection{Integrating out Fermions}
\label{sec-fermionintegration}

In the path integral formalism, the fermionic field variables $\bar \quark_\MI{i}$ and $\quark_\MI{i}$ are Grassman variables, i.e., anticommuting numbers. The integrals over the Grassman degrees of freedom $\bar \quark$ and $\quark$ in the path integral can be evaluated analytically. To this end, consider the generating functional
\begin{equation}
	Z[\bar \eta,\eta,U] = \int d\bar \quark \int dq\ \exp\left( -\sum_{\MI{1},\MI{2}}\ \bar q_\MI{1}\ K_{\MI{1}\MI{2}}[U]\ q_\MI{2} - \sum_\MI{1} \bar \eta_\MI{1} q_\MI{1} - \sum_\MI{1} \bar q_\MI{1} \eta_\MI{1} \right) \ .
\end{equation}
We complete the squares and perform the integrals over the Grassman variables to get
\begin{equation}
	Z[\bar \eta,\eta,U] = \det \left( K[U] \right)\  \exp\left( \sum_{\MI{1},\MI{2}} \bar \eta_\MI{1}\, G_{\MI{1}\MI{2}}[U]\, \eta_\MI{2}\right)\ ,
\end{equation}
where the \terminol{lattice quark propagator} $G_{\MI{1}\MI{2}}[U]$ is the inverse of the fermion matrix:
\begin{equation}
	K_{\MI{1}\MI{2}}[U]\,G_{\MI{2}\MI{3}}[U] = \delta_{\MI{1}\MI{3}}\ .
\end{equation}
The generating function allows us to integrate out the fermions for any expression $O[\bar \quark,\quark,U]$ in the path integral:
\begin{multline}
	O[\bar \quark, \quark, U] \exp\left( -\sum_{\MI{1},\MI{2}}\ \bar q_\MI{1}\ K_{\MI{1}\MI{2}}[U]\ q_\MI{2} \right) = 
	O\left[\frac{\partial}{\partial \eta}, -\frac{\partial}{\partial \bar \eta}, U\right]\ Z[\bar \eta,\eta,U]\ \Big\vert_{\eta=\bar \eta=0} \\
	= \det \left( K[U] \right)\ O\left[\frac{\partial}{\partial \eta}, -\frac{\partial}{\partial \bar \eta}, U\right]\ \exp\left( \sum_\MI{1} \bar \eta_\MI{1}\, G_{\MI{1}\MI{2}}[U]\, \eta_\MI{2}\right) \ \Big\vert_{\eta=\bar \eta=0} \equiv \det \left( K[U] \right)\ {\tilde O}[U]\ .
\end{multline}
With the help of the formula above, we are able to determine a representation ${\tilde O}[U]$ of $O$ free of any fermionic variables. As a simple example, suppose we have $O[\bar \quark, \quark, U] = \quark_\MI{1} \bar \quark_\MI{2}$. Then, according to Grassmann algebra, we obtain
\begin{equation}
	\tilde{O}[U] = -\frac{\partial}{\partial \bar \eta} \frac{\partial}{\partial \eta} Z[\bar \eta,\eta,U]\ \Big\vert_{\eta=\bar \eta=0} = G_{\MI{1}\MI{2}}[U] \equiv 
	\contraction{}{\quark_\MI{1}}{}{\quark_\MI{2}}
	\quark_\MI{1} \quark_\MI{2}\ .
\end{equation}
Integrating out fermions amounts to forming all possible Wick contractions of quark fields in ${O}[\bar \quark, \quark, U]$ and replacing each contracted pair by a propagator $G_{\MI{1}\MI{2}}[U]$.
Finally, the path integral has a form suitable for numerical treatment:
\begin{equation}
	\dlangle O \drangle = \frac{ \int dU\  {\tilde O}[U]\ \det ( K[U] )\ \exp(- S_G^\lat[U]) }{  \int dU\ \det ( K[U] )\ \exp(-S_G^\lat[U])} \ .
	\label{eq-lat-noferm-pathint}	
\end{equation}
Note that up to now, we have restricted ourselves to a single quark flavor. For several quark flavors $\quark=u,d,s,\ldots$, the fermion matrix $K[U]=\mathrm{diag}(K^{(u)}[U],K^{(d)}[U],\ldots)$ is block diagonal, each block being a fermion matrix for a single flavor with components $K^{(\quark)}_{\MI{1}\MI{2}}[U]$. Correspondingly, we can introduce lattice quark propagators $G^{(\quark)}_{\MI{1}\MI{2}}[U]$ for each flavor. 

\subsection{Monte Carlo Calculations}
\label{sec-MCsim}
In eq.\,(\ref{eq-lat-noferm-pathint}), we have to integrate over all independent link variables in our volume. For multidimensional integrals with very many dimensions, \terminol{Monte Carlo} techniques, such as the \terminol{Metropolis algorithm}, become very efficient. At each sampling step, the algorithm produces one \terminol{gauge configuration} $U$, an array of real numbers specifying the values of all the link variables on the lattice. In a simple Monte Carlo algorithm, each gauge configuration is a modification of its preceeding configuration. Not all configurations proposed by the random generator are accepted. The rejection policy is set up in such a way (``detailed balance'') that the resulting sequence of gauge configurations $E_U$, called the \terminol{ensemble}, contains an arbitrary configuration $U$ with a probability proportional to the weight factor in the path integral :
\begin{equation}
	P( U \in E_U ) \propto \det ( K[U] )\ \exp(- S_G^\lat[U]) \ .
\end{equation}
Of course, a requirement for this importance sampling approach is that the weight factor $\det ( K[U] )\ \exp(- S_G^\lat[U])$ is a real and positive number. This is why the Wick rotation to Euclidean space is essential. With an ensemble of $N$ configurations at hand, we can now calculate correlation functions according to
\begin{equation}
	\dlangle O \drangle \approx \frac{1}{N} \sum_{U \in E_U} {\tilde O}[U] \ .
	\label{eq-latexpv}
\end{equation}
For reasons of efficiency, modern lattice calculations typically combine Monte Carlo techniques with deterministic sampling algorithms, such as \terminol{molecular dynamics}. However, the basic definition of lattice expectation values eq.~(\ref{eq-latexpv}) remains unaffected from the particular choice of algorithm.

\subsection{Gauge Fixing}
\label{sec-gfix}

The path integrals eqns.\,(\ref{eq-cont-action}),\,(\ref{eq-cont-eu-act}),\,(\ref{eq-lat-pathint}),\,(\ref{eq-lat-noferm-pathint}) are all of the general form
\begin{equation}
	\dlangle O \drangle \equiv \frac{ \int \mathcal{D}\phi\ O[\phi]\ \theta[\phi] }{\int \mathcal{D}\phi\ \theta[\phi]}\ ,
\end{equation}
where $\phi$ subsumes all fields $\quark$, $\bar\quark$, $\Afield$ and/or $U$ and where the weight factor $\theta[\phi]$ contains the exponentiated action (and, if $\quark$ and $\bar\quark$ are integrated out, the fermion determinant). Gauge fixing can be achieved with a mapping $\mathscr{f}_g$ which transforms a given field configuration $\phi$ in accordance with the gauge transformation rules eqns.\,(\ref{eq-gaugetrans-q})--(\ref{eq-gaugetrans-A}),\,(\ref{eq-gaugetransf-u}) into a new configuration $\phi'$ which fulfills the gauge fixing condition $g$. To this end, the gauge rotation matrix $W(x)\equiv W_g[\phi](x)$ must be chosen appropriately for a given $\phi$. The integration measure $\mathcal{D}\phi$ and the weight factor $\theta[\phi]$ are gauge invariant, in particular $\theta[\phi]=\theta[\mathscr{f}_g[\phi]]$. Therefore, the gauge fixed version of the path integral reads
\begin{equation}
	\dlangle O \drangle_g \equiv \frac{ \int \mathcal{D}\phi\ O[\mathscr{f}_g[\phi]]\ \theta[\phi] }{\int \mathcal{D}\phi\ \theta[\phi]}\ .
\end{equation}
If $O$ is gauge invariant, i.e., if $O[\phi]=O[\mathscr{f}_g[\phi]]$, then gauge fixing has obviously no effect. If, however, $O$ is not gauge invariant, then gauge fixing is mandatory to obtain a meaningful result. This will be the case in sections \ref{sec-martinelli}, \ref{sec-taxidriver} and \ref{sec-divwline}, where we will use the Landau gauge fixing condition $\partial_\muE \Afield_\muE(x) = 0$. This condition becomes $\sum_{x,\muE} \myRe\ \Tr\ U(x,x+\hat\mu) = \mathrm{min}!$ on the lattice \cite{DeGrand:2006zz}. The \toolkit{Chroma} executable has the corresponding minimization algorithm built in, and is able to convert a given gauge configuration $U$ to Landau gauge, i.e., \toolkit{Chroma} provides the mapping $\mathscr{f}_g$.\footnote{The gauge rotations $W_g[U](x)$ determined during the gauge fixing process can be used to produce gauge fixed lattice quark propagators later on.}

\subsection{The lattice as a Regularization Scheme}

Consider an oscillation of the fields with an Euclidean wave vector $k = (k_\Eu{1}, k_\Eu{2}, k_\Eu{3}, k_\Eu{4})$. To the discretized action, wave numbers $|k_\muE| > \pi/a$ are indistinguishable from oscillations with a corresponding wave number in the range $|k_\muE| \leq \pi/a$ (aliasing effect). To phrase it differently,
any mode described by the degrees of freedom of the lattice path integral can be uniquely assigned to a momentum inside the first Brillouin zone, with $-\pi/a < k_\muE \leq \pi/a$. In that sense, the lattice imposes an ultraviolet momentum cutoff $\pi/a$.
Moreover, since we are working with periodic boundary conditions in a box of size $(a\hat L)^3\times(a \hat T)$, wave numbers are multiples of $2\pi/a \hat L$ and $2\pi/a \hat T$, respectively. Thus the ``resolution'' of momenta is limited. As a whole, the lattice provides both an ultraviolet cuttoff (through the lattice spacing) and an infrared cutoff (through finite lattice volume). Consequently, all quantities are finite on the lattice. 

All quantities on the lattice are expressed in terms of dimensionless numbers $\hat \theta(a)$. The corresponding quantity in physical units is obtained according to $\theta^\lat(a) = a^{d_\theta}\, \hat \theta(a)$, where $d_\theta$ is the length dimension of the quantity. Physical observables, e.g., hadron masses,  are not sensitive to the behavior of the theory at very small length scales and converge to their continuum value: $\theta^\lat(a) \rightarrow \theta^\text{cont}$ as $a \rightarrow 0$. There are other quantities which are sensitive the ultraviolet behavior of the theory. They are only well-defined with respect to a given \terminol{renormalization scheme} and \terminol{renormalization scale} or \terminol{renormalization condition}. The renormalization scheme provided by the lattice depends on the details of the lattice action used. It is therefore desirable to translate $\theta^\lat(a)$ to a renormalization scheme in the continuum, such as $\overline{\mathrm{MS}}$. Important examples of renormalization scheme dependent quantities are the quark masses $m_\quark$ and the coupling constant $g$.


\subsection{Determining the Lattice Spacing and Setting Quark Masses}

Notice that after rescaling the fields, the lattice spacing $a$ appears nowhere explicitly in the action.
The lattice spacing is controlled by the coupling constants of the action, in particular by the lattice gauge coupling $g$, or rather $\beta$, in eq.\,(\ref{eq-latgaugeact}).  To make contact to our physical world, we calculate some dimensionful observable on the lattice for a given value of $\beta$ and compare to experimental results. This way, we can determine $a$. We may then adjust the value of $\beta$ until we are close to the desired lattice spacing. 

Since we are going to work with unphysical quark masses, the observable used to determine $a$ should be largely independent of quark masses. A common choice is the static quark potential, which is then compared to phenomenological models describing the spectrum of heavy quarkonia, such as $b \bar b$ states. Some details of this procedure will become clear in section \ref{sec-renstatqpot}, where we are going to calculate the static quark potential in order to renormalize the Wilson line.
The lattice spacing determined this way is subject to statistical uncertainties of the measurement, and, most importantly, to systematic errors inherent to the method.

The lattice quark masses $\hat m_q$ also needs to be tuned. Modern lattice actions typically incorporate the three lightest quarks $u$, $d$, and $s$ as dynamical degrees of freedom. The strange quark mass $m_s$ can be set to an approximately physical value. The light quarks $u$ and $d$ are usually chosen degenerate, $m_u = m_d \equiv m_{ud}$. The most convenient observable to specify the light quark masses is the pion mass $m_\pi$, because, unlike the ``bare'' quark masses $m_{ud}$ on the lattice, it needs no renormalization, yet is very sensitive to $m_{ud}$. In \terminol{chiral effective field theory}, the pion acquires mass only through the explicit breaking of chiral symmetry caused by nonzero quark masses, and it follows the \terminol{Gell-Mann, Oakes and Renner relation} \cite{GellMann:1968rz} $m_\pi^2 \propto m_{ud} + \mathcal{O}(m_{ud}^2)$. The proportionality constant in this relation compensates all renormalization scheme dependence of the quark masses. 

The computational effort in lattice simulations increases drastically for lower quark masses. Only very recently, advanced algorithms and machines made first attempts possible to go down to a realistic pion mass of around $140 \units{MeV}$. However, the ensembles we are going to use for our exploratory calculations feature pion masses no less than $500 \units{MeV}$. With input from such large pion masses, an extrapolation down to the physical pion mass is prone to exhibit a large unknown systematic error. It will be interesting to repeat our calculations at lower pion masses.

Heavier quarks $c$, $b$ and $t$ do not appear in the lattice action, i.e., ``dynamical effects'' of heavy quarks are neglected. Due to the large energy required to create a heavy sea quark pair, such fluctuations can only exist for a very short time, smaller than the lattice spacing. Unless observables involving heavy valence quarks are considered, we can ``integrate out'' heavy quarks. Effectively, the existence of heavy quarks merely amounts to small adjustments of the coupling constants.

%

\subsection{Operator Mixing}
\label{sec-opmix}

For simplicity, let us ignore gauge fields for the moment. Consider the local operators\footnote{In the context of path integrals, they are not operators. We will call them operators none the less - for convenience, and because of their obvious correspondence to operators in the Hamiltonian formalism.} 
\begin{align}
	O_0(a) & = \bar \quark(0)\, \quark(0) \xrightarrow{\text{rescaling}} \frac{1}{a} \bar \quark(0)\, \quark(0) \ , \\
	O_2(a) & = \sum_\muE \bar \quark(0) \left( \quark(-\hat \mu) - 2 \quark(0) + \quark(\hat \mu) \right) \xrightarrow[\text{Taylor expand}]{\text{rescaling}} a \sum_\muE \bar \quark(0) \partial_\muE \partial_\muE \quark(0) + \mathcal{O}(a^3) \ . \label{eq-wilsontermop}
\end{align}
The first operator $O_0$ looks like the mass operator in the action. Switching to dimensionful quark fields and dividing by the cell volume $a^4$ (``rescaling''), we recognize the familiar $1/a$ behavior of the mass term. The second operator looks like the Wilson term eq.\,(\ref{eq-wilsonterm}) and the ``\terminol{na\"ive continuum limit}'' obtained from a Taylor expansion of the fields reveals a relation to the Laplace operator in the continuum. The reason why the Taylor expansion is na\"ive is that there is no one-to-one correspondence between the field configurations on the lattice and continuous fields. This is different from the classical picture, where the lattice can be chosen fine enough to give a rather accurate description of the smooth continuous field. It turns out the Laplace operator appearing on the right hand side of eq.\,(\ref{eq-wilsontermop}) is only part of the continuum operator that corresponds to $O_2$. When we increase the lattice spacing the operator $O_2(a)$ on the fine lattice will have a representation in terms of $O_2(a')$ \emph{and} $O_0(a')$ (and other operators) on the coarse lattice: 
\begin{equation}
	O_2(a) = \mathcal{Z}_{20}(a,a')\,O_0(a') + \mathcal{Z}_{22}(a,a')\, O_2(a') + \ldots \ .
\end{equation}
The $\mathcal{Z}_{ij}$ are \terminol{matching factors}. \terminol{Operator mixing} happens quite generally when switching the renormalization scheme or changing the renormalization scale. The continuum representation of $O_2$ can therefore be specified in the form
\begin{equation}
	O_2(a) \rightarrow \frac{\tilde{\mathcal{Z}}_{20}(a,\mu)}{a} \bar \quark(0)\, \quark(0) + a\,\tilde{\mathcal{Z}}_{22}(a,\mu) \sum_\muE \bar \quark(0) \partial_\muE \partial_\muE \quark(0) + \ldots 
\end{equation}
and the matching factors $\tilde{\mathcal{Z}}_{ij}$ now depend on the lattice action, the lattice spacing $a$, the renormalization scheme in the continuum and the corresponding renormalization scale $\mu$. The mass-like contribution to the continuum representation of $O_2(a)$ is potentially enhanced by a factor $a^{-2}$ with respect to the second derivative term.\footnote{Indeed, for the Wilson fermion action, this means that the Wilson term eq.\,(\ref{eq-wilsonterm}) contributes significantly to the quark mass through an \emph{additive renormalization.}} 

An operator can mix with any other operator that has the same symmetry transformation properties under the symmetries of the action. An important symmetry of the Euclidean continuum action that restricts the number of operators that can mix is the $\mathrm{O}(4)$ rotational invariance. The lattice is not rotationally invariant, but there is a remnant of the $O(4)$ symmetry, the hypercubic group $\mathrm{H}(4)$: The action is invariant under permutation of the axes and under reflections:
\begin{equation}
	(x_\Eu{1},x_\Eu{2},x_\Eu{3},x_\Eu{4}) \rightarrow (\pm x_{\muE_1},\pm x_{\muE_2},\pm x_{\muE_3},\pm x_{\muE_4}) \ .
\end{equation}
Here $(\muE_1,\muE_2,\muE_3,\muE_4)$ is a permutation of $(\Eu{1},\Eu{2},\Eu{3},\Eu{4})$. Since $H(4)$ is less restrictive than $O(4)$, mixing patterns on the lattice are often much more complicated than in the continuum, see, e.g., Ref. \cite{Gockeler:1996mu,Gockeler:2004xba}.

\subsection{Action Improvement}

We have a lot of freedom in setting up the discretized action as an approximation to the continuum action. For example, loops of link variables other than the plaquette also yield the field strength tensor in the continuum limit. The systematic method to design optimized actions is called \terminol{Symanzik improvement}  \cite{Symanzik:1983gh,Symanzik:1983dc}. In essence, the design principle works as follows: All local lattice operators invariant under the desired symmetries of our action can possibly be part of the action. From these, one selects an appropriate finite subset of operators $O_i^\text{lat}$. The improved action is a linear combination of these operators, with coefficients chosen in such a way that spurious operators in the corresponding ``effective'' continuum action cancel up to a certain order in $a$.
However, as we have seen in the previous section, the coefficients appearing in the continuum action cannot be determined from a Taylor expansion of the lattice operators. Rather, lattice perturbation theory or non-perturbative methods must be used.
For more information on action improvement, see, e.g., Ref. \cite{DeGrand:2006zz}.

\subsection{Link Smearing: HYP Blocking}
\label{sec-HYPsmear}

The gauge configurations determined from the Monte Carlo sampling contain a lot of noise, in the form of short distance, high momentum fluctuations. Since we are usually interested in the long distance behavior of the theory, it can be useful to suppress this noise. This can be done by one or several \terminol{smearing} steps. Smeared gauge links, also termed \terminol{fat links}, can be used during the production of gauge configurations as an ingredient to an improved fermion action, or can be used in subsequent steps of the analysis.
The smearing technology applied in this work is a single step of ``\terminol{hypercubic blocking}'', or \terminol{HYP smearing} \cite{Hasenfratz:2001hp}, which is implemented ready to use in the \toolkit{Chroma} library. The fundamental operation of HYP smearing is an \terminol{APE} smearing step \cite{Albanese:1987ds}, where each smeared link variable is formed from the unsmeared link variable and an admixture of staple-shaped gauge links:
\begin{align}
	U_\text{APE}(x,x+\hat \mu) & = \mathrm{Proj}_{\mathrm{SU}(3)} \Big( (1-\alpha) U(x,x+\hat \mu) \nonumber \\
	& + \frac{\alpha}{6} \sum_{\pm \nu \neq \mu} U(x,x+\hat \nu) U(x+\hat \nu,x+\hat \nu+\hat \mu) U(x+\hat \nu+\hat \mu, x+\hat \mu) \Big) \ .
\end{align}
The projection to $\mathrm{SU}(3)$ ensures that the new link variables are again unitary. For HYP blocking, three iterations of APE-like steps are carried out in such a way that the final fat link variable receives contributions from links no farther away than the edges of the surrounding hypercubes.
This way, the smearing operation remains local. Long range properties (universality class of the theory, spectrum, etc.) remain unaffected. 
A disadvantage is the influence on short distance behavior, e.g., of the static quark potential \cite{DeGrand:2006zz}, compare section \ref{sec-renstatqpot} and Refs.~\cite{Albanese:1987ds,Hasenfratz:2001hp}. The smearing strength of HYP blocking is controlled by three parameters. The ones used by LHPC and throughout this work are
$\alpha_1 = 0.75$, $\alpha_2 = 0.6$ and $\alpha_3 = 0.3$.

\section{Nucleon Matrix Elements}

This section will describe the method used to calculate properties of the nucleon on the lattice. 
Since we have used the \toolkit{Chroma} programming library \cite{Edwards:2004sx} and numerical input from the \terminol{Lattice Hadron Physics Collaboration (LHPC)}, we have taken over their conventions and techniques.


\subsection{Baryon Sources and Sinks}

We need to create nucleons on the lattice, although we do not know the precise wave function of the nucleon. The first step is to define a suitable ``interpolating field'', a gauge invariant combination of quark and gluon fields with the same quantum numbers as the nucleon. We shall use 
\begin{equation}
	B_\alpha(t,\vect{P}) \equiv
	\frac{1}{\sqrt{\hat L^3}}\sum_{\vect{x}} e^{-i \vect{P} \cdot \vect{x}} \
	\epsilon_{a b c} u_{a \alpha} (\vect{x}, t) \
	\left( u_b^T(\vect{x},t)\ \GammaDiq\ d_c(\vect{x}, t) \right)
	\label{eq-nuclsource}
	\end{equation}
as a \terminol{nucleon sink} on the lattice. 
Here we have made the Dirac index $\alpha$ and the color indices $a,b,c$ explicit. The nucleon source is located on a time slice $t$ in Euclidean space and has three-momentum $\vect{P}$.
Rather than making the obvious choice $\GammaDiq:=C\gamma_\Eu{5}=\gamma_\Eu{4}\gamma_\Eu{2}\gamma_\Eu{5}$, LHPC uses $\GammaDiq:=C\gamma_\Eu{5}(\Eins + \gamma_\Eu{4})$.
The non-relativistic projection can reduce the number of Dirac indices that need to be processed. The adjoint expression $\bar B_\alpha(t,\vect{P})$ serves as a \terminol{nucleon source}. The quark fields $\quark(\vect{x},t)$ appearing in eq.\,(\ref{eq-nuclsource}) are replaced by ``extended quark fields'', a gauge covariant superposition of quark fields located in the vicinity of $(\vect{x},t)$. This technique of \terminol{source and sink smearing} increases the overlap of the nucleon interpolating field with the nucleon wave function. For details, see Refs.~\cite{Gusken:1989qx,Alexandrou:1990dq,Allton:1993wc,Dolgov:2002zm}. For later convenience, we write source and sink as
\begin{align}
	\overline{B}_\alpha(t,\vect{P}) & =
	\frac{1}{\sqrt{\hat L^3}}\sum_{\vect{x},\MI{1},\MI{2},\MI{3}} e^{i \vect{P} \cdot \vect{x}} \
	\epsilon_{\MI{1}\MI{2}\MI{3}} \ 
	\GammaDiq_{\MI{3}\MI{2}}\ \overline{u}_{\MI{1}} \
	\overline{d}_{\MI{2}}\ \overline{u}_{\MI{3}} \quad
	\left\vert \begin{array}{l}
		\scriptstyle \MI{1},\MI{2},\MI{3}\text{ at }(\vect{x},t) \\ \scriptstyle \text{Dirac}\MI{1} = \alpha 
	\end{array} \right. \ , \nonumber \\
	B_\alpha(t,\vect{P}) & =
	\frac{1}{\sqrt{\hat L^3}} \sum_{\vect{x},\MI{4},\MI{5},\MI{6}} e^{- i \vect{P} \cdot \vect{x}} \
	\epsilon_{\MI{4}\MI{5}\MI{6}} \ 
	\GammaDiq_{\MI{5}\MI{6}}\ u_{\MI{4}} \
	u_{\MI{5}}\ d_{\MI{6}} \quad
	\left\vert \begin{array}{l}
		\scriptstyle \MI{4},\MI{5},\MI{6}\text{ at }(\vect{x},t) \\ \scriptstyle \text{Dirac}\MI{4} = \alpha 
	\end{array} \right. \ ,
	\label{eq-nuclsourcesink}
\end{align}
where it is understood that $\epsilon_{\MI{1}\MI{2}\MI{3}}$ acts only in color space and $\GammaDiq_{\MI{3}\MI{2}}$ only in the space of Dirac indices.

\subsection{The Nucleon Two-Point Function}
\label{sec-nucltwopt}

Two-point correlation functions on the lattice are used for the calculation of hadron masses and will serve us as normalization factors in the computation of nucleon matrix elements. The nucleon two-point correlation function is introduced as
\begin{equation}
	\latcfn^\text{2pt}(t,\vect{P}) = \sum_{\beta\alpha} \GammaTwo_{\beta\alpha}\, \dlangle B_\alpha(t,\vect{P})\, \overline{B}_\beta(0,\vect{P}) \drangle \ ,
	\label{eq-nucltwopoint}
\end{equation}
where the Dirac matrix $\GammaTwo = (\Eins + \gamma_\Eu{4})(\Eins + i \gamma_\Eu{5} \gamma_\Eu{3})/2$ used by LHPC is again a non-relativistic projection, combined with a spin projection. The latter is not needed in principle, but is statistically advantageous in combination with three-point functions. Inserting the source and sink, we get
\begin{align}
	\latcfn^\text{2pt}(t,\vect{P}) & = 
	\sum_{\vect{x},\MI{1},\ldots,\MI{6}} 
	e^{- i \vect{P} \cdot \vect{y}} \
	\epsilon_{\MI{4}\MI{5}\MI{6}}\ 
	\GammaDiq_{\MI{5}\MI{6}}\ 
	\GammaTwo_{\MI{1}\MI{4}}\
	\epsilon_{\MI{1}\MI{2}\MI{3}}\ 
	\GammaDiq_{\MI{3}\MI{2}} 
	\nonumber \\ & \times
	\dlangle u_{\MI{4}} \	u_{\MI{5}}\ d_{\MI{6}}\  
	\overline{u}_{\MI{1}} \	\overline{d}_{\MI{2}}\ \overline{u}_{\MI{3}} \drangle
	\quad
	\left\vert \begin{array}{l}
		\scriptstyle \MI{1},\MI{2},\MI{3}\text{ at }(\vect{x}_\text{src},t_\text{src}) \\ 
		\scriptstyle \MI{4},\MI{5},\MI{6}\text{ at }(\vect{y},t+t_\text{src})
	\end{array}\right. \ ,
\end{align}
where we have already exploited the translation invariance of the expectation value. In the second line of the above expression, we form Wick contractions in order to integrate out the fermions:
\begin{equation}
	\dlangle
		\contraction{}{u_{\MI{4}}}{}{\overline{u}_{\MI{1}}}  u_{\MI{4}}\overline{u}_{\MI{1}}\
		\contraction{}{u_{\MI{5}}}{}{\overline{u}_{\MI{3}}}  u_{\MI{5}}\overline{u}_{\MI{3}}\
		\contraction{}{d_{\MI{6}}}{}{\overline{d}_{\MI{2}}}  d_{\MI{6}}\overline{d}_{\MI{2}}\
		-
		\contraction{}{u_{\MI{4}}}{}{\overline{u}_{\MI{3}}}  u_{\MI{4}}\overline{u}_{\MI{3}}\
		\contraction{}{u_{\MI{5}}}{}{\overline{u}_{\MI{1}}}  u_{\MI{5}}\overline{u}_{\MI{1}}\
		\contraction{}{d_{\MI{6}}}{}{\overline{d}_{\MI{2}}}  d_{\MI{6}}\overline{d}_{\MI{2}}
	\drangle
	\quad
		\left\vert \begin{array}{l}
		\scriptstyle \MI{1},\MI{2},\MI{3}\text{ at }(\vect{x}_\text{src},t_\text{src}) \\ 
		\scriptstyle \MI{4},\MI{5},\MI{6}\text{ at }(\vect{y},t+t_\text{src})
	\end{array}\right. \ .
	\label{eq-nucltwoptcontr}
\end{equation}

\subsection{Point-To-All Propagators}
\label{sec-pttoall}

Each of the contracted pairs in eq.\,(\ref{eq-nucltwoptcontr}) becomes a propagator, see section \ref{sec-fermionintegration}. Since we have $\hat m_u = \hat m_d$, $u$- and $d$-quark propagators are identical. Moreover, notice that all propagators are attached to the source location at $x_\text{src}=(\vect{x}_\text{src},t_\text{src})$. Therefore, it is sufficient to prepare a single type of point-to-all propagator from the inversion
\begin{equation}
	 K^{(u,d)}_{\MI{1}\MI{2}}[U]\,G^{(u,d)}_{\MI{2}\MI{3}}[U] = \delta_{\MI{1}\MI{3}} \qquad \left\vert_{\scriptstyle \MI{3} \text{ at }x_\text{src}} \right. \ .
	\label{eq-propinversion}
\end{equation}
The restriction to a single source location is crucial for the computational feasibility. This way, we only need to perform  $\sim N_c^2 \times 4^2$ inversions of type $\langle\text{sparse matrix}\rangle\times\langle\text{vector}\rangle = \langle\text{vector}\rangle$ for each lattice configuration. The non-relativistic projection reduces the number of inversions further.

The inversion above produces \terminol{forward propagators} $G_{\MI{2}\MI{3}} = \contraction{}{\quark_{\MI{2}}}{}{\overline{\quark}_{\MI{3}}}  \quark_{\MI{2}}\overline{\quark}_{\MI{3}}$, where the second index $\MI{3}$ is fixed to the source location. We can use the relation (based on the so called ``$\gamma_\Eu{5}$-hermiticity'')
\begin{equation}
	\contraction{}{\quark_{\MI{3}}}{}{\overline{\quark}_{\MI{2}}}  \quark_{\MI{3}}\overline{\quark}_{\MI{2}} = G_{\MI{3}\MI{2}} = (\gamma_\Eu{5} G \gamma_\Eu{5})^*_{\MI{2}\MI{3}} 
	\label{eq-backwprop}
\end{equation}
to obtain a \terminol{backward propagator}, which has the first argument fixed at the source location. This relationship works for Wilson-type fermions, including Domain Wall fermions, and can save the large computational costs of further inversions.

\lstset{language=C++,basicstyle=\ttfamily\small,frame=l,basewidth=0.5em,linewidth=0.98\textwidth,xleftmargin=0.02\textwidth}
Once the propagators are prepared, the nucleon two-point function can be evaluated in \toolkit{Chroma} very efficiently, as illustrated in the following code snippet\footnote{Similar code can be found in \lstinline{Chroma::Baryon2PtContractions::sigma2pt}.}
\begin{lstlisting}
LatticePropagator diquark = 
       quarkContract13( prop * BaryonSpinMats::Cg5NR(), 
                        BaryonSpinMats::Cg5NR() * prop );
return trace( BaryonSpinMats::Tmixed() * traceColor( prop * diquark ) ) +
       traceColor( traceSpin( BaryonSpinMats::Tmixed() * prop ) * 
                   traceSpin( diquark ) );
\end{lstlisting}
After projecting out the desired momentum $\vect{P}$, this yields $\latcfn^\text{2pt}(t,\vect{P})$ for all $t$.

\subsection{Transfer Matrix Formalism}
\label{sec-transfmat}
In order to be able to interpret the correlation functions, we switch to the transfer matrix formalism \cite{Luscher:1976ms}. In section \ref{sec-pathintform} we claimed that a lattice expectation value $\dlangle O \drangle$ is related to the vacuum expectation value of $\hat O$. On a periodic lattice of finite extent $T$ in Euclidean time, this is not precisely true. Rather, we obtain a trace. Specifically, for the nucleon two-point function, we obtain if $0 \leq t < T$:
\begin{align}
	\dlangle B_\alpha(t,\vect{P})\, \overline{B}_\beta(0,\vect{P}) \drangle & = 
	\sum_n \bra{n}\ \TFM^{T-t}\ \hat B_\alpha(0,\vect{P})\ \TFM^{t} \ \hat{\overline{B}}_\beta(0,\vect{P})\ \ket{n} \nonumber \\
	& \equiv \BTr\ \TFM^{T-t}\ \hat{B}_\alpha(0,\vect{P})\ \TFM^{t} \ \hat{\overline{B}}_\beta(0,\vect{P})\ .
	\label{eq-twopointtrans}
\end{align}
Here $\lbrace\ket{n}\rbrace$ is a complete set of states which are normalized to unity. The \terminol{transfer matrix} $\TFM$ can be formally written in terms of the Hamilton operator $\hat H$. If the states $\ket{n}$ are eigenstates of the Hamiltonian with energy eigenvalue $E_n$, we may write
\begin{equation}
	\TFM^t \equiv  \exp\left(-\hat H t \right) = \sum_n \ket{n} e^{-E_n t} \bra{n} 
\end{equation}
and substitute the expression on the right at each occurrence of $\TFM$. Suppose now that $T-t \gg t$ is large, such that we can substitute $\TFM^{T-t} \approx \ket{\Omega}\bra{\Omega}$ in eq.~(\ref{eq-twopointtrans}). If $t$ is also large, such that excited states are suppressed, the propagating state between the nucleon interpolating operators must be the nucleon  $\ket{N(P,S)}$. Lighter hadrons or the vacuum are not allowed due to the quantum numbers of the interpolating operators. So we get
\begin{equation}
	\dlangle B_\alpha(t,\vect{P})\, \overline{B}_\beta(0,\vect{P}) \drangle  \approx \sum_S
	\bra{\Omega} \hat{B}_\alpha(0,\vect{P})\ket{N(P,S)}\ \frac{ e^{-E_P t} }{2 E_P} \ \bra{N(P,S)}\hat{\overline{B}}_\beta(0,\vect{P})\ \ket{\Omega}\ .
\end{equation}
The nucleon energy in the denominator appears due to the Lorentz covariant normalization of the nucleon state, $\braKet{N(P,S)}{N(P',S')}{} = 2 E_P\, \delta_{\vect{P}\vect{P}'}\, \delta_{SS'}$. Introducing \terminol{overlaps}
\begin{align}
	\bra{\Omega} \hat{B}_\beta(0,\vect{P}) \ket{N(P,S)} & = \sqrt{Z_N(P)}\, U_\alpha(P,S) \ ,\nonumber \\
	\bra{N(P,S)} \hat{\overline{B}}_\alpha(0,\vect{P}) \ket{\Omega} & = \sqrt{Z_N(P)^*}\, \bar{U}_\alpha(P,S)
\end{align}
the nucleon two-point function eq.\,(\ref{eq-nucltwopoint}) finally reads
\begin{align}
	\latcfn^\text{2pt}(t,\vect{P}) & = \frac{|Z_N(P)|}{2 E_P} e^{-E_P t}\ \Tr_\mathrm{D}\left[\GammaTwo\, \sum_S U(P,S)\, \bar{U}(P,S)\right]\nonumber \\ 
	& = |Z_N(P)|\, e^{-E_P t}\, \frac{\Tr_\mathrm{D} \left\lbrace\GammaTwo\, (-i \slashedE{P} + m_N)  \right\rbrace}{2E_P}
	\ \mathop{=}_{\text{LHPC}}\  |Z_N(P)|\, e^{-E_P t}\, \frac{m_N + E_P}{E_P}\ .
\end{align}
To be precise, we should mention that there is another state that can contribute: the antiparticle of the nucleon's parity partner. This state comes with the overlaps 
\begin{align}
	\bra{\Omega} \hat{B}_\beta(0,\vect{P}) \ket{\bar N'(P,S)} & = \sqrt{Z_{N'}(P)}\, V_\alpha(P,S) \ ,\nonumber \\
	\bra{\bar N'(P,S)} \hat{\overline{B}}_\alpha(0,\vect{P}) \ket{\Omega} & = \sqrt{Z_{N'}(P)^*}\, \bar{V}_\alpha(P,S)	
\end{align}
and thus yields a contribution proportional to $\Tr_\mathrm{D}[\GammaTwo\, \sum_S V(P,S)\, \bar{V}(P,S)]$ in the two-point function. Through our specific choice of $\GammaTwo$, this contribution will be largely suppressed, in particular for small momenta $\vect{P}$. We only see the effect of the nucleon parity partner in the two-point correlator when we approach $t_\text{src}$ from the ``wrong'' side, $t < t_\text{src}$, see Fig.~\ref{fig-nucltwopoint}.

Now, from the slope of  $\ln \latcfn^\text{2pt}(t,\vect{P}) = -E_P t + \text{const}$, we can immediately read off the energy $E_P$ of the nucleon state. For the special case $\vect{P}=0$, we obtain the nucleon mass $m_N$.

\subsection{The Nucleon Three-Point Function}
\label{sec-nuclthreept}

\begin{figure}[tbp]
	\centering
	\includegraphics[scale=1]{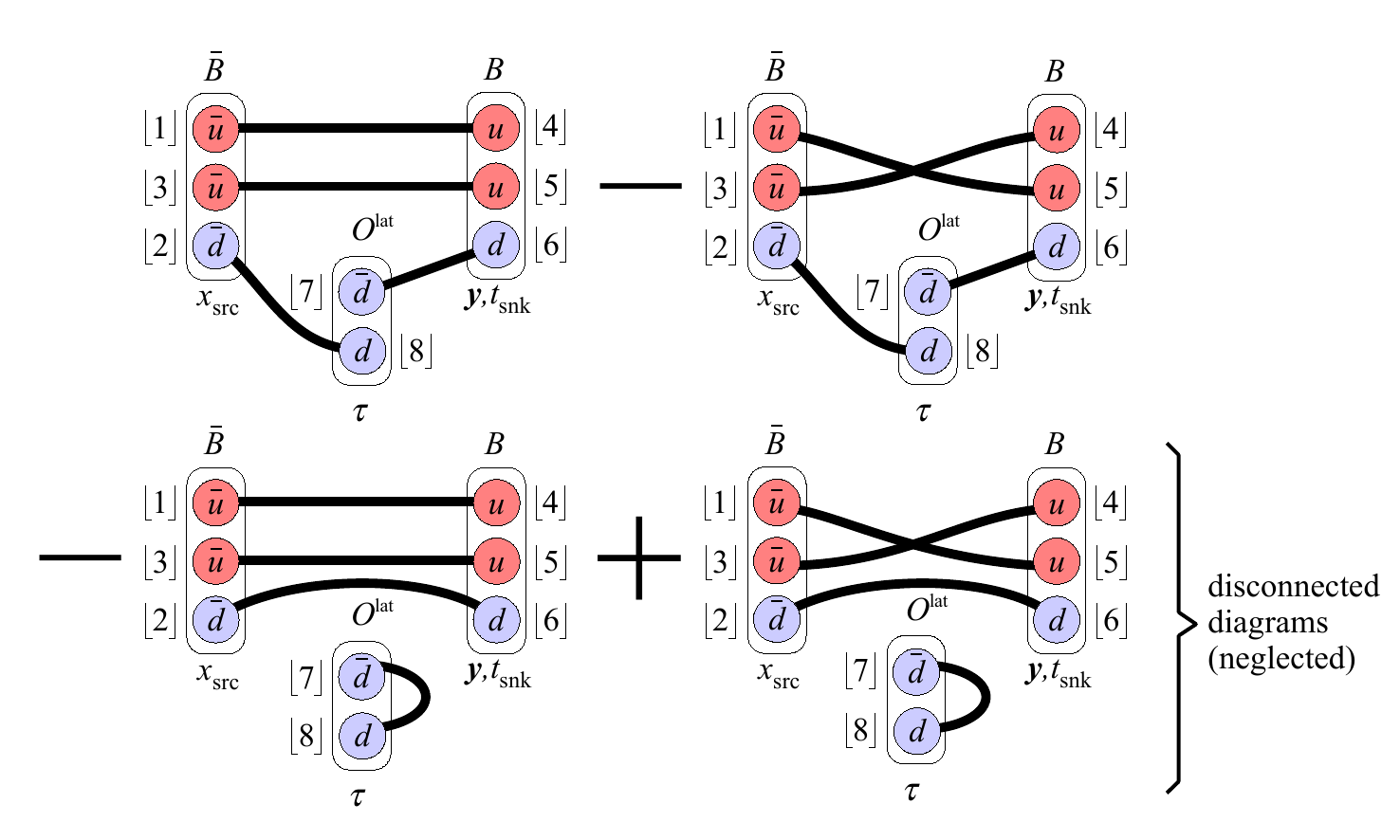}
	\caption[TMDs]{%
		Illustration of diagrams contributing to the nucleon three-point function on the lattice, here for an operator probing $d$-quarks. Each tick line connecting quark fields represents a lattice propagator. The lower two diagrams are neglected in our calculations.
		\label{fig-3ptcontrd}
		}
	\end{figure}

We want to be able to calculate nucleon matrix elements as they appear in eq.\,(\ref{eq-corrssprime}) on the lattice \cite{Martinelli:1988rr,Wilcox:1991cq}. The ``operators'' we encounter are of the form
\begin{equation}
	O_{\GammaOp,\quark}(\mathcal{C}) \ \equiv\ \bar \quark(\elll[\mathcal{C}])\, \GammaOp\ \WlineC{\mathcal{C}}\ \quark(0) \ ,
\end{equation}
where $\elll[\mathcal{C}] \equiv \mathcal{C}(1)-\mathcal{C}(0)$. The quark type $\quark$ can be $u$ or $d$. We use a lattice field combination of the form
\renewcommand{\arraystretch}{0.75}
\begin{align}
	O_{\GammaOp,\quark}^\lat(\mathcal{C}^\lat;z) \ & \equiv\ \bar \quark(\elll[\mathcal{C}^\lat]+z)\, \GammaOp\ \WlineClat{\mathcal{C}^\lat+z}\ \quark(z) \nonumber \\
	& =\ \GammaOp_{\MI{7}\MI{8}}\ \WlineClat{\mathcal{C}^\lat+z}_{\MI{7}\MI{8}}\ \bar \quark_{\MI{7}} \quark_{\MI{8}}
	\quad\left\vert \begin{array}{l} \scriptstyle \MI{7}\text{ at }\elll[\mathcal{C}^\lat] + z \\ \scriptstyle \MI{8}\text{ at }z \end{array} \right.
	\label{eq-olat}
\end{align}
to create analogous operators on the lattice. We have introduced an offset $z$, which leaves the matrix element in eq.~(\ref{eq-corr}) invariant. Here $\WlineClat{\mathcal{C}^\lat+z}$ is a product of connected link variables starting at position $z$ and ending at position $\elll[\mathcal{C}^\lat]+z$, following the convention eq.\,(\ref{eq-lat-gaugelink}). Contracting with the color indices of the two quark fields, it renders $O_{\GammaOp,\quark}^\lat$ gauge invariant. The choice of the link path $\mathcal{C}^\lat$ will be discussed later. However, let us demand from the start that the link path remains on a single time slice, or in other words, that it does not contain any link variables in $\Eu{4}$-direction. We now consider the three-point function
\begin{equation}
	\latcfn^\text{3pt}_{\GammaOp,\quark}(\tau,\vect{P};\mathcal{C}^\lat) = \frac{1}{\hat L^3} \sum_{\vect{z}} \sum_{\beta\alpha} \GammaThr_{\beta\alpha}\, \dlangle B_\alpha(t_\text{snk},\vect{P})\ O_{\GammaOp,\quark}^\lat(\mathcal{C}^\lat;\vect{z},\tau)\ \overline{B}_\beta(t_\text{src},\vect{P}) \drangle\ .
	\label{eq-nuclthreepoint}
\end{equation}
Let us calculate the three-point function for $d$-quarks in the operator. Substituting source, sink and operator, and exploiting translation invariance, we get
\begin{align}
	\latcfn^\text{3pt}_{\GammaOp,d}(\tau,\vect{P};\mathcal{C}^\lat) & = 
	\sum_{\vect{y},\MI{1},\ldots,\MI{8}} 
	e^{- i \vect{P} \cdot \vect{y}} \
	\epsilon_{\MI{4}\MI{5}\MI{6}}\ 
	\GammaDiq_{\MI{5}\MI{6}}\ 
	\GammaThr_{\MI{1}\MI{4}}\
	\epsilon_{\MI{1}\MI{2}\MI{3}}\ 
	\GammaDiq_{\MI{3}\MI{2}}\  
	\nonumber \\ & \times
	\frac{1}{\hat L^3}\sum_{\vect{z}}
	\GammaOp_{\MI{7}\MI{8}}\ 
	\WlineClat{\mathcal{C}^\lat+(\vect{z},\tau)}_{\MI{7}\MI{8}}
	\nonumber \\ & \times
	\dlangle u_{\MI{4}} \	u_{\MI{5}}\ d_{\MI{6}}
	\quad \bar d_{\MI{7}}\ d_{\MI{8}}\quad
	\overline{u}_{\MI{1}} \	\overline{d}_{\MI{2}}\ \overline{u}_{\MI{3}} \drangle
	\quad
	\left\vert \begin{array}{l}
		\scriptstyle \MI{1},\MI{2},\MI{3}\text{ at }x_\text{src} \\ 
		\scriptstyle \MI{4},\MI{5},\MI{6}\text{ at }(\vect{y},t_\text{snk}) \\
		\scriptstyle \MI{7}\text{ at }(\vect{z},\tau),\ \MI{8}\text{ at }({\vect{\elll}[\mathcal{C}^\lat]+\vect{z},\tau})
	\end{array}\right. \ .
\end{align}
There are four possible contractions, as depicted in Fig.~\ref{fig-3ptcontrd}. Two of the contractions are \terminol{disconnected}, i.e., the quark propagators connect the operator with itself rather than with the nucleon. The calculation of disconnected contributions is computationally challenging due to a bad signal to noise ratio, and very expensive, because it requires all-to-all propagators. Meeting these challenges is a hot topic of lattice QCD today. However, throughout this work, we shall ignore disconnected diagrams, assuming that their effect is small. Note that disconnected contributions cancel in \terminol{isovector} quantities, i.e., when we take the linear combination of operators $O_{\GammaOp,u-d}^\lat \equiv O_{\GammaOp,u}^\lat - O_{\GammaOp,d}^\lat$. 

\subsection{Sequential Propagators}
\label{sec-seqprop}

\begin{figure}[tbp]
	\centering
	\includegraphics[scale=1.1]{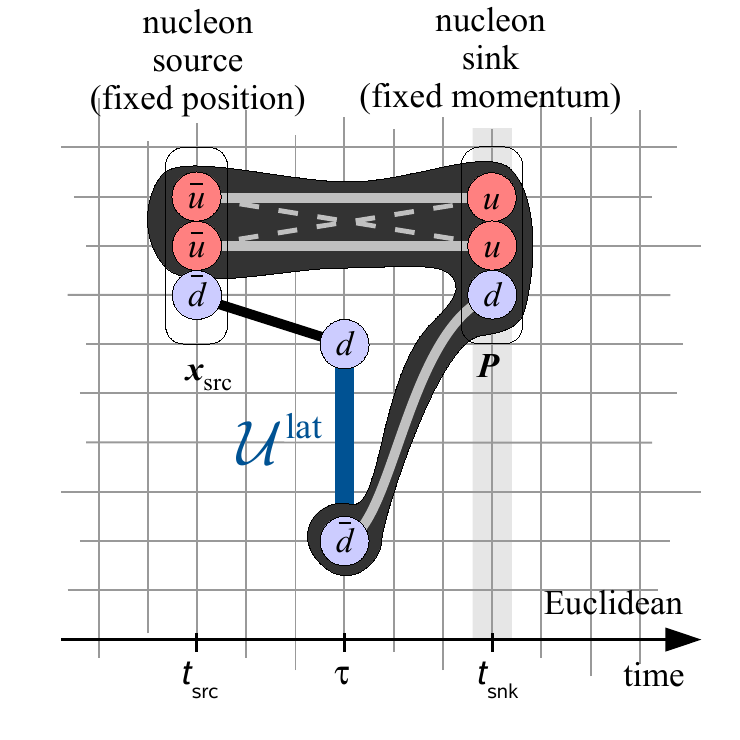}
	\caption[lattice propagator product]{%
		Calculation scheme for nucleon three-point functions on the lattice, here for an operator probing $d$-quarks. Three lattice quark propagators of each of the two contributing diagrams in Fig.~\ref{fig-3ptcontrd} are combined into a single sequential propagator, displayed as the dark blob. To evaluate the three-point function, we calculate the product of the point-to-point propagator between the $d$-quarks, the sequential propagator (for a fixed nucleon three-momentum $\vect{P}$), and a product of link variables from the underlying configuration $\mathcal{U}^\lat$.
		\label{fig-latticepropproduct}
		}
	\end{figure}

Already for the connected diagrams, it looks as though we need an all-to-all propagator: The quark propagator
$\contraction{}{d_{\MI{6}}}{}{\overline{d}_{\MI{7}}}  d_{\MI{6}}\overline{d}_{\MI{7}}$ has neither of its indices $\MI{6}$ and $\MI{7}$ attached to a fixed lattice site. However, in this particular case, there is a trick to avoid all-to-all propagators \cite{Wilcox:1991cq}, see also the appendix of Ref. \cite{Dolgov:2002zm}. The two connected contributions can be brought into the form 
\begin{align}
	\latcfn^\text{3pt}_{\GammaOp,d}(\tau,\vect{P};\mathcal{C}^\lat) & = 
	\sum_{\MI{2},\MI{6},\MI{7},\MI{8}} 
	\frac{1}{\hat L^3}\sum_z
	\GammaOp_{\MI{7}\MI{8}}\ 
	\WlineClat{\mathcal{C}^\lat+(\vect{z},\tau)}_{\MI{7}\MI{8}}
	\nonumber \\ & \times
	\dlangle F_d(x_\text{src},\vect{P},t_\text{snk})_{{\MI{2}}{\MI{6}}}\ 
	\contraction{}{d_{\MI{6}}}{}{\overline{d}_{\MI{7}}}  d_{\MI{6}}\overline{d}_{\MI{7}}\ 
	\contraction{}{d_{\MI{8}}}{}{\overline{d}_{\MI{2}}}  d_{\MI{8}}\overline{d}_{\MI{2}}
	\drangle
	\quad
	\left\vert \begin{array}{l}
		\scriptstyle \MI{2}\text{ at }x_\text{src} \\ 
		\scriptstyle \MI{7}\text{ at }(\vect{z},\tau),\ \MI{8}\text{ at }({\vect{\elll}[\mathcal{C}^\lat]+\vect{z},\tau})
	\end{array}\right.,
\end{align}
\begin{align}
	F_d(x_\text{src},\vect{P},t_\text{snk})_{{\MI{2}}{\MI{6}}} & \equiv 	\sum_{\MI{1},\MI{3},\MI{4},\MI{5}} 
	e^{- i \vect{P} \cdot \vect{y}} \
	\epsilon_{\MI{4}\MI{5}\MI{6}}\ 
	\GammaDiq_{\MI{5}\MI{6}}\ 
	\GammaThr_{\MI{1}\MI{4}}\
	\epsilon_{\MI{1}\MI{2}\MI{3}}\ 
	\GammaDiq_{\MI{3}\MI{2}}\  
	\nonumber \\ & \times
	\dlangle
	\contraction[2ex]{}{u_{\MI{4}}}{\ u_{\MI{5}}\ }{\overline{u}_{\MI{1}}}
	\contraction{u_{\MI{4}}\ }{u_{\MI{5}}}{\ \overline{u}_{\MI{1}}\ }{\overline{u}_{\MI{3}}}
	u_{\MI{4}} \ u_{\MI{5}}\ \overline{u}_{\MI{1}} \ \overline{u}_{\MI{3}}
	-
	\contraction[2ex]{}{u_{\MI{4}}}{\ u_{\MI{5}}\ \overline{u}_{\MI{1}}\ }{\overline{u}_{\MI{3}}}
	\contraction{u_{\MI{4}}\ }{u_{\MI{5}}}{\ }{\overline{u}_{\MI{1}}}
	u_{\MI{4}} \ u_{\MI{5}}\ \overline{u}_{\MI{1}} \ \overline{u}_{\MI{3}}
	\drangle
	\quad
	\left\vert \begin{array}{l}
		\scriptstyle \MI{1},\MI{2},\MI{3}\text{ at }x_\text{src} \\ 
		\scriptstyle \MI{4},\MI{5},\MI{6}\text{ at }(\vect{y},t_\text{snk}) 
	\end{array}\right..
	\label{eq-seqsourced}
	\end{align}
We now introduce the \terminol{sequential propagator} 
\begin{equation}
	S_d(x_\text{src},\vect{P},t_\text{snk})_{\MI{2}\MI{7}} \equiv \sum_{\MI{6}} F_d(x_\text{src},\vect{P},t_\text{snk})_{{\MI{2}}{\MI{6}}}\ 
	\contraction{}{d_{\MI{6}}}{}{\overline{d}_{\MI{7}}}  d_{\MI{6}}\overline{d}_{\MI{7}} 	
	\quad \left\vert_{\scriptstyle \MI{2}\text{ at }x_\text{src}}\right.\ .
\end{equation}
Multiplying the fermion matrix $K^{(d)}_{\MI{7}\MI{9}}$ from the right, we obtain
\begin{equation}
	\sum_{\MI{7}} S_d(x_\text{src},\vect{P},t_\text{snk})_{\MI{2}\MI{7}}\ K^{(d)}_{\MI{7}\MI{9}} = F_d(x_\text{src},\vect{P},t_\text{snk})_{{\MI{2}}{\MI{9}}}\ 	
	\quad \left\vert_{\scriptstyle \MI{2}\text{ at }x_\text{src}}\right.\ .
	\label{eq-seqpropinversion}
\end{equation}
Obviously, the sequential propagator can be computed from a matrix-vector inversion, similar as the point-to-all propagator is calculated in eq.\,(\ref{eq-propinversion}), just with the \terminol{sequential source} $F_d(x_\text{src},\vect{P},t_\text{snk})$ on the right hand side instead of a Kronecker-Delta. Sequential propagators for operators $O_{\GammaOp,u}^\lat$ with $u$ quarks can be set up in an analogous way, only the expression for the sequential source eq.\,(\ref{eq-seqsourced}) differs.
The three-point function now simply reads
\begin{align}
	\latcfn^\text{3pt}_{\GammaOp,\quark}(\tau,\vect{P};\mathcal{C}^\lat) & = 
	\sum_{\MI{2},\MI{7},\MI{8}} 
	\sum_{\vect{z}}
	\GammaOp_{\MI{7}\MI{8}}\ 
	\WlineClat{\mathcal{C}^\lat+(\vect{z},\tau)}_{\MI{7}\MI{8}}
	\nonumber \\ & \times
	\dlangle S_\quark(x_\text{src},\vect{P},t_\text{snk})_{{\MI{2}}{\MI{7}}}\ 
	\contraction{}{d_{\MI{8}}}{}{\overline{d}_{\MI{2}}}  d_{\MI{8}}\overline{d}_{\MI{2}}
	\drangle
	\quad
	\left\vert \begin{array}{l}
		\scriptstyle \MI{2}\text{ at }x_\text{src} \\ 
		\scriptstyle \MI{7}\text{ at }(\vect{z},\tau),\ \MI{8}\text{ at }({\vect{\elll}[\mathcal{C}^\lat]+\vect{z},\tau})
	\end{array}\right. \ .
\end{align}
Notice that the sequential propagator appears like a backward propagator in the expression above, with the \emph{first} argument fixed at the source location. It is common practice to store sequential propagators in the format of forward propagators, so we must apply eq.\,(\ref{eq-backwprop}) before we can use them for our purposes:
\begin{lstlisting}
int G5 = Ns*Ns-1;
LatticePropagator GF = Gamma(opGamma) * fwprop;
LatticePropagator seqpropbackw = Gamma(G5)*seqprop*Gamma(G5);
return localInnerProduct( seqpropbackw, GF );
\end{lstlisting}
The $\toolkit{Chroma}$ code above\footnote{Comparable code can be found in the file \lstinline{lib/meas/hadron/BuildingBlocks_w.cc} of \toolkit{Chroma}.} calculates the three-point function (up to the sum over $z$) for a gauge path of zero length, using a forward propagator \lstinline{fwprop} and a sequential propagator \lstinline{seqprop} loaded from file in advance. The integer value \lstinline{opGamma} allows us to select any Dirac matrix $\GammaOp$ listed in Table \ref{tab-ratios}.

The sequential source technique described above enables us to calculate nucleon matrix elements for any operator insertion $O_{\GammaOp,u/d}^\lat$. Note, however, that separate sequential propagators need to be calculated for each of the quark flavors $u$ and $d$, for each source-sink separation in Euclidean time, for each spin projection $\GammaThr$, and, most importantly, for each nucleon sink momentum $\vect{P}$.

\subsection{Assembling Gauge Paths}

The path $\mathcal{C}^\lat$ appearing in eq.\,(\ref{eq-olat}) is a sequence of adjacent lattice sites starting at the origin. As a compact notation, we specify link paths by a sequence of ``moves'' -- integers in the range $\lbrace-3\ldots-1,1\ldots3\rbrace$. A positive number $\mu$ represents a shift $\hat \mu$, a negative number $-\mu$ describes a shift $-\hat\mu$ in the opposite direction. The paths that our program \toolkit{trans3pt} processes can be specified in an \toolkit{XML} control file. For the assembly of link paths, we used the algorithm in \lstinline{lib/meas/hadron/BuildingBlocks_w.cc}. In essence, the propagator \lstinline{fwprop} in the code snippet above is replaced by a propagator \lstinline{shft_fwprop}, which is initialized with a forward propagator and then iteratively accumulates link paths and shift operations through
\begin{lstlisting}
int mu = spacedir(linkdir,j_decay); // "mu = |linkdir|"
if (linkdir > 0) { // forward shift
   tmpprop = adj( u[ mu ] ) * shft_fwprop;
   shft_fwprop = shift( tmpprop, BACKWARD, mu );
   }
else { // backward shift
   tmpprop = shift( shft_fwprop, FORWARD, mu );
   shft_fwprop = u[ mu ] * tmpprop;
   }
\end{lstlisting}
The program \toolkit{trans3pt} automatically saves some computer time if a link path is an extension of the previous link pattern.

\subsection{Matrix Elements from Ratios}
\label{sec-ratios}

After application of the transfer matrix formalism to the three-point function, the dominant contribution for $0 \ll \tau - t_\text{src} \ll t_\text{snk}-t_\text{src} \ll T$ is 
\begin {align}
	\latcfn^\text{3pt}_{\GammaOp,\quark}(\tau,\vect{P};\mathcal{C}^\lat) & \approx\
	\vert Z_N(P)\vert\ e^{-E_P(t_\text{snk} - t_\text{src})}
	\sum_{S,S'}
	\frac{\overline{U}(P,S)\ \GammaThr\ U(P,S')}{(2E_P)^2}
	\nonumber\\ 
	& \times\  
	\bra{N(P,S')}\, \frac{1}{\hat L^3}\sum_{\vect{z}} O_{\GammaOp,\quark}^\lat(\mathcal{C}^\lat;(\vect{z},0))\, \ket{N(P,S)}\ ,
	\label{eq-threepttransf}
	\end {align}
where we have again used the fact that the contribution from the parity partner of the nucleon is suppressed for our choice of $\GammaThr$. The sum over the offset $\vect{z}$ increases statistics in our computation. The unknown overlaps and the exponential can be cancelled by forming a suitable ratio with the two-point function:
\begin{multline}
	R_{\GammaOp,\quark}(\vect{P};\mathcal{C}^\lat) \equiv \plateau{
	\frac{\latcfn^\text{3pt}_{\GammaOp,\quark}(\tau,\vect{P};\mathcal{C}^\lat)}{\latcfn^\text{2pt}(t_\text{snk}-t_\text{src},\vect{P})} } = \\
	\sum_{S,S'}
	\frac{\overline{U}(P,S)\ \GammaThr\ U(P,S')}{2E_P\ \Tr_\mathrm{D} \left\lbrace\GammaTwo\, (-i \slashedE{P} + m_N)  \right\rbrace}
	\bra{N(P,S')}\, O_{\GammaOp,\quark}^\lat(\mathcal{C}^\lat;0)\, \ket{N(P,S)}\ .
	\label{eq-ratiodef}
\end{multline}
Here $\mathop{\mathrm{plateau}}[\cdot]$ indicates that we are, in principle, referring to the limit $\tau-t_\text{src} \rightarrow \infty$, $t_\text{snk}-\tau \rightarrow \infty$, $T - t_\text{snk} + t_\text{src} \rightarrow \infty$. In practice, we have to live with an approximation, of course. Plotting the ratio for a sufficiently large, fixed source--sink separation as a function of $\tau$, we expect the formation of a plateau between source and sink (Example plots will be shown in Fig.~\ref{fig-plateaus}). We extract the value of $R_{\GammaOp,\quark}(\vect{P};\mathcal{C}^\lat)$ from this plateau region, within which we may assume that contributions from excitations of the nucleon have decayed to a negligible level. A larger source-sink separation $t_\text{src}-t_\text{snk}$ reduces contaminations from excitations, but worsens the signal to noise ratio. 

At this point, it is necessary to establish a relation to renormalized continuum operators $O^\ren$. In general, the continuum operator of interest can be expressed as a linear combination of lattice operators $O_{\GammaOp,\quark}^\lat(\mathcal{C}^\lat)$. The ratio  $R_{O^\ren}(\vect{P})$ for such a renormalized operator $O^\ren$ is then itself a linear combination of ratios $R_{\GammaOp,\quark}(\vect{P};\mathcal{C}^\lat)$, and provides access to the desired continuum matrix elements:
\begin{equation}
	R_{O^\ren}(\vect{P}) =
	\sum_{S,S'}
	\frac{\overline{U}(P,S)\ \GammaThr\ U(P,S')}{2E_P\ \Tr_\mathrm{D} \left\lbrace\GammaTwo\, (-i \slashedE{P} + m_N)  \right\rbrace}
	\bra{N(P,S')}\, O^\ren\, \ket{N(P,S)}\ .
	\label{eq-contratioplat}
\end{equation}
We can now proceed analogous to section \ref{sec-linvamp}, writing the matrix element in the form
\begin{equation}
	\bra{N(P,S')}\, O^\ren\, \ket{N(P,S)} \equiv 
	\bar U(P,S')\ \MDM_{O^\ren}(P)\ U(P,S)\ ,
\end{equation}
where $\MDM_{O^\ren}(P)$ is a Dirac matrix. 
Inserting this into eq.\,(\ref{eq-ratiodef}), and performing the sum over spins, we finally obtain
\begin{equation}
	R_{O^\ren}(\vect{P}) =
	\frac{\Tr_\mathrm{D}\left\lbrace \MDM_{O^\ren}(P)\ (-i \slashedE{P} + m_N)\ \GammaThr\ (-i \slashedE{P} + m_N)\right\rbrace}{2E_P\ \Tr_\mathrm{D} \left\lbrace\GammaTwo\, (-i \slashedE{P} + m_N)  \right\rbrace}\ .
	\label{eq-threeptratio}
\end{equation}

With the continuum parametrization $\MDM_{O^\ren}(P)$ at hand, we gain access to the desired spin channels.
We will make use of this master formula in section \ref{sec-paramconnect} and in equation~(\ref{eq-ratiog4v}). Sample ratio plots will be shown in Fig.~\ref{fig-plateaus}.

\section{The Lattices of the MILC and LHPC Collaborations}
\label{sec-lattices}

The computationally most intensive steps of a lattice simulation are the generation of configurations and the calculation of propagators. 
In this work, we have used configurations and propagators calculated by the MILC and LHPC collaboration. For the purposes of these exploratory studies, we have not selected the largest and most realistic lattices available today. Rather, we have chosen lattices of moderate size with pion masses of around $m_\pi \approx 600 \units{MeV}$, which have allowed us to perform the analysis with reasonable statistics at a computational expense still manageable on the small PC cluster of our theory department.

The gauge configurations we have used are listed in Table~\ref{tab-gaugeconfs}. They have been generated by the MILC collaboration \cite{Ber01,Au04,Bernard:2007ps}. They feature 2+1 dynamical quarks, with the strange quark mass fixed to an approximate physical value. The gauge action is one-loop Symanzik improved (see, e.g., \cite{Alford:1995hw,Bistro} and references in \cite{Ber01}), and the AsqTad fermion action is of the staggered type, built using fat links and with order $a^2$ discretization errors removed at tree level, see Refs.~\cite{Orginos:1999cr,Lepage:1998vj} and references in \cite{Ber01}. Note that the ratio $\hat m_{u,d} / \hat m_{s}$ is constant for the last four lattices listed in Table~\ref{tab-gaugeconfs}. These ensembles serve us to study renormalization of the Wilson line operator in section~\ref{sec-WilsonLineRen}.

The numbers for the lattice spacing $a$ in Table~\ref{tab-gaugeconfs} come from the ``smoothed'' values $r_1/a$ as in Ref.~\cite{Au04}. As conversion factor to physical units we always use the continuum extrapolated value $r_1 = 0.317(7)_\text{sys}(3)_\text{fit}\units{fm}$ given in Ref. \cite{Au04} for all ensembles\footnote{In contrast, the LHPC collaboration typically uses values for $a$ obtained without the continuum extrapolation of $r_1$ but rather from a direct comparison of lattice results to the splittings of the $\Upsilon$ spectrum \cite{Wingate:2003gm,Au04}, which gives $a=0.124\units{fm}$ on the coarse and $a=0.087\units{fm}$ on the fine lattices. However, at present, the lattice spacings for the superfine and supercoarse lattices are only available in terms of $r_1$.}\setcounter{fnnumber}{\thefootnote}. 

\begin{table}[tbp]
	\centering
	\renewcommand{\arraystretch}{1.1}
	\begin{tabular}{|l|l|ll|c|c||c|}
	\hline
	ensemble-ID & $a\units{(fm)}$ & $\hat m_{u,d}$ & $\hat m_{s}$ & $10/g^2$ & $\hat L^3 \times \hat T$\rule{0ex}{1.2em} & $m_\pi^\text{DWF}$ \\
	\hline
	coarse-m050       & $0.1181(17)(27)$  & $0.05$   & $0.05$   & $6.85$  & $20^3 \times 64$  & $797(02)(29)$ \\
	coarse-m030       & $0.1190(16)(27)$  & $0.03$   & $0.05$   & $6.81$  & $20^3 \times 64$  & $621(02)(22)$ \\
	coarse-m020       & $0.1196(15)(27)$  & $0.02$   & $0.05$   & $6.79$  & $20^3 \times 64$  & $514(02)(18)$ \\
	fine              & $0.0854(12)(19)$  & $0.0124$ & $0.031$  & $7.11$  & $28^3 \times 96$  & \\
	superfine         & $0.0601(08)(14)$* & $0.0072$ & $0.018$  & $7.48$  & $48^3 \times 144$ & \\
	supercoarse       & $0.1765(32)(38)$* & $0.0328$ & $0.082$  & $6.485$ & $16^3 \times 48$  & \\ \hline
	\end{tabular}\par\vspace{1ex}
	\renewcommand{\arraystretch}{1.0}
	\caption{Lattice parameters of the MILC gauge configurations \cite{Ber01,Au04,Bernard:2007ps} used in this work. The lattice spacing $a$ in physical units is determined using $r_1$, as explained in the text. The first error quoted for $a$ includes statistical errors and the fit uncertainties specified for $r_1$, the second error stems from the systematic uncertainties of $r_1$. The values marked with an asterisk~* are preliminary \cite{DeTarPriv}. The last column lists the pion masses as determined with the LHPC propagators with domain wall valence fermions \cite{Hagler:2007xi}. The first error is statistical, the second error comes from the conversion to physical units using $a$ as quoted in the table. Note that the pion masses quoted here in physical units differ slightly from those listed in Ref. \cite{Hagler:2007xi}, because we use a different strategy to fix the lattice spacing, see footnote \thefnnumber.
	\label{tab-gaugeconfs}%
	}
\end{table}

We have used the first three lattices of Table~\ref{tab-gaugeconfs} to calculate nucleon matrix elements, with  propagators and sequential propagators from the LHPC collaboration. These propagators have been produced within a ``hybrid action'' approach: The calculation of nucleon matrix elements with staggered fermions is feasible but turns out to be rather complicated due to the unphysical taste degrees of freedom. 
Instead, the approach of the LHPC collaboration is to carry out the inversions eqns. (\ref{eq-propinversion}) and (\ref{eq-seqpropinversion}) in five dimensions with a domain wall fermion matrix $K^{\text{DWF}}_{\MI{1},\MI{2}}[U]$. Even so, the gauge background $U$ needed as input comes from the staggered simulations done by the MILC collaboration. The resulting propagators describe ``domain wall valence quarks on a staggered sea'', i.e. effects of virtual quark loops are included in the staggered formalism. The hybrid action approach is a suitable compromise that enables us to have chirally invariant, doubler-free valence fermions on a computationally affordable gauge background. For a physically meaningfull setup, the valence quark masses should be identical to the quark masses in the staggered sea. To accomplish this, LHPC tunes the valence quark mass parameters until their pion mass agrees with the mass of the lightest state in the pion taste multiplet (``Goldstone pion'') directly obtained with the MILC staggered action \cite{Hagler:2007xi}.
The pion masses after this adjustment are displayed in the last column of Table~\ref{tab-gaugeconfs}.

To reduce computational costs further, the lattices are chopped into two halves of temporal extent $\hat T/2 = 32$. Only every sixth configuration, and alternating temporal halves are taken, reducing autocorrelations to an undetectable level. Noise is reduced by application of HYP-smearing to the gauge configurations before the inversions are performed, see section \ref{sec-HYPsmear}.
The sequential propagators produced by LHPC are available for sink momenta $\vect{P} = 0$ and $\vect{P} = (-1,0,0) \times 2\pi/L$. The latter corresponds to $|\vect{P}|\approx 500\units{MeV}$ and is the lowest non-trivial momentum on these lattices. The source-sink separation is fixed to $\hat t_\text{snk} - \hat t_\text{src} = 10$. 

We should remark here that the point-to-all propagators generated by LHPC are by default calculated with smearing at the source. The sequential propagators have a smeared nucleon source and a smeared nucleon sink; only the free index is point-like. The two-point function in the ratio $R_{\GammaOp,\quark}$ in eq.~(\ref{eq-ratiodef}) must be compatible with this setup, i.e., source and sink must be smeared. Therefore we first need to prepare smeared-smeared point-to-all propagators. This functionality is provided by the \toolkit{Chroma} executable.

\section{Techniques of Statistical Error Estimation}

Powerful theoretical frameworks are available to deal with statistical errors in Monte Carlo data \cite{nla.cat-vn372369,DeGrand:2006zz}. For brevity, we restrict ourselves to a rather informal motivation of the prescriptions we have applied.

The evaluation of $n$ correlation functions $\tilde O_1[U],\ldots,\tilde O_n[U]$ for each of the $N$ gauge configurations $U$ in our ensemble yields \terminol{samples} $s_{ik}$ ($i=1..n$, $k=1..N$) from which we compute our observables. For each $i$, the samples $s_{ik}$ scatter around some ``true'' expectation value $\mathring s_i$. Any of the final observables $\theta$ we want to compute can be expressed as a function of the $s_i$. The ``true'' value is thus $\mathring{\theta} \equiv \theta(\mathring s_1,\ldots,\mathring s_n)$. Of course, the $\mathring s_i$ are not available to us. However, we can calculate an approximate value $\bar \theta \equiv \theta(\bar s_1,\ldots,\bar s_n)$ from the sample means
\begin{equation}
	\bar s_i \equiv \frac{1}{N} \sum_{k=1}^N s_{ik} \ .
	\end{equation}
We also need an estimate of the uncertainty of $\bar \theta$. Calculating the standard error of the sample mean from the scattering of $\theta$ according to 
\begin{equation}
	\sigma_\theta = \sqrt{ \frac{1}{N(N-1)} \sum_{k=1}^N \left( \theta(s_{1k},\ldots,s_{nk}) - \bar \theta \right)^2 } 
	\label{eq-wrongstderr}
	\end{equation}
would be problematic, because the $s_{ik}$ experience large fluctuations, much larger than the expected fluctuations of the $\bar s_i$ around the $\mathring s_i$. Since $\theta$ can behave highly non-linear for such large fluctuations, we would introduce a large uncontrolled bias. The idea of \terminol{resampling} methods is to construct new samples $s_{i(k)}$ that experience fluctuations of similar magnitude and distribution as the $\bar s_i$ around the $\mathring s_i$.

In general, we use the \terminol{jackknife} sampling method by default. In some cases, we opt for the \terminol{bootstrap} method, in particular when input samples from different ensembles (with different lengths $N$) need to be studied in a global analysis.

\subsection{Jackknife Sampling}

Initial \terminol{jackknife samples} are obtained from sample means with one input value missing. Those are used to form jackknife samples of $\theta$:
\begin{equation}
	s_{i(k)} \equiv \frac{1}{N-1} \mathop{\sum_{j=1}}_{j\neq k}^N s_{ij}\,, \qquad
	\theta_{(k)} \equiv \theta(s_{1(k)},\ldots,s_{n(k)})\,.
\end{equation}
First of all, jackknife sampling \cite{Quenouille56,tukey1958bac} is a method to reduce bias in the final estimate $\bar \theta$ . Let us interpret our input samples as random variables $S_{ik}$. The fluctuations in the mean values $\bar S_i$ scale with $N^{-1}$. Thus, in terms of expectation values $E[\cdot]$, it seems appropriate to expand the bias of $\bar \theta = \theta(\bar S_1,\ldots,\bar S_n)$ in powers of $N^{-1}$. The same thing can be done using jackknife samples instead of sample means:
\begin{align}
	E[\theta(\bar S_1,\ldots,\bar S_n)] & = \theta(\mathring s_1,\ldots,\mathring s_n) + \frac{a}{N} + \mathcal{O}\left(\frac{1}{N^2}\right) \ ,\\
	E[\theta(S_{1,(k)},\ldots,S_{n,(k)})] & = \theta(\mathring s_1,\ldots,\mathring s_n) + \frac{a}{N-1} + \mathcal{O}\left(\frac{1}{(N-1)^2}\right) \qquad \forall\, k
\end{align}
with some coefficient $a$. This analysis immediately shows that the \terminol{bias corrected jackknife estimate}
\begin{equation}
	\bar \theta^\text{Jack} \equiv \bar \theta - \frac{N-1}{N} \sum_k \left(\, \theta_{(k)} -   \bar \theta\, \right)
\end{equation}
has remaining bias of order $N^{-2}$. 
Moreover, the variances of the \terminol{jackknife samples} $s_{i(k)}$ are a factor $(N-1)^{-2}$ smaller than those of the $s_{ik}$. This tells us how to modify eq.\,(\ref{eq-wrongstderr}) in order to obtain a viable estimate of the standard deviation of $\bar \theta$, i.e., the \terminol{jackknife error}:
\begin{equation}
	\sigma_\theta^\text{Jack} = \sqrt{ \frac{N-1}{N} \sum_{k=1}^N \left( \theta_{(k)} - \frac{1}{N} \sum_{j=1}^N \theta_{(j)}\, \right)^2 } \ .
	\label{eq-jacknifeerr}
\end{equation}

\subsection{Bootstrap Sampling}
The bootstrap method \cite{Efron1979,nla.cat-vn372369} allows us to form an ensemble of samples whose size $M$ is independent of the number of configurations $N$. In this work, we choose $M=100$ or $M=1000$. Larger values of $M$ give more accurate error estimates but can be computationally costly. The first step of the resampling procedure is to draw $N \times M$ random integers $\alpha_{kl}$ in the range $1..N$. The \terminol{bootstrap samples} are then formed according to
\begin{equation}
	s_{i(l)} = \frac{1}{N} \sum_{k=1}^N s_{i\alpha_{kl}}, \qquad 
	\theta_{(l)} \equiv \theta(s_{1(l)},\ldots,s_{n(l)}) \ .
	\label{eq-bootstrapsamp}
\end{equation}
In the sum above, some samples $s_{ij}$ may appear multiple times, others are left out. In order to handle correlations correctly, it is important that one and the same set of indices $\alpha_{kl}$ is used for each $i$. The distribution of the $s_{i(l)}$ is an approximation of the distribution of the $\bar s_i$. Likewise, we expect the bootstrap samples $\theta_{(l)}$ to be distributed in a similar way as $\bar \theta$ is distributed. To determine an error interval, we simply sort the $\theta_{(l)}$ according to their size, resulting in a sequence $\theta_{(l_1)} \leq \theta_{(l_2)} \leq \ldots \leq \theta_{(l_M)}$. Let us throw away $15.8\%$ of the entries at the start of this sequence, and $15.8\%$ at the end of the sequence. The remaining sequence has a length of about $68.3\%$ of $M$, and the lower and upper values define an error interval with confidence level of about $68.3\%$ (``$1 \sigma$''). 

Bootstrap sampling is particularly useful when combining data of several ensembles $e$, each with a different number of configurations $N_e$. Then each channel $i$ of our input data $s_{ik}$ has been obtained from a certain ensemble $e_i$. Data coming from different ensembles are uncorrelated, so we may determine a new set of indices $\alpha_{kl}^{e}$ for each ensemble. Finally, we obtain resampled data of uniform length $M$ just as in eq.~(\ref{eq-bootstrapsamp}), using the $\alpha_{kl}^{e_i}$ as random indices.

\subsection{Overdetermined Systems, Fits and Correlations}
\label{sec-correl}

In practice, the extraction of an observable $\theta$ often involves solving an overdetermined system of equations. Typically, these equations are of the form $\mathring{s}_i = f_i(a_1,\ldots,a_m)$, with $m<n$ parameters $a_j$. The target observable $\theta$ can be expressed as a function $\theta(a_1,\ldots,a_m)$. Obviously, given approximate input values $\bar s_1,\ldots,\bar s_n$, we cannot find parameters that satisfy all $n$ equations at once. Throughout this work, we will stick to the following strategy:

The parameters $a_j$ are determined by minimizing numerically
\begin{equation}
	\tilde \chi^2 \equiv \sum_{i=1}^n w_i \left( f_i(a_1,\ldots,a_m) - \bar s_i \right)^2\ .
	\label{eq-chisqr}
\end{equation}
Here the choice of the weights $w_i$ involves some arbitrariness. If the $\bar s_i$ are largely uncorrelated, a sensible choice is $w_i = 1/(\Delta s_i)^2$, where $\Delta s_i$ is the jackknife/bootstrap error of $\bar s_i$ itself. If the correlations among the $\bar s_i$ are strong, it would be appropriate to modify eq.~(\ref{eq-chisqr}) such that the correlation matrix can be taken into account. Since the estimated correlation matrix is inaccurate and often close to singular, this approach would require special precautions. For this exploratory study, we will not follow this strategy, but rather adjust the $w_i$ by hand according to simple, problem specific criteria. As a consequence, the minimized value of $\tilde \chi^2$ does not have the strict statistical interpretation as it would have in a proper $\chi^2$ minimization. 

What is important to us is that the minimization yields values for the parameters $\bar a_i$, from which we can determine $\bar \theta = \theta(\bar a_1,\ldots,\bar a_m)$. Regardless of the difficulties with the choice of the $w_i$, this prescription implements $\theta$ as a \terminol{consistent estimator} (see, e.g., \cite{EDJS71}). 

We repeat the minimization eq.\,(\ref{eq-chisqr}) with the entire set of jackknife/bootstrap resampled values $s_{i(l)}$ as input. This can be sped up if we pass the $\bar a_1,\ldots,\bar a_m$ as initial guess to the minimization routine. From the jackknife/bootstrap estimates $\theta_{(l)}$ thus obtained, we determine error intervals as described in the sections above.
These uncertainties are appropriate for the estimator we have used. 

Because of correlations among the input data, the final estimate determined from our fit is statistically not optimal. However, the error bars we calculate with the resampling method take this into account. Their reliability is not compromised by our negligence of correlations in the fit procedure.

\subsection{Autocorrelations}
In the Monte Carlo simulation, the gauge configurations are determined from a Markov process, i.e., new configurations are random modifications of the preceding ones. Therefore, we expect correlations of a sample $s_{ik}$ with preceeding samples $s_{i,k-1}$, $s_{i,k-2}$, $\ldots$. Common techniques to estimate autocorrelations are \terminol{binning} or \terminol{blocking}, or the determination of an \terminol{integrated autocorrelation time} from jackknife samples\footnote{Reliable estimates of autocorrelation times require a large number $N$ of configurations, ideally well beyond 1000.}, see, e.g., \cite{janke10sas,DeGrand:2006zz}. The LHPC collaboration has taken precautions to minimize autocorrelations (see section \ref{sec-lattices}) and has not found autocorrelations of a detectable level in their observables. Therefore, for this study, we assume that autocorrelations can be ignored.

\section{Simulation Setup and Extraction Procedure}
\label{sec-simsetup}

\begin{figure}[p]
  \begin{center}
    \includegraphics{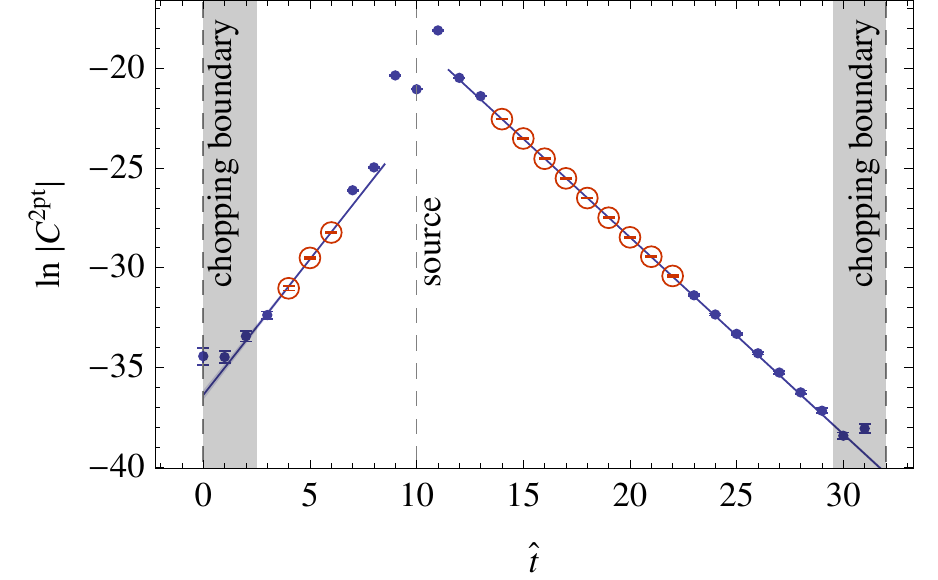}
     \caption{Two-point function for the nucleon on the coarse-m050 ensemble. The circled data points have been used to fit exponential decays. The fit weights have been chosen according to the statistical errors of the individual data points. The decay to the right of the source belongs to the nucleon state. The nucleon mass thus extracted is 
     $m_N = 1.649(08)_\text{stat}(59)_a\units{GeV}$. The data points close to the chopping boundary should not be taken into account.} \label{fig-nucltwopoint}
  \end{center}
\end{figure}

Simulations of nucleon observables are carried out on the coarse lattices, where LHPC propagators are available.
Remember that LHPC applies HYP smearing to the con\-fi\-gu\-ra\-tions and chops them in two halves of temporal extent $\hat T = 32$. The point-to-all propagators have their fixed index (their ``source'') at $\hat t_\text{src} = 10$. These propagators are the only ingredient needed to calculate the nucleon two-point function, which we display in Fig.~\ref{fig-nucltwopoint}.

\begin{figure}[p]
	\centering%
	\hfill%
	\subfloat[][]{%
		\label{fig-plateauPzero}%
		\includegraphics{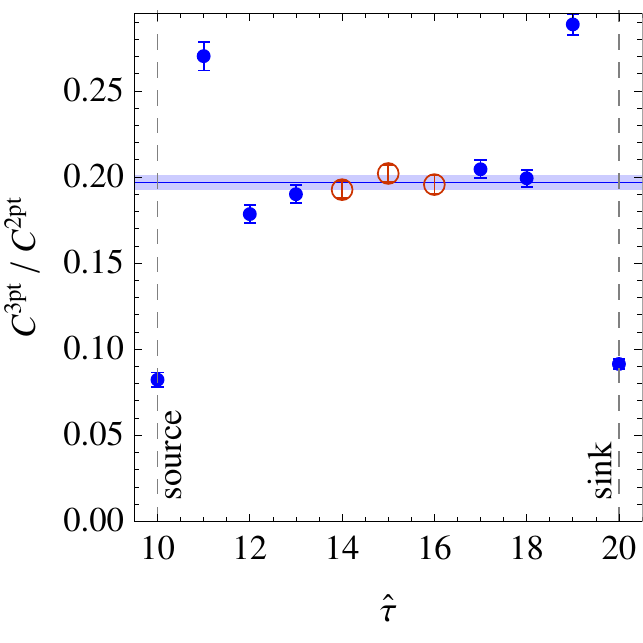}%
		}%
	\hfill%
	\subfloat[][]{%
		\label{fig-plateauPone}%
		\includegraphics{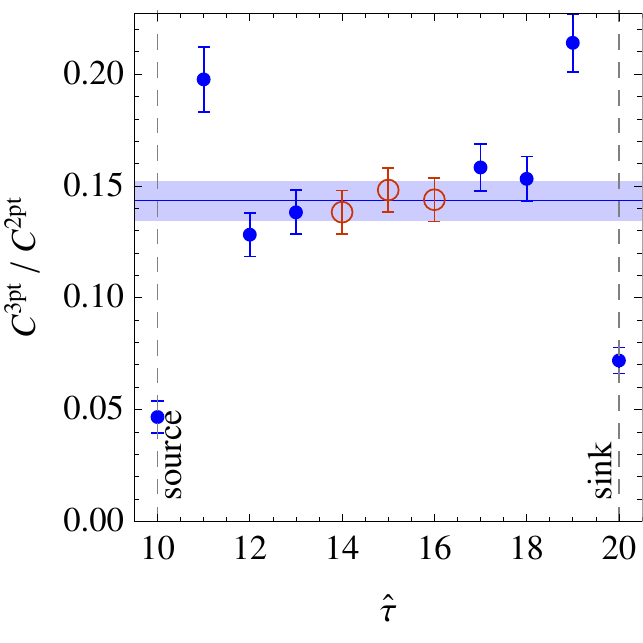}%
		}%
	\hspace*\fill\par%
	\caption[plateaus]{%
		Plateaus plots for the real part of the ratio $\latcfn^\text{3pt}_{\Eu{4},u-d}(\tau,\vect{P};\mathcal{C}^\lat) / \latcfn^\text{2pt}(t_\text{snk}-t_\text{src},\vect{P})$ on the coarse-m050 lattice for the link path $\mathcal{C}^\lat$ given by the moves $1, 2, 1, 1, 2, 1, 1, 2, 1$ as depicted in Fig.~\ref{fig-steplike}. The gauge link was evaluated on the HYP-smeared lattices. The quark separation is $|\vect{\elll}| \approx 6.7 a \approx 0.8 \units{fm}$. We plot the ratio for the two available nucleon momenta: \subref{fig-plateauPzero} $\hat{\vect{P}} = 0$, \subref{fig-plateauPone}  $\hat{\vect{P}} = (-1,0,0)\times 2\pi/\hat L$. The plateau value $R_{\gamma_\Eu{4},u-d}(\vect{P};\mathcal{C}^\lat)$ is extracted from the three circled points and is displayed as a horizontal error band. 
		\label{fig-plateaus}
		}
	\end{figure}

The LHPC sequential propagators we use have the nucleon sink at $\hat t_\text{sink}=20$. The typical appearance of the three-point function is shown in Fig.~\ref{fig-plateaus}. The oscillations of the correlator close to the source are an artifact that can appear when using domain wall fermions. To apply the formalism of section \ref{sec-ratios} we must evaluate the ratio $\latcfn^\text{3pt}_{\GammaOp,\quark}(\tau,\vect{P};\mathcal{C}^\lat) / \latcfn^\text{2pt}(t_\text{snk}-t_\text{src},\vect{P})$ at time slices $\tau$ far enough away from source and sink (and the chopping boundary). We assume here that this is the case for the time slices $\hat \tau = 14, 15, 16$. We obtain the final result for the ratio $R_{\GammaOp}(\vect{P};\mathcal{C}^\lat)$ from an average over the values from these three time slices.\footnote{There are more sophisticated procedures available to extract the ratio value. These take the effect of excited states (and possibly the oscillating states) into account, see, e.g., Ref. \cite{Renner:2007pb}. Using a more sophisticated approach, one may obtain slightly different ratio values and more realistic error bars. Since the primary goal of this work is not a high precision analysis, we stick to the simple averaging procedure.}
Table~\ref{tab-numgaugeconfs} lists the number of configurations we have used on the different ensembles.

\begin{table}[htbp]
	\centering
	\renewcommand{\arraystretch}{1.1}
	\begin{tabular}{|l|l||c|c|}
	\hline
	\multicolumn{2}{|c||}{} &
	\multicolumn{2}{c|}{number of configurations \rule{0ex}{1.2em}} \\
	\hline
	ensemble-ID \rule{0ex}{1.2em} & $m_\pi\units{MeV}$ & HYP-smeared link & unsmeared link \\
	\hline
	coarse-m050     & $\approx 800$ & 425 & 135 \\
	coarse-m030     & $\approx 600$ & 563 & \\ 
	coarse-m020     & $\approx 500$ & 478 & \\
	\hline
	\end{tabular}\par\vspace{2ex}
	\renewcommand{\arraystretch}{1.0}
	\caption{Number of configurations used for our calculations of nucleon three- and two-point functions.
	\label{tab-numgaugeconfs}%
	}
\end{table}

\chapter{\TMDs with Straight Links}
\label{chap-straightlinks}

\label{chap-results}

In this chapter we present first results for \TMDs calculated on the lattice, albeit with a straight gauge link between the gauge fields. As discussed in section \ref{sec-tmddef}, this choice of the Wilson line does not correspond to the situation in scattering experiments. Thus the \TMDs we obtain cannot serve as quantitative predictions for \TMDs used in the description of SIDIS or Drell-Yan experiments. Nevertheless, a straight Wilson line appears to be a good starting point, for several reasons:
\begin{itemize}
	\item The simplest, most obvious way to render the bilocal operator gauge invariant is the straight Wilson line. From a purely theoretical point of view, such operators are appealing in themselves.
	\item We can study a discretized version of the straight gauge link and its properties directly on the lattice.
	\item The parametrization of the matrix elements in terms of Lorentz invariant amplitudes is considerably simpler, see sections \ref{sec-linvamp}.
	\item To our present understanding, \TMDs with straight links have a probability interpretation, see section \ref{sec-probproofstraight}.
\end{itemize}


The fundamental object of interest in this chapter is the quark-quark correlator with a straight Wilson line\footnote{Here and in the following, we omit the index for the quark type $\quark$.}
\begin{align}
	\tilde \Phi^{[\GammaOp]}(\elll,P,S) & =
	\frac{1}{2} \bra{N(P,S)}\ \underbrace{\bar \quark(\elll)\, \GammaOp\ \Wline{\elll,0}\ \quark(0)}_{\displaystyle \equiv O_{\GammaOp}(\elll)}\ \ket{N(P,S)} \nonumber \\ %
	& \equiv \frac{1}{2}\ \bar U(P,S')\, \tilde \MDM_{\GammaOp}(\elll,P)\, U(P,S)\ .
	\label{eq-corrtilde}	
\end{align}
We shall work directly with a discretized version of $O_{\GammaOp}(\elll)$.

\section{Parametrization}

\subsection{Parametrization of the \texorpdfstring{$\elll$}{l}-dependent Correlator}
\label{sec-symtrafo}

We have already discussed a parametrization of the quark-quark correlator in section \ref{sec-linvamp}. There the correlator and amplitudes were $k$-dependent. For our lattice calculation, we need an analogous parametrization of the $\elll$-dependent quantities. 
The symmetry transformation properties of the matrices $\tMDM_{\GammaOp}(\elll,P)$ read
\begin{align}
	&(\dagger): & \left[ \tMDM_\Gamma(\elll,P) \right]^\dagger 
	&= \gamma^0\ \tMDM_{\gamma^0\,\Gamma^\dagger\,\gamma^0} (\mathbf{-}\elll,P)\ \gamma^0 \ ,\\
	&(\mathscr{P}): & \tMDM_\Gamma(\elll,P) 
	&= \gamma^0\ \tMDM_{\gamma^0\,\Gamma\,\gamma^0} (\overline{\elll}, \overline{P})\ \gamma^0\ ,\\
	&(\mathscr{T}): &\left[ \tMDM_\Gamma(\elll,P) \right]^* 
	&= \gamma^5 C\ \tMDM_{C^\dagger \gamma^5\,\Gamma\,\gamma^5 C} (\mathbf{-}\overline{\elll},\overline{P})\ C^\dagger \gamma^5 \ .
	\label{eq-Mtildetimerev}
	\end{align}
It turns out that we can find the structures compliant with the symmetry constraints by replacing $k$ with $i m_N^2 \elll$ in eq.\,(\ref{eq-phitraces})\footnote{and likewise in the $\tMDM$ structures eq.\,(\ref{eq-Mstructs}). For completness, the expressions for the $\tMDM_{\GammaOp}$ are given in eq.\,(\ref{eq-Mtildestructs}) in the appendix.}. We thus obtain
\begin{align}
	\tilde{\Phi}^{[\Eins]}(\elll,P,S) & = 
		2\, m_N\ \tAmp_1 \ ,\nonumber\\
	\tilde{\Phi}^{[\gamma^\mu]}(\elll,P,S) & =
		2\ \tAmp_2\ P^\mu
		+ 2i\,{m_N}^2\ \tAmp_3\ \elll^\mu 
		+ \left[ 2i\,m_N\ \tAmp_{12}\ \epsilon^{\mu \nu \alpha \beta} S_\nu P_\alpha \elll_\beta \right] \ ,\nonumber\\
	\tilde{\Phi}^{[\sigma^{\mu \nu}]}(\elll,P,S) & =
		  \left[ 2i\,m_N\ \tAmp_4\ (P^\mu \elll^\nu - P^\nu \elll^\mu) \right]
		+ 2\ \tAmp_9\ \epsilon^{\mu \nu \alpha \beta} S_\alpha P_\beta \nonumber\\ & \phantom{=\ }
		+ 2i\,{m_N}^2\ \tAmp_{10}\ \epsilon^{\mu \nu \alpha \beta} S_\alpha \elll_\beta  
		- 2\,m_N^2\ \tAmp_{11}\ \epsilon^{\mu \nu \alpha \beta} \elll_\alpha P_\beta (\elll \cdot S)  \ ,\nonumber\\
	\tilde{\Phi}^{[\gamma^\mu \gamma^5]}(\elll,P,S) & = 
		- 2\, m_N\ \tAmp_6\ S^\mu
		- 2i\,m_N\ \tAmp_7\ P^\mu (\elll \cdot S)
		+ 2\,{m_N}^3\ \tAmp_8\ \elll^\mu (\elll \cdot S) \ ,\nonumber\\
	\tilde{\Phi}^{[\gamma^5]}(\elll,P,S) & = \left[ - 2\,{m_N}^2\ \tAmp_5 \ (\elll\cdot S) \right] \ .
	\label{eq-phitildetraces}
	\end{align}
Again, the structures with $\tAmp_4$, $\tAmp_5$ and $\tAmp_{12}$, highlighted with square brackets, are $\mathscr{T}$-odd and vanish for our analysis with straight links.
The amplitudes $\tAmp_i(\elll,\elll \tcdot P)$ appearing in the parametrization above are not constrained to be real valued like their momentum dependent counterparts $A_i(k,k\tcdot P)$. Instead, hermiticity $(\dagger)$ now leads to the following relation:
\begin{equation}
	\left[ \tAmp_i(\elll^2,\elll \tcdot P) \right]^* = \tAmp_i(\elll^2,-\elll \tcdot P)\ .
	\label{eq-tampconstraint}
\end{equation}

\subsection{Amplitudes \texorpdfstring{$\tAmp_i$}{} from the Lattice}
\label{sec-paramconnect}

We now have a continuum parametrization of the $\elll$-dependent matrix elements at hand, and can evaluate our master ratio formula eq.\,(\ref{eq-threeptratio}):
\begin{equation}
	R_{\GammaOp}^\ren(\vect{P},\elll) =
	\frac{\Tr_\mathrm{D}\left\lbrace \tMDM_{\GammaOp}(\elll,P)\ (-i \slashedE{P} + m_N)\ \GammaThr\ (-i \slashedE{P} + m_N)\right\rbrace}{2E_P\ \Tr_\mathrm{D} \left\lbrace\GammaTwo\, (-i \slashedE{P} + m_N)  \right\rbrace}\ .
	\label{eq-ratiorenop}
\end{equation}
For $\tMDM_{\GammaOp}(\elll,P)$ we take the expressions listed in eq.~(\ref{eq-Mtildestructs}), and for  $\GammaTwo$ and $\GammaThr$ we use the LHPC conventions. For any Dirac matrix $\GammaOp$, the equation above enables us to express the ratio $R_{\GammaOp}^\ren(\vect{P},\elll)$ as a combination of amplitudes $\tAmp_i$. We give a list of these expressions for $16$ basic Dirac matrices $\GammaOp$ in Table \ref{tab-ratios} in the appendix. The relations in this table form a system of $16$ equations, which we can solve for the individual amplitudes $\tilde A_i$. However, for this exploratory study, we have only picked a few simple channels, for which we do not need to disentangle the amplitudes:
\begin{itemize}
	\item $\displaystyle 2 \tAmp_2 = R^\ren_{\gamma_\Eu{4}}$\ ,
		\hfill\refstepcounter{equation}\label{eq-ratiogamma4}(\theequation)
	\item $\displaystyle 2 \tAmp_6 = \frac{i E(P)}{m_N} R^\ren_{\gamma_\Eu{3}\gamma_\Eu{5}}$\quad for gauge links with $\vect{\elll}_3=0$\ ,
		\hfill\refstepcounter{equation}(\theequation)
	\item $\displaystyle 2 \tAmp_7 = \frac{-i}{m_N \vect{\elll}_3} R^\ren_{\gamma_\Eu{5}\gamma_\Eu{4}}$\ .
	\hfill\refstepcounter{equation}(\theequation)
\end{itemize}

For the value $E(P)/m_N$ needed to extract $\tilde A_6$, we use the continuum dispersion relation
\begin{equation}
	\frac{E(P)}{m_N} = \frac{\sqrt{\hat m_N^2 + (2\pi/\hat L)^2}}{\hat m_N}\ .
	\label{eq-energyratio}
\end{equation}
Here $\hat m_N$ is determined as described in section~\ref{sec-nucltwopt} and illustrated in Fig.~\ref{fig-nucltwopoint}. We obtain for the ratio above $1.04942(47)_\text{stat}$ on the coarse-m050 lattice, and $1.0713(12)_\text{stat}$ on the coarse-m020 lattice. 
Of course, as an alternative, we can also extract $\hat E(P)$ directly using a two-point function. The value thus obtained for $E(P)/m(N)$ has a larger statistical error, and agrees with the value extracted using eq.~(\ref{eq-energyratio}) almost within error bars.

Note that we will prefer to show results for $m_N \tAmp_7$ rather than the dimensionless amplitude $\tAmp_7$. The calculation of the latter would involve an explicit factor $m_N$, which introduces additional systematic errors and would complicate a chiral extrapolation. For the same reason, we will present results for the corresponding \TMD $g_{1T}$ in terms of $g_{1T}/m_N$ or $g_{1T}|\vprp{k}|/m_N$.

\subsection{From Amplitudes to \TMDs}
\label{sec-amplstraighttmds}

How are the amplitudes $\tAmp_i(\elll,P)$ related to \TMDs? Combining eq.\,(\ref{eq-corr}) and eq.\,(\ref{eq-TMDcorr}), we find that we need to perform the Fourier transformation
\begin{align}
	\Phi^{[\GammaOp]}(x,\vprp{k};P,S) 
	& =  \int \frac{d\elll^-}{2 \pi}\ \frac{d^2 \vprp{\elll}}{(2\pi)^2}\ e^{i( -\elll^- k^+ + \vprp{\elll} \cdot \vprp{k})}\ \tilde \Phi^{[\GammaOp]}(\elll,P,S) \Big \vert_{\elll^+=0}\ ,
\end{align}
where $k^+ = x P^+$, as usual. We rewrite this integral in terms of the Lorentz-invariant quantities $\elll^2$ and $\elll \tcdot P$.
For $\elll^+=0$, we have
\begin{equation}
	\elll \tcdot P = \elll^- P^+\ , \qquad \elll^2 = - \vprp{\elll} \cdot \vprp{\elll} <\ 0 \ .
\end{equation}
Thus we get
\begin{align}
	\Phi^{[\GammaOp]}(x,\vprp{k};P,S) & = \frac{1}{P^+} \int \frac{d(\elll \tcdot P)}{2\pi}\ e^{-i (\elll \cdot P) x}\ \int
	 \frac{d^2 \vprp{\elll}}{(2\pi)^2}\ e^{i \vprp{\elll} \cdot \vprp{k}}\ \tilde \Phi^{[\GammaOp]}(\elll,P,S) \Big \vert_{\elll^+=0}\ .
	\label{eq-phitransformpre}
\end{align}
We substitute $ \vprp{\elll} \cdot \vprp{k} = |\vprp{\elll}|\, |\vprp{k}|\, \cos (\theta) $ and use
\begin{equation}
	\int d^2\vprp{\elll} = \int_0^\infty d\left( \sqrt{ -\elll^2 } \right) \sqrt{ -\elll^2 } \int_0^{2\pi} d \theta  = \frac{1}{2} \int_0^\infty d(-\elll^2) \int_0^{2\pi} d\theta \ .
	\end{equation}
Finally, we get
\begin{align}
	\Phi^{[\GammaOp]}(x, \vprp{k};P,S) & = \frac{1}{P^+} \int \frac{d(\elll \tcdot P)}{2\pi}\ e^{-i (\elll \tcdot P) x}  \int_0^\infty \frac{d(-\elll^2)}{2(2\pi)} \nonumber \\ & \times \int_0^{2\pi} \frac{d\theta}{2\pi}\ e^{i\, \sqrt{-\elll^2}\, |\vprp{k}|\, \cos \theta}\ \tilde \Phi^{[\GammaOp]}(\elll,P,S) \Big \vert_{\elll^+=0}\ .
	\label{eq-phitransform}
\end{align}
Since $\tilde \Phi^{[\GammaOp]}(\elll,P,S)$ is composed of amplitudes $\tilde A_i$, which are functions of $\elll^2$ and $\elll\cdot P$ only, we shall always be able to perform the $\theta$-integral analytically. In the most general case, this can be done as shown in appendix \ref{app-tmdpdfft}. Thus we are left with integrals over the Lorentz-invariant quantities $\elll^2$ and $\elll \cdot P$ only.

\subsubsection{Unpolarized case: \texorpdfstring{$f_1(x,\vprp{k}^2)$}{f1(x,kprp)}}

In the straight link case, $f_1(x,\vprp{k}^2) = \Phi^{[\gamma^+]}(x, \vprp{k};P,S)$ (compare eq\,(\ref{eq-phigammaplus})). With $\elll^+=0$, we get from eq.\,(\ref{eq-phitildetraces}) $\tilde \Phi^{[\gamma^+]}(\elll,P,S) \vert_{\elll^+=0} = 2\, \tAmp_2\, P^+$, which we substitute into eq.\,(\ref{eq-phitransform}):
\begin{align}
f_1(x,\vprp{k}^2) & = \Phi^{[\gamma^+]}(x, \vprp{k};P,S) \nonumber \\ 
& = \frac{1}{P^+} \int \frac{d(\elll \cdot P)}{2\pi}\ e^{-i (\elll \cdot P) x}  \int_0^\infty \frac{d(-\elll^2)}{2(2\pi)} \int_0^{2\pi} \frac{d\theta}{2\pi}\ e^{i\, \sqrt{-\elll^2}\, |\vprp{k}|\, \cos \theta}\ 2\, \tilde{A}_2\ P^+ \nonumber \\
& = \int \frac{d(\elll \cdot P)}{2\pi}\ e^{-i (\elll \cdot P) x}  \int_0^\infty \frac{d(-\elll^2)}{2(2\pi)}\ 2\, \tilde{A}_2\ \int_0^{2\pi} \frac{d\theta}{2\pi}\ e^{i\, \sqrt{-\elll^2}\, |\vprp{k}|\, \cos \theta}\ \nonumber \\
& = \int \frac{d(\elll \cdot P)}{2\pi}\ e^{-i (\elll \cdot P) x}  \int_0^\infty \frac{d(-\elll^2)}{2(2\pi)}\ J_0(\sqrt{-\elll^2}\, |\vprp{k}|)\ 2\,\tilde{A}_2\ .
\label{eq-f1fromtildeA2}
\end{align}
Here $J_0$ is a Bessel function of the first kind. With the notation introduced in appendix \ref{app-tmdpdfft}, we simply write
\begin{align}
\Phi^{[\gamma^+]}(x, \vprp{k};P,S) & = T_{0,0}(x,|\vprp{k}|)\left[\,2\,\tilde{A}_2\,\right]\ .
\end{align}

\subsubsection{Axial vector operator: \texorpdfstring{$g_{1L}(x,\vprp{k})$ and $g_{1T}(x,\vprp{k})$}{g1L and g1T}}

Let us analyze $\Phi^{[\gamma^+ \gamma^5]}(x, \vprp{k})$. The structure we read off from eq.\,(\ref{eq-phitildetraces}) for the contribution from $\tilde A_6$ to $\tilde \Phi^{[\gamma^+ \gamma^5]}$ is:
\begin{equation}
- 2\, m_N\ \tilde{A}_6\ S^+ = - 2\, \Lambda \ \tilde{A}_6\ P^+\ ,
\end{equation}
where we have used the spin conventions eq.\,(\ref{eq-spinparam}) in the appendix. 
Then the corresponding contribution to $\Phi^{[\gamma^+ \gamma^5]}$ is
\begin{align}
\Phi^{[\gamma^+ \gamma^5]}_{(6)}(x, \vprp{k};P,S) & = \Lambda\ T_{0,0}(x,|\vprp{k}|)\left[\,-2\,\tilde{A}_6\,\right]\ .
\end{align}
For $\tAmp_7$, we have
\begin{align}
-2i\,m_N\ \tilde{A}_7\ P^+ (\elll \cdot S) \big \vert_{\elll^+ = 0} =
(-2i\,m_N\ \tilde{A}_7\ P^+)\left( \Lambda\, \elll^-\, \frac{P^+}{m_N} - \vprp{\elll} \cdot \vprp{S} \right)\ . 
\end{align}
Applying the formalism of appendix \ref{app-tmdpdfft}, we obtain
\begin{align}
\Phi^{[\gamma^+ \gamma^5]}_{(7)}(x, \vprp{k};P,S) & = \Lambda\ T_{1,0}(x,|\vprp{k}|)\left[\,-2i\,\tAmp_7\,\right] \nonumber \\
& + \frac{\vprp{k} \cdot \vprp{S}}{m_N}\ T_{0,1}(x,|\vprp{k}|)\left[\,2i\,\tAmp_7\,\right]\ .
\end{align}
Amplitude $\tAmp_8$ drops out since $\elll^+=0$. Comparing to eq.~(\ref{eq-phigammaplusgfive}), we get
\begin{align}
g_{1L}(x,\vprp{k}^2) & = T_{0,0}(x,|\vprp{k}|)\left[\,-2\,\tAmp_6\,\right] + T_{1,0}(x,|\vprp{k}|)\left[\,-2i\,\tilde{A}_7\,\right] 
\ ,\nonumber \\
g_{1T}(x,\vprp{k}^2) & = T_{0,1}(x,|\vprp{k}|)\left[\,2i\,\tAmp_7\,\right]\ .
\end{align}

\subsubsection{Tensor operator}

We now analyze $\Phi^{[\sigma^{i+}]}(x, \vprp{k};P,S)$, for a transverse index $i=1,2$. 
Inserting $\elll^+ = 0$ and $\vprp{P}=0$ into the corresponding line in eq.\,(\ref{eq-phitildetraces}), remembering that $\tAmp_4$ vanishes for straight links,  and using $\epsilon^{i+\alpha\beta} = -g^{\alpha+}\epsilon^{i\beta}_\prp + g^{\beta+}\epsilon^{i\alpha}_\prp$, we get
\begin{align}
\tilde \Phi^{[\sigma^{i+}]}(\elll,P,S) & = 2\, \tilde A_9\, P^+ \epsilon_\prp^{i\alpha} S_\alpha 
- 2 i\, m_N\, \tilde A_{10}\, \Lambda\, P^+\, \epsilon_\prp^{i \beta} \elll_\beta \nonumber \\
& - 2\, m_N^2\, \tilde A_{11}\, P^+\, \epsilon_\prp^{i \alpha} \elll_\alpha \left( \Lambda\, \elll^-\, \frac{P^+}{m_N} - \vprp{\elll} \cdot \vprp{S} \right) \ .
\end{align}
We now apply the formalism of appendix~\ref{app-tmdpdfft} and compare with eq.\,(\ref{eq-phisigmaplusi}). Then
\begin{align}
h_{1T}(x,|\vprp{k}|) & = T_{0,0}(x,|\vprp{k}|)\left[\,2\, \tilde A_9\,\right] \ ,\nonumber \\
h_{1L}^\prp(x,|\vprp{k}|) & = T_{0,1}(x,|\vprp{k}|)\left[\,-2i\, \tilde A_{10}\,\right] 
+ T_{1,1}(x,|\vprp{k}|)\left[\,-2\, \tilde A_{11}\,\right]\ ,\nonumber \\
h_{1T}^\prp(x,|\vprp{k}|) & = T_{0,2}(x,|\vprp{k}|)\left[\,2\, \tilde A_{11}\,\right]\ .
\end{align}

\section{The Discretized Non-Local Operator}

\begin{figure}[tbp]
  \begin{center}
    \includegraphics*{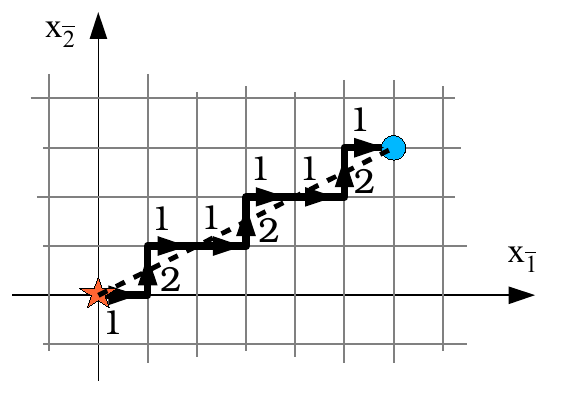}
     \caption{Example of a step-like link path: The straight gauge link in the continuum with $\vect{\elll} = (6,3,0)$ (dashed line) is represented as a product of link variables $U_\muE$ in the directions $\muE = 1, 2, 1, 1, 2, 1, 1, 2, 1$.} \label{fig-steplike}
  \end{center}
\end{figure}

To realize the non-local operator $O_{\GammaOp}(\elll)$ on the lattice, we approximate the Wilson line between the quark fields by a product of connected link variables along a lattice path ${\mathcal{C}^\lat_\elll} = (\elll,x^{(n-1)},x^{(n-2)}\ldots,x^{(1)},0)$. 
We end up with a lattice operator just as in eq.\,(\ref{eq-olat}):
\begin{equation}
	O_{\GammaOp,\quark}^\lat(\mathcal{C}^\lat_\elll;0) \ \equiv\ \bar \quark(\elll)\, \GammaOp\ \WlineClat{\mathcal{C}^\lat_\elll}\ \quark(0) = \bar \quark(\elll)\,U(\elll,x^{(n-1)})\,\cdots\,  U(x^{(1)},0)\,\quark(0)\ .
\end{equation}
If the Wilson line does not coincide with one of the lattice axes, i.e., if it is at an oblique angle, we approximate the straight line by a step-like path as illustrated in Fig.~\ref{fig-steplike}. To this end we programmed a \toolkit{Mathematica} function resembling the Bresenham algorithm \cite{Bresenham:1965}, which lays out the path through lattice sites with minimal distance to the straight line connection. 

We interpret $\elll$ as a finite, physical separation of the quark fields. In the continuum limit $a \rightarrow 0$, the number of link variables required to form the gauge link goes to infinity: $\hat \elll \equiv \elll / a \rightarrow \infty$. The fine-grained details of the discretization prescription $\mathcal{C}^\lat_\elll$ for the link path should not matter, as long as the discretized link can be made to lie arbitrarily close to the straight line by choosing $a$ small enough. If this is maintained, the continuum limit of our lattice gauge link becomes a unitary path ordered product of gauge fields located arbitrarily close to a straight line. In that sense the continuum limit of our lattice operator $O_{\GammaOp,\quark}^\lat(\mathcal{C}^\lat_\elll;0)$ is the continuum operator $O_{\GammaOp}(\elll)$.

Obviously there are many ways we can realize a lattice gauge path close to a straight line. First of all, this raises the question whether the continuum limit is really unique and well-defined. It turns out that the continuum limit of our operator involves a power divergence, proportional to the lattice scale $1/a$. The divergence and the respective ambiguity of the continuum limit must be adequately renormalized, see section \ref{sec-WilsonLineRen}. Together with the renormalization of the quark fields, we will end up with a relation of the form 
\begin{equation}
	O^\ren_{\GammaOp}(\elll) = \renZ^{-1}(-\elll^2;a)\ O_{\GammaOp,\quark}^\lat(\mathcal{C}^\lat_\elll;0)\ .
	\label{eq-renop}
\end{equation}

We will usually assemble our gauge links on a HYP smeared ensemble, so the individual links are in fact fat links. Effectively, the link we construct is a unitarized linear combination of many paths. The HYP smeared gauge links turn out to exhibit almost perfect rotational invariance with respect to $\elll$, indicating that they closely resemble a continuum operator. Without smearing, discretization errors appear more pronounced.

\section{First Observations on the Lattice}
\label{sec-firstobs}

Let us focus on the channel $\GammaOp = \gamma_\Eu{4}$, and ignore the renormalization factor $\renZ(\elll^2;a)$ of eq.~(\ref{eq-renop}) for the moment. From eq.\,(\ref{eq-ratiogamma4}) we get from the \emph{unrenormalized} 
ratio
\begin{equation}
	2 \tAmp_{2}^\unren(\elll^2,\elll \tcdot P) :=  R_{\gamma_\Eu{4}}(\vect{P},\mathcal{C}^\lat_\elll)\ . 
\end{equation}
As mentioned before, the amplitude $\tAmp_2$ is related to the unpolarized \TMD $f_1(x,\vprp{k}^2)$.
Let us begin our explorations on the coarse-m050 lattice, where we get the best statistics. Looking at the sample plateau plots shown in Fig.~\ref{fig-plateaus}, we see a clean signal and surprisingly good statistics considering the rather long gauge link of about $0.8\units{fm}$.

\subsection{The Nucleon at Rest on the Lattice}

If we use the sequential propagators with the nucleon at rest, $\hat{\vect{P}} = \vect{0}$, we can only access the amplitude $\tAmp_{2}^\unren(\elll^2,\elll\tcdot P)$ at $\elll\tcdot P = 0$. Then the results we obtain from the ratio $R_{\gamma_\Eu{4}}(\vect{P},\mathcal{C}^\lat_\elll)$ should only depend on the distance $\elll^2 = -\vect{\elll}^2$ between the quarks; the residual dependence on the details of the link discretization $\mathcal{C}^\lat_\elll$ should be small.

To check this, we generate an initial set of link paths with the algorithm mentionend in the previous section. Firstly, we take link paths aligned with the lattice axes, with a length up to $|\hat{\vect{\elll}}| = 20$ (We have studied link lengths up to $|\hat{\vect{\elll}}| = 40$. Links on the axes that are longer than our lattice size $\hat L = 20$ overlap with their periodic mirror images. We find that the signal remains compatible with $0$ for such very long links). Also, we cover quark separation vectors $\vect{\elll}$ in the first quadrant of the $(\vect{\elll}_1,\vect{\elll}_2)$-plane, as well as two quadrants in the $(\vect{\elll}_1,\vect{\elll}_3)$-plane for quark separations up to $|\hat{\vect{\elll}}|\leq8$. Moreover, we pick certain longer links in the first octant in order to cover more $\elll^2$-values up to $|\hat{\vect{\elll}}|\leq15$. Last but not least, we pick some oblique links with integer lengths.  The choice is illustrated in Fig.~\ref{fig-linkpaths}.

Calculating the ratio $R_{\gamma_\Eu{4}}(\vect{P},\mathcal{C}^\lat_\elll)$ with these gauge paths on the coarse-m050 ensemble, we obtain Fig.~\ref{fig-unrenamp}. For the link paths evaluated on the HYP smeared lattice, we find an almost perfectly smooth dependence on $\elll^2$, even though we include step-like gauge paths. Obviously, the operator constructed from fat links exhibits rotational symmetry to a good approximation. This corroborates the notion that the discretized operator with an extended gauge link is an approximation to the straight link continuum operator. The situation is a bit worse for the non-smeared gauge background, where the gauge links at oblique angles systematically yield somewhat lower values. This error can be reduced considerably with our taxi driver correction, see section \ref{sec-taxidriver}. Notice, however, that the amplitude decays much more quickly with increasing $|\vect{\elll}|$ on the unsmeared ensemble than on the smeared one. The Gaussian-like shape emerging on the unsmeared lattice is much narrower than on the smeared one. We will address this issue in section \ref{sec-WilsonLineRen}, when we discuss renormalization.

\subsection{The Nucleon at Non-Zero Lattice Momentum}

\begin{figure}[tbp]
	\centering%
	\subfloat[][]{%
		\label{fig-linkpaths}%
		\includegraphics{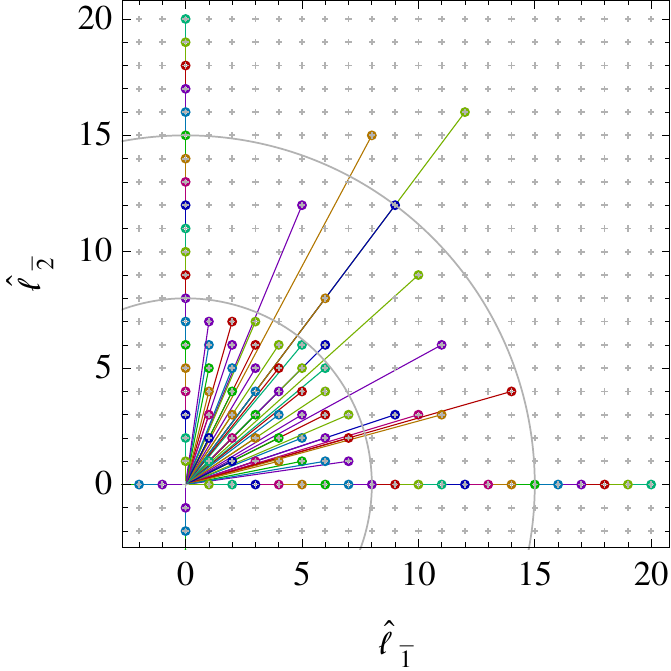}%
		}\hfill%
	\subfloat[][]{%
		\label{fig-lppoints}%
		\includegraphics{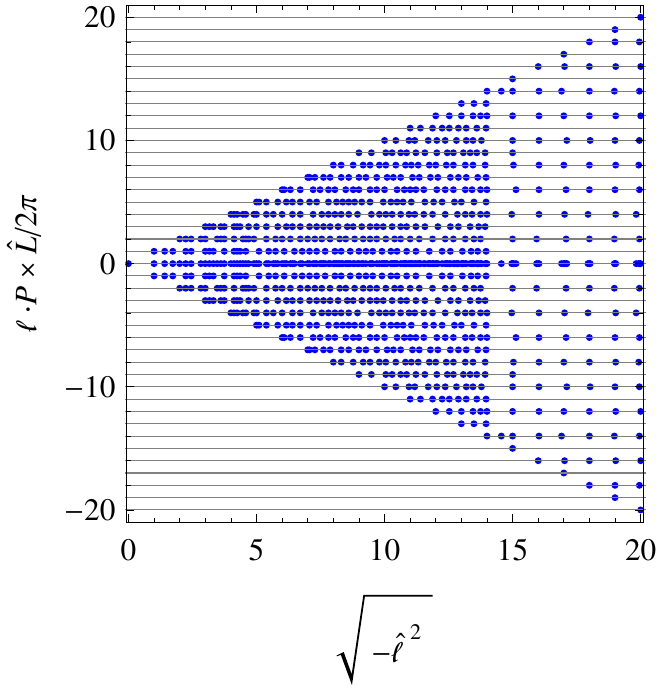}%
		}\par%
	\caption[Selection of link paths.]{%
		Selection of link paths.\par
		\subref{fig-linkpaths}
			Illustration of the set of quark separations $\elll$ chosen for the analysis at $\hat{\vect{P}}=\vect{0}$. Only the gauge paths with $\vect{\elll}_3=0$ are shown.\par			
		\subref{fig-lppoints}
			For the analysis at $\vect{P}=(-1,0,0)\times2\pi/L$, we choose the quark separations $\elll$ in such a way that the $(\elll^2,\elll\tcdot P)$-plane is approximately covered evenly in the accessible region.%
		\label{fig-linkselection}
		}
\end{figure}

\begin{figure}[tbp]
	\centering%
	\subfloat[][]{%
		\label{fig-unrenamp-sm}%
		\includegraphics[width=0.85\textwidth,clip=true]{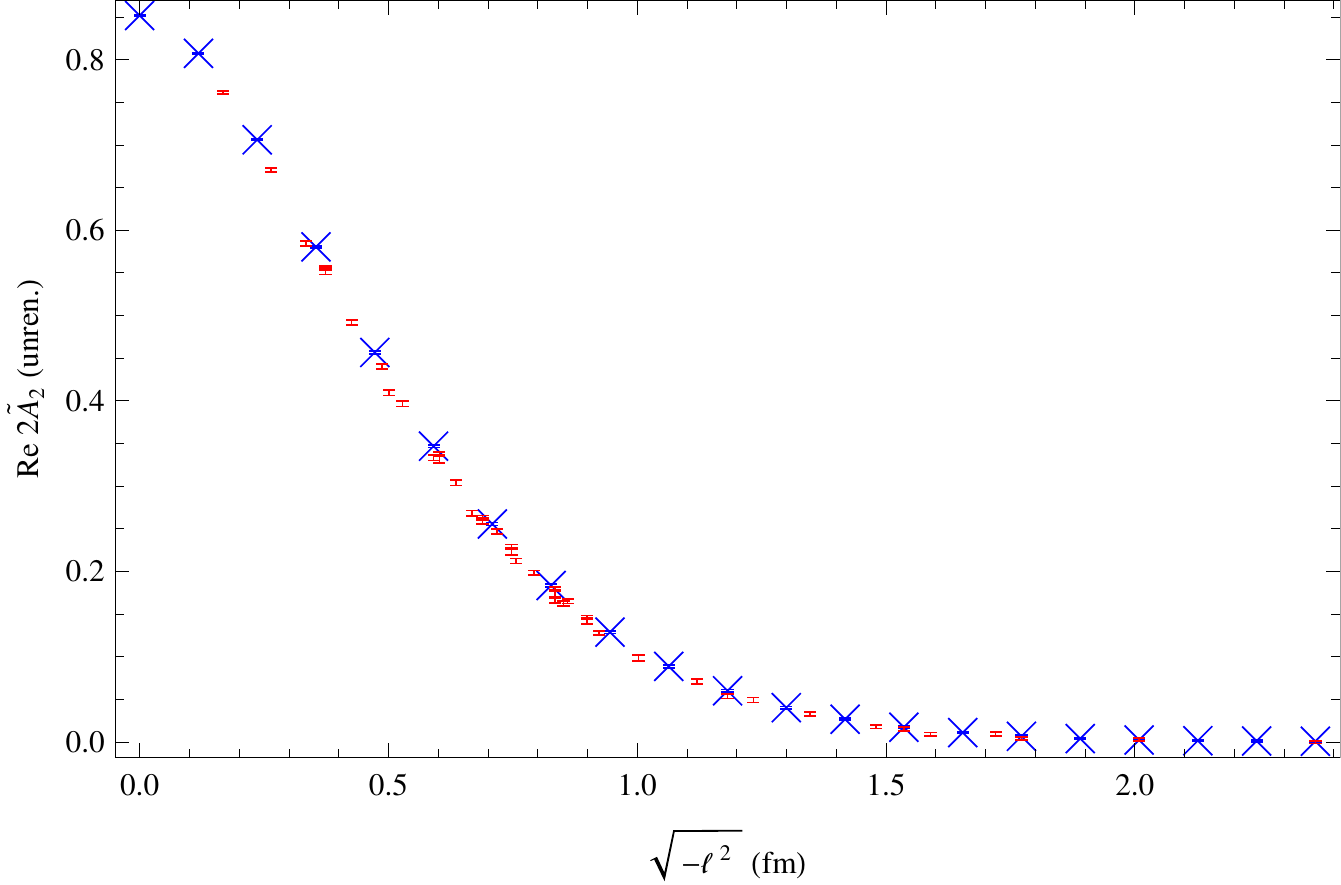}%
		}\\ \vspace{1ex}%
	\subfloat[][]{%
		\label{fig-unrenamp-unsm}%
		\includegraphics[width=0.85\textwidth,clip=true]{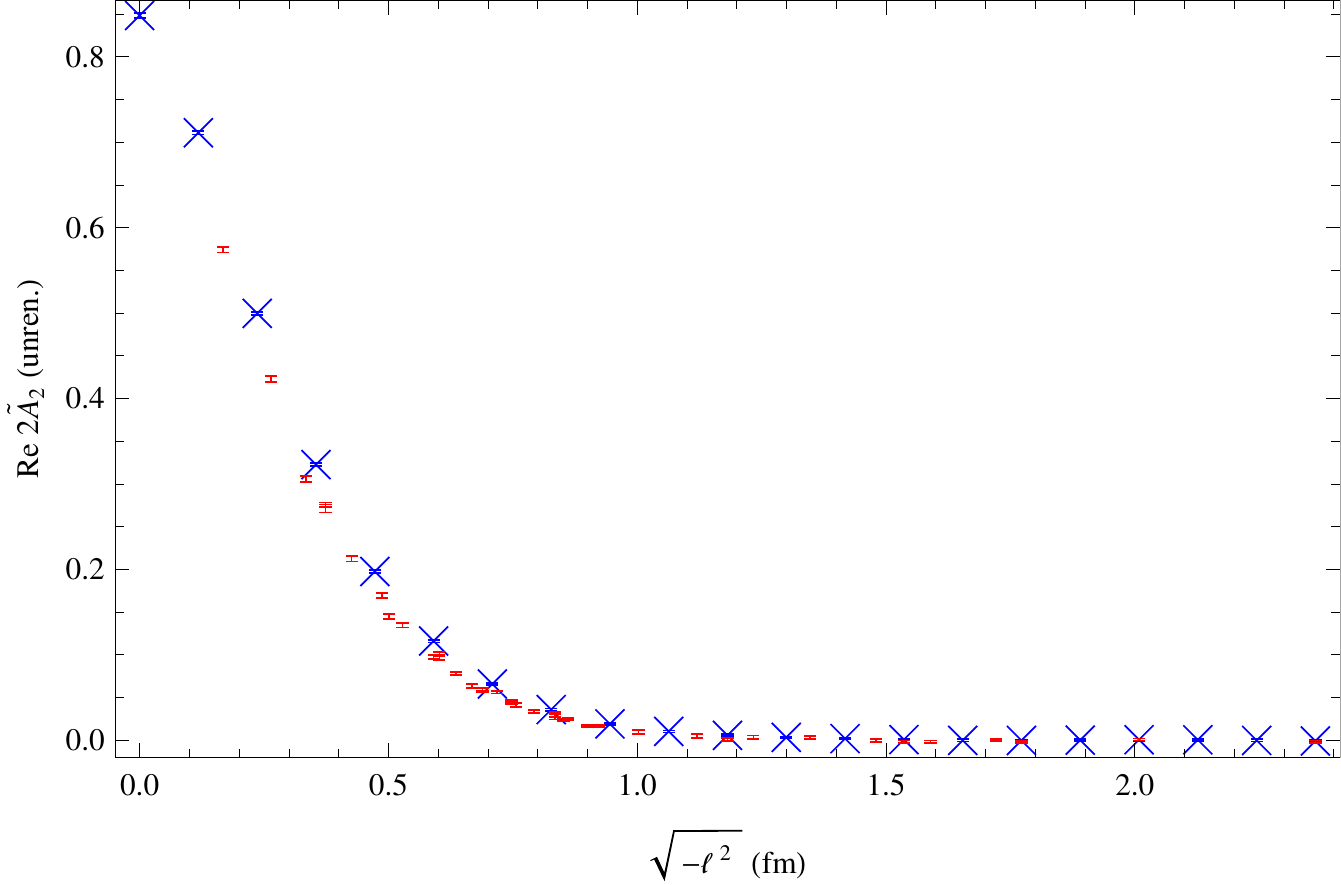}%
		}\par\vspace{1ex}%
	\caption[Unrenormalized Amplitudes]{%
		Unrenormalized amplitude $2 \tAmp_{2,u-d}^\unren(\elll^2,0)$ obtained directly from the ratio $R_{\gamma_\Eu{4},u-d}(\vect{P},\mathcal{C}^\lat_\elll)$, for $\vect{P}=\vect{0}$ on the coarse-m050 lattice. For clarity, we have averaged over link paths of the same shape (equivalent paths under $\mathrm{H}(4)$ transformations, see section \ref{sec-opmix}). Link paths coinciding with the lattice axes are marked with a blue cross; the red error bars belong to link paths at oblique angles.\\
		\subref{fig-lpsurface-re} gauge links evaluated on the HYP smeared lattice, \\			
		\subref{fig-lpsurface-im} gauge links evaluated without smearing.%
		\label{fig-unrenamp}
		}
\end{figure}

\begin{figure}[tbp]
	\centering%
	\begin{overpic}[width=0.85\textwidth,clip=true]{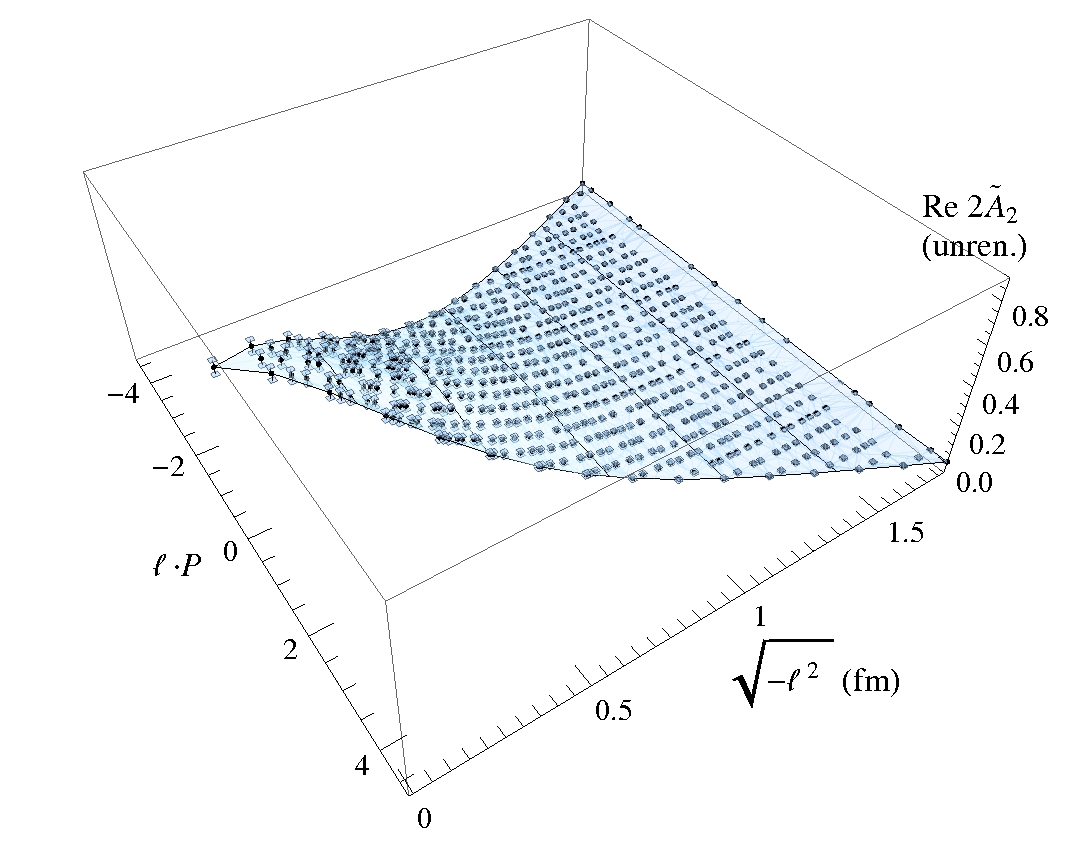}
		\put(0,73){\subfloat[][]{\label{fig-lpsurface-re}\hphantom{(m)}}}
	\end{overpic}\par
	\begin{overpic}[width=0.85\textwidth,clip=true]{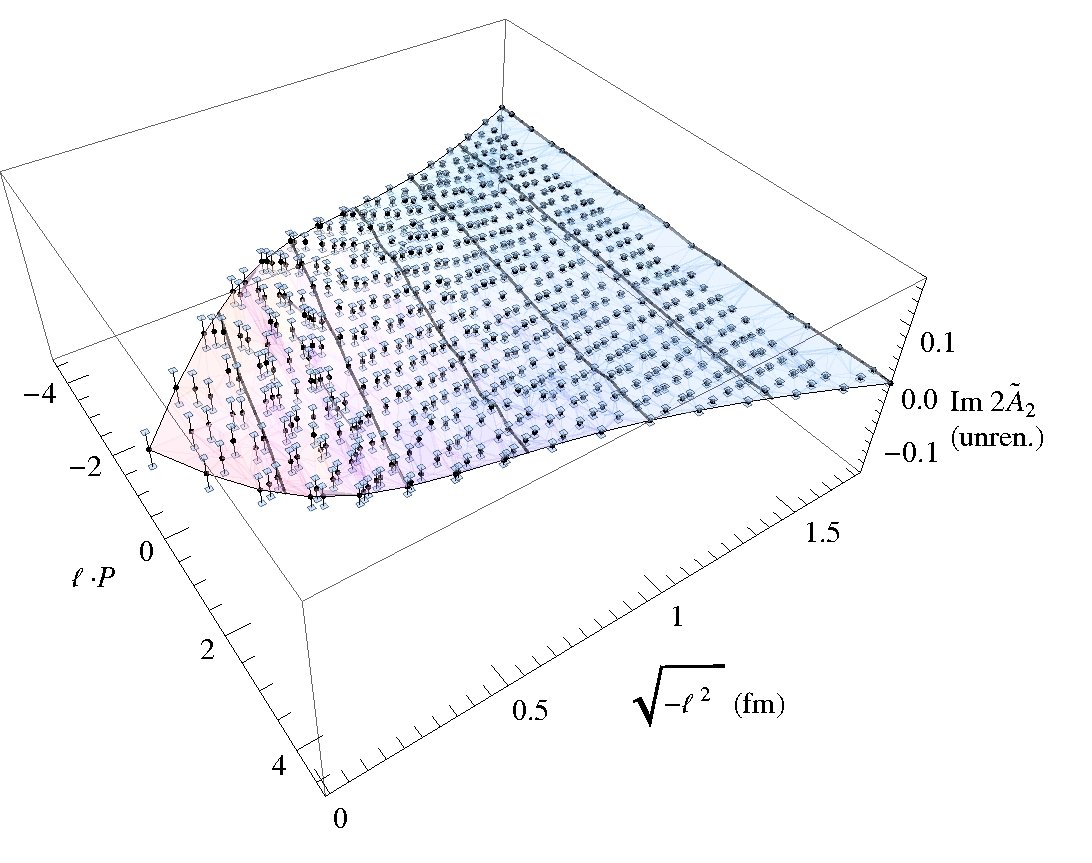}
		\put(0,73){\makebox{\subfloat[][]{\label{fig-lpsurface-im}\hphantom{(m)}}}}
	\end{overpic}\par
	\caption[3D plots]{%
		The unrenormalized amplitude $\tAmp_{2}^\unren(\elll^2,\elll\tcdot P)$ obtained directly from the ratio $R_{\gamma_\Eu{4},u-d}(\tau,\vect{P},\mathcal{C}^\lat_\elll)$ using the sequential propagators with $\vect{P}=(-1,0,0)\times2\pi/L$ on the coarse-m050 ensemble and applying HYP smearing to the gauge fields. Statistical error bars are shown as small squares floating above and below the interpolating surface.\\
		\subref{fig-lpsurface-re}~real part,
		\subref{fig-lpsurface-im}~imaginary part.%
		\label{fig-lpsurface}
		}
\end{figure}

Let us also have a look at the results for $R_{\gamma_\Eu{4}}(\vect{P},\mathcal{C}^\lat_\elll)$ at $\vect{P}=(-1,0,0)\times2\pi/L$. Here we can calculate the amplitude $\tAmp_{2}^\unren(\elll^2,\elll\tcdot P)$ at non-zero values of $\elll \tcdot P$. However, since we cannot implement link-paths with a Minkowski-time component $\elll^0 \neq 0$, we are confined to the region
\begin{equation}
	\elll^2 = - \vect{\elll}^2 \leq 0, \qquad
	| \elll \tcdot P | \leq |\vect{P}|\, \sqrt{-\elll^2}\ .
	\label{eq-accessiblelp}
\end{equation}
The above equation cuts out a wedge-shaped region in the $(\elll^2,\elll\tcdot P)$-plane, compare Figure~\ref{fig-lppoints}. Only this region is accessible to us on the lattice. The wedge has an opening angle proportional to $\vect{P}$, i.e., our investigations are limited by the highest nucleon momentum available to us.
In order to be able to explore the accessible region thoroughly, we choose additional gauge links. Restricting ourselves to $\vect{\elll}_2,\vect{\elll}_3\geq 0$, we have determined a set of gauge paths such that the $(\elll^2,\elll \tcdot P)$-plane is densely covered, see Fig.~(\ref{fig-lppoints}).  Performing the analysis for this extended set of over 1000 link paths on the HYP smeared ensemble results in the plots of Fig.~\ref{fig-lpsurface}. The dominant feature of the real part is the Gaussian-like decay with $\sqrt{-\elll^2}$. The $\elll \tcdot P$-dependence is rather weak. The situation is different for the imaginary part. Here the amplitude fulfills within statistics nicely the constraint $\myIm\,\tAmp_{2}^\unren(\elll^2,\elll\tcdot P) = -\myIm\,\tAmp_{2}^\unren(\elll^2,-\elll\tcdot P)$ following from eq.\,(\ref{eq-tampconstraint}). The $\elll \tcdot P$-dependence is related to the $x$-dependence of the \TMDs, and we will be concerned with it in detail in section~\ref{sec-xdep}.

\subsection{Restrictions from the Euclidean Lattice}
\label{sec-euclidrestric}

Figures \ref{fig-lppoints} and \ref{fig-lpsurface} give us a vivid impression of the limited range of $\elll^2$ and $\elll \tcdot P$ accessible on the Euclidean lattice according to eq.\,\ref{eq-accessiblelp}. What are the respective implications for the calculation of \TMDs? From the formalism developed in section \ref{sec-amplstraighttmds} we have learned that the Fourier transform of our amplitudes $\tAmp_i(\elll,\elll \tcdot P)$ to momentum space always involves an integral over $\elll^2$ from $-\infty$ to $0$, and an integral with respect to $\elll \tcdot P$ over the whole real axis, as in, e.g., eq.\,(\ref{eq-f1fromtildeA2}). Clearly, the lattice calculations cannot provide data in the whole integration range. 

Nevertheless, the amplitudes $\tAmp_i(\elll,\elll \tcdot P)$ at $\elll \tcdot P = 0$ give access to the first Mellin moments we have introduced in eq.\,(\ref{eq-deffirstmellin}). As we will show in more detail in section \ref{sec-mellin}, we can carry out the Fourier transformation with respect to $\elll^2$ and get first Mellin moments such as $f_1^{(1)}(\vprp{k}^2) \equiv \Phi^{[\gamma^+](1)}(\vprp{k};P,S)$. The first Mellin moment contains information about the probability densities of quarks with respect to transverse momentum $\vprp{k}$ alone. For example, in the case of $f_1^{(1)}(\vprp{k}^2)$, we obtain the difference between the unpolarized quark and antiquark density. We can also learn something from the observable $\elll\tcdot P$-dependence of the amplitudes from the lattice, if we are willing to make additional assumptions, see section \ref{sec-xdep}.

However, without further assumptions, we cannot directly determine the $x$-dependence of quark densities. In particular, it is interesting to observe that PDFs, i.e., the $\vprp{k}$-integrated distributions, are inaccessible to us. For example, from equation\,(\ref{eq-f1fromtildeA2}), we get\footnote{We remind the reader that, without special precautions, $f_1(x)$ is strictly speaking not the $\vprp{k}$-integral of $f_1(x,\vprp{k})$, see the discussion starting in section \ref{sec-corr}.}
\begin{equation}
	\text{``  }\int d^2 \vprp{k}\ f_1(x,\vprp{k})\text{ ''} = f_1(x) = 
	\int \frac{d(\elll \cdot P)}{2\pi}\ e^{-i (\elll \cdot P) x}\ 2\,\tilde{A}_2(\elll^2,\elll \tcdot P) \Big\vert_{\elll^2=0}\ .
\end{equation}
Clearly, we have no freedom to vary $\elll \tcdot P$ for $\elll^2=0$ on the lattice. Nevertheless, it is possible to calculate Mellin moments of PDFs, which corresponds to an expansion in terms of local operators.

\section{Renormalization of the Gauge Link}
\label{sec-WilsonLineRen}

As mentioned before, the differences between Fig. \ref{fig-unrenamp-sm} and \ref{fig-unrenamp-unsm} indicate a strong renormalization scheme dependence of our operators. What are the renormalization properties of the Wilson line?

\subsection{The Wilson Line in the Continuum}
\label{sec-wlinerencont}

Let $\mathcal{C}$ be a continuously differentiable (``smooth''), non-overlapping contour of total length $l[C]$. In the continuum, the Wilson line along such a path is renormalized according to \cite{Dotsenko:1979wb,Arefeva:1980zd,Craigie:1980qs,Dorn:1986dt}
\begin{equation}
	\Wlineren{\mathcal{C}} = \renZ_z^{-1} \exp\left( -\delta m\, l[\mathcal{C}] \right) \Wline{\mathcal{C}}\ .
	\label{eq-linkrencont}
\end{equation}
Here $\renZ_z$ and $\delta m$ are renormalization constants.\footnote{This formula can be derived on very general grounds using the auxiliary $z$-field method \cite{Gervais:1979fv} and BRS transformations \cite{Becchi:1974md}. Wave function renormalization of the auxiliary field $z$ is the origin of the renormalization constant $\renZ_z^{-1}$.} The exponential factor in the equation above is an example of a \terminol{power divergence}, i.e., an ultraviolet divergence which behaves like a power of the renormalization scale. For example, for the regularization prescription $1/x^2 \rightarrow 1/(x^2+a^2)$, having an inherent renormalization scale $\sim a^{-1}$, Dorn \cite{Dorn:1986dt} finds at one loop order
$\delta m \propto g^2 / a + \mathcal{O}(g^4)$. In lattice gauge theory, which corresponds to a cutoff scheme with cutoff scale $a^{-1}$, perturbation theory gives a result of the same form, see section \ref{sec-perturbren} below. The power divergence also appears in other schemes, such as the Pauli-Villars scheme \cite{Pauli:1949zm} and cutoff schemes in general. In dimensional regularization, $\delta m$ is zero, but \terminol{renormalon ambiguities} appear, compare, e.g., Ref.~\cite{Pineda:2005kv}.

Each ``disruption'' in the smooth Wilson line gives rise to another renormalization factor. Consider Wilson lines with a finite number $n_\text{cusps}[\mathcal{C}]$ of cusps (i.e., points where the contour is not continuously differentiable). Then the generalized formula for the renormalization of the Wilson line reads
\begin{equation}
	\Wlineren{\mathcal{C}} = \renZ_z^{-1} \exp\left( -\delta m\, l[\mathcal{C}] - \sum_{i=1}^{n_\text{cusps}[\mathcal{C}]} \nu(\theta_i) \right) \Wline{\mathcal{C}}\ ,
	\label{eq-linkrencontcor}
\end{equation}
where the renormalization constants $\nu(\theta_i)$ depend on the opening angles $\theta_i$ of the cusps. On the other hand, for the color trace of a closed contour $\mathcal{C}_\text{loop}$, the multiplicative renormalization with $\renZ_z^{-1}$ does not appear:
\begin{equation}
	\Tr_c\ \Wlineren{\mathcal{C}_\text{loop}} = \exp\left( -\delta m\, l[\mathcal{C}_\text{loop}]  - \sum_{i=1}^{n_\text{cusps}[\mathcal{C}_\text{loop}]} \nu(\theta_i) \right)\ \Tr_c\ \Wline{\mathcal{C}_\text{loop}}\ .
	\label{eq-linkrenloop}
\end{equation}
For the operator $O_{\GammaOp}(\elll)$ defined in eq.\,(\ref{eq-corrtilde}), renormalization factors arise at the end points of the Wilson line from quark field renormalization and the quark -- gauge link vertex:
\begin{equation}
	O^\ren_{\GammaOp}(\elll) = \underbrace{\renZ_\psi^{-1}\, \renZ_{(\psi z)}^2\, \renZ_z^{-1}}_{\displaystyle \renZ^{-1}_{\psi,z}}\ \exp\left( -\delta m\, l[\mathcal{C}]  \right)\ O_{\GammaOp}(\elll)\ . 
	\label{eq-opren}
\end{equation}
For a straight gauge link, $l[\mathcal{C}]=\sqrt{-\elll^2}$, so the renormalized operator is of the form we have stated in eq.~(\ref{eq-renop}). It is of primary importance to get the length dependent renormalization with $\delta m$ under control, and we shall focus on this issue on the following pages. We will discuss the multiplicative renormalization factor $\displaystyle \renZ^{-1}_{\psi,z}$ in section \ref{sec-mellin}.

\subsection{Renormalization Conditions}

The renormalization constant $\delta m$ cannot be determined unambiguously without an additional \terminol{renormalization condition}. To see this, let us write all $a$-dependences explicitly, and let us make the replacement
\begin{equation}
	\delta m(a) \rightarrow \delta m_\text{old}(a) = \delta m(a) + \delta m_0
	\label{eq-renambig}
\end{equation}
in eq.\,(\ref{eq-linkrencont}). Then 
\begin{equation}
	\Wlineren{\mathcal{C}} = \renZ_z^{-1}(a)\ \exp\left( -\delta m(a)\, l[\mathcal{C}] - \delta m_0\, l[\mathcal{C}] \right)\ \Wline{\mathcal{C}}(a)\ .
	\label{eq-deltam0}
\end{equation}
The second $l$-dependent term is not $a$-dependent, so it can be absorbed in the definition of the renormalized Wilson line, and we get
\begin{equation}
	\WlineI{^\text{ren2}}{\mathcal{C}} = \renZ_z^{-1}(a)\ \exp\left( -\delta m(a)\, l[\mathcal{C}] \right)\ \Wline{\mathcal{C}}(a)
\end{equation}
in a new renormalization scheme labelled ``ren2''. Obviously $\delta m(a)$ is determined only up to a scale independent constant. First of all, we note that the dimensionless quantity $\Delta \hat m(a)$ defined by
\begin{equation}
	\Delta \hat m(a) \equiv a^2 \frac{d}{da} \delta m(a) = a \frac{d}{d \ln a} \delta m(a)
	\label{eq-Deltama}
\end{equation}
is free of the aforementioned ambiguity and thus adequate to specify results in a renormalization scheme independent way. For the comparison of ensembles with two different lattice spacings $a_1$ and $a_2$, it will be useful to approximate the derivative in the definition above by a finite difference. Here we choose to discretize the logarithmic derivative, i.e., the relative change in $a$:
\begin{equation}
	\Delta \hat m(a_1,a_2) \equiv \sqrt{a_1 a_2}\, \frac{\delta m(a_2) - \delta m(a_1)}{\ln(a_2/a_1)} \approx \Delta \hat m(a)\ .
	\label{eq-Deltamaa}
\end{equation}
To fix a value for $\delta m(a)$, we need to specify a renormalization condition, i.e., we must provide some piece of information that uniquely defines the numerical values of the renormalized Wilson line at lengths $l[\mathcal{C}] \gg a$. In sections \ref{sec-renstatqpot} and \ref{sec-martinelli}, we explore two such renormalization conditions. 


\subsection{Perturbative Link Renormalization}
\label{sec-perturbren}

\begin{figure}[tbp]
	\centering%
	\hfill
	\subfloat[][]{%
		\rule{0pt}{3cm}\includegraphics{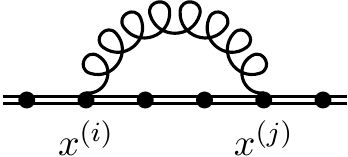}
		\label{fig-sunset}
		}\hfill%
	\subfloat[][]{%
		\rule{0pt}{3cm}\includegraphics{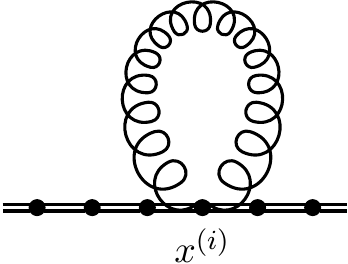}
		\label{fig-tadpole}
		}\hfill\rule{0pt}{0pt}\par%
	\caption[Sunset and tadpole]{%
		Leading loop diagrams for gauge links in lattice perturbation theory.\par
		\subref{fig-sunset}~Sunset diagram, \subref{fig-tadpole}~tadpole diagram.
		}
\end{figure}

It is instructive to follow Refs. \cite{Eichten:1989kb,Boucaud:1989ga,Maiani:1991az} and to use lattice perturbation theory at leading order to calculate the renormalization constant $\delta m$ for the discretized Wilson line
\begin{equation}
	\WlineClat{\mathcal{C}^\lat} = U(x^{(0)},x^{(1)})\, U(x^{(2)},x^{(3)})\, \cdots\, U(x^{(n-1)},x^{(n)})\ .
\end{equation}
Before we start, we need to introduce a lattice gluon propagator. To this end, the lattice gauge action $S^\lat_G[U]$ is expanded to lowest order in terms of $\Afield$-fields using $U(x,x+\hat \mu) = \exp(i g a \Afield_\muE(x) )$. Adding a gauge fixing term $S^\lat_\text{gf}[U]$ (necessary in perturbation theory) and expressing everything in momentum space, the action can be brought into the form
\begin{equation}
	S^\lat_G[U] + S^\lat_\text{gf}[U] = \frac{1}{2} \int_{-\pi/a}^{\pi/a} \frac{d^4 k}{(2\pi)^4} \sum_{\muE\nuE a}
	\tilde \Afield_\muE^a(k)\,\frac{1}{a^2} \left( \tilde D_{\muE\nuE}(k) - \left(1-\frac{1}{\xi}\right) \hat{\kappa}_\muE \hat{\kappa}_\nuE \right) \tilde \Afield_\nuE^a(-k)\ ,
	\label{eq-definversegluprop}
\end{equation}
where $\hat \kappa_\muE \equiv 2 \sin(a k_\muE /2 )$ and where $\xi$ is the gauge fixing parameter. 
The expression for $\tilde D_{\muE\nuE}(k)$ is specific to the gluon action used.
Now the gluon propagator $\tilde G_{\nuE\Eu{\rho}}^{ab} \equiv \delta_{ab} \tilde G_{\nuE\Eu{\rho}}$ is defined through the inversion
\begin{equation}
	\sum_\nuE \frac{1}{a^2} \left( \tilde D_{\muE\nuE}(k) - \left(1-\frac{1}{\xi}\right) \hat{\kappa}_\muE \hat{\kappa}_\nuE \right) \tilde G_{\nuE\Eu{\rho}}(k) = \delta_{\muE\Eu{\rho}}
\end{equation}
and the Fourier transformed propagator is defined by
\begin{equation}
	G^{a b}_{\muE \nuE}(x) = \delta^{a b} \int_{-\pi/a}^{\pi/a} \frac{d^4 k}{(2\pi)^4} \exp(i k_\Eu{\rho} x_\Eu{\rho})\ \tilde G_{\muE\nuE}(k)
	= \delta^{a b} \int_{-\pi}^{\pi} \frac{d^4 \hat{p}}{(2\pi\, a)^4} \exp(i \hat{p}_\Eu{\rho} x_\Eu{\rho}/a)\ \tilde G_{\muE\nuE}(\hat{p}/a)\ ,
\end{equation}
where we have introduced the dimensionless momentum variable $\hat{p}=ak$.
Let us now expand the gauge link in terms of $\Afield$-fields. For simplicity, we pick a path on one of the lattice axes, i.e., we set $x^{(n)} = n \hat{\mu}$. We get 
\begin{align}
	\dlangle \WlineClat{\mathcal{C}} \drangle &=  \Eins 
	\underbrace{- g^2 a^2 \sum_{i=0}^{n-2} \sum_{j=i+1}^{n-1} \Afield_{\muE}(x^{(i)}) \Afield_{\muE}(x^{(j)})}_{\displaystyle \text{``sunset'', see Fig.~\ref{fig-sunset}}} 
	\underbrace{- \frac{1}{2} g^2 a^2 \sum_{i=0}^{n-1} \Afield_{\muE}(x^{(i)})\,\Afield_{\muE}(x^{(i)})}_{\displaystyle \text{``tadpole'', see Fig.~\ref{fig-tadpole}}}
	 + \mathcal{O}(g^4)  \nonumber \\
	&=
	\Eins - \frac{1}{2} g^2 a^2 \sum_{i,j=0}^{n-1} G_{\muE \muE}^{a b}(x^{(j)}-x^{(i)})\, T^a\, T^b + \mathcal{O}(g^4)\ ,
\end{align}
where $T^a$, $T^b$ are the generators of $\mathrm{SU}(3)_c$. 
In terms of the gluon propagator in momentum space, we are able to perform the sums over $i$ and $j$, and we obtain
\begin{align}
	\dlangle \WlineClat{\mathcal{C}^\lat} \drangle 
	& =  \Eins - \Eins\,\frac{C_F g^2}{2} \int_{-\pi}^{\pi} \frac{d^4 \hat{p}}{(2\pi)^4}
	\frac{1- \cos(\hat{p}_\muE l[\mathcal{C}^\lat]/a)}{1-\cos(\hat{p}_\muE)} \frac{\tilde G_{\muE\muE}(\hat{p}/a)}{a^2} + \mathcal{O}(g^4) \ ,	
\end{align}
where $l[\mathcal{C}^\lat] = a n$ is the length of the gauge link and $\Eins C_F = \sum_a T^a T^a = (4/3)\Eins$.

Taking into account that $\tilde G_{\muE\muE}(\hat{p}/a)/a^2$ is $a$-independent, we take the derivative of the above expression with respect to $a$ in order to isolate the divergent part:
\begin{align}
	\frac{d}{d a} \dlangle \WlineClat{\mathcal{C}^\lat} \drangle 
	& = \frac{\Eins C_F g^2 l[\mathcal{C}^\lat]}{2 a^2 } \int_{-\pi}^{\pi} \frac{d \hat{p}_{\muE}}{(2\pi)^4}
	\frac{\sin(\hat{p}_\muE l[\mathcal{C}^\lat]/a)}{1-\cos(\hat{p}_\muE)} 
	\int_{-\pi}^{\pi} \frac{d^3 \hat{p}}{(2\pi)^3} \frac{\tilde G_{\muE\muE}(\hat{p}/a)}{a^2} + \mathcal{O}(g^4)\ ,	
\end{align}
where the three dimensional integration is carried out over all indices other than $\muE$. The integrand of the $\hat p_\muE$-integral is a representation of the $\delta$-functional, so that we get for $a \rightarrow 0$
\begin{align}
	\frac{d}{d a} \dlangle \WlineClat{\mathcal{C}^\lat} \drangle 
	& = \frac{\Eins C_F g^2 l[\mathcal{C}^\lat]}{2 a^2 } \int_{-\pi}^{\pi} \frac{d^3 \hat{p}}{(2\pi)^3} \frac{\tilde G_{\muE\muE}(\hat{p}/a)}{a^2}\Big\vert_{\hat{p}_\muE = 0} + \mathcal{O}(g^4)	\ .
\end{align}
Thus we find
\begin{equation}
	\dlangle \WlineClat{\mathcal{C}^\lat} \drangle 
	= \dlangle  \Eins - g^2 \left( \frac{l[\mathcal{C}^\lat]}{a} \frac{C_F}{2} \int_{-\pi}^{\pi} \frac{d^4 \hat{p}}{(2\pi)^3} 
	\frac{\tilde G_{\muE\muE}(\hat{p}/a)}{a^2} \Big\vert_{\hat{p}_\mu = 0}
	+ \text{non-divergent} \right) \drangle + \mathcal{O}(g^4)\ .
\end{equation}
From this we identify the renormalization constant $\delta m$:
\begin{equation}
	\delta m = \frac{g^2}{a}\ \underbrace{\frac{C_F}{2} \int_{-\pi}^{\pi} \frac{d^4 \hat{p}}{(2\pi)^3} 
	\frac{\tilde G_{\muE\muE}(\hat{p}/a)}{a^2} \Big\vert_{\hat{p}_\mu = 0}}_{\displaystyle \equiv X}\ +\ \text{const.}
	\label{eq-deltamperturb}
\end{equation}
Using eq.\,(\ref{eq-Deltama}), this corresponds to $\Delta \hat m = - g^2 X$. Indeed, we find that $\delta m$ grows linearly with the cutoff $a^{-1}$.

We can also take HYP smearing of the gauge fields into account. Following Ref.~\cite{DeGrand:2002va}, the effect of smearing can be implemented in our one-loop calculation by replacing the gauge field by a smeared one:
\begin{equation}
	\Afield_\muE(x) \rightarrow \Afield^\text{sm}_\muE(x) = \sum_{y,\nuE} h_{\muE\nuE}(y)\, \Afield_\nuE(x+y)\ .
\end{equation}
Here, the coefficients $h_{\muE\nuE}(y)$ are specific to the smearing procedure. Effectively, this just means that we have to replace the gluon propagator eq.\,(\ref{eq-deltamperturb}) according to 
\begin{equation}
	\tilde G_{\muE\nuE}(k) \longrightarrow \tilde G_{\muE \nuE}^{(\text{sm})}(k) = \sum_{\muE'\nuE'} \tilde h_{\muE\muE'}(-k)\, \tilde G_{\muE'\nuE'}(k)\, \tilde h_{\nuE'\nuE}(-k)\ .
\end{equation}
We have evaluated equation\,(\ref{eq-deltamperturb}) numerically, both for the smeared and the unsmeared case, using the following ingredients, which we have listed for convenience in section \ref{sec-pertingred}:
\begin{itemize}
	\item the inverse gluon propagator of the MILC action \cite{Bistro},
	\item the parameters $c_1$, $c_2$ and $c_3$ of the MILC action, where we have used the values for $u_0$ listed next to the unsmeared ensembles in Table~(\ref{tab-gaugeconfs}),
	\item the HYP smearing coefficients $\tilde h_{\muE\nuE}(k)$ \cite{DeGrand:2002va} with the values $\alpha_i$ specified in section \ref{sec-HYPsmear},
	\item for the coupling $g$, the bare lattice coupling listed in Table \ref{tab-gaugeconfs}.
\end{itemize}
The results are independent of the gauge fixing parameter $\xi$ and are listed in Table~(\ref{tab-gaugeconfs}). 
As an interesting qualitative feature, notice that $\Delta \hat m$ is much smaller on the smeared ensembles.
Note that the MILC gluon propagator reduces to the Wilson gluon propagator if we set $c_i = 0$. In this case, we get $X = -0.168487$, in agreement with Ref. \cite{Eichten:1989kb}.

Two improvements of the calculation discussed above are possible:
Firstly, instead of using the bare lattice coupling, there are more sophisticated ways to adjust the coupling $g$. Secondly, we could go to higher loop order. Both of these strategies have been followed, e.g., in Ref. \cite{Martinelli:1998vt}. 
However, our aim in the sections to follow will be to examine non-perturbative methods to fix $\delta \hat m$. What we have gained from the perturbative calculation is the confidence that our discretized operator will indeed need a length dependent renormalization of the same form as derived for the the continuum operator, and the insight that gluon self energy diagrams are responsible for the divergence proportional to the cutoff $a^{-1}$.

\newcommand{\hlraise}[1]{\smash{\raisebox{0.6em}{#1}}}
\begin{sidewaystable}[htbp]{
	\centering
	\small
	\renewcommand{\arraystretch}{1.1}
	\begin{tabular}{|ll||l|l|l||r|l|l||r||l||r|l|l|}
	\hline
	& & \multicolumn{3}{|c||}{perturbative} & 
	\multicolumn{3}{|c||}{Wilson line} & 
	\multicolumn{5}{|c|}{Wilson loops} \\ \hline
	& & & & & \multicolumn{3}{|c||}{at $l=0.5\units{fm}$} & & scaling & 
	\multicolumn{3}{|c|}{$V_\text{string}$} \\
	\hline
	ensemble \rule{0ex}{1.2em} & & 
	$u_0$ & $-X$ &$\Delta \hat m(a)$ & 
	$N_\text{cnf}$ & $-\delta \hat m$ & $\Delta \hat m(a_1,a_2)$ &
	$N_\text{cnf}$ & $\Delta \hat m(a_1,a_2)$ & 
	$\hat t_\mmin$ & $-\delta \hat m$ & $\Delta \hat m(a_1,a_2)$ \\
	\hline
	\hline
	(Wilson)       & smeared & & 0.04866 &        &     &                       &          &     &                 & & & \\
	coarse-m050    & smeared & & 0.04518 & 0.0659 &  34 & 0.27002(60)$^\dagger$ &          &     &                 & & & \\
	coarse-m030    & smeared & & 0.04516 & 0.0663 &     &                       &          &     &                 & & & \\
	super coarse   & smeared & & 0.04501 & 0.0694 & 188 & 0.30048(25)           &                      & 200 &                 &  4 & 0.1043(94) &\\ 
	coarse-m020    & smeared & & 0.04516 & 0.0665 &  74 & 0.27363(40)$^*$       & \hlraise{0.2186(14)} & 264 & \hlraise{0.227} &  6 & 0.1553(47)$^*$ & \hlraise{0.265(25)} \\
	fine           & smeared & & 0.04526 & 0.0637 &  30 & 0.24802(47)           & \hlraise{0.1850(17)} &  79 & \hlraise{0.186} &  8 & 0.1639(35)     & \hlraise{0.186(18)} \\
	super fine     & smeared & & 0.04535 & 0.0606 &   9 & 0.22268(47)           & \hlraise{0.1634(20)} &  15 & \hlraise{0.164} & 10 & 0.1578(17)     & \hlraise{0.145(11)} \\
	\hline                                                       
	(Wilson)       &  &        & 0.1685  &        &     &                       &          &     &                 & & &\\  
	coarse-m050    &  & 0.8707 & 0.1361  & 0.1987 & 184 & 0.51544(29)$^\dagger$ &          &     &                 & & &\\
	coarse-m030    &  & 0.8696 & 0.1360  & 0.1997 &     &                       &          &     &                 & & &\\
	super coarse   &  & 0.8558 & 0.1348  & 0.2077 & 178 & 0.54951(28)           &                      & 178 &                 &  3 & 0.361(60) &\\ 
	coarse-m020    &  & 0.8688 & 0.1359  & 0.2002 & 182 & 0.52189(29)$^*$       & \hlraise{0.4668(11)} & 182 & \hlraise{0.47}  &  4 & 0.397(35)$^*$ & \hlraise{0.48(17)} \\
	fine           &  & 0.8788 & 0.1368  & 0.1924 &  67 & 0.48417(31)           & \hlraise{0.3918(13)} &  67 & \hlraise{0.38}  &  4 & 0.382(10)     & \hlraise{0.341(94)} \\
	super fine     &  & 0.8881 & 0.1376  & 0.1840 &   9 & 0.44263(69)           & \hlraise{0.3458(26)} &   9 & \hlraise{0.32}  &  5 & 0.361(11)     & \hlraise{0.311(44)} \\
	\hline                     
	\end{tabular}\par\vspace{1ex}
	\renewcommand{\arraystretch}{1.0}
	}\caption{Renormalization constants for the Wilson line, obtained using several different methods. The values for $\Delta m(a_1,a_2)$ have been determined from two ensembles and are placed between the two corresponding lines. All errors in the table are purely statistical. From the comparison of smeared and unsmeared ensembles according to section \ref{sec-discerr}, we estimate a systematic uncertainty of about $\pm 0.024$ for the values marked with an asterisk (*), and about $\pm 0.041$ for the values marked with a dagger ($\dagger$). The number of configurations used for the studies with Wilson lines and Wilson loops is displayed in the columns $N_\text{cnf}$.}
	\label{tab-renconst}%
\end{sidewaystable}

\subsection{Wilson Loops: A Study on Multiple Scales}
\label{sec-wloopscaling}

A Wilson loop
\begin{equation}
	W(r,t;a) \equiv \frac{1}{N_c} \Tr_c\ \WlineClat{\mathcal{C}^\lat_{\hat r,\hat t}} 
\end{equation}
is the color trace of a closed gauge path $\mathcal{C}^\lat_{\hat r,\hat t}$ describing a rectangle of dimensions $r \times t$ (in physical units) on the lattice. The major advantage of using Wilson loops for the determination of the renormalization constant $\delta m(a)$ is their gauge invariance. 

Can we determine the renormalization constants from the scaling behavior of the Wilson loop?
According to the continuum formalism eq.\,(\ref{eq-linkrenloop})
\begin{equation}
	W^\ren(r,t) = \exp\left( - \delta m(a) l - 4 \nu_\prp(a) \right) W(r,t;a)\ ,
	\label{eq-wloopren}
\end{equation}
where we abbreviate $2(r+t) \equiv l$ and the renormalization constant for $90^\circ$ corners with $\nu_\prp(a)\equiv\nu(90^\circ;a)$. Demanding that $W^\ren(r,t)$ be the same at two different lattice spacings $a_1$ and $a_2$, we get
\begin{equation}
	\ln \left( \frac{\dlangle W(r,t,a_2) \drangle}{\dlangle W(r,t,a_1) \drangle} \right) =
	l \left\{\delta m(a_2)-\delta m(a_1)\right\} + 4\,\left\{\nu_\prp(a_2)-\nu_\prp(a_1)\right\}\ .
\end{equation}
Obviously, we can determine the difference of renormalization constants at different lattice spacings.
Let us define a quantity which can be expressed in terms of $\Delta \hat m(a_1,a_2)$ as introduced in eq.\,(\ref{eq-Deltamaa}):
\begin{equation}
	Y_\text{scal}(r,t;a_1,a_2)\equiv\frac{\sqrt{a_1 a_2}}{l\, \ln(a_2/a_1)}\ln \left( \frac{\dlangle W(r,t,a_2) \drangle}{\dlangle W(r,t,a_1) \drangle} \right) =
	\Delta \hat m(a_1,a_2) + \frac{4 \sqrt{a_1 a_2}}{l} \Delta \nu_\prp(a_1,a_2) \ ,
	\label{eq-scaling}
\end{equation}
where $\Delta \nu_\prp(a_1,a_2) \equiv \lbrace \nu_\prp(a_2)-\nu_\prp(a_1)\rbrace/\ln(a_2/a_1)$.

\begin{figure}[tbp]
	\centering%
	\subfloat[][]{%
		\label{fig-scaling-overview-one}%
		\includegraphics[clip=true]{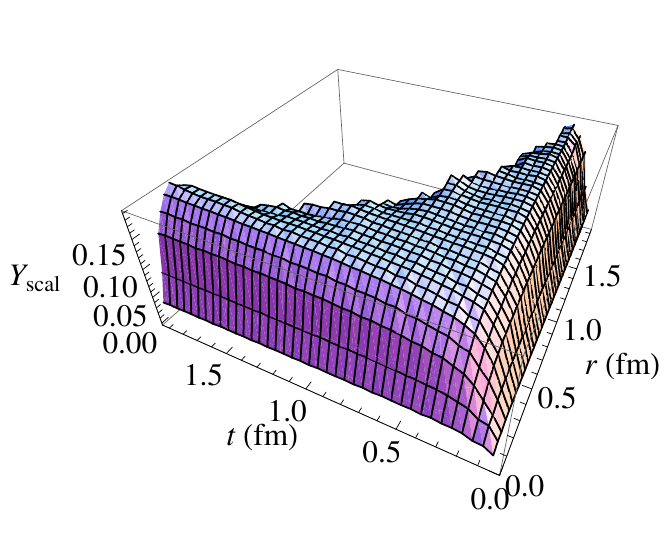}%
		}\hfill%
	\subfloat[][]{%
		\label{fig-scaling-overview-two}%
		\includegraphics[clip=true]{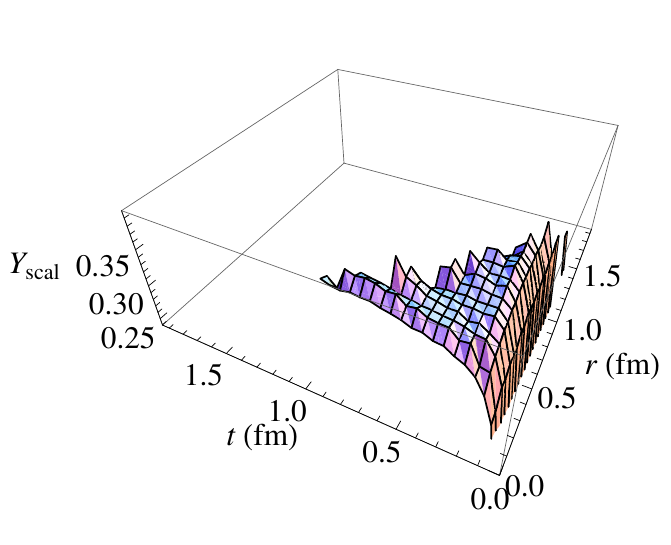}%
		}\\%
	\vspace{2em}
	\subfloat[][]{%
		\label{fig-scaling-yfits}%
		\includegraphics[clip=true]{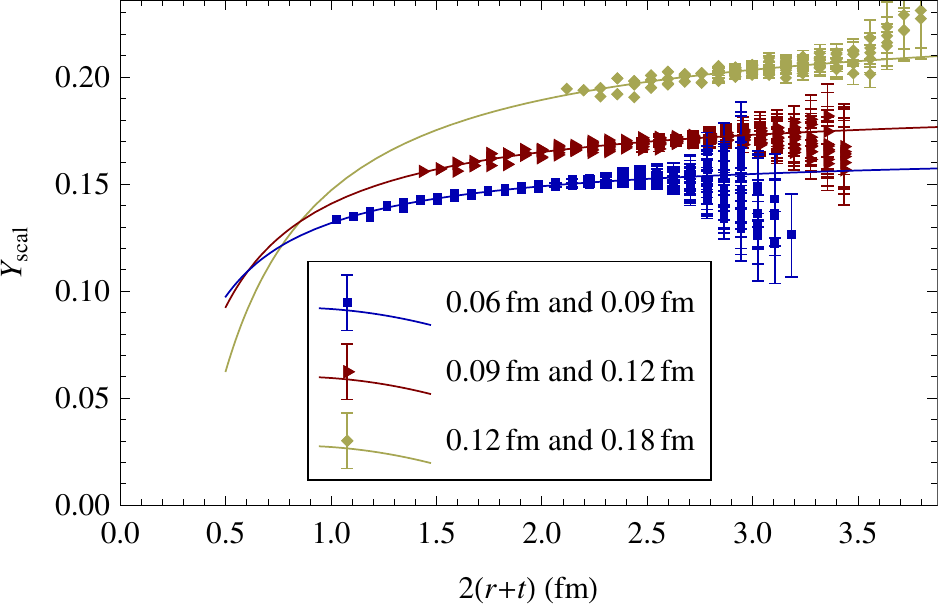}%
		}\par%
	\vspace{2em}
	\caption[scaling plots]{%
		\subref{fig-scaling-overview-one} and \subref{fig-scaling-overview-two}: $Y_\text{scal}(r,t;a_1,a_2)$ evaluated from the ratio of equally sized Wilson loops on the coarse-m020 and fine ensemble. We only include points with errors below $\pm0.02$. \subref{fig-scaling-overview-one}: with HYP smearing, \subref{fig-scaling-overview-two} without HYP smearing; statistics breaks down much earlier.	\par
		\subref{fig-scaling-yfits}: $Y_\text{scal}(r,t;a_1,a_2)$ from pairs of HYP-smeared ensembles at $\hat m_{u,d}= 0.4\hat m_s$. Selected data points fulfill $r,t \geq 3 a$ ($a$ taken from the larger lattice) and have statistical errors below $\pm0.02$. The fits are performed with the parametrization given in eq.~(\ref{eq-scaling}).%
		\label{fig-scaling}
		}
\end{figure}

To study the quantity $Y_\text{scal}(r,t;a_1,a_2)$ on the lattice, we have evaluated Wilson loops on ensembles which differ only in their lattice spacings. From the lattices listed in Table~\ref{tab-gaugeconfs}, we selected the supercoarse, coarse-m020, fine and superfine ensembles, which all have approximately the same physical strange quark mass and a light to strange quark mass ratio $\hat m_{u,d} = 0.4 m_s$.
Using planar Wilson loops, only integer dimensions $\hat r \times \hat t$ are available on the lattice, making it difficult to compute loops of the same physical size for different lattice spacings. To overcome this, we interpolate $\ln \dlangle W(r,t;a)\drangle$ linearly in the $(r, t)$-plane. Typical results for $Y_\text{scal}(r,t;a_1,a_2)$ are shown in Fig.~\ref{fig-scaling-overview-one} and \ref{fig-scaling-overview-two}. Clearly, for larger loops, $Y_\text{scal}(r,t;a_1,a_2)$ approaches a plateau value. 

We now evaluate $Y_\text{scal}(r,t;a_1,a_2)$ on a grid of points in the $(r,t)$-plane. We restrict ourselves to the region $r,t\geq 3\,\mathrm{max}(a_1,a_2)$, where the linear interpolation of $\ln \dlangle W(r,t;a)\drangle$ works reasonably well. Also, we reject points with statistical errors of more than $\pm0.02$. The results, plotted with respect to $l = 2(r+t)$, are shown for the smeared ensembles in Fig.~\ref{fig-scaling-yfits}. As they should, data points with the same $l$ coincide, even if they differ in $r$ and $t$. Using eq.\,(\ref{eq-scaling}), we can now determine $\Delta \hat m(a_1,a_2)$ and $\Delta \nu_\prp(a_1,a_2)$ from fits to data from pairs of ensembles with similar lattice spacing. (In this case, we take uniform fit weights for all input data.) The results for $\Delta \hat m(a_1,a_2)$ are given in the column labelled ``scaling'' in Table~\ref{tab-renconst}. We do not quote uncertainties. 

Notice that the numbers for $\Delta \hat m(a_1,a_2)$ we get are much larger than the values $\Delta m(a)$ determined perturbatively at similar lattice spacings. We conclude that our simple perturbative calculation gives very inaccurate results. However, some qualitative features agree: $\Delta \hat m$ increases with the lattice spacing, and smearing reduces $\Delta \hat m$ significantly.

In the following sections, we will discuss methods that allow us to fix the renormalization constant $\delta m$ for a given ensemble. The numbers we have obtained for $\Delta \hat m(a_1,a_2)$ from the scaling behavior of Wilson lines can serve as a valuable cross check of these methods.

\subsection{Renormalization with Wilson Lines}
\label{sec-martinelli}

\begin{figure}[tbp]
	\centering%
	\subfloat[][]{%
		\label{fig-YlineRaw}%
		\includegraphics[clip=true]{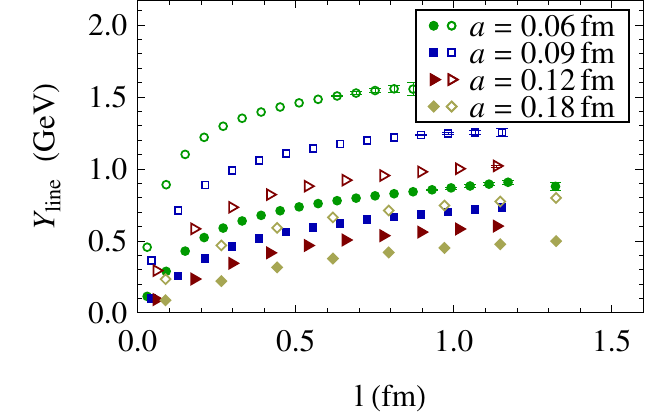}%
		}\hfill%
	\subfloat[][]{%
		\label{fig-YlineInfRen}%
		\includegraphics[clip=true]{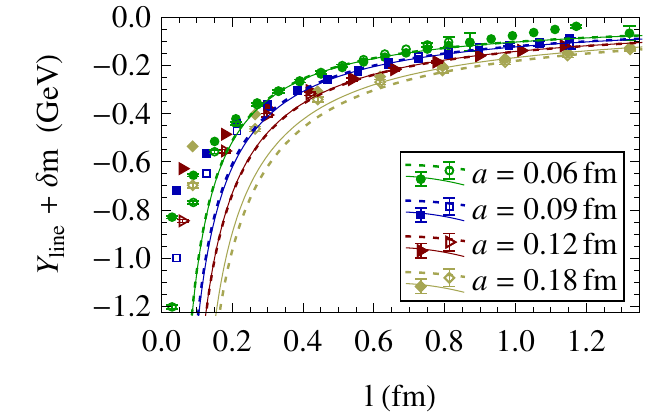}%
		}\par%
	\vspace{2em}
	\subfloat[][]{%
		\label{fig-YlinePtRen}%
		\includegraphics[clip=true]{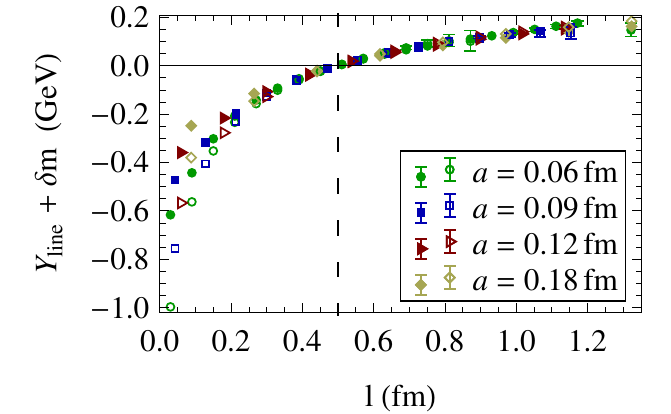}%
		}\hfill%
	\subfloat[][]{%
		\label{fig-YlineBazRen}%
		\includegraphics[clip=true]{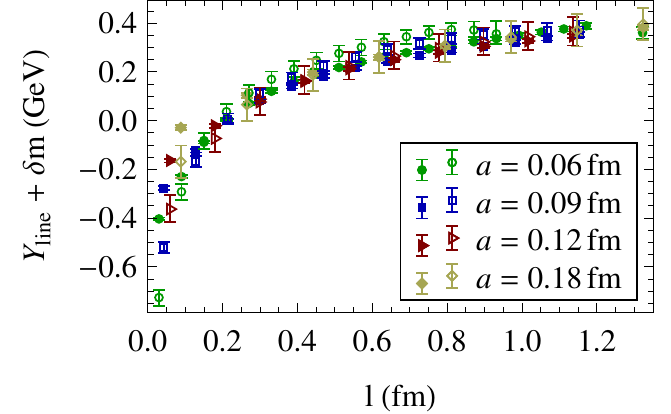}%
		}\par%
	\vspace{2em}
	\subfloat[][]{%
		\label{fig-YlineBazRen-sm}%
		\includegraphics[clip=true]{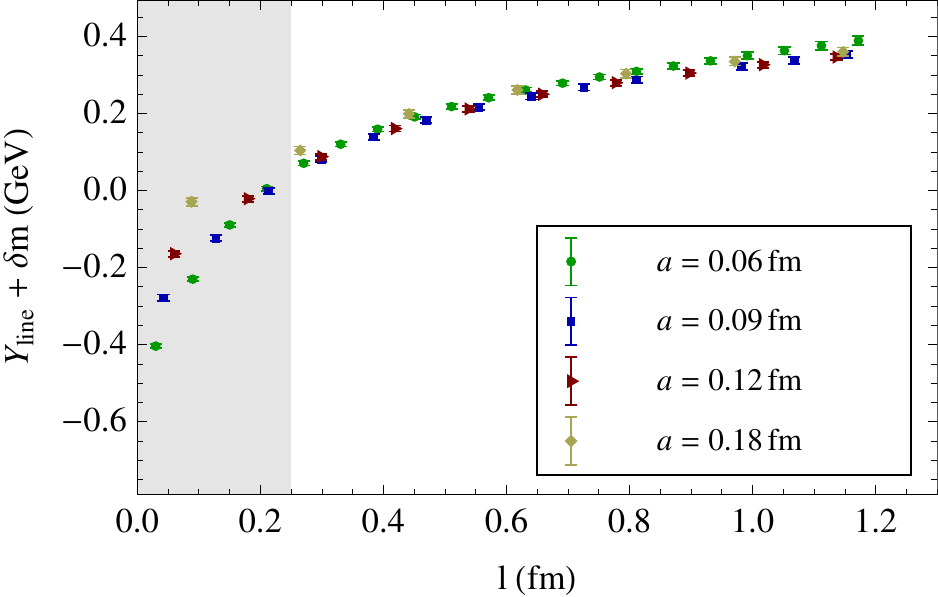}%
		}\par%
	\vspace{2em}
	\caption[Yline]{%
		The length derivative of the straight Wilson line in Landau gauge $Y_\text{line}(l)$ from the four ensembles at $\hat m_{u,d} = 0.4 \hat m_s$, with HYP smearing (filled symbols) and without smearing (open symbols).
		\subref{fig-YlineRaw} unaltered data, \subref{fig-YlineInfRen} fitted and renormalized based on an assumption about the asymptotic behavior, \subref{fig-YlinePtRen} renormalized by imposing a renormalization condition at $l=0.5\units{fm}$, \subref{fig-YlineBazRen} renormalization based on the string potential, see section \ref{sec-renstatqpot}, \subref{fig-YlineBazRen-sm} same as the previous item, but for the smeared ensembles only. The data in the shaded region is clearly affected by strong lattice cutoff effects.
		\label{fig-Yline}
		}
\end{figure}

Martinelli and Sachrajda \cite{Martinelli:1995vj,Crisafulli:1995pg} suggest to analyze Wilson lines along straight contours $\mathcal{C}_l$ of length $l$ as follows 
\begin{equation}
	Y_\text{line}(l) \equiv - \frac{d}{dl} \ln \Tr_c\, \dlangle \WlineC{\mathcal{C}_l} \drangle = 
	- \delta m - \frac{d}{dl} \ln \Tr_c\, \dlangle \Wlineren{\mathcal{C}_l} \drangle
	\approx - \frac{1}{a} \ln \frac{\Tr_c\, \dlangle \WlineClat{\mathcal{C}^\lat_{l+a/2}} \drangle}{\Tr_c\, \dlangle \WlineClat{\mathcal{C}^\lat_{l-a/2}} \drangle}\ .
	\label{eq-Ylinedef}
\end{equation}
The expectation values of open Wilson lines are not gauge invariant; the equation above is therefore only valid in the context of gauge fixed ensembles. Here we choose Landau gauge fixing, see section \ref{sec-gfix}. Our results for $Y_\text{line}(l)$ are plotted in Fig.~\ref{fig-YlineRaw}. We have selected data with errors of less than $0.02/a$ obtained with straight link paths on the axes of length $l\leq L/2$. According to the above equation, renormalization of the Wilson lines manifests itself as a shift of $Y_\text{line}(l)$ by a value $\delta m$, which can be different for each ensemble. Thus we should shift the data points of each ensemble up or down until they all agree with each other. The agreement can be optimized at a certain value of $l$, i.e., at a certain renormalization point. 

For example, a suggestion in Ref. \cite{Martinelli:1995vj} is to adjust the curves at $l \rightarrow \infty$, imposing the renormalization condition $\lim_{l \rightarrow \infty} \frac{d}{dl} \ln \Tr_c\,\Wlineren{\mathcal{C}_l} = 0$. Note that the authors point out that the existence of this limit is theoretically not proven. Fitting the form $a Y_\text{line}(l) = - \delta \hat m - \gamma \ln(1 + a/(l-a/2))$ (as suggested in Ref. \cite{Crisafulli:1995pg}) to data with $l \geq 5a$, we arrive at Fig.~\ref{fig-YlineInfRen}. In this plot, we have already offset the data by $\delta m$, so that the plotted fit functions approach zero at infinity. The fit is not very stable, and the values $\Delta m$ we obtain from this method are not in good agreement with our scaling analysis in the previous section. 

Therefore, let us try another renormalization point, and demand $\frac{d}{dl} \ln \Tr_c\,\Wlineren{\mathcal{C}_l} = 0$ at $l=0.5\units{fm}$, a scale at which we have accurate data. Using simple linear interpolation between data points at $l$-values above and below $0.5\units{fm}$, we can determine $\delta \hat m$ and arrive at Fig.~\ref{fig-YlinePtRen}. Numbers are listed in the column labelled ``Wilson line at $l=0.5\units{fm}$'' in Table~\ref{tab-renconst}. We have also listed the resulting values $\Delta \hat m(a_1,a_2)$ between adjacent ensembles, and find very good agreement with the scaling analysis of the previous section.

In Fig.~\ref{fig-YlineBazRen}, we show $Y_\text{line}(l)$ with offsets $\delta m$ determined from renormalization with the string potential, as described in the following section. In our case, this method obviously does not perform very well on the unsmeared ensembles: As we shall see, this problem is not unexpected and could be resolved with more input configurations. Therefore, we restrict ourselves to the smeared ensembles in Fig~\ref{fig-YlineBazRen-sm}.

Looking, in particular, at Fig.~\ref{fig-YlinePtRen} and Fig.~\ref{fig-YlineBazRen-sm}, we find that the data of different ensembles can be brought to almost perfect agreement for Wilson lines of lengths greater than about $0.3 \units{fm}$. This observation corroborates our notion that the corresponding lattice operators approximate continuum operators with straight Wilson lines, and thus share their renormalization properties. On the other hand, it is obvious that our data exhibits substantial lattice cuttoff effects for Wilson lines of lengths $\lesssim 0.25\units{fm}$. For such short Wilson lines the continuum inspired renormalization prescription fails: the data points of different ensembles do not just differ by a constant shift $\delta m$. Hence we will discard data for operators with links shorter than $0.25\units{fm}$ in our analysis of nucleon observables. 

Moreover, we learn that is important to impose a renormalization prescription which is sensitive to the data in the region where statistical and systematic errors are both small: The gauge link should be long enough, so that finite-$a$-effects are small, and yet not too long, so that finite volume effects and statistical uncertainties are acceptable.

Finally, we draw again attention to the fact that different renormalization conditions may produce values for $\delta m$ that can differ by a renormalization scale independent constant. This is the reason why the plots in Fig.~\ref{fig-YlineInfRen}--\ref{fig-YlineBazRen} feature different offsets in the ordinates.

\subsection{Discretization Errors estimated with Wilson Lines}
\label{sec-discerr}

A hint about the size of lattice cutoff effects can be obtained from the comparison of smeared and unsmeared ensembles. To this end, let us take a look at
\begin{equation}
	\Delta[ \delta m] \equiv \left[ Y^\text{sm}_\text{line}(l_1) - Y^\text{unsm}_\text{line}(l_1) \right] - \left[ Y^\text{sm}_\text{line}(l_2) - Y^\text{unsm}_\text{line}(l_2) \right]\ .
\end{equation}
In each of the square brackets we calculate the difference between the renormalization constants $\delta m$ on the smeared and unsmeared ensembles, albeit at two different renormalization points $l_1$ and $l_2$. If we had no lattice cutoff effects, we should obtain exactly the same result at both renormalization points, and $\Delta[ \delta m]$ would be zero. 
To be able to evaluate $Y^\text{sm}_\text{line}(l)$ for any $l$ on a given ensemble, we generate spline interpolations.
Setting $l_1 = 0.25 \units{fm}$ and $l_2 = 0.75 \units{fm}$, we obtain 
$\Delta[ \delta m ] = 0.027(11) \units{GeV}$,
$\Delta[ \delta m ] = 0.0308(26) \units{GeV}$,
$\Delta[ \delta m ] = 0.0384(14) \units{GeV}$,
$\Delta[ \delta m ] = 0.04711(56) \units{GeV}$
on the superfine, fine, coarse-m020 and supercoarse ensembles, respectively. These numbers suggest that $\Delta[\delta m]$ extrapolates linearly to $0$ at $a=0$.

For our calculations of nucleon structure in later sections, we will work with data obtained on the coarse ensemble with quark separations ranging from about $0.25\units{fm}$ to $1.2\units{fm}$. So let us set $l_1 = 0.25\units{fm}$ and $l_2 = 1.2 \units{fm}$. 
Within our limited statistics, we cannot determine $\Delta[ \delta m ]$ with reasonable precision on the two finest lattices for these choices of $l_1$ and $l_2$. However, on the coarse-m020 ensemble, we obtain $\Delta[ \delta m] = 0.0393(88) \units{GeV}$, or, in lattice units, $\Delta[ \delta \hat m] = 0.0238(54)$. This number will serve as an estimate of systematic uncertainties from lattice cutoff effects. We also calculate this error on the coarse-m050 ensemble, where we obtain $\Delta[ \delta \hat m] = 0.0405(56)$. 
The uncertainties thus obtained should not be ignored; they are in general much larger than uncertainties from other sources. 


\subsection{Renormalization based on the Static Quark Potential}
\label{sec-renstatqpot}

\subsubsection{The Principle}
\label{sec-renstatqpotprinciple}

The two Wilson lines running parallely in $t$ direction in a Wilson loop can be interpreted as propagators of static quarks $\Quark$ and $\bar \Quark$, compare section \ref{sec-auxfields} and textbooks such as Refs. \cite{Roth,DeGrand:2002va}. The other two Wilson lines are then regarded as a gauge invariant source of a $\Quark\bar \Quark$ pair separated by a distance $r$. Just as we did it for the nucleon in section \ref{sec-transfmat}, we can extract the energy of the $\Quark\bar \Quark$ system from the slope of the logarithmic correlator at large $t$. We call this energy \terminol{static quark potential}
\begin{equation}
	V(r;a) \equiv - \lim_{t \rightarrow \infty} \frac{\partial}{\partial t} \ln \dlangle W(r,t;a) \drangle \ .
\end{equation}
Replacing $V(r;a)$ and $W(r,t;a)$ by renormalized quantities, and substituting eq.\,(\ref{eq-wloopren}), we obtain
\begin{align}
	V^\ren(r) = 2\, \delta m(a) - \lim_{t \rightarrow \infty} \frac{\partial}{\partial t} \ln  \dlangle W(r,t;a) \drangle 
	& = 2\, \delta m(a) + V(r;a) \ , \\
	&\ \hat{=}\ 2\, m^\ren_Q - E_\text{bind}(r)\ ,
	\label{eq-potren}
\end{align}
which has an interpretation in terms of a binding energy $E_\text{bind}(r)$ and the mass of the static quark $m^\ren_Q$ in the last line of the above expression. We observe that a change in the renormalization condition for the Wilson line $\delta m(a) \rightarrow \delta m(a)+\delta m_0$ is equivalent to a finite quark mass renormalization $m^\ren_Q \rightarrow m^\ren_Q + \delta m_0/2$.

We can fix $\delta m(a)$ by imposing a renormalization condition on $V^\ren(r)$. We want the condition to be insensitive to our determination of the lattice scale $a$. One successful approach \cite{Cheng:2007jq,Petrov:2007ug,BazavovPriv,Bazavov:2009zn} makes use of the observation that the static potential at large distances $r$ can be very well described by the formula 
\begin{equation}
	V^\ren_\text{string}(r) = \sigma r - \frac{\pi}{12\,r} + C^\ren
	\label{eq-potstring}
\end{equation}
derived in Ref. \cite{Luscher:1980fr}. The formula\footnote{The excellent agreement of $V^\ren_\text{string}(r)$ with lattice data at large $r$ is shown in Fig.~1 of Ref.~\cite{Cheng:2007jq}.} originates from a small-$\hbar$ approximation based on the assumption that the Wilson loop satisfies the equation of motion of a quantized string \cite{Nambu:1978bd}. Note that an exact result for the string potential is also available \cite{Arvis:1983fp}, which shows that the string model certainly cannot describe the static quark potential for smaller values of $r$. In the expression above, the \terminol{string tension} $\sigma$ is a fundamental constant, while $C^\ren$ is the renormalization constant related to the self energy of the Wilson loop. It turns out that the lattice data can be fitted very well (in terms of dimensionless quantities) to a function of the form
\begin{equation}
	a V(r;a) \approx \hat V_\text{fit}(r;a) = \hat \sigma_a \hat r - \frac{\alpha_a}{\hat r} + \hat C_a
	\label{eq-potfit}
\end{equation}
with the fit parameters $\hat \sigma_a$, $\alpha_a$ and $\hat C_a$. Let us set the lattice potential and the string potential equal at some fixed $r=r_\text{match}$, which we choose large but still in the range where lattice data is available. Taking $\sigma = \hat \sigma_a / a^2$ and putting together eqns.~(\ref{eq-potren}), (\ref{eq-potstring}) and (\ref{eq-potfit}), we get
\begin{equation}
	\hat V^\ren_\text{fit}(r_\text{match}) \mathop{=}^! \hat V^\ren_\text{string}(r_\text{match}) \quad \Rightarrow \quad
	2\, \delta \hat m(a) = \frac{a}{r_\text{match}}\left(\alpha_a - \frac{\pi}{12}\right) + (a C^\ren - \hat C_a)\ .
\end{equation}
Here the lattice spacing $a$ appears explicitly in combination with $r_\text{match}$, which we must specify in physical units in order to implement the renormalization condition scale-in\-de\-pen\-dent\-ly. It is useful to eliminate this appearance of $a$ in favor of the \terminol{Sommer scale} $r_0$, which is defined by the condition \cite{So93}
\begin{equation}
	\left. r^2 \frac{\partial V(r;a)}{\partial r} \right|_{r=r_0} = 1.65\ .
	\label{eq-sommerscale}
\end{equation}
Effective potentials reproducing the experimental spectrum of heavy quarkonia ($b\bar b$ bound states, etc.) show that $r_0 \approx 0.5 \units{fm}$.\footnote{Note that the MILC collaboration prefers to modify the condition above, replacing the constant $1.65$ by $1$. The corresponding scale $r_1=0.317(10)\units{fm}$ \cite{Au04} has been used to determine the lattice spacings quoted in Table \ref{tab-gaugeconfs}.} Using our parametrization eq.~(\ref{eq-potfit}), the Sommer scale in lattice units is $\hat r_0 = \sqrt{(1.65-\alpha_a)/\hat \sigma_a}$. Substituting $a=r_0/\hat r_0$ we finally obtain
\begin{equation}
	2\, \delta \hat m(a) = \left(\frac{r_0}{r_\text{match}}\right) \sqrt{\frac{\hat \sigma_a}{1.65-\alpha_a}} \left(\alpha_a - \frac{\pi}{12}\right) - \hat C_a + a C^\ren \ .
	\label{eq-stringren}
\end{equation}
In order to be able to use this equation to determine $2\, \delta \hat m(a)$, we need to make a choice regarding $C^\ren$. This choice is part of the definition of our renormalization condition and is related to the arbitrary shift $\delta m_0$ in eq.\,(\ref{eq-deltam0}). We will follow Ref. \cite{Cheng:2007jq} and choose $C^\ren = 0$. The right hand side is now free of any explicit $a$-dependence. Together with the convention $r_\text{match}/r_0 = 1.5$ of Ref. \cite{Cheng:2007jq}, the equation above forms a practical and robust renormalization prescription. 

The approach discussed above may be interpreted in the following way:
Matching to the string potential, we anticipate that the static potential will converge to a straight line $\sigma r + C^\ren$ for large $r$.\footnote{Just from the symmetry of the Wilson loop under exchange of temporal and spatial extent, it can be shown that the Wilson loop is bounded from above by a linear function in $r$ \cite{Seiler:1978ur}.} By setting $C^\ren = 0$, we demand that this line run through the origin, see the dashed straight line $\sigma r$ in Fig.~\ref{fig-StringRenSm}. 


\subsubsection{Implementation and Results}
\label{sec-implstaticpot}

\begin{figure}[tbp]
	\centering%
	\hfill
	\subfloat[][]{%
		\includegraphics[scale=0.5]{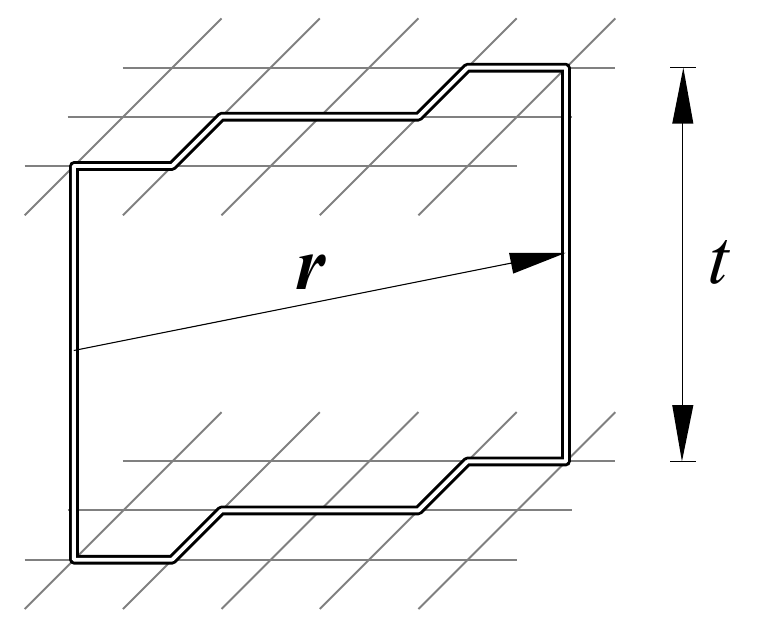}
		\label{fig-obliquewloop}
		}\hfill%
	\subfloat[][]{%
		\rule{0pt}{2.8cm}\includegraphics{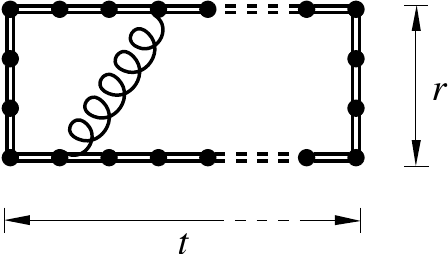}
		\label{fig-wloopglueexch}
		}\hfill\rule{0pt}{0pt}\par%
	\caption[Rectangular Wilson Loop]{%
		Rectangular Wilson loop in the calculation of the static quark potential.\par
		\subref{fig-obliquewloop}~Wilson loop with step-like, spatial connections at an oblique angle.\par
		\subref{fig-wloopglueexch}~Gluon exchange between the temporal Wilson lines in leading order lattice perturbation theory.
		\label{fig-wlooppot}
		}
\end{figure}

Apart from planar Wilson loops, let us consider Wilson loops with oblique, step like gauge paths in the spatial direction, see Fig.~\ref{fig-obliquewloop}. The temporal Wilson lines of these loops will then be separated by a lattice vector $\vect{r}$ in the spatial hyperplane. To calculate the static quark potential $\hat V(r)$, we pick a vector $\vect{r}$ and select Wilson loops with a temporal extent that is at least $t_\text{min}$ (see Table \ref{tab-renconst}) and at most $\hat T/2$.
Assuming that the ground state dominates for these temporal extents, we can perform a fit of the form
\begin{equation}
	\dlangle W(\vect{r},t;a) \drangle = b_{\vect{r}}\, \exp\left(- \hat t\, \hat v_{\vect{r}} \right)\ ,
\end{equation}
where $b_{\vect{r}}$ and $\hat v_{\vect{r}}$ are fit parameters. We set the fit weights to the bootstrap errors of the individual input data points. 

The values $\hat v_{\vect{r}}$ obtained in the fits correspond to the potential at the given $r=|\vect{r}|$. 
At small $|\vect{r}|$, the potential thus determined exhibits considerable breaking of rotational invariance, which is a clear sign of discretization errors. However, these errors can be considerably reduced with a perturbative correction \cite{Michael:1992nj,Bali:2000gf}. Calculating the gluon exchange between the temporal Wilson line segments in the loop (cf. Fig~\ref{fig-wloopglueexch}) to leading order in lattice perturbation theory similarly as in section \ref{sec-perturbren}, one obtains
\begin{equation}
	\lim_{t\rightarrow \infty} \frac{\partial}{\partial \hat t} \ln \dlangle W(r,t;a)\drangle \approx C_F g^2  \int_{-\pi}^{\pi} \frac{d^3 \hat p}{(2\pi)^3} \left. \exp(i \hat p_\muE \hat r_\muE) \frac{\tilde G_{\Eu{4}\Eu{4}}(\hat p/a)}{a^2} \right|_{\hat p_\Eu{4} = 0}
	\equiv \frac{C_F g^2}{4\pi} \left[ \frac{1}{\vect{\hat r}} \right]\ .
\end{equation}
The corresponding calculation in continuum perturbation theory simply produces the Coulomb potential $C_F g^2/(4\pi |\vect{\hat r}|)$. Indeed, $[1/\vect{\hat r}]$ approaches $1/\hat r$ with increasing $\hat r$  very quickly, see appendix \ref{sec-glucorr} for details. To apply the correction in practice, choosing the appropriate coupling $g$ is non-trivial, so the usual method is to determine a strength parameter $\lambda_a$ for the lattice artefacts along with the parameters $\hat \sigma_a$, $\hat C_a$ and $\alpha_a$ in a fit to the potential. In this fit, the values $\hat{v}_{\vect{r}}$ determined above serve as input data. The fit constraits are of the form
\begin{equation}
	\hat v_{\vect{r}} = \hat \sigma_a \hat r - \frac{\alpha_a}{\hat r} + \hat C_a - \lambda_a \left(\, \left[ \frac{1}{\vect{\hat r}} \right]-\frac{1}{\hat r}\,\right)\ .
	\label{eq-potfitcorr}
\end{equation}
Again, we choose fit weights according to the statistical errors of the input data. Taking the fit results thus obtained, we can determine the renormalization constant $\delta \hat m(a)$ according to eq.~(\ref{eq-stringren}). The parameter $\lambda_a$ does not enter the renormalization condition. However, it turns out that the additional degree of freedom introduced through the perturbative correction is very important for the quality of the fit eq.~(\ref{eq-potfitcorr}). Only with the corrections, we can use data points $\hat v_{\vect{r}}$ down to $\hat r = \sqrt{2}$. For details regarding the corrections, see section~\ref{sec-glucorr} in the appendix. As a consistency check, we also calculate the lattice scale $a$ determined from the fits.
On the smeared lattices, we obtain deviations below $3\%$ with respect to the numbers obtained from the MILC collaboration. Systematic errors could be further reduced by increasing the minimal size of the Wilson loops, which would however demand a larger number of input configurations. We have estimated these systematic errors for $\delta m$ by varying the minimal size of the Wilson loops in the fits and get errors that are negligible compared to the systematic uncertainties determined from the smeared--unsmeared comparison in section \ref{sec-discerr}. The final results for $\delta m$ are quoted in Table~\ref{tab-renconst}. 

\begin{figure}[tbp]
	\centering%
	\subfloat[][]{%
		\label{fig-StringRenSm}%
		\includegraphics[clip=true]{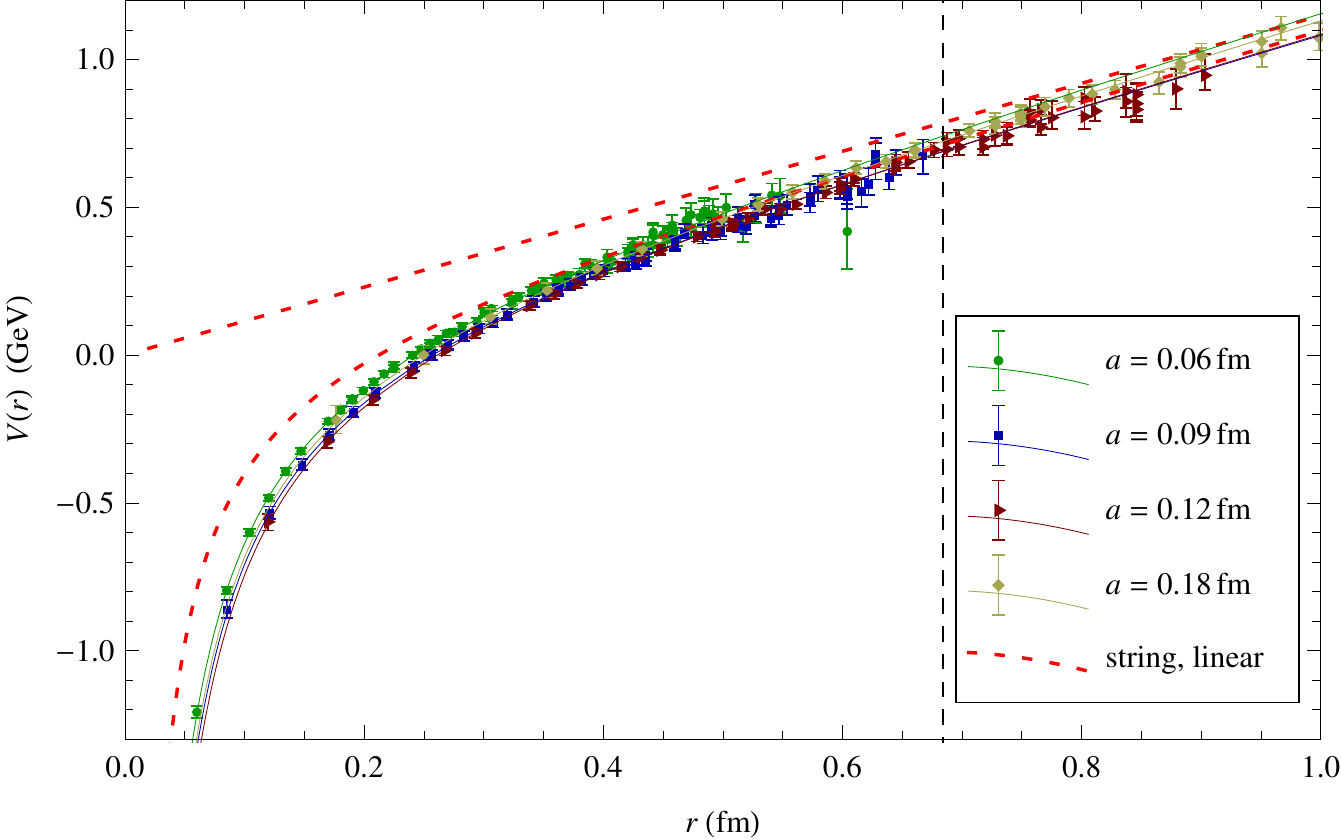}%
		}\par%
	\vspace{1em}
	\subfloat[][]{%
		\label{fig-StringRenUnsm}%
		\includegraphics[clip=true]{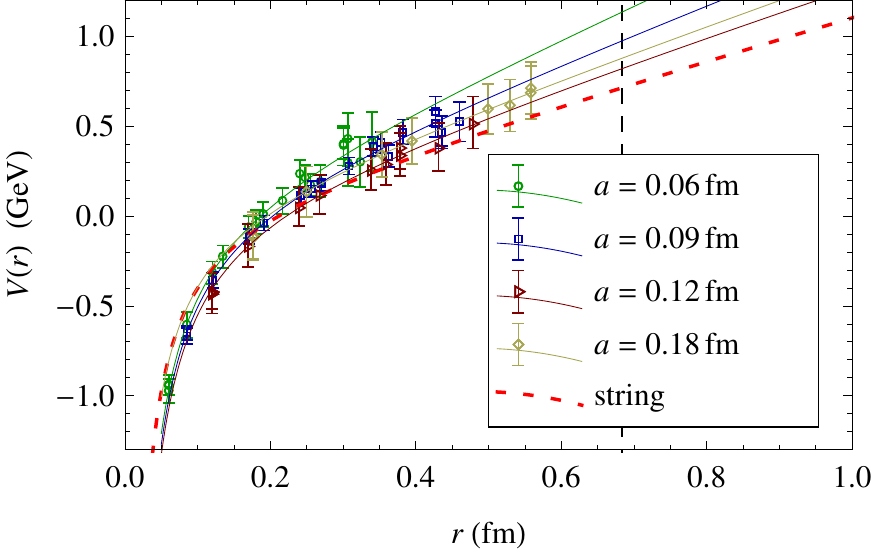}%
		}\par%
	\vspace{1em}
	\caption[Renormalization Using The String Potential]{%
		Renormalization using the string potential. The data points shown have been shifted by the renormalization constants determined from the fits and have been corrected for known discretization errors by subtracting $\lambda_a \left(\,[ 1/\vect{\hat r}] - 1/\hat r \,\right)$. The thin solid lines through the data points correspond to fits of the form eq.~(\ref{eq-potfit}). For the shown string potential $V^\ren_\text{string}$ from eq.~(\ref{eq-potstring}), we take $C^\ren=0$ and an average $\sigma$ determined from the fits to the smeared ensembles. The vertical dashed line marks approximately $1.5r_0$, where the matching to the string has been performed. \par
		\subref{fig-StringRenSm} HYP smeared ensembles. We also show the linear part $\sigma r$ of the potential for an averaged numerical value of $\sigma$. \par
		\subref{fig-StringRenUnsm} Unsmeared ensembles. The statistics we have accumulated here is very limited; the small values of $\hat t_\mmin$ we have used to get some coarse results may entail significant systematic uncertainties. For the red dashed line showing the string potential, we took the same parameter values as in the smeared case.
		\label{fig-StringRen}
		}
\end{figure}

Figure \ref{fig-StringRenSm} shows the static quark potential after renormalization with the method described above on the HYP smeared ensembles. The data points of the different ensembles lie very close to each other. This is a clear indication that the method works: The renormalized static quark potential from the lattice is unique; in particular it is independent of the lattice spacing. Without renormalization, we would see large offsets between the lattice results for the static quark potential on the different ensembles.

Our results are worse on the unsmeared ensembles, see Fig.~\ref{fig-StringRenUnsm}. However, it is obvious that our calculations for the unsmeared ensembles could be easily improved by taking more configurations (to improve statistics) and by raising the minimal temporal extent of the Wilson loops $\hat t_\mmin$ (to reduce systematic uncertainties).

\subsection{Taxi Driver Correction}
\label{sec-taxidriver}

\begin{figure}[tbp]
	\centering%
	\subfloat[][]{%
		\label{fig-taxiUnsmSpline}%
		\includegraphics[clip=true]{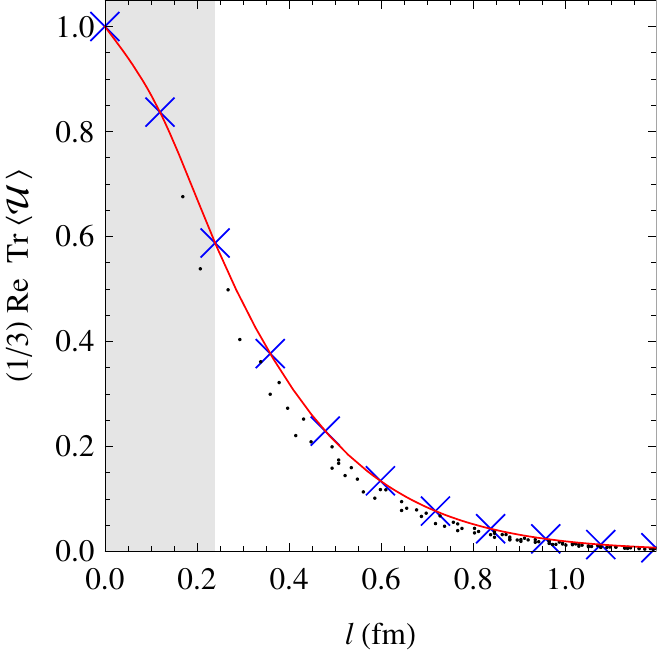}%
		}\hfill%
	\subfloat[][]{%
		\label{fig-taxiUnsmCorr}%
		\includegraphics[clip=true]{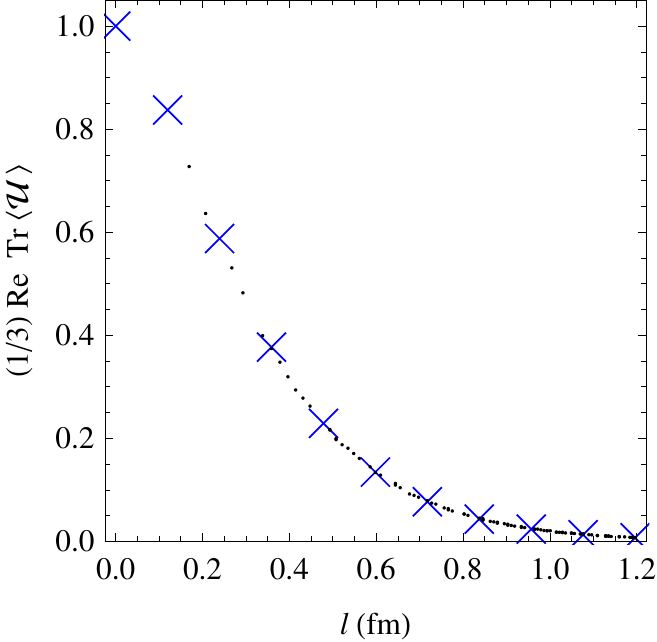}%
		}\par%
	\vspace{2em}
	\subfloat[][]{%
		\label{fig-taxiSmSpline}%
		\includegraphics[clip=true]{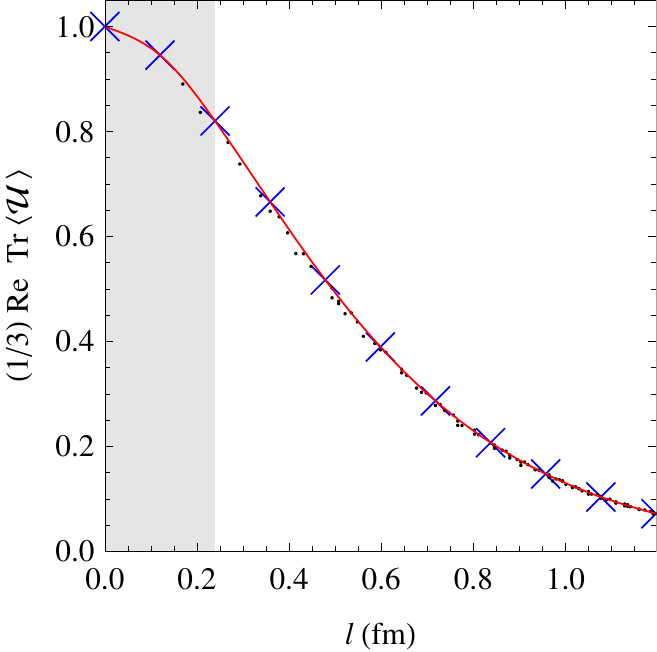}%
		}\hfill%
	\subfloat[][]{%
		\label{fig-taxiSmCorr}%
		\includegraphics[clip=true]{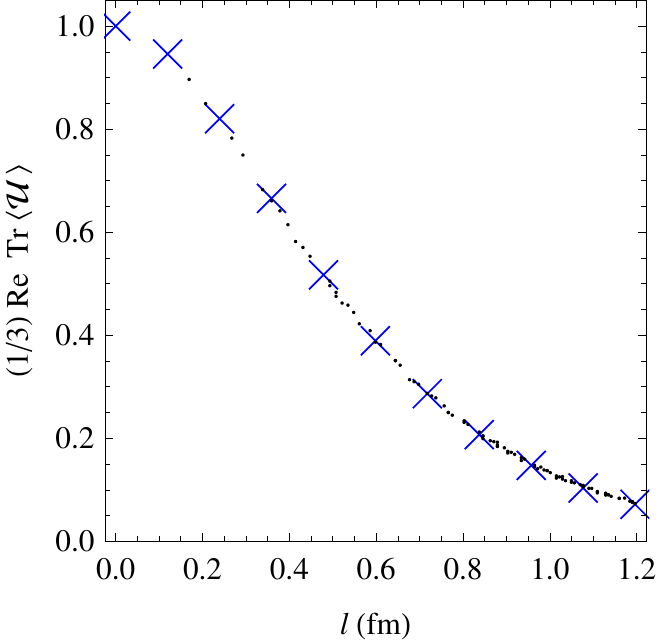}%
		}\par%
	\vspace{2em}
	\caption[Taxi Driver Correction]{%
		Taxi driver correction: We plot the unrenormalized expectation value of the Wilson line $\frac{1}{N_c} \myRe\, \Tr_c\, \dlangle \WlineI{^\lat}{\mathcal{C}^\lat} \drangle$ on the coarse-m020 ensemble. The blue crosses indicate results from straight link paths, the black dots are obtained with step-like link paths. Error bars are smaller than the symbols in this plot. \par
		\subref{fig-taxiUnsmSpline}, \subref{fig-taxiUnsmCorr}: without smearing,\\
		\subref{fig-taxiSmSpline}, \subref{fig-taxiSmCorr}: evaluated on the HYP smeared ensemble.\par
		In \subref{fig-taxiUnsmSpline} and \subref{fig-taxiSmSpline}, we show the uncorrected data, and the smooth curve is a natural spline interpolating the blue crosses. The shaded area is the region excluded from the fit that determines $\delta \hat m_\text{taxi}$ and $\delta \nu_\text{taxi}$. In \subref{fig-taxiUnsmCorr} and \subref{fig-taxiSmCorr} we plot the data after the taxi driver correction.
		\label{fig-taxi}
		}
\end{figure}

In the previous sections, we have discussed ways to determine a constant $\delta m$ which can be used to renormalize \emph{straight} Wilson lines on the lattice. For a straight link path $\mathcal{C}^\lat_{\elll}$ connecting two points separated by a vector $\elll$, the length is related to the number of link variables by $|\vect{\elll}| = l[\mathcal{C}^\lat] = a\, n_\text{links}[\mathcal{C}^\lat]$. What about the step-like paths that we use to discretize Wilson lines at oblique angles? We have already observed in Fig.~\ref{fig-unrenamp} that operators with step-like link paths produce expectation values that lie systematically lower than those of operators with straight link paths of the same length $|\vect{\elll}|$, especially on the unsmeared ensembles. Like a taxi driver navigating on a rectangular grid of streets and avenues, the link path on the lattice has a longer length than the continuum Wilson line we wanted to model: $a\, n_\text{links}[\mathcal{C}^\lat] > |\vect{\elll}|$. The link path has the shape of a polygon, which approximates a straight, smooth Wilson line and has an infinite density of kinks in the limit $a \rightarrow 0$. In the continuum, Wilson lines of this shape have already been discussed in Ref.~\cite{Craigie:1980qs}. They are renormalized like the smooth Wilson line they approximate, but the renormalization constant $\delta m$ receives an extra contribution. Inspired by this continuum result, we may conjecture that the step-like Wilson line on the lattice can be improved by a correction factor:
\begin{equation}
	\WlineI{^\lat_\text{corr}}{\mathcal{C}^\lat_\elll} = \exp\left( -(n_\text{links}[\mathcal{C}^\lat]-|\vect{\elll}|/a)\, \delta \hat m_\text{taxi} - n_\text{cusps}[\mathcal{C}^\lat]\,\nu_\text{taxi} \right)\ \WlineI{^\lat}{\mathcal{C}^\lat_\elll}\ .
	\label{eq-taxicorr}
\end{equation}
Here $n_\text{cusps}[\mathcal{C}^\lat]$ is the number of $90^\circ$ angles in the link path $\mathcal{C}^\lat$, and 
$\delta \hat m_\text{taxi}$ and $\nu_\text{taxi}$ are constants, which we simply determine from a fit, based on the requirement that the corrected operators should give expectation values that depend smoothly on $|\vect{\elll}|$. To this end, we take expectation values of Wilson lines 
\begin{equation}
	\frac{1}{N_c} \myRe\, \Tr_c\, \dlangle \WlineI{^\lat}{\mathcal{C}^\lat} \drangle
\end{equation}
calculated on a Landau gauge fixed ensemble. We interpolate the results for straight link paths by a natural spline, as shown in Fig.~\ref{fig-taxiUnsmSpline} and \ref{fig-taxiSmSpline}. Next, we apply the correction eq.~(\ref{eq-taxicorr}) to the Wilson lines with step-like link paths, adjusting $\delta m_\text{taxi}$ and $\nu_\text{taxi}$ until the mean squared distance between these data points and the spline becomes minimal. The fit weights are chosen according to the bootstrap errors of the individual data points. In the fit, we restrict ourselves to data points with $2a < |\vect{\elll}| \leq L/2$ in order to exclude the regions where strong lattice cutoff effects and finite volume effects may be expected. The fit results are listed in Table~\ref{tab-taxi}.

The corrected data points, shown in Fig.~\ref{fig-taxiUnsmCorr} and \ref{fig-taxiSmCorr}, do not lie perfectly on the spline within their tiny errors. Obviously, the prescription eq.~(\ref{eq-taxicorr}) cannot remove all artefacts created by the use of step-like link paths. Nevertheless, the improvement is significant, in particular on the unsmeared ensembles. For example, the weighted root mean square distance to the spline (the square root of ``$\chi^2$'') is reduced by a factor of almost 14 on the unsmeared coarse-m020 ensemble. The correction seems to work well for the whole range of $|\vect{\elll}|$. Even the two data points in the region excluded from the fit in Fig.~\ref{fig-taxiUnsmSpline} are visibly improved in Fig.~\ref{fig-taxiUnsmCorr}. On the smeared ensembles, where the data points lie close to the spline from the start, the improvement is much smaller.

We conjectured in Ref. \cite{Musch:2008jd} that the constant $\delta\hat m_\text{taxi}$ determined from the fit to the spline might be used for the overall renormalization of the Wilson line, i.e., that the entire renormalization of the lattice operator depends on the number of link variables and corners only:
\begin{equation}
	\text{``}\quad \WlineI{^\ren}{\mathcal{C}_\elll} = \exp\left( -n_\text{links}\, \delta \hat m_\text{taxi} - n_\text{cusps}[\mathcal{C}^\lat]\,\nu_\text{taxi} \right)\ \WlineI{^\lat}{\mathcal{C}^\lat_\elll}\quad \text{''.}
	\label{eq-taxiren}
\end{equation}
The numerical results of our studies at several lattice spacings are not in support of this conjecture. In particular, the values $\Delta \hat m(a_1,a_2)$ obtained from $\delta\hat m_\text{taxi}$ disagree with our results from section \ref{sec-wloopscaling}. They are too small by a factor of more than 2 on the unsmeared ensembles, and by roughly an order of magnitude on the smeared ensembles. It appears that the taxi driver correction merely addresses a certain kind of discretization error. On top of the taxi driver correction, an $|\vect{\elll}|$-dependent renormalization as discussed in the previous sections is mandatory.

\begin{table}
	\centering
	\renewcommand{\arraystretch}{1.1}
	\begin{tabular}{|ll||l|l|}
	\hline
	ensemble  \rule{0ex}{1.2em} & & $\delta m_\text{taxi}$ & $\nu_\text{taxi}$ \\
	\hline \hline                     
   super coarse   & smeared & -0.0300(12)  & 0.00728(45) \\
   coarse-m020    & smeared & -0.0223(12)  & 0.00624(52) \\
   fine           & smeared & -0.02217(93) & 0.00735(57) \\
   super fine     & smeared & -0.02180(91) & 0.00762(53) \\
   \hline                    
   super coarse   &  &  -0.2294(15)  & 0.04700(65) \\
   coarse-m020    &  &  -0.20664(72) & 0.04756(36) \\
   fine           &  &  -0.18511(61) & 0.04387(31) \\
   super fine     &  &  -0.1804(13)  & 0.04682(76) \\
	\hline
	\end{tabular}\par\vspace{1ex}
	\renewcommand{\arraystretch}{1.0}
	\caption{Parameters determined for the taxi driver correction.}
	\label{tab-taxi}%
\end{table}

\section{Dividing Amplitudes by the Wilson Line}
\label{sec-divwline}

Consider the operator
\begin{equation}
	O^\text{div}_{\GammaOp}(\elll) \equiv 
	\frac{O_{\GammaOp}(\elll)}{\frac{1}{N_c}\myRe\, \Tr_c\, \dlangle \Wline{0,\elll} \drangle} = 
	\frac{\bar \quark(\elll)\, \GammaOp\ \Wline{\elll,0}\ \quark(0)}{\frac{1}{N_c}\myRe\, \Tr_c\, \dlangle \Wline{0,\elll} \drangle} \ ,
\end{equation}
where the expectation value of the Wilson line in the denominator requires some gauge fixing. We choose the Landau gauge.
According to eqns.~(\ref{eq-linkrencont}) and (\ref{eq-opren}), the operator defined above is renormalized according to
\begin{equation}
	O^{\text{div},\ren}_{\GammaOp}(\elll) = \renZ_\psi^{-1}\, \renZ_{(\psi z)}^2\ O^\text{div}_{\GammaOp}(\elll)\ .
\end{equation}
The renormalization factor $\exp(-\delta m\,l[\mathcal{C}])$ cancels. Thus the unrenormalized amplitudes explored in section \ref{sec-firstobs} divided by the Wilson line expectation value
\begin{equation}
	 \tAmp_{i}^\text{div}(\elll^2,\elll \tcdot P) \equiv \frac{\tAmp_{i}^\unren(\elll^2,\elll \tcdot P)}{\frac{1}{N_c}\myRe\, \Tr_c\, \dlangle \Wline{0,\elll} \drangle}
\end{equation}
are quantities that need no $\elll$-dependent renormalization. In section \ref{sec-selfenergy} the square root of a Wilson loop serves as a subtraction factor with the same function as the Wilson line here. The obvious disadvantage of using the Wilson line as subtraction factor is that we introduce dependence on the gauge fixing condition. 

\begin{figure}[tbp]
	\centering%
	\subfloat[][]{%
		\label{fig-ampdiv-m050umd2}%
		\begin{overpic}[clip=true]{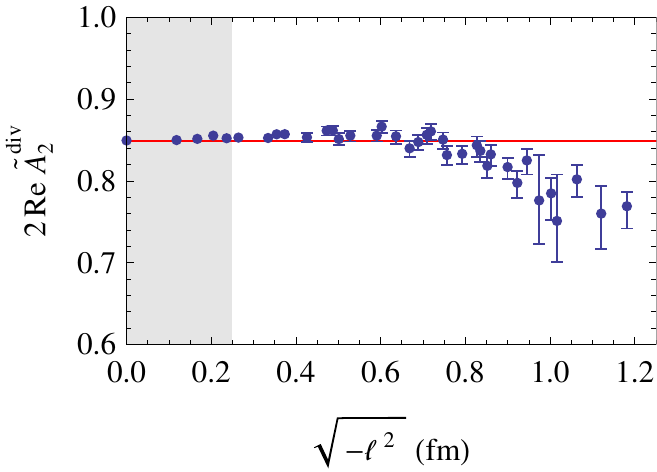}%
			\put(23,23){$\tilde A_2$, u-d quarks, coarse-050}
		\end{overpic}
		}\hfill%
	\subfloat[][]{%
		\label{fig-ampdiv-m050umd6}%
		\begin{overpic}[clip=true]{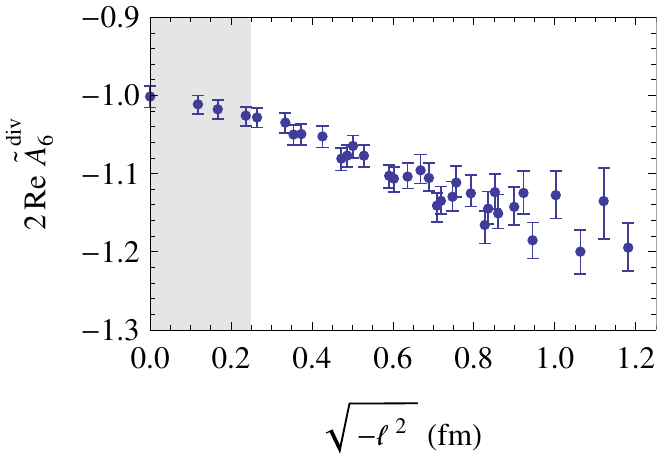}%
			\put(26,23){$\tilde A_6$, u-d quarks, coarse-050}
		\end{overpic}
		}\par%
	\vspace{1em}
	\subfloat[][]{%
		\label{fig-ampdiv-m050u2}%
		\begin{overpic}[clip=true]{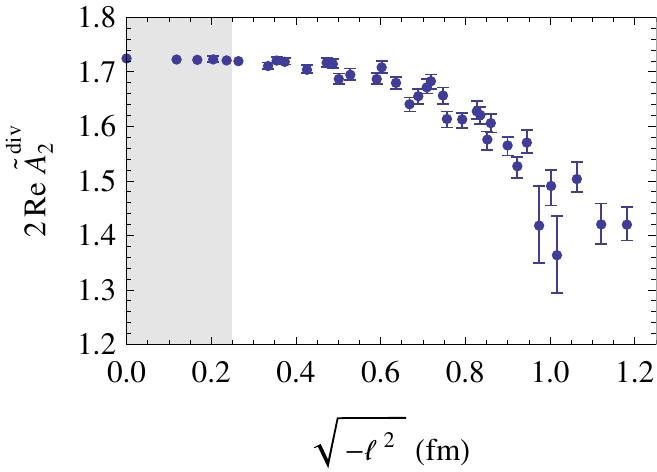}%
			\put(23,23){$\tilde A_2$, u quarks, coarse-050}
		\end{overpic}
		}\hfill%
	\subfloat[][]{%
		\label{fig-ampdiv-m050u6}%
		\begin{overpic}[clip=true]{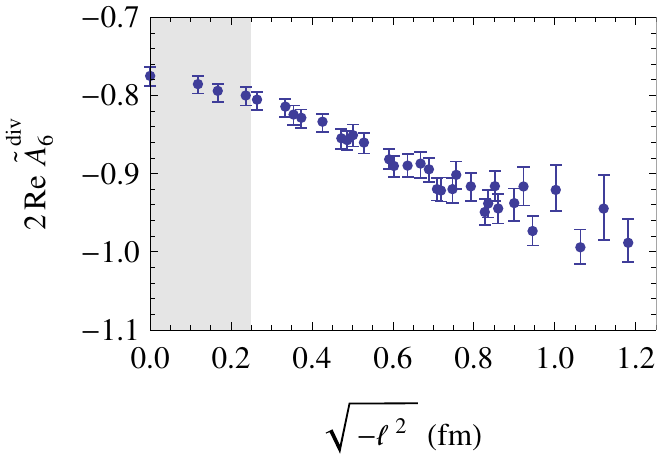}%
			\put(26,23){$\tilde A_6$, u quarks, coarse-050}
		\end{overpic}
		}\par%
	\vspace{1em}
	\subfloat[][]{%
		\label{fig-ampdiv-m020u2}%
		\begin{overpic}[clip=true]{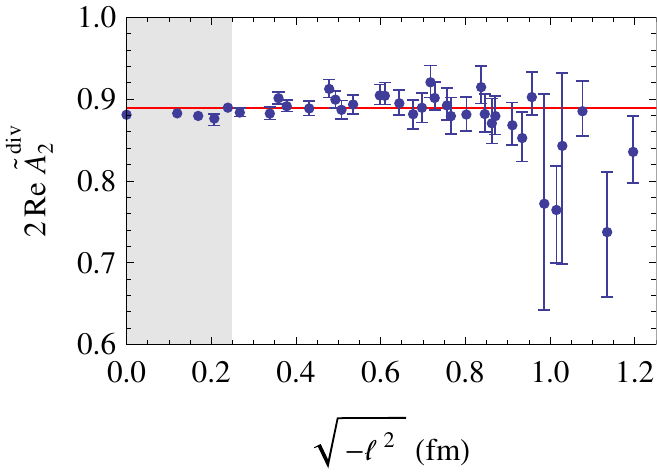}%
			\put(23,23){$\tilde A_2$, u-d quarks, coarse-020}
		\end{overpic}
		}\hfill%
	\subfloat[][]{%
		\label{fig-ampdiv-m020u7}%
		\begin{overpic}[clip=true]{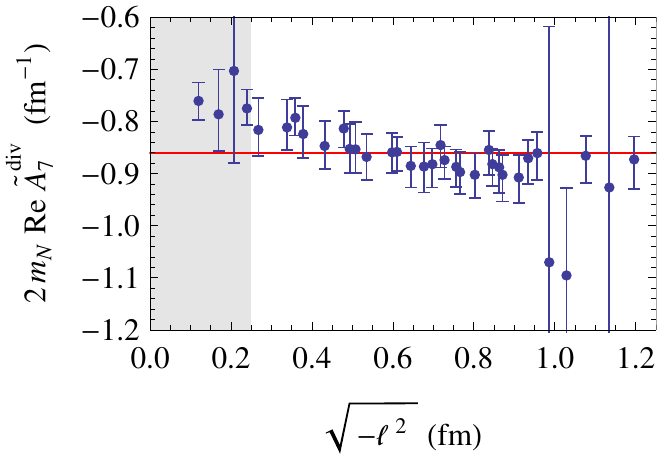}%
			\put(26,23){$\tilde A_7$, u-d quarks, coarse-020}
		\end{overpic}
		}\par%
	\vspace{1em}
	\caption[Amplitudes divided by Wilson lines]{%
		The divided Amplitudes $\tilde A_i^\text{div}(\elll^2,\elll\tcdot P=0)$ evaluated on the HYP smeared ensembles. In some of the plots, red horizontal lines have been inserted at an average value to guide the eye. Note that statistical errors are smaller on the coarse-m050 ensemble. The amplitude $\tilde A_7$ has been calculated only for the coarse-m020 ensemble up to now. (The shaded region indicates that we expect significant lattice cutoff effects in the amplitudes below about $0.25 \units{fm}$.) 
		\label{fig-ampdiv}
		}
\end{figure}

Figure~\ref{fig-ampdiv} shows some examples of amplitudes divided by Wilson lines, plotted as a function of $\sqrt{-\elll^2}$. The amplitudes have been obtained from ratios as described in section \ref{sec-paramconnect}. Surprisingly, the data shows only weak dependence on $\sqrt{-\elll^2}$. It appears that the characteristic Gaussian-like decay we saw in the unrenormalized amplitudes, e.g., in Fig.~\ref{fig-unrenamp}, has been cancelled in the ratio with the Wilson line. Some of the divided amplitudes are almost constant within the whole range of $\elll^2$ accessible to us. 
These results suggest that the typical Gaussian-like shape of the unrenormalized amplitudes is not a feature characteristic of the nucleon. Instead, the Gaussian decay seems to be largely driven by the Wilson line, which already shows this behavior when evaluated with Landau gauge fixing in the vacuum.
This indicates that the $\elll^2$-dependence of our straight link correlator is predominantly given by the dynamics of the gauge field. 

\section{The First Mellin Moment}
\label{sec-mellin}

\begin{sidewaysfigure}[tbp]
	\captionsetup[subfigure]{justification=raggedright,singlelinecheck=false}
	\centering
	\begin{overpic}{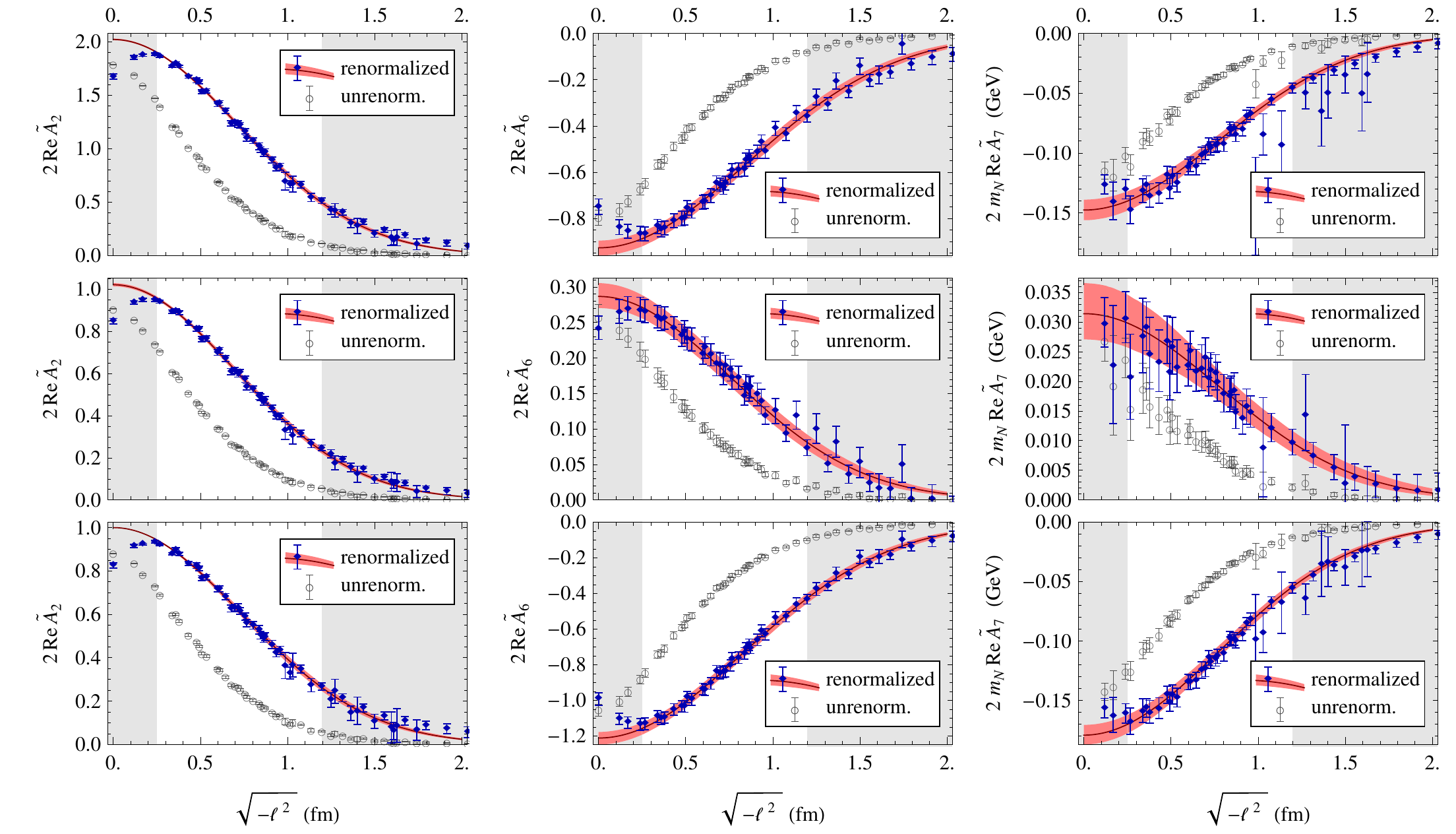}
		\put(8.5,40){\subfloat[][{up}]{\hspace*{2cm}} }
		\put(42,51.5){\subfloat[][{up}]{\hspace*{2cm}} }
		\put(75,51.5){\subfloat[][{up}]{\hspace*{2cm}} }
		\put(8.5,23){\subfloat[][{down}]{\hspace*{2cm}} }
		\put(42,23){\subfloat[][{down}]{\hspace*{2cm}} }
		\put(75,23){\subfloat[][{down}]{\hspace*{2cm}} }
		\put(8.5,6.5){\subfloat[][{u-d}]{\hspace*{2cm}} }
		\put(42,18){\subfloat[][{u-d}]{\hspace*{2cm}} }
		\put(75,18){\subfloat[][{u-d}]{\hspace*{2cm}} }
	\end{overpic}\par
	\caption[Amplitudes]{%
		The amplitudes $\tilde A_i(\elll^2,\elll\tcdot P=0)$ evaluated with straight Wilson lines on the HYP smeared ensemble coarse-m020. For renormalization, we have used $\delta \hat m$ obtained with the string potential, and $\renZ^{-1}_{\psi,z}$ has been determined from the Gaussian fit requiring $\tAmp_{2,u-d}(0,0)=1$. The shaded regions have been excluded from the fits since they are affected by lattice cutoff effects (below $0.25 \units{fm}$) and possibly by finite volume effects  (above $L/2$, as a conservative estimate). Errors and error bands are statistical.
		\label{fig-ampmellin}
		}
	\end{sidewaysfigure}

In section \ref{sec-euclidrestric} we showed that we cannot calculate the amplitudes for all $(\elll \tcdot P)$-values that enter the Fourier transformation to momentum space. However, if we restrict ourselves to the first Mellin moment as defined in section \ref{sec-defmellin}, the Fourier transform with respect to $\elll$ is restricted to $\elll^+=\elll^-=0$, i.e. we only need the amplitudes for $\elll^2 < 0$ and $\elll \cdot P = 0$. We plot the real parts of the amplitudes $\tAmp_2$, $\tAmp_6$ and $\tAmp_7$ at $\elll \tcdot P = 0$ evaluated on the HYP smeared coarse-m020 ensemble in Fig.~\ref{fig-ampmellin}. 
(The imaginary part of these amplitudes vanishes at $\elll \tcdot P = 0$ due to eq.\,(\ref{eq-tampconstraint}).)
In the following, we explain how we arrive at the renormalized amplitudes and the Gaussian parametrization shown in this figure.

\subsection{Gaussian Parametrization}

In Ref.~\cite{Musch:2007ya}, we showed that a type of Gaussian fit function can successfully describe our results for the unrenormalized amplitude $\tAmp_2(\elll^2,0)$ from the lattice (compare Fig.~\ref{fig-unrenamp-sm}) and enabled us to make the Fourier transform to momentum space.
The Gaussian ansatz has been widely used and has a number of virtues \cite{Collins:2005ie}. A look at first Mellin moments obtained from models also suggest that a Gaussian function might be a good starting point, compare Fig.~\ref{fig-diquark}. Now that we know more about the renormalization of our operators, we still find the Gaussian parametrization useful, although there are certain limitations, which we shall address in the section \ref{sec-critgauss}. The simple Gaussian fit function we presently use for our amplitudes reads
\begin{equation}
	\tAmp_{j}(\elll^2,0) \approx \frac{1}{2} c_j \exp\left( - \frac{(-\elll^2)}{\sigma_j^2}\right) = \frac{1}{2} c_j \exp\left( - \frac{\vprp{\elll}^2}{\sigma_j^2}\right)\ ,
	\label{eq-gaussamps}
\end{equation}
where $c_j$ and $\sigma_j$ are the fit parameters, the latter one characterizing the width of the amplitude:
\begin{equation}
	\langle \vprp{\elll}^2 \rangle_{\tAmp_j}^{-1/2} \equiv \sqrt{ \frac{\int d\vprp{\elll}^2\ \vprp{\elll}^2\ \tAmp_j(-\vprp{\elll}^2,0)}{\int d\vprp{\elll}^2\ \tAmp_2(-\vprp{\elll}^2,0)} } = \sigma_j\ .
\end{equation}
Let us now study \TMDs based on the Gaussian parametrization.
In terms of the amplitudes $\tAmp_j$, the first Mellin moment defined in eq.~(\ref{eq-deffirstmellin}) is of the form 
\begin{align}
	\Phi^{[\GammaOp](1)}(\vprp{k};P,S) & = \frac{1}{P^+} \int \frac{d^2\vprp{\elll}}{(2\pi)^2}\ e^{i \vprp{k}\cdot \vprp{\elll} } \ \sum_{\mathclap{j,n_1,\ldots,n_{m_j}}}\ a^{(\GammaOp)}_{j,n_1,\ldots,n_{m_j}}\ \vect{\elll}_{\prp n_1} \cdots\, \vect{\elll}_{\prp n_{m_j}}\ 2 \tAmp_j(\elll^2,0) \ ,
\end{align}
where the $a^{(\GammaOp)}_{j,n_1,\ldots,n_{m_j}}$ are appropriate coefficients that can be read off from the parametrization eq.\,(\ref{eq-phitildetraces}). Inserting the Gaussian model function eq.\,(\ref{eq-gaussamps}), we can readily perform the Fourier transform and get
\begin{align}
	\Phi^{[\GammaOp](1)}(\vprp{k};P,S) & = \sum_{\mathclap{j,n_1,\ldots,n_{m_j}}} a^{(\GammaOp)}_{j,n_1,\ldots,n_{m_j}} \left(\frac{i \vect{k}_{\prp n_1} \sigma_j^2}{2} \right) \cdots \left(\frac{i \vect{k}_{\prp n_{m_j}} \sigma_j^2}{2} \right)
	\frac{c_j\, \sigma_j^2}{4\pi P^+} \exp\left( \frac{-\vprp{k}^2}{(2/\sigma_j)^2}\right)\ .
\end{align}
By comparison with eqns.\,(\ref{eq-phigammaplus}) -- (\ref{eq-phisigmaplusi}), we can easily identify the first Mellin moments of \TMDs. In particular, we find
\begin{align}
	\Phi^{[\gamma^+](1)}(\vprp{k};P,S) & = \frac{c_2\, \sigma_2^2}{4\pi} \exp\left( \frac{-\vprp{k}^2}{(2/\sigma_2)^2} \right) \nonumber \\
	& = f_1^{(1)}(\vprp{k})  \label{eq-mellinunpol} \ ,\\
	\Phi^{[\gamma^+\gamma^5](1)}(\vprp{k};P,S) & = 
	- \Lambda \frac{c_6\, \sigma_6^2}{4\pi} \exp\left( \frac{-\vprp{k}^2}{(2/\sigma_6)^2} \right) 
	- \frac{\vprp{k} \cdot \vprp{S}}{m_N}\, \frac{c_7\,m_N^2\,\sigma_7^4}{8\pi} \exp\left( \frac{-\vprp{k}^2}{(2/\sigma_7)^2} \right) \nonumber \\
	& = \Lambda\, g^{(1)}_{1L}(\vprp{k}^2) + \frac{\vprp{k} \cdot \vprp{S}}{m_N}\ g^{(1)}_{1T}(\vprp{k}^2)\ .
	\label{eq-mellinpol}
\end{align}

\subsection{Renormalized Amplitudes from Gaussian Fits}

According to eq.~(\ref{eq-opren}), the renormalized amplitudes are given by
\begin{equation}
	\tAmp_j(\elll^2,0) = \renZ^{-1}_{\psi,z}\ \exp\left( -\delta m\, \sqrt{-\elll^2} \right)\ \tAmp^\text{unren}_j(\elll^2,0) \ .
\end{equation}
In a first step, we renormalize our data with respect to the $\elll^2$-dependent term, using the value $\delta m$ obtained with the help of the static potential, as described in section \ref{sec-renstatqpot}. We also include the taxi driver correction from section \ref{sec-taxidriver}, although the effect is small on the smeared ensembles. Next, we fit Gaussians of the form eq.~(\ref{eq-gaussamps}) to the data, from which we obtain the fit parameters $c_j^\text{unren}$ and $\sigma_j$. The coefficients $c_j^\text{unren}$ do not include multiplicative renormalization with $\renZ^{-1}_{\psi,z}$, yet.
For the Gaussian fit, we exclude data points with $\sqrt{-\elll^2} \leq 0.25 \units{fm}$, because gauge links of such short lengths are subject to significant lattice cutoff effects, which is an observation we made looking at Fig.~\ref{fig-YlineBazRen-sm}. We also exclude data points with $\sqrt{-\elll^2} \geq L/2$, i.e., for operators larger than half the box size of the periodic lattice. For such operators, the separation between ``copies'' on the periodic lattice can become smaller than the extent of the non-local operators themselves. This might introduce finite volume effects. However, we remark that the results would not change much if we were to include the data points with $\sqrt{-\elll^2} \geq L/2$ in the fits. In the fit procedure, we adjust the fit weights to the statistical errors of the individual data points.

In the last step, we address the multiplicative renormalization factor $\renZ^{-1}_{\psi,z}$. Here we explicitly make use of the assumption that $f_{1,\quark}(x,\vprp{k})$ has an interpretation as the density of quarks of flavor $\quark$ in the proton, as discussed in \ref{sec-lfquant}. From this assumption follows 
\begin{equation}
	\int d^2 \vprp{k} \int dx\ f_{1,\quark}(x,\vprp{k}) = \left\lbrace \begin{array}{lcl} 2 & : & \quark=u \\ 1 & : & \quark = d \end{array} \right. \ ,
	\label{eq-quarkcount}
\end{equation}
where the right hand side specifies the number of valence quarks in the proton. Sea quark contributions cancel in the integration over all $x$, because the antiquark densities are given by $f_{1,\bar \quark}(x,\vprp{k}) = - f_{1,\quark}(-x,\vprp{k})$ \cite{Tangerman:1994eh}.
For the determination of $\renZ^{-1}_{\psi,z}$, we take the results for $\quark=u-d$, where disconnected diagrams in our three-point function cancel. Thus, in terms of amplitudes, we simply demand
\begin{equation}
	2 \tAmp_{2,u-d}(0,0) = 1\ .
	\label{eq-multrencond}
\end{equation}
In practice, we determine $\renZ^{-1}_{\psi,z}$ and the renormalized coefficients $c_j$ according to 
\begin{equation}
	\renZ_{\psi,z} := c_{2,u-d}^\text{unren}, \qquad c_j := \renZ^{-1}_{\psi,z}\ c_j^\text{unren}\ .
\end{equation}
Note that we deliberately do not use the data point at $\elll^2=0$ for renormalization. As mentioned before, we exclude it from the fit, along with all other results obtained with gauge links shorter than about two lattice spacings. The corresponding operators must be regarded as local operators; their renormalization properties are different and can be treated with other techniques than the ones discussed in this work, see, e.g., Ref. \cite{Martinelli:1994ty,Gockeler:1998ye}. For local operators, the multiplicative renormalization constants depend on $\GammaOp$. In contrast, in the non-local case we consider here, the constant $\renZ^{-1}_{\psi,z}$ is universally applicable regardless of $\GammaOp$, see section 4.2 of Ref. \cite{Dorn:1986dt}.

Numerical results for the Gaussian fits are given in Tables \ref{tab-amp2results}, \ref{tab-amp6results} and \ref{tab-amp7results}. In Table \ref{tab-amp2results}, the values for $c_2$ show that the condition eq.\,(\ref{eq-quarkcount}) is approximately fulfilled in the $u$ and $d$ channel using the renormalization constant $\renZ^{-1}_{\psi,z}$ determined from the $u-d$ channel. This is an indication that contributions of disconnected diagrams are negligible at the present level of accuracy for the observables we study here.

\subsection{A Critical Look at the Renormalization and Fit Prescription}
\label{sec-critgauss}

\begin{figure}[tbp]
	\centering%
	\begin{overpic}[trim=0 0 0 20,clip=true]{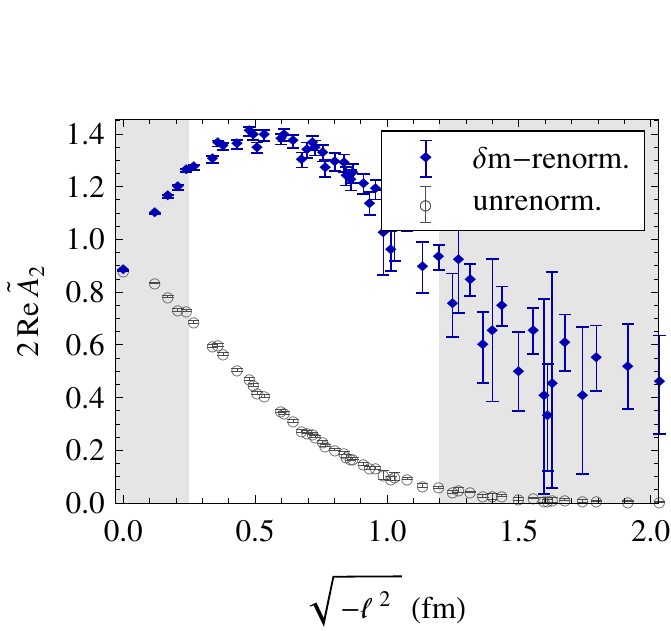}%
		\put(20,22){u-d}
		\end{overpic}
	\caption{%
		Amplitude $\tilde A_2(\elll^2,\elll\tcdot P=0)$ evaluated with straight Wilson lines on the HYP smeared ensemble coarse-m020, with a different renormalization condition as in Fig.~\ref{fig-ampmellin}: We take $\delta \hat m$ from our renormalization procedure with Wilson lines on the Landau gauge fixed ensemble at a renormalization point $l=0.5\units{fm}$ as described in section \ref{sec-martinelli}. The data is not renormalized with respect to the multiplicative factor $\renZ^{-1}_{\psi,z}$.
		\label{fig-ampmellinl0p5ren}
		}
\end{figure}

A serious issue in the above procedure is the choice of the renormalization condition for the Wilson line. In Fig.~\ref{fig-ampmellin} we have renormalized with the static potential and $C^\ren = 0$ as described in section \ref{sec-renstatqpot}. The picture changes drastically if we choose another renormalization condition. Just to show the effect, we plot in Fig.~\ref{fig-ampmellinl0p5ren} data which has been renormalized with $\delta m$ obtained from the Wilson lines in a Landau gauge fixed ensemble, as described in  \ref{sec-martinelli}. This corresponds to the choice $C^\ren \approx 0.2 \units{GeV}$ in terms of renormalization with the static potential. The renormalized data now keeps rising out to $\sqrt{-\elll^2} \approx 0.6\units{fm}$. Clearly, a Gaussian fit would not work in this case, unless we are willing to exclude a much larger range of data points from the fit. Is there a physical interpretation of the renormalization condition, i.e. the value $C^\ren$? Can we establish a relation to a continuum factorization scale, or to a continuum renormalization scheme and scale? These are important questions that have to be addressed in future studies.

Another issue concerns the high-$\vprp{k}$-behavior of \TMDs we already encountered in sections \ref{sec-relationPDFs} and  \ref{sec-probprob}.
The Fourier transformed amplitude, $f_1^{(1)}(\vprp{k})$, is in turn a Gaussian function, see eq.\,(\ref{eq-mellinunpol}).
Obviously, a Gaussion does not adequately describe the large-$\vprp{k}$-behavior $\sim 1/\vprp{k}^2$ predicted by perturbation theory. 
On the other hand, the $\vprp{k}$-integral in our normalization equation eq.~(\ref{eq-quarkcount}) is now well-defined, even without a $|\vprp{k}|$-cutoff. With the Gaussian approach, no explicit cutoff is needed. Instead, the Gaussian ansatz itself functions as a regularization prescription. 

At the level of amplitudes, the asymptotic behavior $f_1^{(1)}(\vprp{k}) \approx b/\vprp{k}^2$ translates into a contribution of the form $- 2\pi b\, \ln( \sqrt{-\elll^2} / l_0 )$ to $\tAmp_2(\elll^2,0)$ at small $\sqrt{-\elll^2}$, see section \ref{sec-highkt}. This contribution diverges at $\elll=0$. Note that our lattice results do not exhibit the divergence at $\elll=0$ because the lattice imposes a momentum cutoff. Instead, we observe effects of the lattice cutoff, which is the reason why we exclude data points with $\sqrt{-\elll^2} \lesssim 0.25\units{fm}$. Our Gaussian fit constitutes an interpolation that smoothly bridges the excluded region at small $\sqrt{-\elll^2}$. Roughly speaking, with our fit prescription, we ``do not resolve'' short range behavior at $\sqrt{-\elll^2} \lesssim 0.25\units{fm}$. Correspondingly, we expect that the \TMDs we calculate become unreliable for $\vprp{k} \gtrsim 1/0.25\units{fm} \approx 0.8 \units{GeV}$.

In the future, we should investigate whether we can detect, isolate and interpret the onset of the logarithmic short-range behavior in our amplitudes from the lattice. Special attention will have to be payed to lattice cutoff effects, which also set in at small $\sqrt{-\elll^2}$. Fit functions that exhibit the correct behavior in the perturbative regime will enable us to treat the high-$\vprp{k}$ behavior in a more systematic way.



\subsection{Comparing a Smeared and an Unsmeared Ensemble}

\begin{figure}[tbp]
	\centering%
	\begin{overpic}[clip=true]{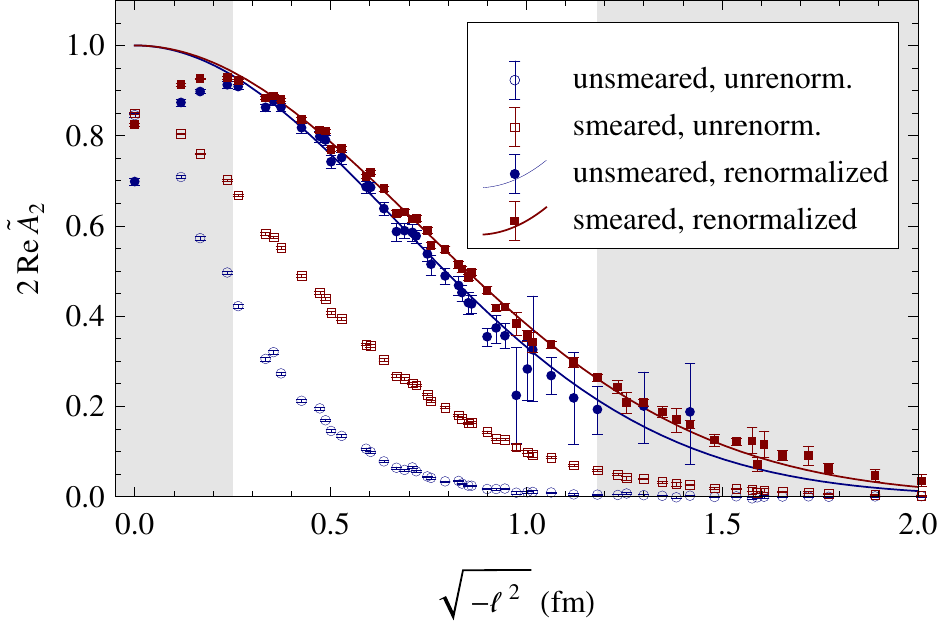}%
		\put(15,16){u-d}
		\end{overpic}
	\caption{%
		Amplitude $\tilde A_2(\elll^2,\elll\tcdot P=0)$ evaluated with straight Wilson lines for $u{-}d$ quarks on the ensemble coarse-m050. We compare the results for gauge links evaluated on the unsmeared gauge configurations to the results obtained with HYP smearing. We take $\delta m$ from our renormalization procedure with Wilson lines on the Landau gauge fixed ensemble at a renormalization point $l=0.5\units{fm}$ as described in section \ref{sec-martinelli} and add $0.1945 \units{GeV}$ to reproduce approximately the renormalization condition from the static quark potential with $C^\ren = 0$. 
		\label{fig-rensmunsm}
		}
\end{figure}

In Fig.~\ref{fig-unrenamp} we observed a big difference in the widths of the unrenormalized amplitudes $\tAmp_2^\unren$ on the smeared and unsmeared coarse-m050 ensemble. Does the renormalization procedure outlined here eliminate these differences? Up to now, we do not have renormalization constants from the static quark potential available for the coarse-m050 ensemble, which is the only ensemble for which we have calculated correlators with unsmeared gauge links. To be able to make the comparison nonetheless, let us use $\delta m$ as determined from the Wilson line at the renormalization point $0.5\units{fm}$ and add a constant $0.1945 \units{GeV}$. This will approximately correspond to renormalization with the static potential at $C^\ren = 0$. We show the results in Fig.~\ref{fig-rensmunsm}. Smeared and unsmeared results agree much better, but there is still an observable difference. The inverse widths of the Gaussian fits to the renormalized data are $4/\sigma_2^2 = 0.4146(85)_\text{stat} \units{GeV}$ for the unsmeared, and $4/\sigma_2^2 = 0.3872(34)_\text{stat} \units{GeV}$ for the smeared ensemble. However, the discrepancy is no larger than we would expect it to be from our study of Wilson lines in section~\ref{sec-discerr}: With the help of eq.~(\ref{eq-widtherrprop}), we obtain a systematic error in $4/\sigma_2^2$ of $\pm 0.037\units{GeV}$.

\subsection{\TMDs and Densities from the Gaussian Parametrization}
\label{sec-gaussresults}

\begin{sidewaysfigure}[tbp]
	\captionsetup[subfigure]{justification=raggedright,singlelinecheck=false}
	\centering
	\begin{overpic}{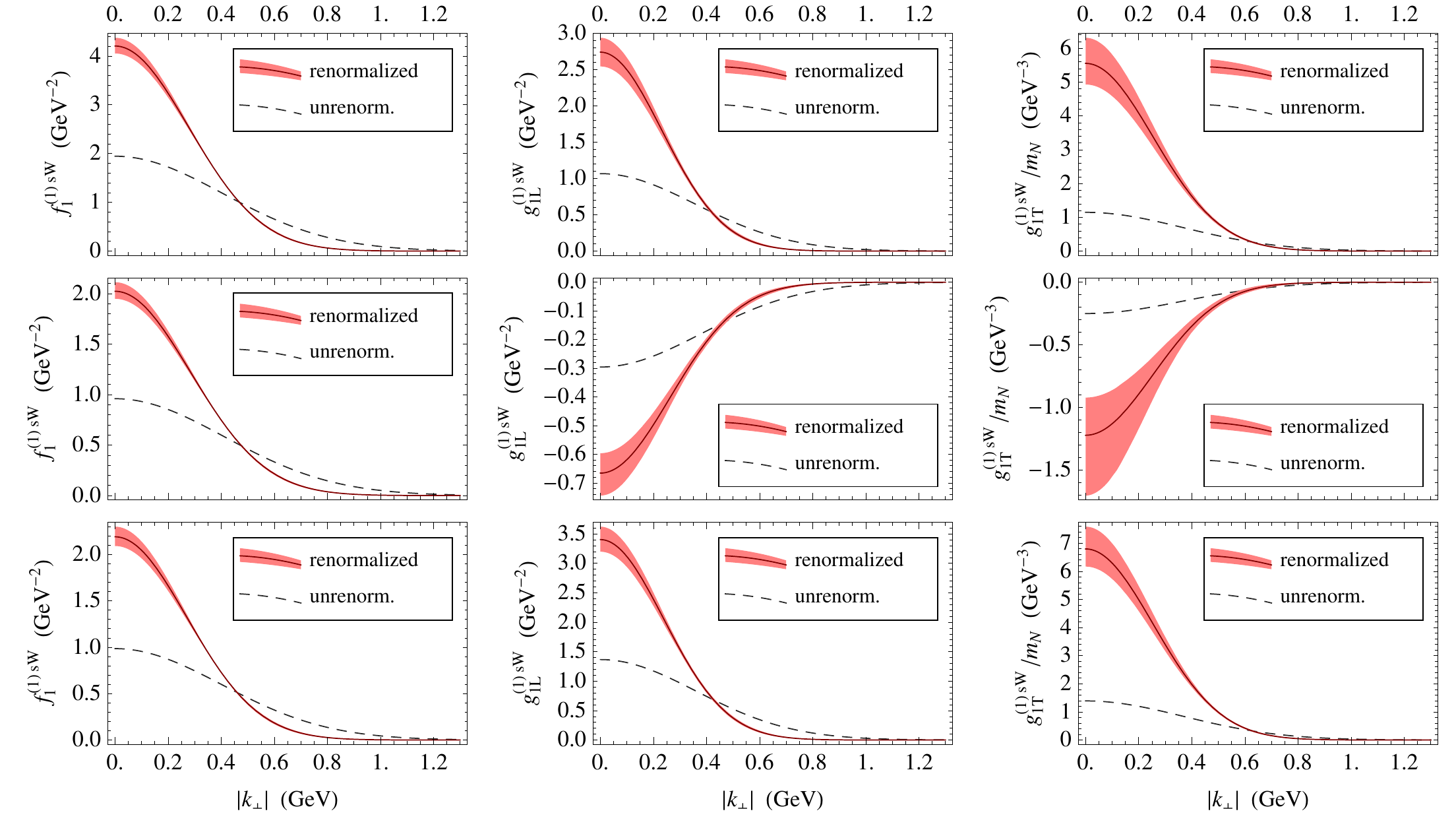}
		\put(8.5,40){\subfloat[][{up}]{\hspace*{2cm}} }
		\put(59,40){\subfloat[][{up}]{\hspace*{2cm}} }
		\put(93,40){\subfloat[][{up}]{\hspace*{2cm}} }
		\put(8.5,23){\subfloat[][{down}]{\hspace*{2cm}} }
		\put(58,34.5){\subfloat[][{down}]{\hspace*{2cm}} }
		\put(92,34.5){\subfloat[][{down}]{\hspace*{2cm}} }
		\put(25.5,6.5){\subfloat[][{u-d}]{\hspace*{2cm}} }
		\put(59,6.5){\subfloat[][{u-d}]{\hspace*{2cm}} }
		\put(93,6.5){\subfloat[][{u-d}]{\hspace*{2cm}} }
	\end{overpic}\par
	\caption[TMDs]{%
		The first Mellin moment of \TMDs $f_1^{(1)\text{sW}}$, $g_{1L}^{(1)\text{sW}}$ and $g_{1T}^{(1)\text{sW}}$ as obtained from the amplitudes plotted in Fig~\ref{fig-ampmellin}. The index ``sW'' is a reminder that these \TMDs have been obtained with straight Wilson lines and with a particular choice for the renormalization condition and for the fit model function. The ``unrenormalized'' \TMDs plotted here are from fits to the unrenormalized data but are also normalized to fulfill exact quark counting in the u-d channel. The error bands include statistical errors only.
		\label{fig-tmdsmellin}
		}
	\end{sidewaysfigure}

\begin{figure}[tbp]
	\centering%
	\subfloat[][]{%
		\label{fig-bachhineq-u}%
		\begin{overpic}[clip=true]{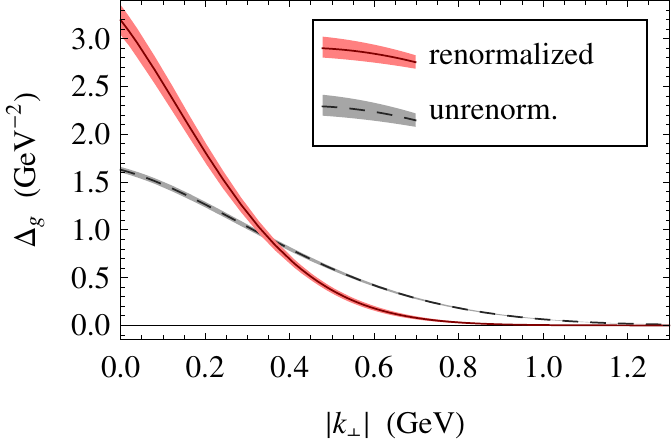}%
			\put(74,36){u quarks}
		\end{overpic}
		}\hfill%
	\subfloat[][]{%
		\label{fig-bachhineq-d}%
		\begin{overpic}[clip=true]{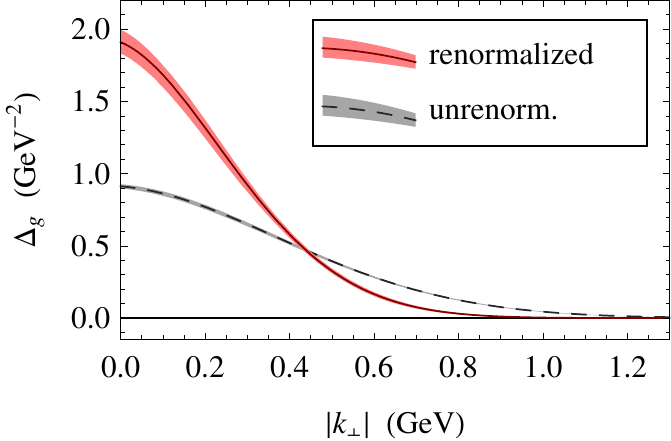}%
			\put(74,36){d quarks}
		\end{overpic}
		}\par%
	\caption[Bacchetta inequality]{%
		A numerical study of bounds on \TMDs: On the HYP-smeared coarse-m020 ensemble, we test whether the function $\Delta_g(|\vprp{k}|)$ defined in the text is positive. The error bands include statistical errors only. Renormalization is based on the static quark potential with the renormalization condition $C^\ren = 0$. 
		\label{fig-bachhineq}
		}
\end{figure}

Notwithstanding the issues of section \ref{sec-critgauss}, we give an interpretation of our Gaussian fit results in terms of \TMDs and quark densities. From eqns.\,(\ref{eq-mellinunpol}) and (\ref{eq-mellinpol}) we readily read off $f_1^{(1)}$, $g_{1L}^{(1)}$ and $g_{1T}^{(1)}$ in terms of the fit parameters and plot them in Fig.~\ref{fig-tmdsmellin}. In our plots, we label our profile functions with an extra superscript ``sW'', to remind the reader of the fact that these \TMDs have been obtained with straight Wilson lines and are therefore not strictly identical to the \TMDs defined and used in the literature and for the description of, e.g., SIDIS.


As discussed in \ref{sec-defmellin}, our first Mellin moments are combinations of quark and antiquark densities.
Presumably, the contribution from antiquarks is small, so it is interesting to test if our first Mellin moments numerically fulfill bounds analogous to those derived for \TMDs in Ref. \cite{Bacchetta:1999kz}.
Consider the bounds on $g_{1T}$ given by eq.~(\ref{eq-g1Tbounds}). In Fig.~\ref{fig-bachhineq}, we plot the function
\begin{equation}
	\Delta_g(|\vprp{k}|) \equiv \sqrt{ \left( f_1^{(1)\text{sW}}(\vprp{k}^2) \right)^2 
	- \left( g_{1L}^{(1)\text{sW}}(\vprp{k}^2) \right)^2 }  -
	\left| \frac{\vprp{k}}{m_N}\, g_{1T}^{(1)\text{sW}}(\vprp{k}^2) \right|
\end{equation}
with respect to $|\vprp{k}|$ on the coarse-m020 ensemble for $u$ and $d$-quarks, and find that it is positive.
Thus, on a numerical level, we find that the first Mellin moment $g_{1T}^{(1)\text{sW}}(\vprp{k}^2)$ complies with a similar bound as the corresponding $x$-dependent \TMD.
 
In the following, we want to give our Mellin moments an interpretation in terms of densities of quarks (minus antiquarks). In particular, we consider
\begin{align}
	\rho_{UU}(\vprp{k}) & \equiv \Phi^{[\gamma^+](1)}(\vprp{k};P,S)  \nonumber \\
		& = f_1^{(1)}(\vprp{k}^2) \ ,\\
	\rule{0pt}{2em}%
	\rule{0pt}{2em}%
	\rho_{TL}(\vprp{k};\vprp{S},\lambda) & \equiv \frac{1}{2}\,\Phi^{[\gamma^+](1)}(\vprp{k};P,S) + \frac{\lambda}{2}\,\Phi^{[\gamma^+\gamma^5](1)}(\vprp{k};P,S) \big \vert_{S=S_\prp} \nonumber \\
	& = \frac{1}{2} f_{1}^{(1)}(\vprp{k}^2) + \frac{\lambda}{2} \frac{\vprp{k} \cdot \vprp{S}}{m_N}\ g_{1T}^{(1)}(\vprp{k}^2) \ .
\end{align}
Here $\rho_{TL}(\vprp{k};\vprp{S},\lambda)$ is defined in analogy to eq.\,(\ref{eq-polarizedquarkdens}). 

\begin{figure}[p]
	\centering
	\begin{overpic}[width=0.85\textwidth,trim=10 20 20 20,clip=true]{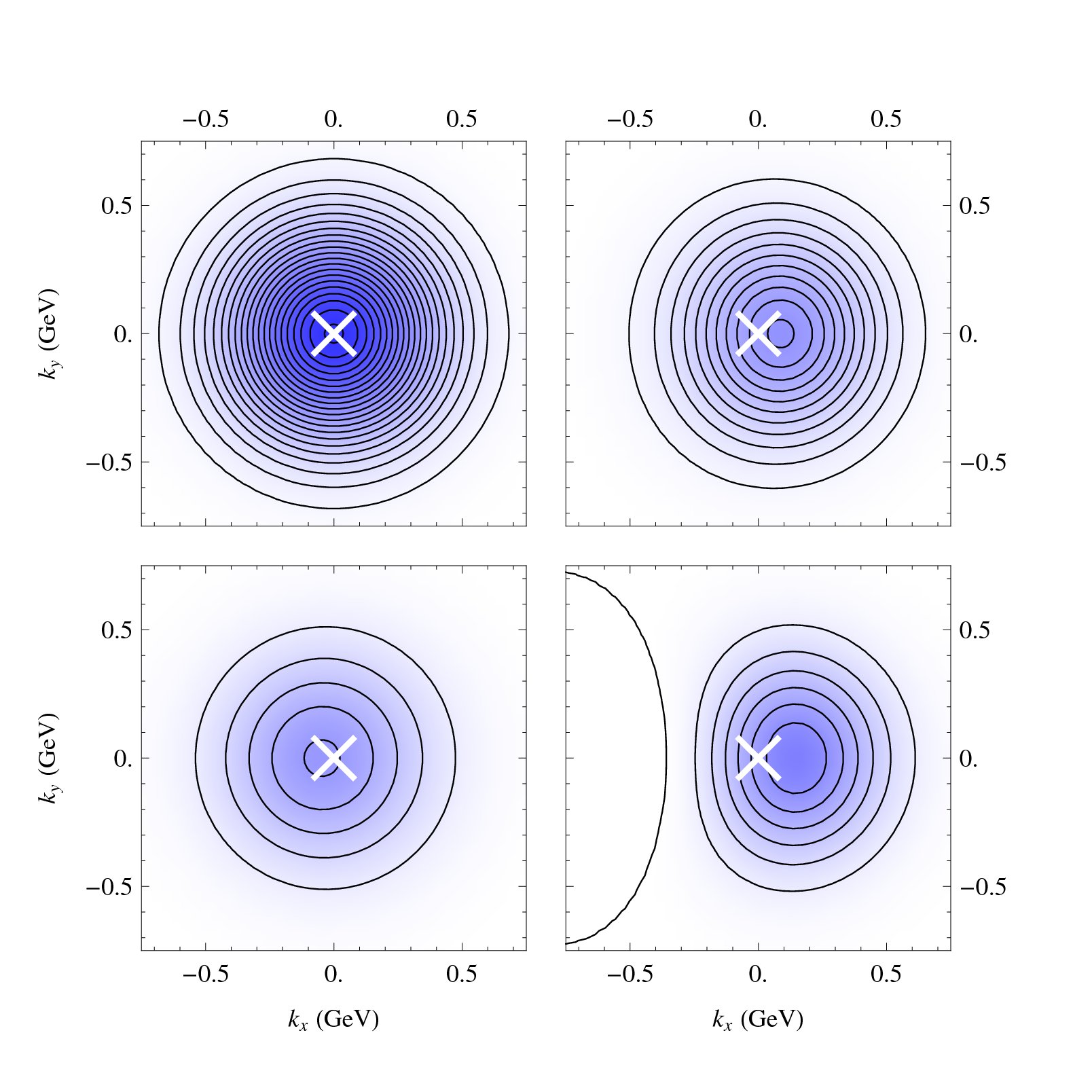}
		\put(12.5,51){\subfloat[][]{\label{fig-unpolDens}\hphantom{(m)}}}
		\put(44.5,86){up}
		\put(55,51){\subfloat[][]{\label{fig-polDensU}\hphantom{(m)}}}
		\put(87,86){up}
		\put(55,78.5){\includegraphics[width=1.6cm]{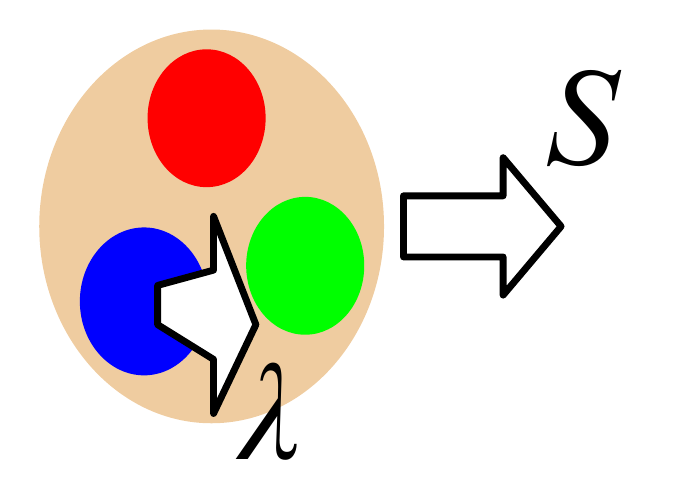}}
		\put(12.5,8.5){\subfloat[][]{\label{fig-polDensD}\hphantom{(m)}}}
		\put(41,42.5){down}
		\put(12.5,36.5){\includegraphics[width=1.6cm]{figs/PolarizedNucleon2.pdf}}
		\put(55,8.5){\subfloat[][]{\label{fig-polDensUminusD}\hphantom{(m)}}}
		\put(86,42.5){u-d}
		\put(55,36.5){\includegraphics[width=1.6cm]{figs/PolarizedNucleon2.pdf}}
		\end{overpic}\par
	\vspace{2em}
	\caption[polarized]{%
		Quark density plots in the transverse momentum plane at $m_\pi = 500\units{MeV}$ obtained from the Gaussian fits to the amplitudes $\tAmp_2(\elll^2,0)$ and $\tAmp_7(\elll^2,0)$ as depicted in Fig.~\ref{fig-ampmellin}. We highlight the origin with a white cross.\par
		\subref{fig-unpolDens}\mquad%
			Density of up quarks (minus antiquarks) $\rho_{UU,u}^{\text{sW}}(\vprp{k}) = f_{1,u}^{(1)\text{sW}}(\vprp{k})$.\par
		\subref{fig-polDensU}\mquad%
			Density $\rho_{TL,u}^{\text{sW}}(\vprp{k};\vprp{S},\lambda)$ of up quarks (minus antiquarks) with positive helicity $\lambda=1$ (i.e., with spin pointing in $z$-direc\-tion) in a nucleon polarized in transverse x-direction $\vprp{S} = (1,0)$. The upper left inset symbolizes the nucleon with its spin vector pointing to the right, and the probed quark with its spin pointing towards the reader. \par%
		\subref{fig-polDensD}
			Same as in \subref{fig-polDensU} but for down quarks. \par%
		\subref{fig-polDensD} Same as in \subref{fig-polDensU} for up quarks minus down quarks. The deformation appears amplified. \par%
		\label{fig-dens}
		}
	\end{figure}

The unpolarized density $\rho_{UU}$ is the density of quarks (minus antiquarks) in a nucleon with respect to the intrinsic transverse momentum of the quarks, averaged over the nucleon spin. We plot this density in the two-dimensional transverse momentum plane in Fig.~\ref{fig-unpolDens}. The distribution is axially symmetric. \par

Particularly interesting is $\rho_{TL}$, the transverse momentum dependent density of quarks (minus antiquarks) with definite helicity in a transversely polarized nucleon, compare section \ref{sec-poldens}. 
We plot this distribution for up quarks in Fig.~\ref{fig-polDensU}. For an intuitive interpretation, imagine a proton is moving towards us, the observer. The proton is polarized in transverse direction, with the spin vector pointing in positive x-direction, $\vprp{S} = (1,0)$. We now measure the probability distribution of $u$-quarks with spins pointing towards us (more precisely, with positive helicity $\lambda=1$). The density of quarks in the transverse momentum plane thus obtained appears deformed -- it is no longer axially symmetric.  The peak of the density is shifted to the right, to a positive value of $\vect{k}_{\prp 1}$. The reason for this deformation is that $g_{1T}$ is non-zero. For the corresponding $d$-quark distribution plotted in Fig.~(\ref{fig-polDensD}), the peak is shifted to the opposite direction, a result of the sign change in $\tAmp_7$ or rather $g_{1T}^{(1)}$, see Figs. \ref{fig-ampmellin} and \ref{fig-tmdsmellin}. The deformation appears amplified if we plot $\rho_{TL}(\vprp{k};\vprp{S},\lambda)$ for $u-d$ quarks as in Fig.~\ref{fig-polDensUminusD}. 

Within the formalism of Ref.~\cite{Miller:2007ae}, the deformations we observe in $\rho_{TL}$ are evidence for the ``non-spherical shape of the nucleon''. The densities plotted in Fig.~\ref{fig-dens} are reminiscent of the spin densities in the spatial transverse plane presented in Ref. \cite{Gockeler:2006zu}. In contrast to the study at hand, these densities have been obtained from GPDs, which have an interpretation in terms of impact parameter dependent distributions \cite{Diehl:2005jf}. Concerning the comparison to impact parameter $\vprp{b}$-dependent densities, the distribution $g_{1T}^{(1)}(\vprp{k}^2)$ is somewhat peculiar. It describes a correlation between transverse nucleon spin, quark helicity and transverse momentum of the form $\lambda \vprp{S} \tcdot \vprp{k}$.  A corresponding impact parameter dependent distribution with a correlation of the form $\lambda \vprp{S} \tcdot \vprp{b}$ does not exist \cite{Diehl:2005jf}. Another peculiarity of $g_{1T}^{(1)}(\vprp{k}^2)$ and its associated density $\rho_{TL}$ is the shift of quark density in the direction of the transverse nucleon spin. In contrast, the densities plotted in Ref. \cite{Gockeler:2006zu} feature an asymmetry perpendicular to the spin vector, due to a correlation in the form of a vector product $\vprp{S}\times\vprp{b}$.


To give a quantitative description of our findings, let us calculate some $\vprp{k}$-moments:
\begin{align}
	\left\langle \vprp{k}^2 \right\rangle_{\rho_{UU}} & \equiv \frac{ \int d^2 \vprp{k}\ \vprp{k}^2\ \rho_{UU}(\vprp{k})}{\int d^2 \vprp{k}\ \rho_{UU}(\vprp{k})} \ ,\nonumber \\
	\left\langle \vprp{k} \right\rangle_{\rho_{TL}} & \equiv \frac{ \int d^2 \vprp{k}\ \vprp{k}\ \rho_{TL}(\vprp{k};\vprp{S},\lambda)}{\int d^2 \vprp{k}\ \rho_{TL}(\vprp{k};\vprp{S},\lambda)} \ ,\nonumber \\
	\left[\frac{g_A}{g_V}\right]_\text{TMD} & = \frac{ \int d^2 \vprp{k}\ \Phi^{[\gamma^+\gamma^5](1)}(\vprp{k};P,S)\big\vert_{\Lambda=1,\ \vprp{S}=0}}{\int d^2 \vprp{k}\ \Phi^{[\gamma^+](1)}(\vprp{k};P,S)} \ .
	\label{eq-ktmoms}
\end{align}
Here $g_A/g_V$ is a quantity that can be obtained without any reference to transverse momentum dependence. The nucleon vector coupling constant $g_V$ is given by the number of valence quarks (i.e., $2$ in the $u$-- and $1$ in the $d$-- or $(u{-}d)$--channel). The nucleon axial vector coupling constant $g_A$ for $u{-}d$ quarks has been determined experimentally with high accuracy from neutron $\beta$-decay: $\left[g_A/g_V\right]^\text{phys}_{u-d} = 1.2695(29)$ \cite{Amsler:2008zzb}.
Using the Gaussian parametrization of the amplitudes, we find
\begin{align}
	\left\langle \vprp{k}^2 \right\rangle_{\rho_{UU}} & = (2/\sigma_2)^2  \ , \nonumber \\
	\left\langle \vprp{k} \right\rangle_{\rho_{TL}} & = - \lambda \vprp{S} \frac{m_N c_7}{c_2}  \ ,\nonumber \\
	\left[ \frac{g_A}{g_V} \right]_\text{TMD} & = - \frac{c_6}{c_2}\ .
	\label{eq-ktmomentsfromgauss}
\end{align}
Again, we remind the reader that the integrals in eq.\,(\ref{eq-ktmoms}) only exist if the distributions decay sufficiently fast with $\vprp{k}$. With the Gaussian parametrization, this is guaranteed, but the perturbative high-$|\vprp{k}|$ behavior is not reproduced correctly. 

Numerical results for the $\vprp{k}$-moments are included in Tables \ref{tab-amp2results}, \ref{tab-amp6results} and \ref{tab-amp7results}. Our results renormalized with the static quark potential and $C^\ren=0$ yield $\langle \vprp{k}^2 \rangle_{\rho_{UU}} \approx 0.4\units{GeV}$, which is of the same order of magnitude as the experimental estimate $0.5\units{GeV}$ of Ref.~\cite{Anselmino:2005nn} mentioned in section \ref{sec-cahn}. However, a serious quantitative comparison of our results for the RMS transverse momentum to phenomenological estimates is not justified at this stage, primarily for the following reasons:
Firstly, in contrast to $[g_A/g_V]_\text{TMD}$ and $\langle \vprp{k} \rangle_{\rho_{TL}}$, the RMS transverse momentum $\langle \vprp{k}^2 \rangle_{\rho_{UU}}$ is very sensitive to $\delta \hat m$, and thus to the choice of the renormalization condition. Moreover, we have employed a simplified contour for the Wilson line and pion masses much larger than the physical ones. 

The numbers in Table \ref{tab-amp7results} for $\left\langle \vprp{k} \right\rangle_{\rho_{TL}}$ are in line with what we saw in Figs.~\ref{fig-polDensU} and \ref{fig-polDensD}: Helicity polarized quarks in a transversely polarized nucleon carry a non-zero average transverse momentum. This average transverse momentum shift differs in sign for $u$-- and $d$--quarks. Specifically, with renormalization from the static quark potential, we find a shift of $\langle \vprp{k} \rangle_{\rho_{TL}} = (73 \pm 5) \lambda \vprp{S} \units{MeV}$ for up quarks, and a shift of about half the magnitude, $\langle \vprp{k} \rangle_{\rho_{TL}} = (-31 \pm 5) \lambda \vprp{S} \units{MeV}$, for down quarks.

For $u-d$ quarks, we find $[g_A/g_V]_\text{TMD} = 1.21 \pm 4$ on the coarse-m020 ensemble with renormalization from the static quark potential, see Table \ref{tab-amp6results}. We find agreement within errors with the value $g_A/g_V = 1.173(29)$ determined on the same ensemble in Ref. \cite{Edwards:2005ym}. Notice that our result is not so far from the experimental value. A quantitative comparison to experiment can of course only be made with an extrapolation to the physical pion mass, as performed, e.g., in the above reference \cite{Edwards:2005ym}. We rate the successful determination of $g_A/g_V$ as an important crosscheck of our methods.


\begin{sidewaystable}[tbp]
	\centering
	\renewcommand{\arraystretch}{1.1}
	\begin{tabular}{|c|c|c||l||l|l|l||l|}
	\hline
	ens. & quarks & $-\delta \hat m$ &
	$\chi^2$ &
	$\renZ_{\psi,z}$ & $c_2$ & $\sigma_2\ (\mathrm{fm})$ &
	$\langle \vprp{k}^2 \rangle^{1/2}_{\rho_{UU}}\ (\mathrm{GeV})$ \\
	\hline \hline
	m020 & $u-d$ & 0.1553(47)  & 0.6         & 1.063(13)   & 1           & 1.035(25)   & 0.381(10)(22)(14) \\
	"    & $u$   &  "          & 0.7         & "           & 2.0212(74)  & 1.009(19)   & 0.391(08)(22)(14) \\
	"    & $d$   &  "          & 0.6         & "           & 1.0211(77)  & 0.985(20)   & 0.401(08)(22)(14)  \\
	\hline \hline
	m020 & $u-d$ & 0           & 3.2         & 0.7676(43)  & 1           & 0.6942(58)  & 0.5685(47) \\
	"    & $u$   & "           & 9.1         & "           & 2.0223(68)  & 0.6861(34)  & 0.5752(28) \\
	"    & $d$   & "           & 7.5         & "           & 1.0253(68)  & 0.6766(39)  & 0.5833(34) \\
	\hline
	m030 & $u-d$ & 0           & 7.4         & 0.7544(30)  & 1           & 0.6960(38)  & 0.5670(31)  \\
	\hline
	m050 & $u-d$ & 0           & 10          & 0.7423(21)  & 1           & 0.6876(31)  & 0.5740(26)\\
	\hline
	\end{tabular}\par\vspace{1ex}
	\renewcommand{\arraystretch}{1.0}
	\caption{Numbers obtained from the Gaussian fits for $\tAmp_2$, determined on the HYP smeared coarse ensembles. The value for $\delta \hat m$ in the upper section has been determined using the static quark potential, see section \ref{sec-renstatqpot}. The renormalization constant $\renZ_{\psi,z}$ is determined from $u-d$ quarks and applied globally. Due to correlations, the value $\chi^2$ (per degree of freedom) specified here gives only a crude orientation about the fit quality, see section \ref{sec-correl}. The meaning of the errors is, in sequence, 
	(1) statistical error, determined from 1000 bootstrap samples,
	(2) systematic error determined from the smeared--unsmeared comparison according to section~\ref{sec-discerr} and eq.~(\ref{eq-widtherrprop})
	and (3) uncertainty from the determination of the lattice spacing.}
	\label{tab-amp2results}%
\end{sidewaystable}






\begin{table}[tbp]
	\centering
	\renewcommand{\arraystretch}{1.1}
	\subfloat[][]{
	\label{tab-amp6results}
	\begin{tabular}{|c|c|c||l||l|l||l|}
	\hline
	ens. & quarks & $-\delta \hat m$ &
	$\chi^2$ &
	$c_6$ & $\sigma_6\ (\mathrm{fm})$ &
	$g_A/g_V$ \\
	\hline \hline
	m020 & $u-d$ & 0.1553(47)  & 0.2         &           -1.211(36)  & 1.173(33)   & \phantom{-}1.211(36)(14) \\
	"    & $u$   &  "          & 0.5         &           -0.927(33)  & 1.202(39)   & \phantom{-}0.458(16)(06) \\
	"    & $d$   &  "          & 0.9         & \phantom{-}0.287(18)  & 1.066(52)   &           -0.281(18)(03) \\
	\hline
	\hline
	m020 & $u-d$ & 0           & 1.9         &           -1.124(34)  & 0.7713(72)  & \phantom{-}1.124(34) \\
	"    & $u$   & "           & 1.8         &           -0.860(31)  & 0.7789(94)  & \phantom{-}0.425(15) \\
	"    & $d$   & "           & 1.0         & \phantom{-}0.269(17)  & 0.734(19)   &           -0.262(17) \\
	\hline
	m030 & $u-d$ & 0           & 4.3         &           -1.149(20)  & 0.7638(40)  & \phantom{-}1.149(20) \\
	\hline
	m050 & $u-d$ & 0           & 9.6         &           -1.123(13)  & 0.7541(34)  & \phantom{-}1.123(13) \\
	\hline
	\end{tabular}\par}\vspace{2em}
	\subfloat[][]{
	\label{tab-amp7results}
	\begin{tabular}{|c|c|c||l||l|l||l|}
	\hline
	ens. & quarks & $-\delta \hat m$ &
	$\chi^2$ &
	$c_7$ & $\sigma_7\ (\mathrm{fm})$ &
	$\langle \vprp{k} \rangle_{\rho_{TL}}\ (\mathrm{GeV} \times \lambda \vprp{S})$ \\
	\hline \hline
	m020 & $u-d$ & 0.1553(47)  & 0.4         &           -0.1087(50) & 1.096(31)   & \phantom{-}0.1793(83)(30)(63)  \\
	"    & $u$   &  "          & 0.6         &           -0.0895(53) & 1.094(37)   & \phantom{-}0.0730(42)(13)(26)  \\
	"    & $d$   &  "          & 1.1         & \phantom{-}0.0190(29) & 1.103(97)   &           -0.0307(48)(06)(11)  \\
	\hline \hline
	m020 & $u-d$ & 0           & 0.7         &           -0.0970(46) & 0.7586(95)  & \phantom{-}0.1600(76)  \\
	"    & $u$   & "           & 0.8         &           -0.0796(47) & 0.760(12)   & \phantom{-}0.0649(38)  \\
	"    & $d$   & "           & 1.1         & \phantom{-}0.0169(27) & 0.765(36)   &           -0.0271(44)  \\
	\hline
	\end{tabular}\par}\vspace{2em}
	\renewcommand{\arraystretch}{1.0}
	\caption{Numbers obtained from the Gaussian fits for $\tAmp_6$ and $\tAmp_7$, determined on the HYP smeared coarse ensembles. The value for $\delta \hat m$ in the upper sections has been determined using the static quark potential, see section \ref{sec-renstatqpot} in the text. The renormalization constant $\renZ_{\psi,z}$ is taken from the analysis of $\tAmp_{2,u-d}$. Due to correlations, the value $\chi^2$ (per degree of freedom) specified here gives only a crude orientation about the fit quality, see section \ref{sec-correl}. The meaning of the errors is, in sequence, 
	(1) statistical error, determined from 1000 bootstrap samples,
	(2) crude estimate of systematic errors from the smeared--unsmeared comparison of section~\ref{sec-discerr},
	and (3) uncertainty from the determination of the lattice spacing.}
	\label{tab-amp6amp7results}%
\end{table}

\subsection{Ratios of Amplitudes}
\label{sec-ampratios}

\begin{table}[tbp]
	\centering
	\renewcommand{\arraystretch}{1.1}
	\begin{tabular}{|c|c||l|l|}
	\hline
	\rule{0pt}{1.2em}%
	ens. & quarks & 
	$g_A/g_V$ &
	$\left[\langle \vprp{k} \rangle_{\rho_{TL}}\right]_\text{reg}\ (\mathrm{GeV} \times \lambda \vprp{S})$ \\
	\hline \hline
	m020 & $u-d$ & \phantom{-}1.216(48)  & \phantom{-}0.162(19)(06) \\
	"    & $u$   & \phantom{-}0.468(21)  & \phantom{-}0.072(10)(03)    \\
	"    & $d$   &           -0.256(23)  &           -0.017(12)(01) \\
	\hline \hline
	m030 & $u-d$ & \phantom{-}1.199(30)  & \\
	\hline
	m050 & $u-d$ & \phantom{-}1.193(22)  & \\
	\hline
	\end{tabular}\par\vspace{1em}
	\renewcommand{\arraystretch}{1.0}
	\caption{Numbers obtained from the renormalization scale and scheme independent ratios $\tAmp_i/\tAmp_2$, determined on the HYP smeared coarse ensembles. The first error is statistical and has been determined using 1000 bootstrap samples, the second error comes from the determination of the lattice spacing $a$. Unspecified systematic uncertainties also arise from the polynomial extrapolation of the ratios of amplitudes down to $\elll^2=0$.}
	\label{tab-ampratioresults}%
\end{table}

\begin{figure}[p]
	\captionsetup[subfigure]{justification=raggedright,singlelinecheck=false}
	\centering
	\begin{overpic}{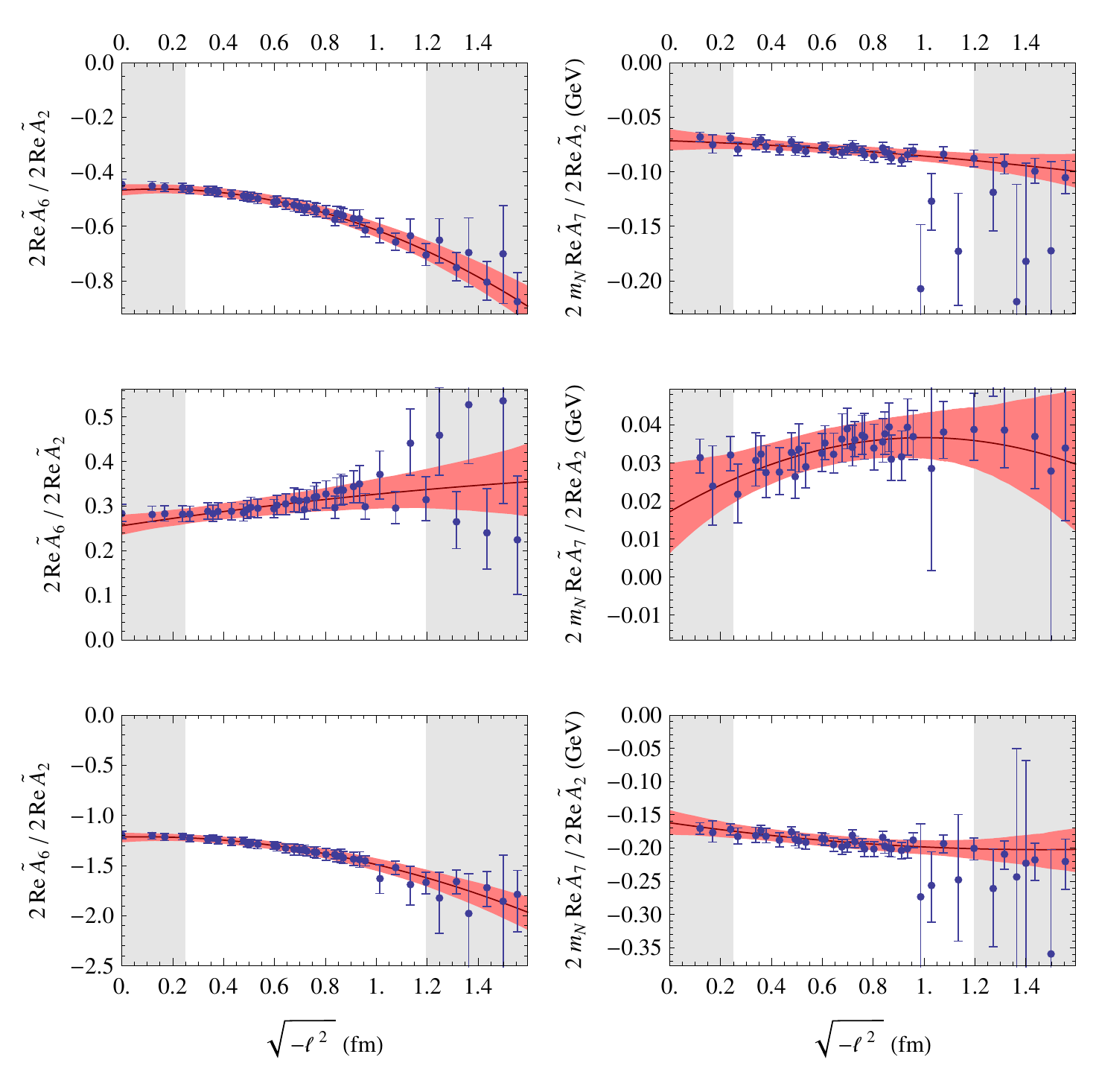}
		\put(13,89){\subfloat[][{up}]{\hspace*{2cm}} }
		\put(63,89){\subfloat[][{up}]{\hspace*{2cm}} }
		\put(13,59){\subfloat[][{down}]{\hspace*{2cm}} }
		\put(63,59){\subfloat[][{down}]{\hspace*{2cm}} }
		\put(13,29){\subfloat[][{u-d}]{\hspace*{2cm}} }
		\put(63,29){\subfloat[][{u-d}]{\hspace*{2cm}} }
	\end{overpic}\par
	\vspace{2em}
	\caption[TMDs]{%
		Ratios of amplitudes $\tAmp_i(\elll^2,\elll\cdot P=0)$ to $\tAmp_2(\elll^2,\elll\cdot P=0)$ for straight Wilson lines calculated on the HYP smeared coarse-m020 ensemble. These quantities need no renormalization. The shaded regions in the range $\sqrt{-\elll^2} \lesssim 0.25\units{fm}$ and for $\sqrt{-\elll^2} \gtrsim L/2$ give a conservative estimate where we expect lattice cutoff effects and finite volume effects. 
		\label{fig-ampratiosmellin}
		}
	\end{figure}
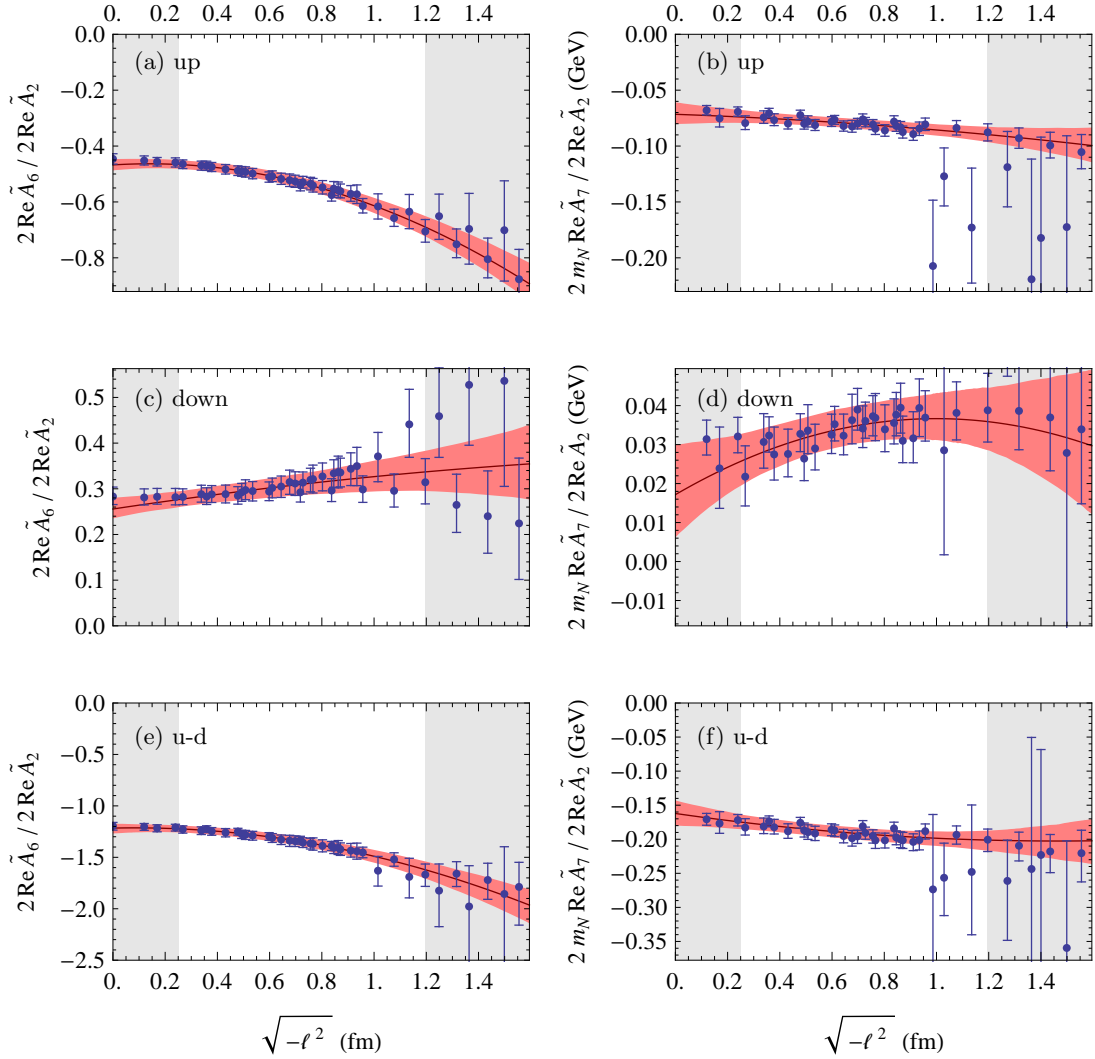

Observables which can be obtained from ratios of amplitudes $\tAmp_i$ are particularly attractive, because renormalization factors cancel entirely:
\begin{equation}
	\frac{\tAmp_i(\elll^2,0)}{\tAmp_2(\elll^2,0)} = \frac{\tAmp^\unren_i(\elll^2,0)}{\tAmp^\unren_2(\elll^2,0)}\ .
	\label{eq-ampratios}
\end{equation}
The unrenormalized amplitudes can be directly obtained from the unrenormalized ratios, provided $\elll^2$ is in the range where discretization artefacts are small enough. We plot ratios of amplitudes in Fig.~\ref{fig-ampratiosmellin}.

What can we read off from these amplitudes? The formalism in appendix \ref{sec-kprpmomentsformalism} shows how $\vprp{k}$-moments can be expressed in terms of the amplitudes $\tAmp(\elll^2,0)$ at $\elll^2=0$. Concerning the observables of the previous section, we readily obtain
\begin{align}
	\left[ \left\langle \vprp{k}^2 \right\rangle_{\rho_{UU}} \right]_\text{reg} & = -4\, \frac{\frac{\partial}{\partial(-\elll^2)}\,\tAmp_2(\elll^2,0)\vert_{\elll=0}}{\tAmp_2(0,0)} \label{eq-rmsktfromratio} \ , \\
	\left[ \frac{g_A}{g_V} \right]_\text{TMD} & = - \frac{\tAmp_6(0,0)}{\tAmp_2(0,0)} \label{eq-gafromratio} \ ,\\
	\left[ \left\langle \vprp{k} \right\rangle_{\rho_{TL}} \right]_\text{reg} & = - \lambda \vprp{S} \frac{m_N \tAmp_7(0,0)}{\tAmp_2(0,0)} \ . \label{eq-meanktfromratio}
\end{align}
In eqns.\,(\ref{eq-rmsktfromratio}) and (\ref{eq-meanktfromratio}), the subscript ``reg'' means that we have omitted certain terms that vanish if we make the additional assumption that derivatives of $\tAmp_i(\elll^2,0)$ are finite at $\elll^2=0$. The omitted terms are higher derivatives of the amplitudes multiplied by factors of $\vprp{\elll}$ or $\elll^2$, compare the right hand side of eq.~(\ref{eq-ktmomentsfromamps}) in the appendix.
Incidentally, the assumption of finite derivatives at $\elll^2=0$ is true for Gaussian amplitudes, so that we recover eq.\,(\ref{eq-ktmomentsfromgauss}) upon substitution of the Gaussian parametrizations.

The solid curves and error bands in Fig.~\ref{fig-ampratiosmellin} have been obtained from fits to polynomials of second order, where only the data for $0.25 \units{fm} \leq \sqrt{-\elll^2} \leq L/2$ has been used as input, and where the fit weights have been chosen according to the statistical errors of the individual data points. Let us make an extrapolation of the ratios $\tAmp_i(\elll^2,0)/\tAmp_2(\elll^2,0)$ down to $\elll^2=0$ based on the polynomial fits. This way, we obtain numbers for $[g_A/g_V]_\text{TMD}$ and $[ \left\langle \vprp{k} \right\rangle_{\rho_{TL}} ]_\text{reg}$, which we list in Table \ref{tab-ampratioresults}. The numbers qualitatively agree with the results from the Gaussian fits. Small differences are expected, because the ratio method and the Gaussian parametrization represent two different ways to extrapolate down to $\elll^2=0$. In that sense, these differences give a crude estimate for a systematic uncertainty. The numbers for $[g_A/g_V]_\text{TMD}$ in the $(u{-}d)$--channel agree within their (relatively large) statistical errors with the results for $g_A$ quoted in Ref. \cite{Edwards:2005ym}, although we note that they are systematically higher.

We now understand why $g_A/g_V$ and $\left\langle \vprp{k} \right\rangle_{\rho_{TL}}$ turned out to be insensitive to $\delta \hat m$ in the previous chapter: They can be obtained from ratios of amplitudes. It will be interesting to study the small-$\elll^2$-behavior of the amplitudes and its effect on the quantity $\langle \vprp{k} \rangle_{\rho_{TL}}$.


%

\subsection{Chiral Extrapolation}

\begin{figure}[tbp]
	\centering
	\includegraphics{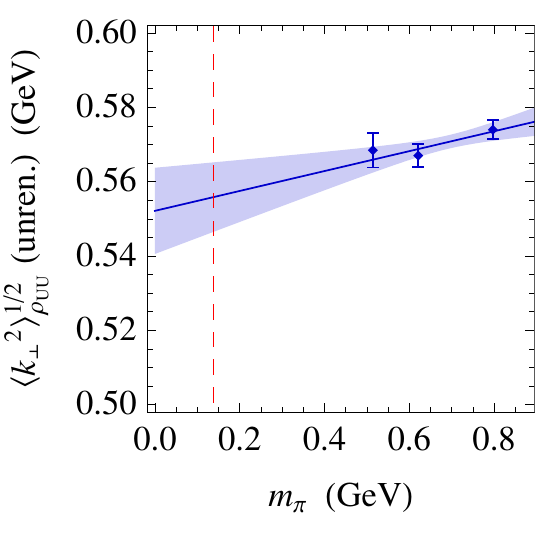}
	\caption[Chiral extrapolation]{%
		Linear chiral extrapolation of the unrenormalized RMS width of the transverse momentum distribution $\left\langle \vprp{k}^2 \right\rangle_{\rho_{UU}}$ obtained from the coarse ensembles. The error bars and the error band correspond to the statistical errors. We indicate the location of the physical pion mass by a dashed vertical line. 
		\label{fig-rmskt-extrapol}
		}
	\end{figure}

To make physical predictions, an extrapolation to physical quark masses is needed. For some quantities, such as $g_A$, sophisticated extrapolation formulae have been worked out in chiral effective field theory. Using a simple linear extrapolation instead introduces large unquantifiable systematic errors but may serve as a first guess. Let us look at quantities specifically related to \TMDs. With the data presently available from our analyses, we can only investigate the quark mass dependence of the RMS width of the transverse momentum distribution $\left\langle \vprp{k}^2 \right\rangle_{\rho_{UU}}$ as we obtain it from the Gaussian fit. Up to now, we have not worked out the renormalization constant $\delta \hat m$ for the coarse-m030 and coarse-m050 ensembles, but it is plausible that they will not depend strongly on the sea quark masses. Let us take the values $\left\langle \vprp{k}^2 \right\rangle_{\rho_{UU}}^{1/2}$ for $u{-}d$ with $\delta \hat m = 0$ from Table~\ref{tab-amp2results} and extrapolate them linearly down to the physical pion mass. We obtain $\left\langle \vprp{k}^2 \right\rangle_{\rho_{UU}}^{1/2} = 0.5558(94)_\text{stat} \units{GeV}$ (with a $\chi^2$ per degree of freedom of $0.7$). The extrapolation is illustrated in Fig.~\ref{fig-rmskt-extrapol}. It turns out that the extrapolated value hardly differs from our input at large $m_\pi$. For the high pion masses in our study, the RMS transverse momentum shows practically no sensitivity to the light quark masses. However, we should be aware of the fact that a much more pronounced quark mass dependence may set in at lower pion masses\footnote{as is expected in, e.g., the extrapolation of $g_A$ \cite{HPW03,Khan:2006de,Procura:2006gq,Edwards:2005ym} and of $\langle x \rangle$ \cite{Dorati:2007pv}.}.

\section{Dependence on the Longitudinal Momentum Fraction \texorpdfstring{$x$}{x}}
\label{sec-xdep}

\begin{figure}[tbp]
	\centering
	\includegraphics[width=0.7\textwidth]{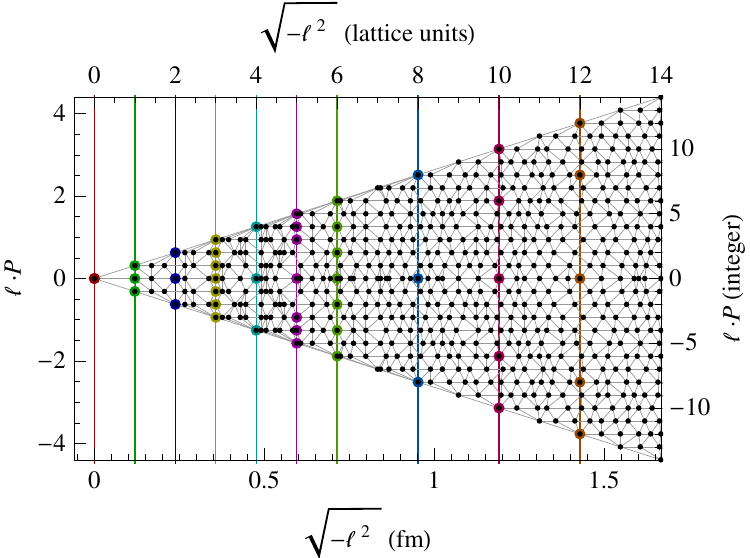}
	\caption[TMDs]{%
		Illustration of our procedure to obtain cuts at constant $\elll^2$. The black dots are points in the $(\elll^2,\elll \tcdot P)$-plane where lattice data is available. The vertical lines are selected values of $\elll^2$ for which we plot  $\tilde A_2(\elll^2,\elll \tcdot P)$ in Fig.~\ref{fig-amp2ldPcutsunren}. The colored dots correspond to the lattice data points shown in these figures. To obtain intermediate values, we have interpolated linearly on the $(\elll^2,\elll \tcdot P)$-plane, using the triangulation shown as a grey mesh.
		\label{fig-ldPcutsselection}
		}
	\end{figure}

\begin{figure}[tbp]
	\centering%
	\subfloat[][]{%
		\label{fig-amp2ldPcutsRe}%
		\includegraphics[clip=true,width=0.7\textwidth]{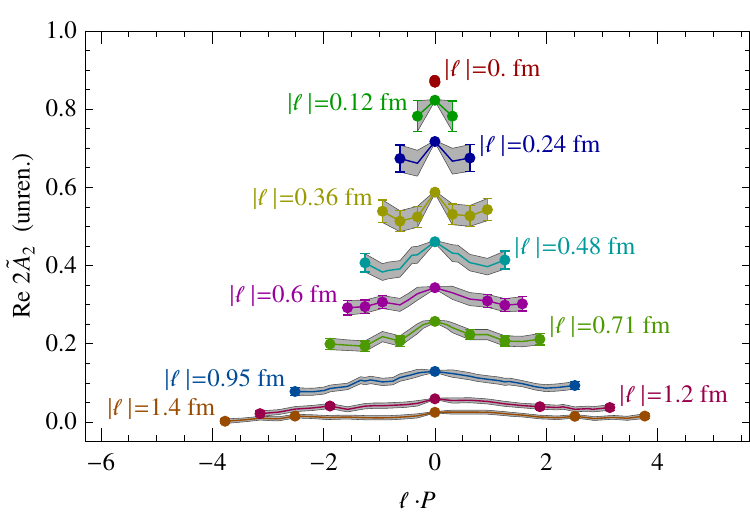}%
		}\par%
	\vspace{1em}
	\subfloat[][]{%
		\label{fig-amp2ldPcutsIm}%
		\includegraphics[clip=true,width=0.7\textwidth]{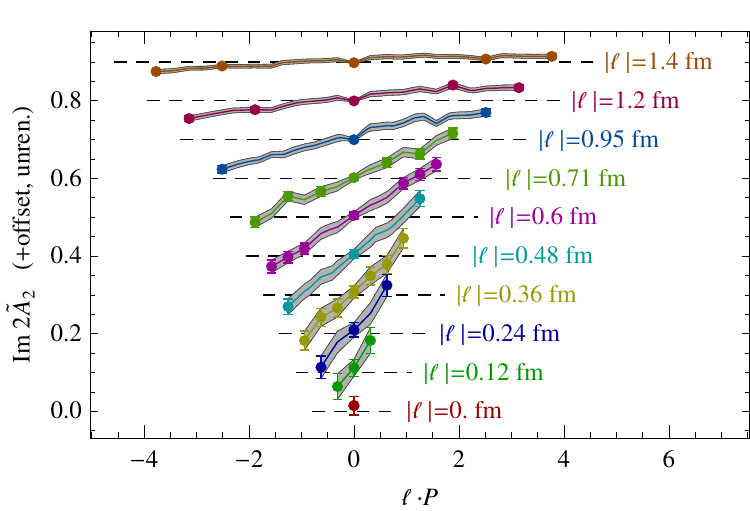}%
		}\par%
	\vspace{1em}
	\caption[$\elll \tcdot P$ cuts]{%
		Cuts through the unrenormalized amplitude $\tAmp_2(\elll^2,\elll\tcdot P)$ for a selection of fixed values $\sqrt{-\elll^2}=|\vect{\elll}|$ on the smeared coarse-m030 ensemble for $u-d$ quarks. The error band between the data points has been obtained from linear interpolation as illustrated in Fig.~\ref{fig-ldPcutsselection}.\quad \subref{fig-amp2ldPcutsRe} Real part, \quad \subref{fig-amp2ldPcutsIm} Imaginary part. For the sake of clarity, we have added offsets in the ordinate in steps of $0.1$. The dashed lines indicate the respective zero lines. 
		\label{fig-amp2ldPcutsunren}
		}
\end{figure}

In the previous section we have studied lattice data for $\elll \tcdot P=0$. Let us now explore the $(\elll \tcdot P)$-dependence of $\tilde A_2(\elll^2,\elll \tcdot P)$. The first integral in the last line of eq.\,(\ref{eq-f1fromtildeA2}) shows that it is related to the dependence of $f_1^\lat(x,\vprp{k})$ on the longitudinal momentum fraction $x$ via a Fourier transformation. 


We gave a three-dimensional overview of the available data for this amplitude in Fig.~\ref{fig-lpsurface}. Let us now work on the coarse-m030 ensemble, where we have rather good statistics.
In order to be able to plot $\tilde A_2(\elll^2,\elll \tcdot P)$ as a function of $\elll \cdot P$ at fixed values of $\elll^2$, we interpolate linearly between our data points as illustrated in Fig.~ \ref{fig-ldPcutsselection}. This enables us to take a closer look at the $(\elll \tcdot P)$-dependence in Fig.~\ref{fig-amp2ldPcutsunren}. First of all, the symmetry of the plots clearly confirms the relation $[ \tAmp_i(\elll^2,\elll \tcdot P) ]^* = \tAmp_i(\elll^2,-\elll \tcdot P)$ we found in section \ref{sec-symtrafo}. 
Moreover, we recognize a general suppression of the amplitude with increasing $\elll^2$. 
Apart from this suppression, and apart from finer structures which might be statistical fluctuations or discretization errors from the steplike linkpaths, the $(\elll \tcdot P)$--behavior follows a general trend independent of $\elll^2$: In the real part, we notice a slight curvature downward. The imaginary part shows strong $(\elll \tcdot P)$-dependence in the form of an almost linear rise.

\subsection{A Normalized Amplitude}

\begin{figure}[tbp]
	\centering%
	\subfloat[][]{%
		\label{fig-amp2normRe}%
		\includegraphics[clip=true,width=0.7\textwidth]{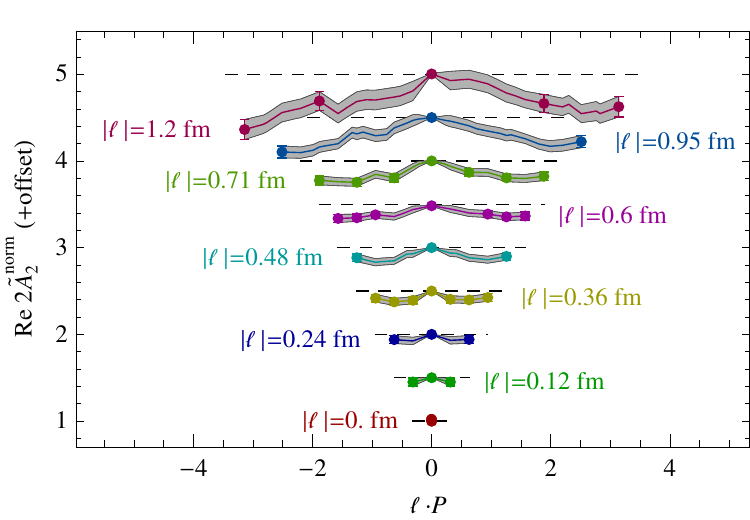}%
		}\par%
	\vspace{1em}
	\subfloat[][]{%
		\label{fig-amp2normIm}%
		\includegraphics[clip=true,width=0.7\textwidth]{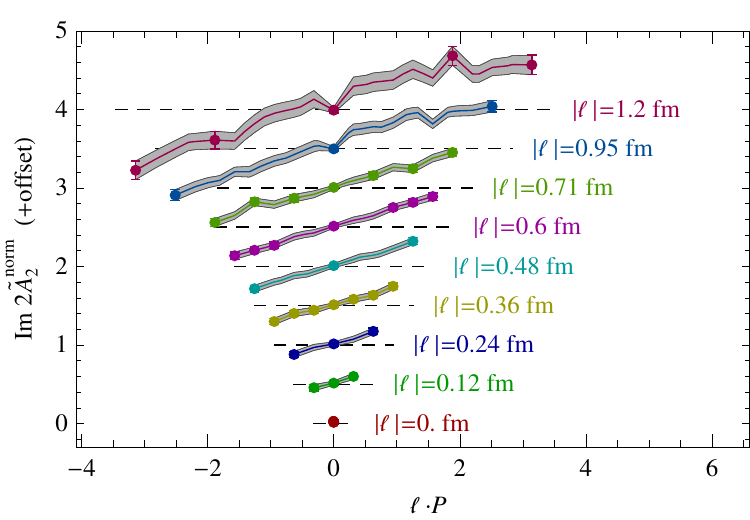}%
		}\par%
	\vspace{1em}
	\caption[$\elll \tcdot P$ cuts]{%
		Cuts through the normalized amplitude $\tAmp_2^\norm(\elll^2,\elll\tcdot P)$ for a selection of fixed values $\sqrt{-\elll^2}=|\vect{\elll}|$ on the smeared coarse-m030 ensemble for $u-d$ quarks. The error band between the data points has been obtained from linear interpolation as illustrated in Fig.~\ref{fig-ldPcutsselection}. For the sake of clarity, we have added offsets in the ordinate in steps of $0.5$. \par \subref{fig-amp2ldPcutsRe} Real part. The dashed lines are plotted at an ordinate of $1$, shifted by the respective offset. \par \subref{fig-amp2ldPcutsIm} Imaginary part. The dashed lines indicate the respective zero lines.
		\label{fig-amp2norm}
		}
\end{figure}

The observations above have inspired us to study the normalized amplitude
\begin{equation}
	\tAmp_2^\norm(\elll^2,\elll \tcdot P) \equiv \frac{\tAmp_2(\elll^2,\elll \tcdot P)}{\tAmp_2(\elll^2,0)}
	= \frac{\tAmp^\text{unren}_2(\elll^2,\elll \tcdot P)}{\tAmp^\text{unren}_2(\elll^2,0)}\ .
	\label{eq-normamp}
\end{equation}
We point out that the quantity defined above needs no renormalization, just as the ratios of amplitudes in eq.\,(\ref{eq-ampratios}). At present, we obtain $\tAmp_2^\norm(\elll^2,\elll \tcdot P)$ from 
\begin{equation}
	\tAmp_2^\norm(\elll^2,\elll \tcdot P) = \frac{R_{\gamma_\Eu{4}}(\vect{P},\mathcal{C}^\lat_\elll)}{I(\elll^2)}\ .
\end{equation}
Here $R_{\gamma_\Eu{4}}(\vect{P},\mathcal{C}^\lat_\elll)$ is the (unrenormalized) ratio defined in eq.~(\ref{eq-ratiodef}). The function $I(\elll^2)$ is a smooth interpolation of $R_{\gamma_\Eu{4}}(\vect{P},\mathcal{C}^\lat_\elll)$ at $\elll \tcdot P = 0$, where we have chosen straight link paths $\mathcal{C}^\lat_\elll$ on the lattice axes. In the future, we should use the ratio
\begin{equation}
	\tAmp_2^\norm(\elll^2,\elll \tcdot P) = \frac{R_{\gamma_\Eu{4}}(\vect{P},\mathcal{C}^\lat_\elll)}{R_{\gamma_\Eu{4}}(\vect{0},\mathcal{C}^\lat_\elll)}\ .
\end{equation}
The expected advantage of the latter is an optimal cancellation of statistical fluctuations and residual discretization errors from the step-like link paths. Up to now, data for $R_{\gamma_\Eu{4}}(\vect{0},\mathcal{C}^\lat_\elll)$ has not been calculated for the full set of link paths, so we must content ourselves with the interpolation method. 

We plot $\tAmp_2^\norm(\elll^2,\elll \tcdot P)$ as a function of $\elll \tcdot P$ for fixed $\elll^2$ in Fig.~\ref{fig-amp2norm}. It appears as though the normalized amplitude is largely $\elll^2$-independent. In the real part, we observe a slight curvature; the amplitude bends down with increasing $|\elll \tcdot P|$. A much cleaner picture emerges for the imaginary part. Here the amplitude basically describes a straight line through the origin, with a slope that appears to be $\elll^2$-independent. 

\begin{figure}[tbp]
	\centering%
	\subfloat[][]{%
		\label{fig-amp2lindepRe}%
		\includegraphics[clip=true,width=0.49\textwidth]{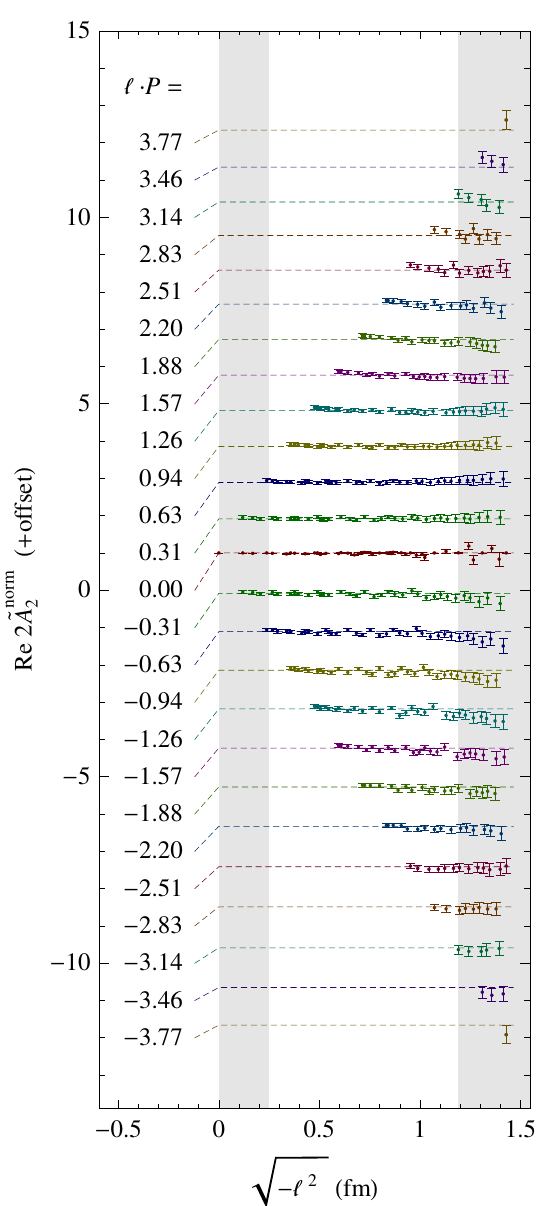}%
		}\hfill%
	\subfloat[][]{%
		\label{fig-amp2lindepIm}%
		\includegraphics[clip=true,width=0.49\textwidth]{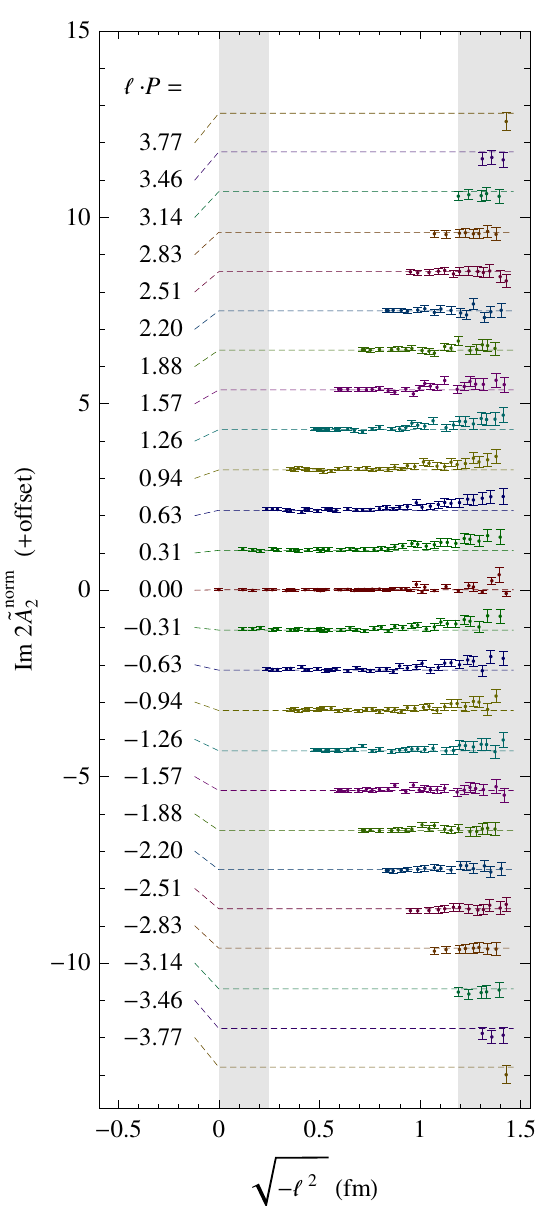}%
		}\par%
	\vspace{1em}
	\caption[Checking $\elll^2$-independence]{%
		Testing $\elll^2$-independence on the smeared coarse-m030 ensemble for $u-d$ quarks: We plot our data for $\tAmp_2^\norm(\elll^2,\elll \tcdot P)$ with offsets $(\elll \tcdot P)\hat L/2\pi$ in the ordinate. To guide the eye, we display the average of $\tAmp_2^\norm(\elll^2,\elll \tcdot P)$ at constant $\elll \tcdot P$ as horizontal dashed lines. In the averaging procedure for the horizontal lines, we also include data points with $\elll \tcdot P$ of opposite sign, taking into account the property $[\tAmp_2^\norm(\elll^2,\elll \tcdot P)]^* = \tAmp_2^\norm(\elll^2,-\elll \tcdot P)$.
		The magnitude of the amplitude relative to the offset can be read off from the inclined dashed lines on the left. The regions with a gray background are more likely to be affected by significant lattice artefacts.
		\label{fig-amp2lindep}
		}
\end{figure}

The values for $\elll \tcdot P$ we can access on the lattice are integer multiples of $2\pi/\hat L$. In Fig.~\ref{fig-amp2lindep} we plot $\tAmp_2^\norm(\elll^2,\elll \tcdot P)$ for each of these values as a function of $\elll^2$. We do not see any significant dependence of $\tAmp_2^\norm(\elll^2,\elll \tcdot P)$ on $\elll^2$. In conclusion, within the range of available lattice data and within our level of precision it is appropriate to write
\begin{equation}
	\tAmp_2^\norm(\elll^2,\elll \tcdot P) \approx \tAmp_2^\norm(\elll \tcdot P)
\end{equation}
and, making use of definition eq.\,(\ref{eq-normamp}),
\begin{equation}
	\tAmp_2(\elll^2,\elll \tcdot P) \approx \tAmp_2^\norm(\elll \tcdot P)\ \tAmp_2(\elll^2,0)\ .
	\label{eq-ampfactorization}
\end{equation}
In other words, the amplitude $\tAmp_2(\elll^2,\elll \tcdot P)$ approximately factorizes into an $\elll^2$-- and an $(\elll \tcdot P)$-dependent part.

\subsection{Factorization Hypothesis}

What might be the implication of this observation in terms of \TMDs?
Suppose eq.\,(\ref{eq-ampfactorization}) were true in the entire domain of $\elll^2$ and $\elll \tcdot P$. Inserting this hypothetical factorization into eq.\,(\ref{eq-f1fromtildeA2}), we obtain
\begin{align}
\text{`` }f_1(x,\vprp{k}^2)
& = \int \frac{d(\elll \tcdot P)}{2\pi}\ e^{-i (\elll \cdot P) x}\ \tAmp_2^\norm(\elll \tcdot P) \ \int_0^\infty \frac{d(-\elll^2)}{2(2\pi)}\ J_0(\sqrt{-\elll^2}\, |\vprp{k}|)\ 2\,\tAmp_2(\elll^2,0) \nonumber \\
& = \underbrace{ \int \frac{d(\elll \tcdot P)}{2\pi}\ e^{-i (\elll \cdot P) x}\ \frac{\tAmp_2(0,\elll \tcdot P)}{\tAmp_2(0,0)} }_{\displaystyle f_1(x)/\tAmp_2(0,0) }\ \underbrace{\int_0^\infty \frac{d(-\elll^2)}{2(2\pi)}\ J_0(\sqrt{-\elll^2}\, |\vprp{k}|)\ 2\,\tAmp_2(\elll^2,0)}_{\displaystyle  f_1^{(1)}(\vprp{k})} \text{ ''.}
\label{eq-f1factorization}
\end{align}
The two integrations can now be carried out independently, so that the factorization assumption of $\tAmp_2(\elll^2,\elll \tcdot P)$ translates into a factorization of the quark density 
\begin{equation}
	\text{`` } f_1(x,\vprp{k}^2) = \frac{f_1(x)}{\mathcal{N}}\ f_1^{(1)}(\vprp{k}) \text{ ''}
	\label{eq-f1factorized}
\end{equation}
into an $x$- and a $\vprp{k}$-dependent part. The latter is just the first Mellin moment discussed in section~\ref{sec-mellin}. 
The $x$-dependent part is identified as the regular PDF $f_1(x)$ divided by the number of valence quarks $\mathcal{N} \equiv \tAmp_2(0,0)$. 

At this point, some serious words of caution are in order. The observations we make with limited statistical significance and within the limited range of $\elll^2$ and $\elll \tcdot P$ on the lattice certainly do not justify a claim that $\tAmp_2(\elll^2,\elll \tcdot P)$ factorizes strictly in the entire domain. Moreover, the amplitude $\tAmp_2^\norm(\elll^2,\elll \tcdot P)$ needs no renormalization, while, in contrast, $f_1(x)$ is a renormalized quantity. Obviously, somewhere in our calculation eq.\,(\ref{eq-f1factorization}) we have ignored necessary renormalization steps. As discussed in section section \ref{sec-critgauss}, we expect that in the continuum the amplitude $\tAmp_2(\elll^2,\elll \tcdot P)$ diverges for $\elll^2 \rightarrow 0$. Thus we may reason that the expression $\tAmp_2(0,\elll \tcdot P)/\tAmp_2(0,0)$ is only well defined within a suitable regularization prescription, which then carries over to $f_1(x)$. 

At the present stage, eq.~(\ref{eq-f1factorization}) shows us that there is a qualitative analogy between the observed factorization of $\tAmp_2(\elll^2,\elll \tcdot P)$ and a factorization assumption of the form of eq.\,(\ref{eq-f1factorized}). The latter has been frequently used as a working hypothesis in phenomenological applications, in particular in combination with the Gaussian parametrization of $f_1^{(1)}(\vprp{k})$, see, e.g., the analysis of Ref.~\cite{Anselmino:2005nn} discussed in section \ref{sec-cahn}. Our lattice data provides some evidence that the factorization assumption is justified as an approximation, at least in a certain kinematical region, with the caveat that we have used straight Wilson lines in our correlators. 

In Fig.~\ref{fig-amp2normldP}, we plot our lattice data for $\tAmp_2^\norm(\elll^2,\elll \tcdot P)$ in stripes of constant $\elll \tcdot P$, omitting data with $\sqrt{-\elll^2} < 0.25\units{fm}$ to be on the safe side concerning lattice cutoff effects. Small offsets on the abscissa enable us to verify again that there is no strong $\elll^2$-dependence at a given $\elll \tcdot P$.
The error bands superimposed on the data have been obtained with polynomial fits which correspond to a parametrization of $\tAmp_2^\norm$ as
\begin{equation}
	\tAmp_2^\norm(\elll^2,\elll \tcdot P) = 1 + i (\elll \tcdot P)\, b_2 - \frac{1}{2} (\elll \tcdot P)^2\, b_3 - \frac{i}{6} (\elll \tcdot P)^3\, b_4 + \frac{1}{24} (\elll \tcdot P)^4\, b_5 \ .
\end{equation}
Again, the fit weights have been chosen according to the statistical errors of the individual data points.
The values we obtain from our fits in the $u-d$ channel (where $\mathcal{N}=1$) are
\begin{align}
	b_2 & = 0.242(19)\ , &
	b_3 & = 0.226(49)\ , &
	b_4 & = 0.019(18)\ , &
	b_5 & = 0.157(60)\ .
\end{align}

From the factorization assumption eq.~(\ref{eq-f1factorization}) follows 
\begin{multline}
	\int dx\ x^{n} f_1(x)  = \mathcal{N} \int \frac{d(\elll \tcdot P)}{2\pi}\ \tAmp_2^\norm(\elll \tcdot P)\ \frac{\partial^n}{\partial (-i \elll \tcdot P)^n}\ e^{-i (\elll \cdot P) x} \\
	= \mathcal{N} (-i)^n \frac{\partial^n}{\partial (\elll \tcdot P)^n}  \tAmp_2^\norm(\elll \tcdot P) \Big \vert_{\elll \tcdot P = 0} = \mathcal{N} b_{n+1} \ .
\end{multline}
Thus, if we could take eq.~(\ref{eq-f1factorization}) literally, the $b_n$ would be Mellin moments of $f_1(x)$. 
An analysis of $x$-moments based on renormalized local operators on the same ensemble \cite{Hagler:2007xi} finds $\int dx\ x\, f_{1,u{-}d}(x) = 0.226(4)$, which is quite similar to our $b_2$, however, our $b_3$ differs strongly from their result $\int dx\ x^2\, f_{1,u{-}d}(x) = 0.074(5)$. 

\subsection{Qualitative Comparison to PDFs from Phenomenology}

Taking eq.~(\ref{eq-f1factorization}) literally once more, we can invert the Fourier transformation and get
\begin{align}
	\text{`` } \tAmp_{2,\quark}^\norm(\elll \tcdot P) & = \mathcal{N}^{-1} \int_{-1}^1 dx\ e^{i (\elll \cdot P) x}\ f_{1,\quark}(x) \nonumber \\ & =  
	\mathcal{N}^{-1} \int_{-1}^1 dx\ e^{i (\elll \cdot P) x}\ \left\lbrace \begin{array}{lcl} \phantom{-}f_{1,\quark}(x) & : & x\geq 0 \\ -f_{1,\bar \quark}(-x) & : & x<0 \end{array} \right. 
	\text{ '',}
\end{align}
where we have made use of the support properties of $f_1(x)$. Let us take the phenomenologically determined \toolkit{CTEQ5M} parametrization \cite{CTEQ5} of $f_{1,u-d}(x)$ and $f_{1,\bar u-\bar d}(x)$ at a scale $Q^2 = (2 \units{GeV})^2$ as input to the above equation. An unofficial release of the \toolkit{CTEQ5M} distributions is available as a \toolkit{Mathematica} input file. We have evaluated the above integral numerically, excluding a small $x$--interval $[-0.00001,0.00001]$ to avoid numerical problems. The results are shown as red dashed curves in Fig.~\ref{fig-amp2normldP}. At a qualitative level, there is an obvious similarity to the lattice data. 

\subsection{Qualitative Comparison to the Diquark Model}

\begin{figure}[tbp]
	\centering%
	\begin{overpic}[clip=true,width=0.97\textwidth]{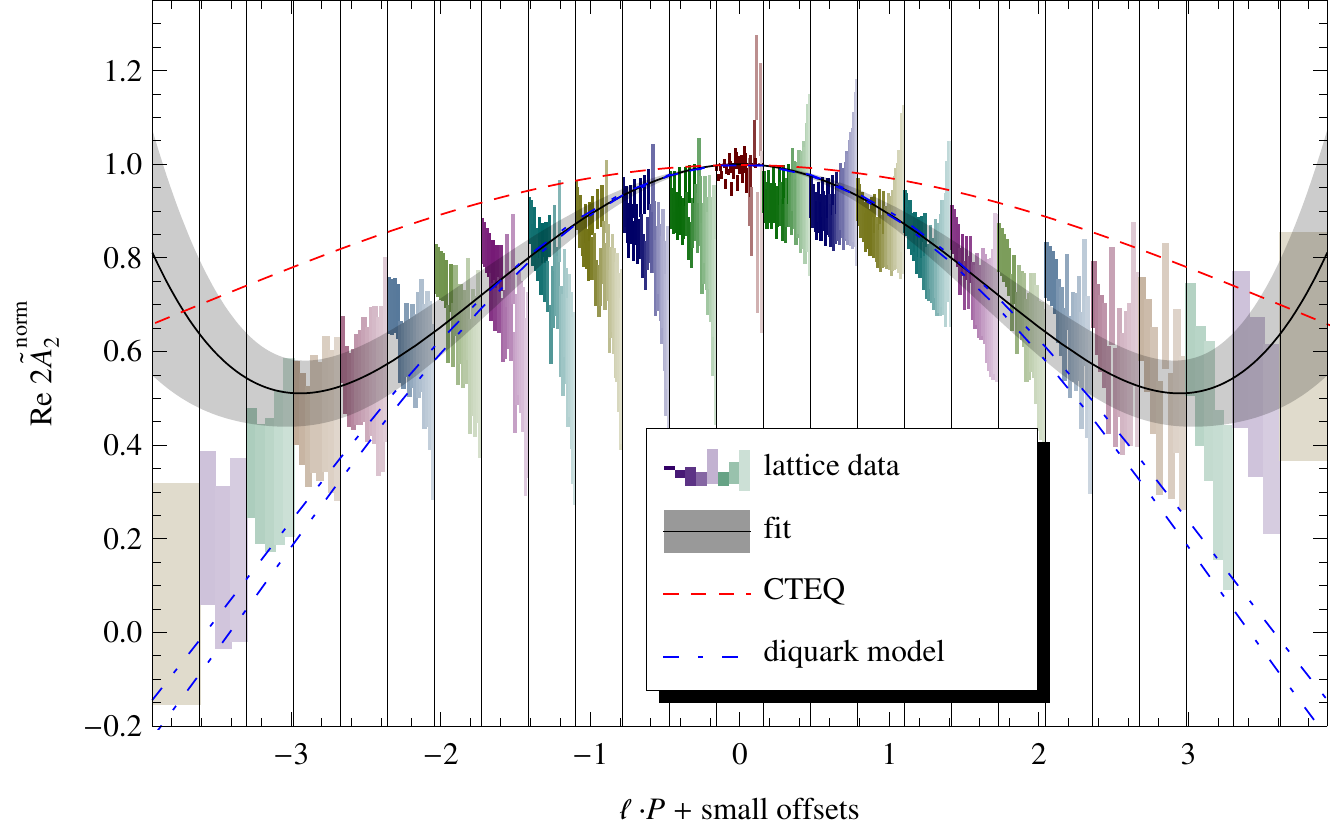}%
		\put(0,57){\subfloat[][]{\label{fig-amp2normldPRe}\hphantom{(m)}}}
	\end{overpic}\par\vspace{5mm}
	\begin{overpic}[clip=true,width=0.97\textwidth]{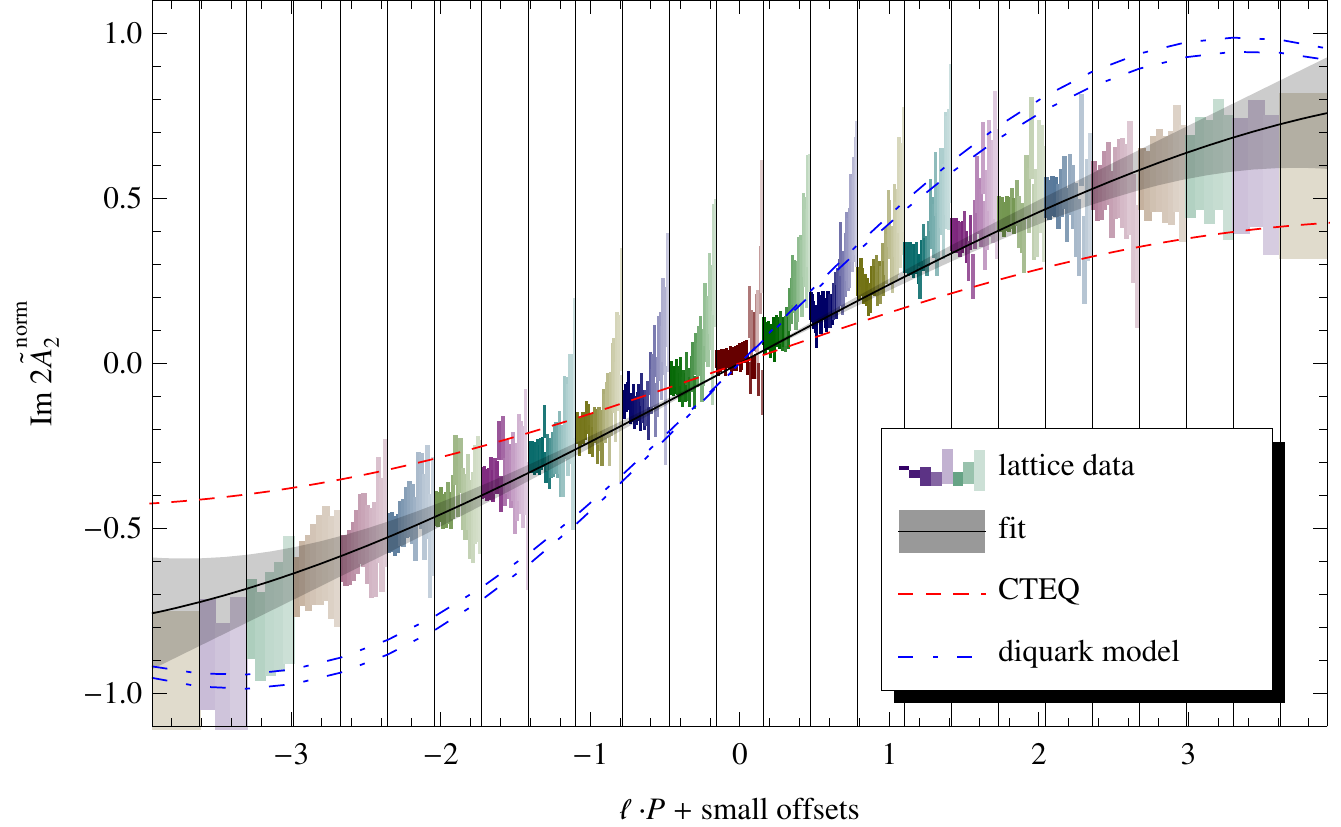}%
		\put(0,57){\subfloat[][]{\label{fig-amp2normldPIm}\hphantom{(m)}}}
	\end{overpic}\par
	\caption[factorization plot]{%
		Results for $\tAmp_2^\norm(\elll^2,\elll \tcdot P)$ obtained on the smeared coarse-m030 ensemble for $u-d$ quarks. The statistical error bars of the lattice data are shown as colored rectangles; data with larger statistical errors are drawn with lighter colors. In each vertical stripe we show lattice data for one value of $\elll \tcdot P$; data points obtained for larger values of $\sqrt{-\elll^2}$ are drawn further to the right inside the stripe. We have omitted data points with $\sqrt{-\elll^2} < 0.25\units{fm}$ because of possible lattice cutoff effects. The curves labeld ``CTEQ'' and ``diquark model'' are explained in the text and are only intended to provide a qualitative comparison.
		\label{fig-amp2normldP}
		}
\end{figure}

The model of section \ref{sec-models} provides us with an explicit expression for the quark distribution $f_1(x,\vprp{k})$ valid for $x>0$. We can convert this into an amplitude $\tAmp_2$ via the inverse Fourier transformation 
\begin{equation}
	\tAmp_2(\elll^2,\elll \tcdot P) = \int_0^1 dx\ e^{i (\elll \cdot P) x}  \int_0^\infty d|\vprp{k}|\  2\pi|\vprp{k}| J_0(\sqrt{-\elll^2}\, |\vprp{k}|)\ f_1(x,\vprp{k}^2)
\end{equation}
provided $f_1(x,\vprp{k}^2)$ decays asymptotically fast enough with $\vprp{k}$ (otherwise, some regularization procedure is required, e.g., via a cutoff). Let us take the parameter set from section \ref{sec-models} in its original form with $\alpha = 2$, as in Ref.~\cite{Jakob:1997wg}. In this case, we can evaluate the above integrals numerically without a cutoff in $|\vprp{k}|$.
From the resulting amplitude $\tAmp_2(\elll^2,\elll \tcdot P)$, we calculate $\tAmp_2^\norm(\elll^2,\elll \tcdot P)$ according to its definition eq.\,(\ref{eq-normamp}). We plot it in Fig.~\ref{fig-amp2normldP} with respect to $\elll \tcdot P$ for two different values of $\elll^2$, namely $\sqrt{-\elll^2} = 0$ and $\sqrt{-\elll^2} = 1 \units{fm}$.

First of all, we find qualitative agreement between the model and lattice data. What is perhaps even more astonishing is that the model amplitude $\tAmp_2^\norm(\elll^2,\elll \tcdot P)$ is very similar for $\sqrt{-\elll^2} = 0$ and $\sqrt{-\elll^2} = 1 \units{fm}$. We conclude, that the factorization assumption on the level of amplitudes $\tAmp_2(\elll^2,\elll \tcdot P)$ is also a good approximation for the scalar diquark model within the region of $(\elll^2,\elll\tcdot P)$ accessible to us on the lattice.

\chapter{Work in Progress}
\label{chap-ideas}
In this chapter, we would like to present some preliminary studies and ideas that may help us in the future. In particular, we will address the question whether we can go beyond straight Wilson lines in our operator, with the aim to calculate ``realistic'' \TMDs, as they occur in the description of scattering experiments such as SIDIS.

\section{Auxiliary Fields}
\label{sec-auxfields}

The operator $\bar \quark(\elll)\, \GammaOp\ \WlineC{\mathcal{C}_\elll}\ \quark(0)$ is non-local. This complicates the probability interpretation and renormalization of our quark-quark correlator. Therefore, it can be useful to give the gauge link an alternative interpretation as a propagator of some auxiliary field, which produces the gauge link when the auxiliary field is integrated out. We draw attention to the fact that such an auxiliary field formalism (``$z$-field formalism'') has already been used in the literature to derive the renormalization properties we make use of in section~\ref{sec-wlinerencont}. 
Moreover, the idea of integrating out a fermion is somehow linked to the motivation for the introduction of the Wilson line in the first place: As we discussed section \ref{sec-linkrole}, the longitudinal Wilson line effectively describes interactions with a fast propagating parton. In the following, we show why auxiliary field techniques (and effective theories) can become interesting for us.

\subsection{Heavy Particles}

That heavy auxiliary fields can be advantageous in the context of hadron structure calculations on the lattice has already been put forward in Ref.~\cite{Detmold:2005gg}. The heavy quark action is derived from the QCD fermion Lagrange density for a single heavy quark flavor:
\begin{equation}
	\mathcal{L}_F[\bar \Psi,\Psi,\Afield] = \Psi(x) ( i \slashed{D} - m_h ) \Psi(x)\ .
	\label{eq-QCDheavyfermion}
\end{equation}
Here $m_h$ is the mass of the heavy quark field $\Psi(x)$. Due to its large inertia, the motion of the heavy quark is characterized by small fluctuations around the classical path. Let us decompose its momentum according to $p = m_\Quark v + k$, where $v$ is a fixed velocity four vector ($v\tcdot v = 1$) and where $m_\Quark \approx m_h$. The choice of $m_\Quark$ and $v$ are arbitrary to a certain extent, as long as the remaining dynamical momentum $k$ is small compared to $m_\Quark$. One now decomposes $\Psi(x)$ such that 
\begin{equation}
	\frac{1+\slashed{v}}{2} \Psi(x) = \exp( -i\,m_\Quark\,v \tcdot x )\ h(x)\,, \qquad
	\frac{1-\slashed{v}}{2} \Psi(x) = \exp( -i\,m_\Quark\,v \tcdot x )\ H(x)\,.
	\label{eq-heavyfieldspinproj}
\end{equation}
with projections $(1\pm\slashed{v})/2$. The phase factor absorbs oscillations due to the motion with velocity $v$. Making use of the classical equations of motion for the field $H(x)$, we arrive at the HQET Lagrangian \cite{Georgi:1990um,Neubert:1993mb}
\begin{equation}
	\mathcal{L}_\text{HQET}[\bar h,h,\Afield] = \bar h(x)\, \left( i v\tcdot D - \frac{(\slashed{D}_T)^2}{2 m_\Quark} - m_R \right)\, h(x) + \mathcal{O}(1/m_\Quark^2)\ .
	\label{eq-hqetaction}
\end{equation}
Here $m_R \equiv m_h - m_\Quark$ is the residual mass \cite{Falk:1992fm} and $D_T^\mu \equiv D^\mu - (v \tcdot D) v^\mu$. Choosing $v$ in temporal direction and going to Euclidean space, we get
\begin{equation}
	\mathcal{L}_\text{NR0}^\text{E}[\bar h,h,\Afield] = \bar h(x)\, \left( -D_\Eu{4} - \frac{\vect{D} \tcdot \vect{D}}{2 m_\Quark} + m_R \right)\, h(x)\ , 
	\label{eq-hqetactioneuclnr}
\end{equation}
where we have omitted a spin dependent relativistic effect which is part of $\slashed{D}_T \slashed{D}_T$, and where $\vect{D}$ refers to the three spatial components of $D$. 
In the \terminol{static limit} $m_\Quark \rightarrow \infty$, the second derivative term vanishes. A simple discretization of the static action on a lattice $\mathbb{L}$ of infinite volume then reads
\begin{equation}
	S_\text{stat}^\text{lat}[\bar h, h,U] = \sum_{x \in \mathbb{L}} \bar h(x)\, \left[h(x) - U(x,x+\hat 4) h(x+\hat 4) \right] + \bar h(x) \hat m_R h(x)\ .
\end{equation}
A solution for the propagator of a static quark is thus 
\begin{align}
	\contraction{}{h(x)}{}{\bar h(y)} h(x) \bar h(y) & = (1+ \hat m_R)^{y_\Eu{4}-x_\Eu{4}-1}\ \nonumber \\ & \times\ 
	\left\lbrace \begin{array}{lcl} U(x,x+\hat 4) U(x+\hat 4, x+2\,\hat 4) \cdots U(y-\hat 4,y) & : &x_\Eu{j}=y_\Eu{j},\  y_\Eu{4} > x_\Eu{4} \\
   \Eins & : & x = y \\ 0 & : & \text{otherwise} \end{array} \right. \ .
	\label{eq-staticquarkprop}
\end{align}
The quark can only propagate in one direction in Euclidean time. The propagator becomes a straight gauge link in $\hat 4$-direction, once the quark fields are integrated out \cite{Eichten:1987xu}.\footnote{We set the fermion determinant of the heavy quark action to one (``quenched approximation''), i.e., we neglect virtual heavy quark loops.} We have already alluded to this formalism in section~(\ref{sec-renstatqpotprinciple}), when we interpreted the temporal line segments of Wilson loops as propagators of static quarks. Note that $(1+ \hat m_R)^{y_\Eu{4}-x_\Eu{4}} \approx \exp( \hat m_R (y_\Eu{4}-x_\Eu{4}) )$. This factor resembles the Wilson line self energy $\exp(-\delta m \sqrt{-\elll\tcdot\elll})$. 

\begin{figure}[tbp]
	\centering%
	\subfloat[][]{%
		\label{fig-hoppingaxis}%
		\includegraphics[clip=true,width=0.49\textwidth]{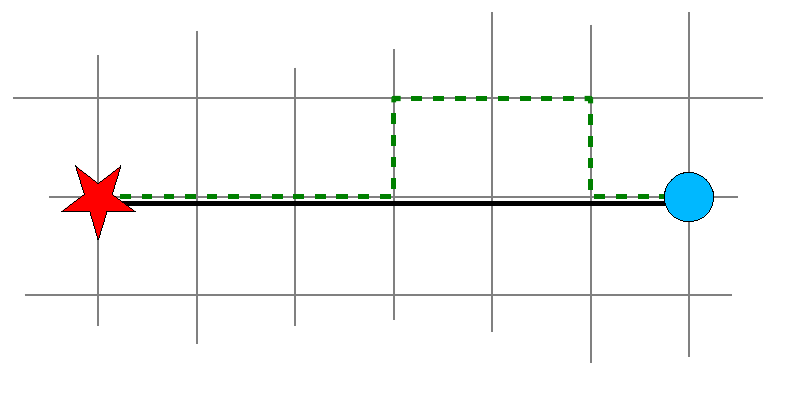}%
		}\hfill%
	\subfloat[][]{%
		\label{fig-hoppingoblique}%
		\includegraphics[clip=true,width=0.49\textwidth]{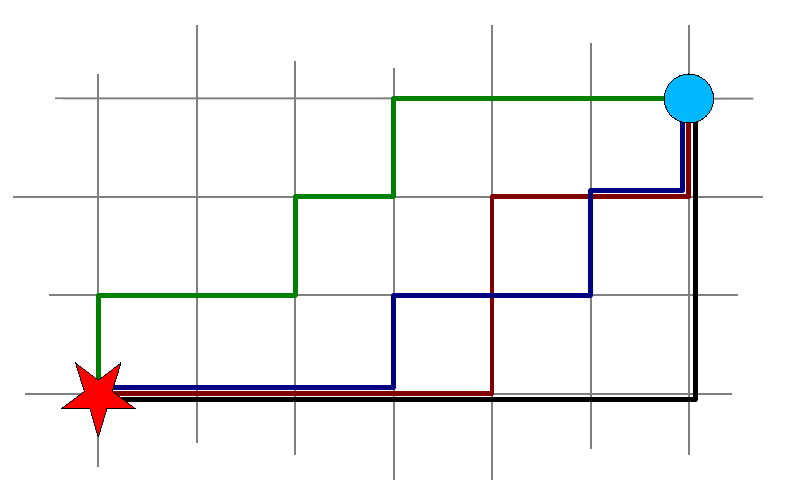}%
		}\par
	\subfloat[][]{%
		\label{fig-hoppingshort}%
		\includegraphics[clip=true,width=0.49\textwidth]{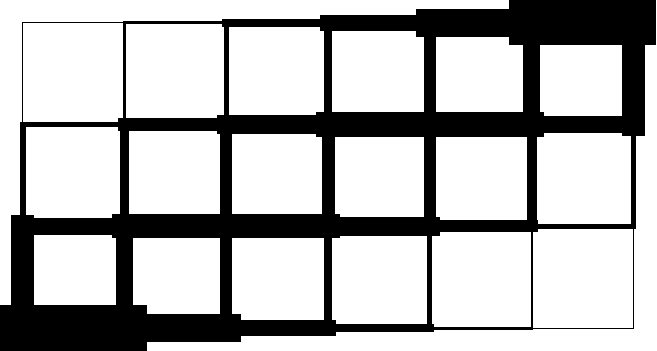}%
		}\hfill%
	\subfloat[][]{%
		\label{fig-hoppingshort2}%
		\includegraphics[clip=true,width=0.49\textwidth]{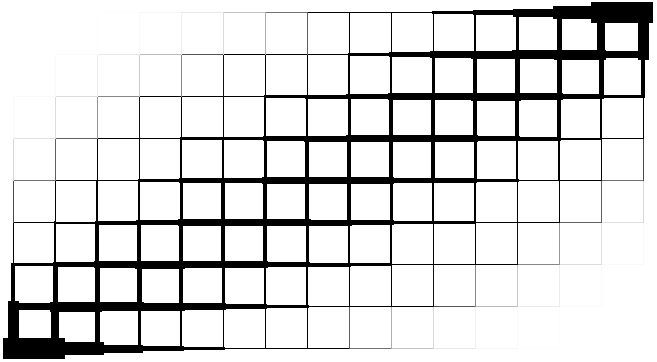}%
		}\par%
	\caption[Hopping Parameter Expansion]{%
		Hopping parameter expansion for a spatial propagator in our heavy particle action.
		\subref{fig-hoppingaxis} Leading link path and example of a subleading link path for propagation along a spatial lattice axis.
		\subref{fig-hoppingoblique} Examples of leading link paths for propagation at an oblique angle.
		\subref{fig-hoppingshort}, \subref{fig-hoppingshort2} Superposition of all leading link paths for propagation at an oblique angle. The line thickness of a each link variable is proportional to the number of times the respective link variable occurs in the whole set of link paths.
		\label{fig-hopping}
		}
\end{figure}

For our quark--quark correlator with the straight Wilson line, we would like to find an action that produces the straight link between any two points separated by a spatial vector $\elll$.
Ideally, this action will be independent of $\elll$.
In pursuit of such an action, let us study eq.~(\ref{eq-hqetactioneuclnr}) again. This time, we also take the term $\vect{D}^2/2m_\Quark$ into account. We remark that the whole expression eq.~(\ref{eq-hqetactioneuclnr}) is just the leading order Lagrangian in the NRQCD formalism \cite{Caswell:1985ui,Bodwin:1994jh}, which has a different power counting scheme than HQET (compare, e.g., Ref. \cite{Luke:1996hj}). A simple discretization is
\begin{align}
	S_\text{NR0}^\text{lat}[\bar h,h,U] & = S_\text{stat}^\text{lat}(\bar h,h,U) -  
	\frac{1}{2 \hat m_\Quark} \sum_{x \in \mathbb{L}} \sum_{\pm\muE = 1}^{3} \big[  \bar h(x)\, U(x,x+\hat \mu)\, h(x+\hat \mu) - \bar h(x) h(x) \big] \nonumber \\
	& = \sum_{x\in\mathbb{L}} \sum_{y\in\mathbb{L}} \bar h(x) K(x,y) h(y) \ ,
\end{align}
where 
\begin{align}
	K(x,y) = \delta_{x,y} \underbrace{\left( 1 + \hat m_R + \frac{3}{\hat m_Q} \right)}_{\displaystyle \equiv \rho} - \delta_{x,y-\hat 4}\, U(x, y) - \frac{1}{2 \hat m_\Quark} \underbrace{\sum_{\pm \muE = 1}^{3} \delta_{x,y-\hat \mu}\, U(x,y) }_{\displaystyle \equiv M(x,y)}\ .
\end{align}
The propagator $G(z,x) = \contraction{}{h(z)}{}{\bar h(x)} h(z) \bar h(x)$ fulfills
\begin{align}
	\delta_{z,y} \mathop{=}^! \sum_{x \in \mathbb{L}} G(z,x) K(x,y) & = 
	\rho G(z,y) - G(z,y-\hat 4) U(y-\hat 4, y) \nonumber \\ & - \frac{1}{2 \hat m_\Quark} \mathop{\sum_{x \in \mathbb{L}}}_{\smash{x_\Eu{4} = y_\Eu{4}}} G(z,x) M(x,y)\ .
\end{align}
Let us study the restriction of the above equation to $z_\Eu{4} = y_\Eu{4}$. If we do not allow the particle to propagate backward in Euclidean time (analogous to eq.\,(\ref{eq-staticquarkprop})), then $G(z,y-\hat 4) = 0$. We can now solve for $G$ (in matrix notation):
\begin{equation}
	\Eins = G \left(\rho\, \Eins - \frac{1}{2 \hat m_\Quark} M\right) \qquad \Rightarrow \qquad 
	G = \frac{1}{\rho} \sum_{j=0}^{\infty} \left( \frac{1}{2\,\rho\, \hat m_Q} M \right)^j\ .
\end{equation}
Now, consider the propagator of the particle from the origin to $\elll$, with $\elll_\Eu{4} = 0$. 
Higher powers of $M$ are suppressed by powers of $1/\hat m_\Quark$ in the series above. 
The matrix $M$ connects a given lattice site with the six neighboring lattice sites in the same time slice. 
The expansion in powers of $M$ is called a \terminol{hopping parameter expansion}, see, e.g., Ref. \cite{Roth}.
The leading non-vanishing contribution to the propagator has $\lceil \elll \rceil \equiv \sum_{\muE=1}^3 | \elll_\muE |$ factors of $M$:
\begin{align}
	\contraction{}{h(\elll)}{}{\bar h(0)} h(\elll) \bar h(0) & = \rho^{-1} {(2\rho \hat m_\Quark)}^{-\lceil \elll \rceil} \left\{  \qquad\sum_{\mathclap{z^{(1)},\ldots,z^{(\lceil \elll \rceil-1)}}} M(x,z^{(\lceil \elll \rceil-1)}) \cdots M(z^{(2)}\!\!,z^{(1)}) M(z^{(1)}\!\!,0) + \mathcal{O}(\hat m_\Quark^{-1}) \right\} \nonumber \\
	 & = \rho^{-1} {(2\rho \hat m_\Quark)}^{-\lceil \elll \rceil} \left\{ 
	\sum_{\mathcal{C}^\lat} \WlineClat{\mathcal{C}^\lat}  \left| \begin{array}{l}\mathcal{C}^\lat \text{ to } \elll \\ n_\text{links}[\mathcal{C}^\lat] = \lceil \elll \rceil \end{array} \right. + \mathcal{O}(\hat m_\Quark^{-1}) 
	\right\} \ .
	\label{eq-hprop}
\end{align}
If $\elll$ is chosen to lie on one of the lattice axes, we immediately see that the leading order contribution is the shortest possible gauge path from $0$ to $\elll$: a straight link. This leading contribution comes with a factor $\sim \hat m_\Quark^{-|\vect{\elll}|}$; again this factor reminds us of the Wilson line self-energy $\exp(-\delta m |\vect{\elll}|)$. The next order introduces a dent in the path, see Fig.~\ref{fig-hoppingaxis}. The dent is suppressed by $\mathcal{O}(m_\Quark^{-2})$, but there are many possible locations along the path where the dent can be inserted, and the dent can have different lengths. If the vector $\elll$ is at an oblique angle compared to the lattice axes, many link paths contribute to the leading order, see illustration Fig.~\ref{fig-hoppingoblique}. Link paths running close to the geometrically direct connection dominate combinatorially, as illustrated in Figs.~\ref{fig-hoppingshort} and \ref{fig-hoppingshort2}. Comparing the two figures, we note that the resemblance to a direct connection intensifies on a finer lattice. However, the continuum limit $a \rightarrow \infty$ cannot be taken within our picture, because the hopping parameter expansion is only applicable (probably as an asymptotic series) on a not too fine lattice, where $a m_\Quark \gg 1$. Nevertheless, we conclude that at leading order in $\hat m_\Quark^{-1}$ for not too small $|\vect{\elll}|$ the propagator effectively approximates a straight Wilson line from $0$ to $\elll$.\footnote{Note that our ``leading contribution'' might not be the leading one if we do not truncate our heavy quark action in the beginning.} It seems that for a given $m_\Quark$, this Wilson line has a certain ``thickness'', in the sense that gauge links with small dents deviating from the direct path contribute (no matter how fine we choose the lattice). The larger we choose $m_\Quark$, the ``thinner'' the line will get. In the continuum formalism, paths with tiny wiggles can be absorbed into the renormalization constant $\delta m$ of a straight Wilson line, see Ref. \cite{Craigie:1980qs}, so we see another correlation between $m_\Quark$ and $\delta m$. In summary, we rewrite eq.\,(\ref{eq-hprop}) symbolically as 
\begin{align}
	\text{`` }\contraction{}{h(\elll)}{}{\bar h(0)} h(\elll) \bar h(0) \approx \mathcal{U}^{\text{thick}}[\elll,0]\ \tilde m_\Quark^{-|\vect{\elll}|}\text{  '',}
	\label{eq-hpropsymb}
\end{align}
where $\mathcal{U}^{\text{thick}}[\elll,0]$ represents the combination of Wilson lines from $0$ to $\elll$, and where the factor $m_\Quark^{-|\vect{\elll}|}$ reminds us that there is a suppression factor that exponentiates with the distance $|\vect{\elll}|$.

How can we make use of these observations? 
With respect to the action $S[\bar q,q,U] + S_\text{NR0}[\bar h,h,U]$, we can consider the correlator
\begin{equation}
	\Phi_h(\vprp{k},P,S)  = \int \frac{d^2 \vprp{\elll}}{(2\pi)^2 P^+} \ 
	e^{-ik \cdot \elll}\ 
	\frac{1}{2} \bra{N(P,S)}\ \bar \quark(\elll) \Gamma^{h\dagger} h(\elll)\ \bar h(0) \Gamma^h \quark(0)\ \ket{N(P,S)} \big \vert_{\elll\pm=0}
	\label{eq-NRcorr}
\end{equation}
(in analogy to the first Mellin moment in eq.\,(\ref{eq-deffirstmellin})). Here the Dirac structure $\GammaOp$ of our original correlator eq.\,(\ref{eq-corr}) is encoded in $\Gamma^{h}$. More research is needed to understand the role of $\Gamma^{h}$, and the role of spin in general in this context\footnote{Note that the heavy quark fields $h(x)$ involve a spin projection, see eq.\,(\ref{eq-heavyfieldspinproj}). This restricts valid choices of $\Gamma^{h}$.}. 
We now integrate out the heavy degrees of freedom. This replaces $h(\elll) \bar h(0)$ in the expression above by the heavy quark propagator $\contraction{}{h(\elll)}{}{\bar h(0)} h(\elll) \bar h(0)$. As we have demonstrated, this propagator translates into a Wilson line, at least in an approximate sense. 
Within the symbolical notation of eq.\,(\ref{eq-hpropsymb}), we thus obtain
\begin{align}
	\text{`` } \Phi_h(\vprp{k},P,S)  & \approx \int \frac{d^2 \vprp{\elll}}{(2\pi)^2 P^+} \ 
	e^{-ik \cdot \elll}\ 
	\frac{1}{2} \bra{N(P,S)}\ \bar \quark(\elll)\ \Gamma^{h\dagger}\Gamma^h \nonumber \\ 
	& \times \mathcal{U}^{\text{thick}}[\elll,0]\ \tilde m_\Quark^{-|\vect{\elll}|}\  \quark(0)\ \ket{N(P,S)} \big \vert_{\elll^+=\elll^-=0} \text{ ''.}
\end{align}
Similar as in section \ref{sec-probproofmech}, but without the difficulties with the gauge link, we can rewrite eq.~(\ref{eq-NRcorr}) as
\begin{equation}
	\Phi_h(\vprp{k},P,S)  = \frac{1}{P^+} \sum_n\  
	\Big\vert\ \bra{n}\ \bar h(0) \Gamma^h \quark(0)\ \ket{N(P,S)} \Big\vert^2 \ \delta^{(2)}(\vect{p}_{n\prp} + \vprp{k}) \ \geq 0
\end{equation}
for a complete set of states $\ket{n}$. 
This is the probability that a fictitious interaction converts a quark into an $h$-particle such that the final state $\ket{n}$ carries transverse momentum $-\vprp{k}$. 

\subsection{How Auxiliary Fields may help us}

The preceding section shows that we can build correlators that look quite similar to our original one in  eq.~(\ref{eq-corr}) once the auxiliary fields are integrated out. However, heavy particles are not the only type of auxiliary particles that might be interesting to us. Soft collinear effective theory (SCET), for example, describes very \emph{fast} rather than very heavy quarks, and can generate the Wilson lines out to infinity that appear in SIDIS factorization. In general, the auxiliary field and action need not have a direct interpretation in the context of the actual physical process we want to study (e.g., SIDIS). The primary goal is to design well-defined observables that provide us with information about the transverse momentum distributions of quarks inside the nucleon. As demonstrated above, such an observable can have a probability interpretation. 
Moreover, we may learn more about the meaning of the renormalization condition.
Last but not least, we might be able to construct improved lattice operators with reduced discretization artefacts, profiting, e.g., from existing experience with heavy quarks on the lattice.

\section{\TMDs with Extended Links}
\label{sec-staples}

The quark--quark correlator defining \TMDs in the context of scattering experiments contains a Wilson line with sections running close to the $\hat n_-$ direction, see chapter \ref{chap-dis}. Can we construct such correlators on the lattice?

\subsection{Accessible Link Directions on the Lattice}

\begin{figure}[tbp]
	\centering%
	\subfloat[][]{%
		\label{fig-largemomentum1}%
		\includegraphics[clip=true,width=0.49\textwidth]{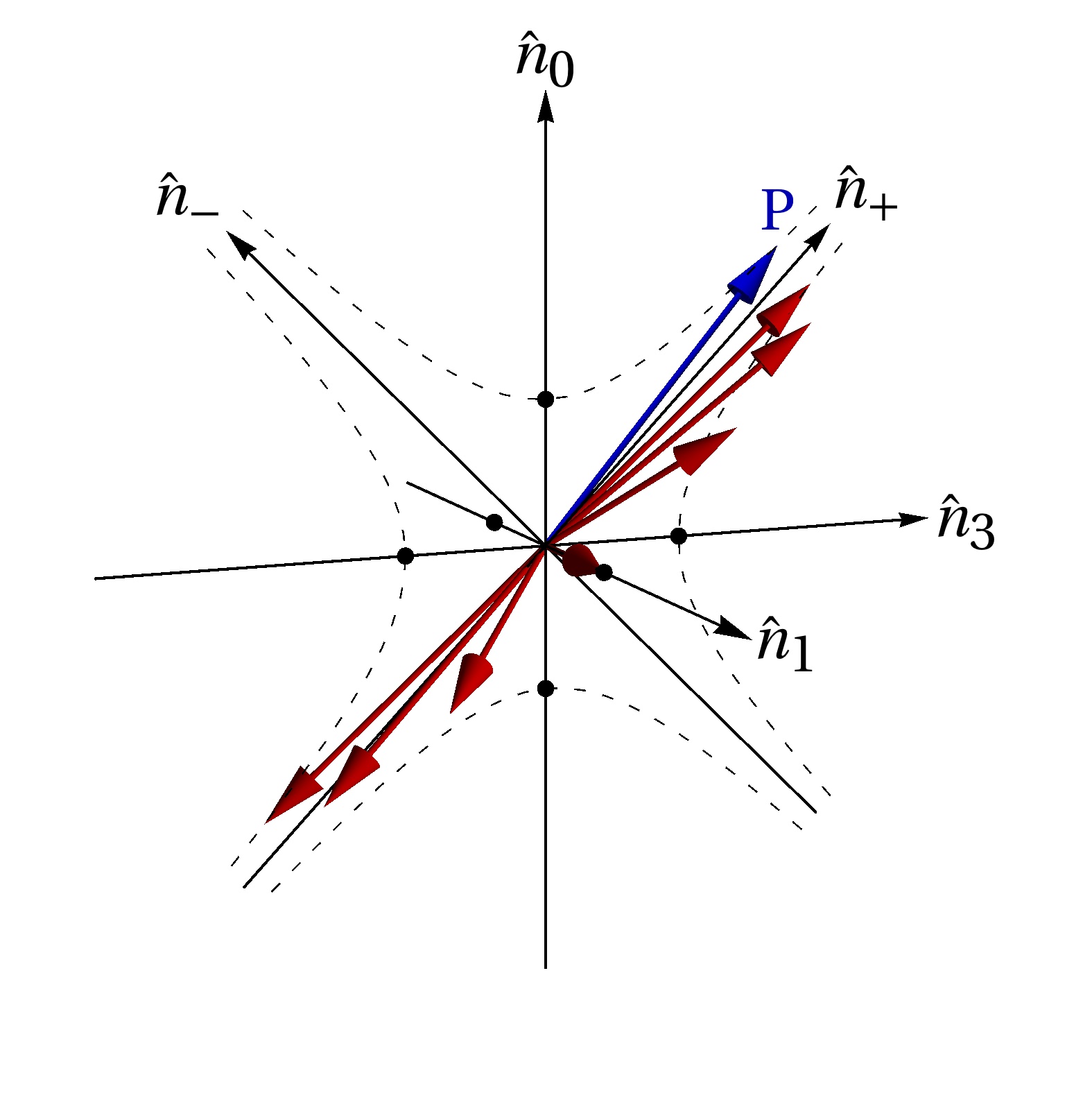}%
		}\hfill%
	\subfloat[][]{%
		\label{fig-largemomentum2}%
		\includegraphics[clip=true,width=0.49\textwidth]{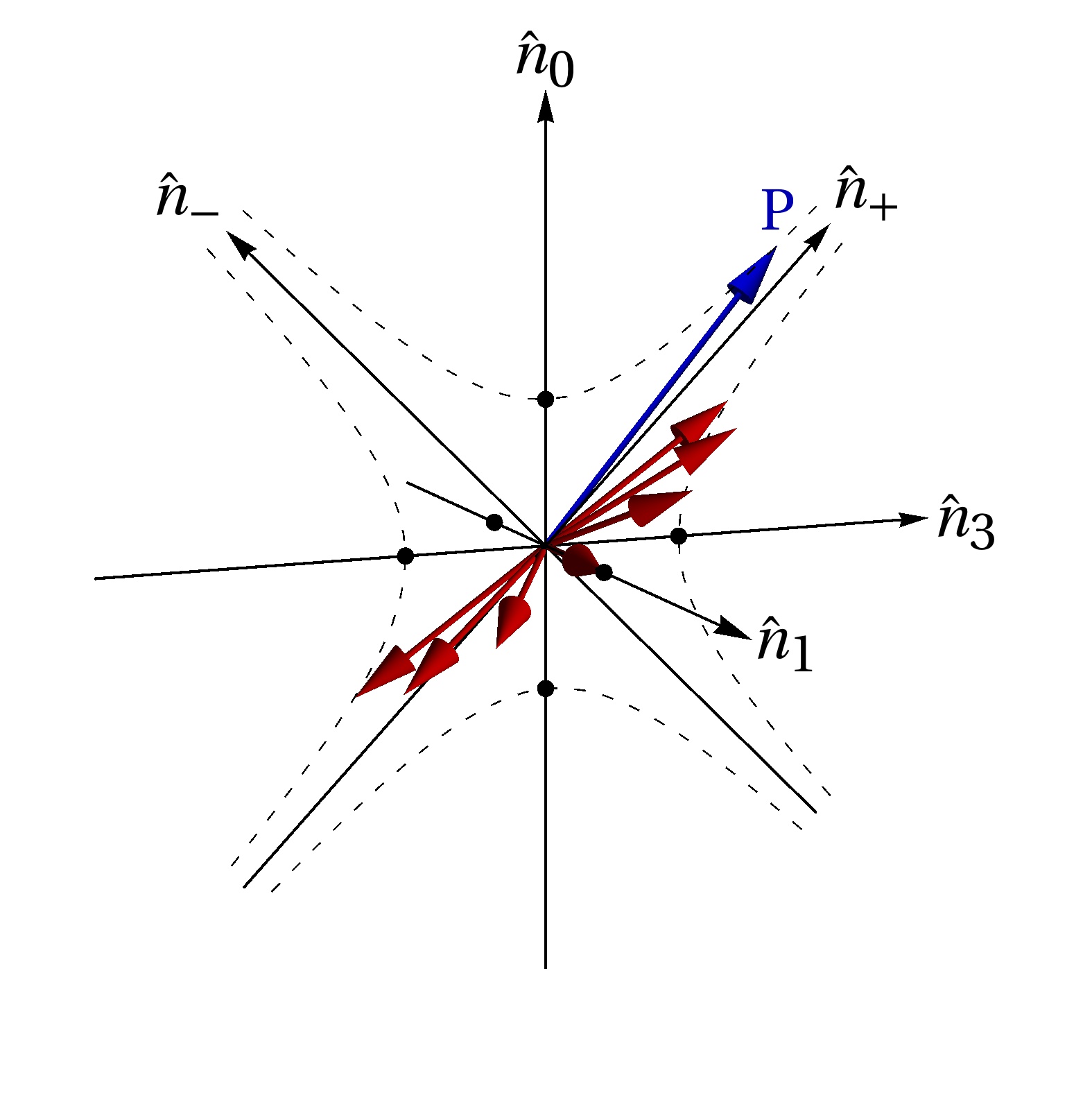}%
		}\par
	\caption[Large Momentum Frame]{%
		Boosting the frame at rest relative to the lattice to the large momentum frame. In this illustration, the frame of reference is fixed by the requirement that $P^+ = 2 \sqrt{2} m_N$. The vector labelled ``$P$'' depicts the nucleon momentum. The other vectors represent $v$ for $\theta = 0^\circ$, $30^\circ$, $60^\circ$, $90^\circ$, $120^\circ$, $150^\circ$, and $180^\circ$. The dashed hyperbolae connect points of equal proper length $\sqrt{-(x^0)^2 + (x^3)^2} = w$ and of equal invariant mass $\sqrt{(x^0)^2 - (x^3)^2} = m_N$. 
		\subref{fig-largemomentum1} Nucleon at rest on the lattice, $P^3_\lat = 0$.
		\subref{fig-largemomentum2} Nucleon moving with $P^3_\lat = m_N/2$ with respect to the lattice.
		\label{fig-largemomentum}
		}
\end{figure}

Let us first visualize the situation. Suppose the operator we want to use to probe the nucleon contains a straight Wilson line whose direction is given by the Minkowski vector $v$. We want to implement this Wilson line directly on the lattice. In a frame of reference at rest relative to the lattice, $v$ cannot have a Minkowski time component, $v^0_\lat \neq 0$. (Neither do we consider Wilson lines with an extent in Euclidean time direction, because they complicate the interpretation in terms of the transfer matrix formalism.) For the purpose of our discussion, let us restrict ourselves to a vector of the form $ v_\lat = w \sin(\theta) \hat n_1 +  w \cos(\theta) \hat n_3 $, and consider a lattice nucleon momentum 
$P_\lat = \sqrt{m_N^2+(P^3_\lat)^2}\, \hat n_0 + P^3_\lat \hat n_3$.

The quantities we want to calculate on the lattice have an interpretation in a frame of reference where the momentum of the nucleon is very large, so let us boost in 3-direction to a fixed large nucleon momentum $P^+$. 
Figure~\ref{fig-largemomentum} shows what happens to the vector $v$ under this boost. For illustration purposes, we have chosen $P^+ = 2 \sqrt{2} m_N$ (i.e., a Lorentz factor $\gamma = 2$ with respect to the nucleon rest frame).
In Fig.~\ref{fig-largemomentum1}, we start out with a nucleon at rest on the lattice, $P^3_\lat = 0$, while in Fig.~\ref{fig-largemomentum2}, we set $P^3_\lat = m_N/2$, which is roughly comparable to our situation with the LHPC sequential propagators, where $|\vect{P}| \approx 500 \units{MeV}$.
Under the boost, the nucleon momentum vector moves to the right along the dashed hyperbola at the top. Other vectors (except for purely transverse ones) are also pulled towards the $\hat n_+$ axis. 
Of particular interest is the vector $v$ for $\cos(\theta) = 1$, whose arrow head touches the dashed hyperbola to the right. Obviously, by changing $P^3_\lat$, we can move the arrow head up and down along the hyperbola. 
If $P^3_\lat$ is of the order of $P^+/\sqrt{2}$, the arrow is approximately horizontal, parallel to the $\hat n_3$ axis. If we were to increase $P^3_\lat$ to a value of the order $(P^+)^2/m_N$, the arrow head would move all the way down and appear close to the $\hat n_-$ axis. Of course, we can never reach $v \propto \hat n_-$ exactly.

The direction $v$ of the Wilson line relative to the nucleon momentum can be described by the Lorentz-invariant quantity $\zeta \equiv (P \cdot v)^2/|v^2| \sim (P^+)^2/2$ we already encountered in section \ref{sec-rapidity}. In the limit of lightlike $v$, $\zeta$ becomes infinite. On the lattice, $\zeta$ is limited by the largest attainable lattice nucleon momentum: $\sqrt{\zeta} \leq |\vect{P}_\lat|$.


\subsection{Staple-Shaped Wilson lines on the Lattice}
\label{sec-stapleresults}

\begin{figure}[tbp]
	\centering%
	\hfill%
	\subfloat[][]{%
		\label{fig-staplelink}%
		\includegraphics[clip=true]{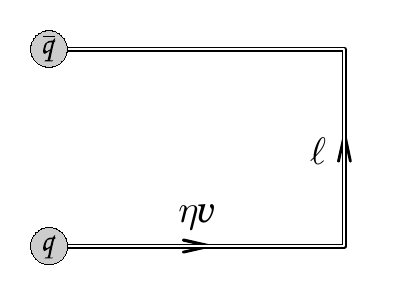}%
		}\hfill%
	\topalignbox{\includegraphics{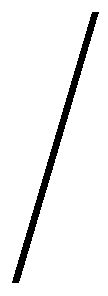}}\hfill%
	\subfloat[][]{%
		\label{fig-stapleloop}%
		\includegraphics[clip=true]{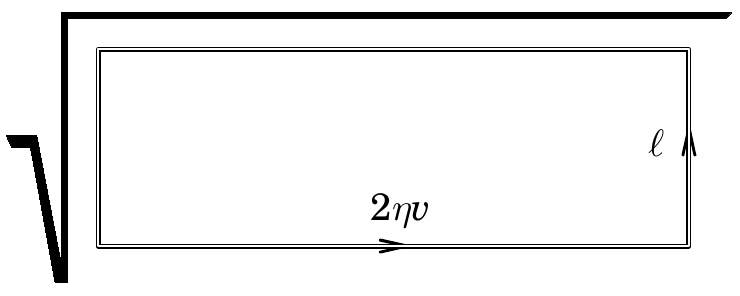}%
		}\hfill\vphantom{M}\par
	\caption[Staple Link]{%
		\subref{fig-staplelink} Staple shaped gauge link,
		\subref{fig-stapleloop} Wilson loop subtraction factor.
		\label{fig-staple}
		}
\end{figure}

We have performed some preliminary studies on the lattice with staple shaped gauge links of the form $\mathcal{C}_{\elll,v} = \Wline{\elll,\eta v+\elll,\eta v,0}$ as depicted in Fig.~\ref{fig-staplelink}. Again, we have implemented them directly as a product of link variables $\WlineClat{\mathcal{C}^\lat_{\elll,\eta v}}$. We have chosen a quark separation in $2$-direction, $\elll = \hat \elll_\Eu{2}\, \hat 2$, and a staple vector $v$ in 1-direction, $v \propto \hat 1$. In combination with the lattice nucleon momentum $\vect{P} = (2\pi/L)(-1,0,0)$, this choice enables us to work at non-zero $v\tcdot P$. Let us keep $v$ normalized to $v^2 = -1$.  With the replacement $k \rightarrow i m_N^2 \elll$, we deduce from eq.~(7) of Ref. \cite{Goeke:2005hb} that for our particular choice of $v$ and $\elll$, the ratio $R^\ren_{\gamma_\Eu{4}}$ should depend on the amplitudes $\tAmp_2$, $\tAmp_{12}$ and $\tilde B_8$ for $\eta \rightarrow \infty$. At finite $\eta$, let us denote the amplitudes by $\tilde a_i(\elll^2,\elll \tcdot P,\eta v\tcdot \elll, -\eta^2, \eta v \tcdot P)$ and $\tilde b_i(\elll^2,\elll \tcdot P, \eta v\tcdot \elll, -\eta^2, \eta v \tcdot P)$. 
In analogy to section \ref{sec-symtrafo} and eq.~(\ref{eq-ratiorenop}), we obtain the ratio $R^\ren_{\gamma_\Eu{4}}$ via structures $\tMDM$. For our test case, the relevant $\tMDM$-structures read
\begin{align}
	\tMDM_{\gamma^\mu}(\elll,P,\eta v) 
		&= \frac{2}{m_N}\ \tilde a_2\ P^\mu 
		 + 2 i\ \tilde a_{12}\ \epsilon^{\mu \nu \alpha \beta}  P_\alpha \elll_\beta\ \gamma_\nu \gamma^5
		 + 2 i m_N^2\ \tilde b_8\    \epsilon^{\mu \nu \alpha \beta}  \elll_\alpha \eta v_\beta\ \gamma_\nu \gamma^5
		 + \ldots \ .
\end{align}
In the limit of large $\eta$, we expect the structures of the above equation to converge to finite values. The amplitudes we want to extract are then
\begin{align}
	\tAmp_{2} \left(\elll^2,\elll \tcdot P,\frac{v \tcdot \elll}{|v \tcdot P|},\zeta^{-1},\mathrm{sgn}(v \tcdot P)\right)\ &\equiv\ \lim_{\eta \rightarrow \infty} \tilde a_2(\elll^2,\elll \tcdot P,\eta v \tcdot \elll,-\eta^2, \eta v \tcdot P) \ ,\\
	\tAmp_{12} \left(\elll^2,\elll \tcdot P,\frac{v \tcdot \elll}{|v \tcdot P|},\zeta^{-1},\mathrm{sgn}(v \tcdot P)\right)\ &\equiv\ \lim_{\eta \rightarrow \infty} \tilde a_{12}(\elll^2,\elll \tcdot P,\eta v \tcdot \elll,-\eta^2, \eta v \tcdot P) \ ,\\
	\tilde B_{8} \left(\elll^2,\elll \tcdot P,\frac{v \tcdot \elll}{|v \tcdot P|},\zeta^{-1},\mathrm{sgn}(v \tcdot P)\right)\ &\equiv\ \lim_{\eta \rightarrow \infty} (\eta v \tcdot P)\ \tilde b_{8}(\elll^2,\elll \tcdot P,\eta v \tcdot \elll,-\eta^2, \eta v \tcdot P) \ .
\end{align}
Note that $\tilde b_{8}$ must vanish at least as fast as $1/\eta$ to keep $\tilde B_8$ finite.
Taking into account the transformation properties of the staple-like like link path  eq.\,(\ref{eq-stapletrafoprops}), the constraints on $\tMDM$ from discrete symmetries eq.\,(\ref{eq-Mtildetimerev}) must now be generalized to 
\begin{align}
	&(\dagger): & \left[ \tMDM_\Gamma(\elll,P,v) \right]^\dagger 
	&= \gamma^0\ \tMDM_{\gamma^0\,\Gamma^\dagger\,\gamma^0} (\mathbf{-}\elll,P,v)\ \gamma^0 \ ,\\
	&(\mathscr{P}): & \tMDM_\Gamma(\elll,P,v) 
	&= \gamma^0\ \tMDM_{\gamma^0\,\Gamma\,\gamma^0} (\overline{\elll}, \overline{P},\overline{v})\ \gamma^0 \ ,\\
	&(\mathscr{T}): &\left[ \tMDM_\Gamma(\elll,P,v) \right]^* 
	&= \gamma^5 C\ \tMDM_{C^\dagger \gamma^5\,\Gamma\,\gamma^5 C} (\mathbf{-}\overline{\elll},\overline{P},-\overline{v})\ C^\dagger \gamma^5 \ .
	\label{eq-Mtildenewprops}
	\end{align}
From hermiticity $(\dagger)$ follows for the amplitudes $\tAmp_i$ (and analogously for the $\tilde B_i$)
\begin{equation}
	\left[ \tAmp_{i} \left(\elll^2,\elll \tcdot P,\frac{v \tcdot \elll}{|v \tcdot P|},\zeta^{-1},\mathrm{sgn}(v \tcdot P)\right) \right]^* = \tAmp_{i} \left(\elll^2,-\elll \tcdot P,-\frac{v \tcdot \elll}{|v \tcdot P|},\zeta^{-1},\mathrm{sgn}(v \tcdot P)\right)\ .
	\label{eq-staplehermprop}
\end{equation}
The time reversal operation $(\mathscr{T})$ implies
\begin{align}
	\left[ \tAmp_{2} \left(\elll^2,\elll \tcdot P,\frac{v \tcdot \elll}{|v \tcdot P|},\zeta^{-1},\mathrm{sgn}(v \tcdot P)\right) \right]^* 
	& = \phantom{-}\tAmp_{2} \left(\elll^2,-\elll \tcdot P,\frac{v \tcdot \elll}{|v \tcdot P|},\zeta^{-1},-\mathrm{sgn}(v \tcdot P)\right) \ ,\nonumber \\
	\left[ \tAmp_{12} \left(\elll^2,\elll \tcdot P,\frac{v \tcdot \elll}{|v \tcdot P|},\zeta^{-1},\mathrm{sgn}(v \tcdot P)\right) \right]^* 
	& = -\tAmp_{12} \left(\elll^2,-\elll \tcdot P,\frac{v \tcdot \elll}{|v \tcdot P|},\zeta^{-1},-\mathrm{sgn}(v \tcdot P)\right) \ ,\nonumber \\
	\left[ \tilde B_{8} \left(\elll^2,\elll \tcdot P,\frac{v \tcdot \elll}{|v \tcdot P|},\zeta^{-1},\mathrm{sgn}(v \tcdot P)\right) \right]^*  
	& = -\tilde B_{8} \left(\elll^2,-\elll \tcdot P,\frac{v \tcdot \elll}{|v \tcdot P|},\zeta^{-1},-\mathrm{sgn}(v \tcdot P)\right) \ .
\end{align}
In combination with eq.~(\ref{eq-staplehermprop}), this confirms that $\tAmp_{2}$ is a $\mathscr{T}$-even amplitude, $\tAmp_{2}(\ldots,1)=\tAmp_{2}(\ldots,-1)$, while $\tAmp_{12}$ and $\tilde B_{8}$ are $\mathscr{T}$-odd, $\tAmp_{12}(\ldots,1)= -\tAmp_{12}(\ldots,-1)$ and $\tilde B_8(\ldots,1)= -\tilde B_{8}(\ldots,-1)$, as stated in Ref. \cite{Goeke:2005hb}.\footnote{Note that the third argument of the amplitudes also changes sign under combined $(\dagger)$ and $(\mathscr{T})$.}

In our specific case, $v\tcdot \elll = 0$ and $\elll \tcdot P = 0$, so according to eq.~(\ref{eq-staplehermprop}), all our amplitudes will be real valued. From our master formula eq.\,(\ref{eq-threeptratio}) and with the LHPC choice of spin projectors, we obtain the ratio
\begin{equation}
	R^\ren_{\gamma_\Eu{4}}(\vect{P},\elll,\eta v) = 2 \tilde a_2 - \frac{2 i m_N \elll_\Eu{2}}{E(P)} \left( P_\Eu{1}\, \tilde a_{12} + m_N^2\, \eta v_\Eu{1}\,\tilde b_8\right) \ .
	\label{eq-ratiog4v}
\end{equation}

Our test calculation has been carried out on the 425 HYP smeared gauge configurations of the coarse-m050 ensemble. In Figure~\ref{fig-staplereg4}, we plot the real part of the unrenormalized ratio $R_{\gamma_\Eu{4}}(\vect{P},\mathcal{C}^\lat_{\elll,\eta v})$ for our choices of $\elll$ and $v$. Clearly, the ratio goes to zero for large $|\eta v \tcdot P|$. By now we know that this is the expected behavior: the gauge link on the lattice decays with increasing length. Partly, this decay can be attributed to a renormalization factor $\exp(2 \delta m |\eta|)$. We realized that we can cancel all renormalization factors introduced by the gauge link if we divide by the square root of a vacuum expectation value of a rectangular Wilson loop of dimensions $2|\eta|\times\sqrt{-\elll^2}$, see illustration Fig.~\ref{fig-stapleloop} and the discussion in  section \ref{sec-selfenergy} and Ref.~\cite{Collins:2008ht}. The quantity
\begin{equation}
	R^{\text{divW}}_{\gamma_\Eu{4}}(\vect{P},\mathcal{C}^\lat_{\elll,v}) \equiv \frac{R_{\gamma_\Eu{4}}(\vect{P},\mathcal{C}^\lat_{\elll,\eta v})}{\sqrt{W(2|\eta|,\sqrt{-\elll^2})}} 
\end{equation}
does not need renormalization up to the quark field renormalization factor $\renZ_\psi$. We plot the real part of $R^{\text{divW}}_{\gamma_\Eu{4}}$ in Fig.~\ref{fig-staplereg4divw}. For short quark separations $\elll$, the divided ratio appears to reach a plateau value, just at the border to the gray shaded regions. According to eq.~(\ref{eq-ratiog4v}), this plateau value gives access to $2 \tilde A_2$  (divided by the Wilson loop factor):
\begin{equation}
	\myRe\ R^{\text{divW}}_{\gamma_\Eu{4}}(\vect{P},\mathcal{C}^\lat_{\elll,\eta v}) \xrightarrow{ |\eta v \tcdot P| \text{ large}}
	\tAmp_{2}^\text{divW} (\elll^2,0,0,\zeta^{-1},\pm 1)\ ,
\end{equation}
where, for our parameters, $\zeta^{1/2} = |v\tcdot P| = 0.52 \units{GeV}$. Inside the shaded regions, the Wilson loop overlaps with itself on the periodic lattice ($2|\eta| \geq L = 20$). Obviously, data points inside the shaded area must be discarded. For longer quark separations $\elll$, statistics breaks down before we can discern a plateau. However, we clearly observe that the divided amplitude $\tAmp_{2}^\text{divW}$ rises with increasing quark separation. Whether this divided amplitude gives access to useful, well-defined \TMDs requires further studies.

\subsection{Ratios of Amplitudes, \texorpdfstring{$\mathscr{T}$}{T}-odd Effects from the Lattice}
\label{sec-oddratio}

Already in section \ref{sec-ampratios} we noted that ratios of amplitudes are theoretically attractive, since they are inherently independent of the lattice renormalization scheme and scale. (That said, we should keep in mind that we found sizeable lattice cutoff effects for quark separations $\elll$ shorter than about $2a$.) In the context of our test calculation with extended links, let us look at the quantity
\begin{align}
	R_\text{odd}(\vect{P},\mathcal{C}^\lat_{\elll,v}) & \equiv 
	- \frac{E(P)}{m_N\,\elll_\Eu{2}\,P_\Eu{1} }
	\frac{\myIm\ R_{\gamma_\Eu{4}}(\vect{P},\mathcal{C}^\lat_{\elll,\eta v})}{\myRe\ R_{\gamma_\Eu{4}}(\vect{P},\mathcal{C}^\lat_{\elll,\eta v})} = 
	\frac{ \tilde a_{12} + (m_N^2 \eta v_\Eu{1}/P_\Eu{1})\,\tilde b_8}{\tilde a_2 } \nonumber \\ 
	& \xrightarrow{\pm \eta v\tcdot P \text{ large}}
	\frac{ \tAmp_{12}(\elll^2,0,0,\zeta^{-1},\pm 1) + (m_N/P_\Eu{1})^{2}\,\tilde B_8(\elll^2,0,0,\zeta^{-1},\pm 1)}{\tAmp_2(\elll^2,0,0,\zeta^{-1},\pm 1) } \ ,
\end{align}
where again $\zeta^{1/2} = 0.52 \units{GeV}$, and where $(m_N/P_\Eu{1})^2 \approx 9.9$ for the study at hand.
The quantity above allows us to have a first glimpse at $\mathscr{T}$-odd effects from the lattice.
The signal we observe for $R_\text{odd}$ contains a contribution from the Sivers function $f_{1T}^\prp(x,\vprp{k}^2)$ via the amplitude $\tAmp_{12}$, compare section \ref{sec-todd}. The values $\tAmp_{12}(\ldots, 1)$ and $\tAmp_{12}(\ldots,-1)$, differing only in sign, belong to different experimental setups, e.g., SIDIS versus Drell-Yan.

\begin{figure}[tbp]
	\hbox to \linewidth{\smash{\topalignbox{\subfloat[][]{%
		\label{fig-staplereg4}\hphantom{(M)}%
		}}}\hphantom{M}\hfill%
	\topalignbox{\includegraphics[clip=true]{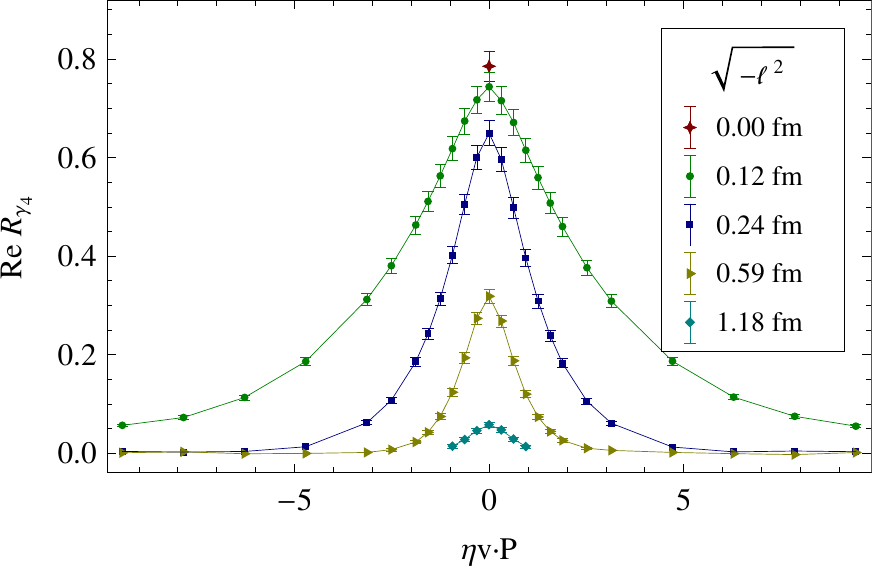}}%
	\hfill\hphantom{M}}\par\vspace{2em}%
	\hbox to \linewidth{\smash{\topalignbox{\subfloat[][]{%
		\label{fig-staplereg4divw}\hphantom{(M)}%
		}}}\hphantom{M}\hfill%
	\topalignbox{\includegraphics[clip=true]{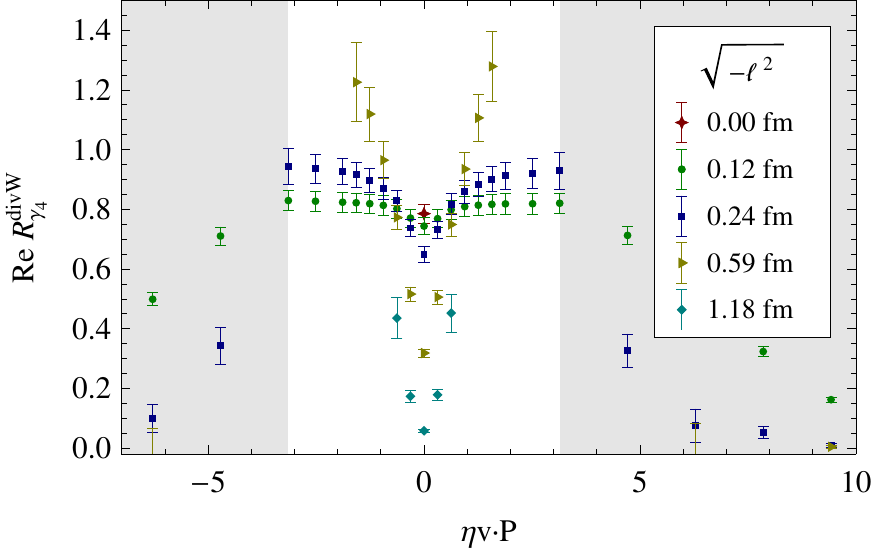}}%
	\hfill\hphantom{M}}\par\vspace{2em}%
	\hbox to \linewidth{\smash{\topalignbox{\subfloat[][]{%
		\label{fig-stapleimg4diva2}\hphantom{(M)}%
		}}}\hphantom{M}\hfill%
	\topalignbox{\includegraphics[clip=true]{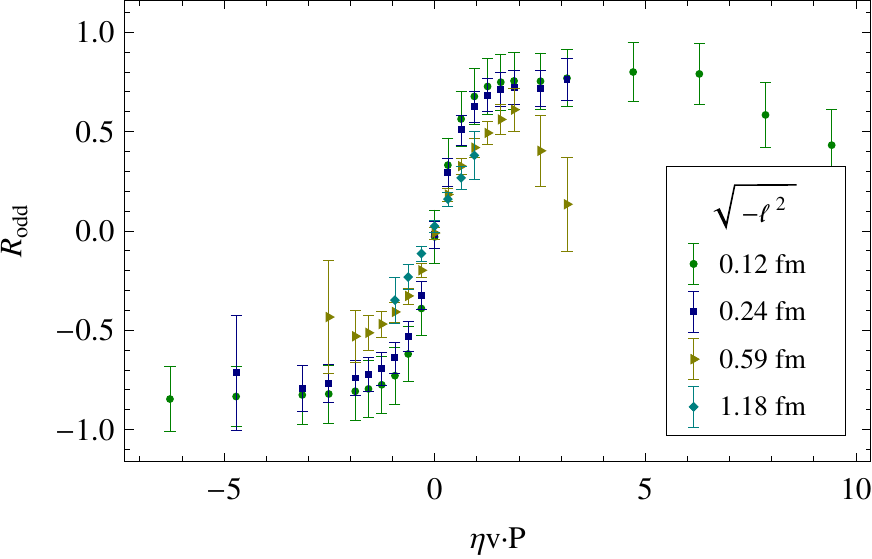}}%
	\hfill\hphantom{MM}}\par\vspace{1em}
	\caption[Results for Staple Link]{%
		Results from a test calculation with staple-shaped link paths on the smeared coarse-m050 ensemble. Details regarding the setup are described in the text.\par
		\subref{fig-staplereg4} Real part of the unrenormalized ratio $R_{\gamma_\Eu{4}}(\vect{P},\mathcal{C}^\lat_{\elll,\eta v})$. \par
		\subref{fig-staplereg4divw} Real part of the unrenormalized ratio divided by a Wilson loop subtraction factor. In the shaded region, the Wilson loop overlaps with its periodic image on the lattice.\par
		\subref{fig-stapleimg4diva2} Relative size of a linear combination of $\mathscr{T}$-odd amplitudes to $\tAmp_2$, obtained from a ratio of imaginary and real part of $R_{\gamma_\Eu{4}}(\vect{P},\mathcal{C}^\lat_{\elll,\eta v})$.
		\label{fig-stapleres}
		}
\end{figure}

Fig.~\ref{fig-stapleimg4diva2} displays first encouraging results for $R_\text{odd}$. 
Firstly, the signal is clearly odd in $\eta v \tcdot P$, i.e., the lattice reproduces the expected $\mathscr{T}$-odd effects. The data points for $\sqrt{-\elll^2}=0.12\units{fm}$ and $0.24\units{fm}$ reach a stable plateau once $|\eta v \tcdot P|$ becomes greater than about $2$. At a quark separation of $5a \approx 0.59\units{fm}$, it still looks like the data reaches a plateau value before statistics breaks down, and one may read off an estimate for the relative size of the linear combination $\tAmp_{12} + (m_N/P_\Eu{1})^{2}\,\tilde B_8$ compared to $\tAmp_{2}$. 

Considering these results of our test calculation, it looks like it is possible to use lattice QCD for the calculation of ``realistic'' \TMDs, with gauge links as they occur in the description of experimental scattering processes such as SIDIS. Limitations are of technical, not of principal nature: the small nucleon momenta $|\vect{P}|$ available in present simulations restrict the evolution parameter $\zeta$ to rather small values, and the proliferation of statistical noise as we increase the staple extent $\eta$ confines us to small quark separations. Certainly, a smart choice of directions $v$, $\elll$, $\vect{P}$ and the spin projection matrices $\GammaThr$ is required to disentangle the 32 independent amplitudes $\tAmp_i$ and $\tilde B_i$ \cite{Goeke:2005hb} from the real and imaginary parts of the 16 channels $\GammaOp$. As long as some open questions concerning the renormalization remain, we may resort to the calculation of ratios of amplitudes. Such ratios enable us to characterize the relative size of spin dependent phenomena (compare section \ref{sec-ampratios}) or, as in the present case, $\mathscr{T}$-odd effects. 


\captionsetup[subfigure]{position=bottom}
\captionsetup[subfigure]{position=top}

\chapter{Conclusion}
\label{chap-conclusion}

\section{Summary}

This work has been a first exploration of concepts, techniques, prospects and limitations for the calculation of transverse momentum dependent quark distribution functions (\TMDs) with lattice QCD. Comprehensive test calculations, mostly with a simplified operator (straight Wilson line), have produced encouraging first results and have enabled us to study renormalization properties of the relevant non-local operators in practice. We summarize the results as follows:

\begin{itemize}
	\item We can directly implement non-local operators $\bar q(\elll) \WlineC{\mathcal{C}_\elll} q(0)$ on the lattice, by assembling the gauge link $\WlineC{\mathcal{C}_\elll}$ as a product of link variables connecting the two quark fields. 
	\item A gauge link of length $l$ introduces a power divergence of the form $\exp(- \delta \hat m(a)\, l / a)$. Based on an analysis of smeared and unsmeared ensembles with four different lattice spacings, we confirm this behavior for our lattice operators with gauge links longer than about two lattice spacings. We have tested and compared methods to determine the renormalization constant $\delta \hat m$. Two different non-perturbative methods, based on the static quark potential and on Wilson lines, prove to be successful and consistent with each other. The value $\delta \hat m$ is defined unambiguously only with respect to a renormalization condition. 
	\item Fixing a renormalization condition, we are able to specify our results in terms of amplitudes $\tAmp_i$ unambiguously and independent of the lattice scheme and scale, up to a global multiplicative constant $\renZ^{-1}_{\psi,z}$. The \TMDs are related to the corresponding amplitudes $\tAmp_i$ by a Fourier transformation. 
	\item Restrictions from the Euclidean metric of the lattice preclude us from calculating the $x$-dependence of \TMDs directly. However, we do have access to Mellin moments. The first Mellin moment is just the $x$-integral of \TMDs and describes the $\vprp{k}$-dependent distribution of quarks irrespective of their longitudinal momentum.
 	\item \label{it-gaussfits} From Gaussian fits to our amplitudes, we have calculated the first Mellin moment of selected \TMDs, namely $f_1^{(1)\text{sW}}(\vprp{k})$, $g^{(1)\text{sW}}_{1L}(\vprp{k})$ and $g^{(1)\text{sW}}_{1T}(\vprp{k})$. Note that:
	\begin{itemize}
	\item We have connected the quark fields in our correlators with a straight Wilson line. The \TMDs we get (tagged with a superscript ``sW'') are therefore not strictly identical to those defined and used in the literature and for the description of, e.g., SIDIS. 
	\item In order to obtain unambiguous results, we need to fix a renormalization condition for $\delta \hat m$. We choose a condition based on the static quark potential ($C^\ren = 0$, see section \ref{sec-WilsonLineRen}). The precise meaning of the renormalization condition in the context of factorization theorems for \TMDs has still to be worked out.
	\item We work at a pion mass of about $500\units{MeV}$. 
	\item To be able to do the Fourier transform, we fit Gaussian functions to our amplitudes, excluding input data for quark separations smaller than $0.25 \units{fm} \approx 2 a$ because of lattice cutoff effects. 
	\item 
	The resulting Gaussian \TMDs clearly cannot reproduce the large-$\vprp{k}$ behavior expected from perturbation theory, which is related to a singular dependence on $\elll$ at short distances.
	\end{itemize}
	\item We interpret our results in terms of quark densities as functions of the intrinsic transverse momentum of quarks inside the nucleon. The following observations are based on $f_1^{(1)\text{sW}}(\vprp{k})$, $g^{(1)\text{sW}}_{1L}(\vprp{k})$ and $g^{(1)\text{sW}}_{1T}(\vprp{k})$ determined under the conditions and subject to the limitations mentioned above:
	\begin{itemize}
		\item The unpolarized quark density in an unpolarized nucleon is axially symmetric in the transverse momentum plane. The width of the distribution, given by the root mean square transverse momentum, is $\langle \vprp{k}^2 \rangle^{1/2}_{\rho_{UU}} = (381 \pm 28) \units{MeV}$ for $u-d$ quarks. This number is very sensitive to the renormalization condition we have imposed, but seems to be rather insensitive to the quark masses on our ensembles with pion masses ranging from about $500$ to $800\units{MeV}$.
		\item The density of longitudinally polarized quarks in a transversely polarized nucleon is no longer axially symmetric but appears deformed. This is due to a sizable contribution from $g^{(1)\text{sW}}_{1T}$. The peak of the density is shifted along the direction of the transverse nucleon spin vector. The direction of the shift is opposite for up and down quarks. The magnitude of the deformation can be characterized by the average transverse momentum, which we determine to be $\langle \vprp{k} \rangle_{\rho_{TL}} = \left[ (73 \pm 5) \units{MeV} \right] \lambda \vprp{S}$ for up quarks, and $\langle \vprp{k} \rangle_{\rho_{TL}} = \left[ (-31 \pm 5) \units{MeV} \right] \lambda \vprp{S}$ for down quarks, all at a pion mass of about $500\units{MeV}$. These results are rather insensitive to the renormalization condition.
		Numbers compatible to the ones quoted above have been obtained with an alternative method, which involves extrapolation of a ratio of amplitudes that does not need renormalization.
		\item As a cross check, we compute the axial vector coupling constant $g_A$ from the $\vprp{k}$-integral of $g^{(1)}_{1L}(\vprp{k})$, and find reasonable agreement with results in the literature extracted from the same ensembles. 
	\end{itemize}
	\item The dependence on the longitudinal momentum fraction $x$ is encoded in the $\elll \tcdot P$-dependence of our amplitudes. Analyzing the renormalization scheme independent, normalized amplitude $\tAmp_2(\elll^2,\elll \tcdot P)/\tAmp_2(\elll^2,0)$, we observe factorization in $\elll^2$ and $\elll \tcdot P$ within our statistical and kinematical limits. This corresponds to the factorization $f^{\text{sW}}_1(x,\vprp{k}) \approx f_1(x) f_1^{(1)\text{sW}}(\vprp{k}) / \mathcal{N}$. The results for our normalized amplitude are in qualitative agreement with a phenomenological parametrization of $f_1(x)$ and with a diquark model.
	\item We carry out first studies with extended, staple-shaped gauge links, as they appear in the description of scattering experiments like SIDIS or the Drell-Yan process. The attainable evolution parameter $\zeta\equiv(P \cdot v)^2/|v^2|$ is limited by the maximum nucleon momentum $|\vect{P}|$ achievable on the lattice. It is important to cancel the length dependent renormalization factors associated to the gauge link. For a time-reversal odd ratio of amplitudes, our test calculation yields a non-zero signal of encouraging quality.
\end{itemize}


\section{Open Questions and Future Projects}

Our study motivates a number of future research topics:
\begin{itemize}
	\item So far, we have studied only a small selection of \TMDs. A large number of other structures with interesting density interpretations can be readily analyzed with the techniques described in this work. 
	\item Can we detect the short distance behavior predicted by perturbation theory in our amplitudes obtained from the lattice?
	\item In order to remove the power divergence associated to the gauge link, we have introduced a renormalization condition. How can we interpret this renormalization condition in terms of a continuum renormalization or factorization scale?
	\item Can effective field theories or auxiliary field techniques give us hints regarding these issues?
	\item It will be interesting to expand our test calculations with staple-shaped gauge links. Are the evolution equations in $\zeta$ applicable to lattice results? Until all issues of renormalization are resolved, one may resort to ratios of amplitudes, which should allow us to estimate the relative sizes of spin dependent or time-reversal-odd effects.
	\item In particular in connection with ``realistic'', staple-shaped gauge links, how should we set up a Wilson loop subtraction factor to produce \TMDs that have a well-defined meaning as probability densities or as a non-perturbative ingredient to a factorized cross-section of a high-energy scattering process?
	\item Can we improve our lattice operators by perturbatively motivated correction terms, similar as in the case of the static potential?
\end{itemize}

\newpage
\section{Résumé}

Within this thesis, we have explored ways to calculate intrinsic transverse momentum distributions in hadrons with lattice QCD. There are remaining challenges to design well-defined observables that can describe these distributions. Using a simplified definition, we provide for the first time an insight into transverse momentum dependent quark distributions of the nucleon from a model independent calculation within full QCD. As an example, we determine the strength of an axially asymmetric deformation of a spin-selective quark distribution inside the spin-polarized nucleon. Unlike most lattice calculations of hadron structure, we directly implement non-local operators and remove their length dependent divergence non-perturbatively. First test calculations indicate that lattice QCD has the potential to go beyond the simplified operator definition. It is well conceivable that the lattice perspective, being truly non-perturbative from the start and offering a numerical test bed, will be helpful in developing a conceptually improved definition. Considering our initial success, we are optimistic that lattice QCD will become an important tool for the calculation of intrinsic transverse momentum dependent distributions of quarks.




\begin{appendices}

\chapter{Conventions and Useful Relations}
\section{Wilson Lines}
\label{sec-WLines}

Let the continuous, piecewise differentiable function $\mathcal{C}(\lambda)$, defined for $0\leq \lambda \leq 1$, pa\-ra\-me\-trize a path from $x$ to $y$, i.e., $\mathcal{C}_{x,y}(0) = x$, $\mathcal{C}_{x,y}(1) = y$. We define a \terminol{Wilson line} (also called \terminol{gauge link}) along this path according to
\begin{equation}
	\WlineC{\mathcal{C}} \equiv \mathcal{P} \exp \left( -i\,g\int_{\mathcal{C}}\ d\xi^\mu\ \Afield_\mu(\xi) \right) = \mathcal{P} \exp \left( -i\,g\int_0^1 d\lambda\ \mathcal{C}'^\mu(\lambda)\ \Afield_\mu(\mathcal{C}(\lambda)) \right)\ .
	\label{eq-WLineAlongPath}
\end{equation}
Here $\mathcal{C}'(\lambda) = d\mathcal{C}(\lambda)/d\lambda$ is the derivative of the parametrization of the path. The path ordering symbol $\mathcal{P}$ indicates that the fields $\Afield_\mu(\mathcal{C}(\lambda))$ are sorted with increasing $\lambda$ from left to right after applying the definition of the matrix exponential. A Wilson line along a straight path from $x$ to $y$ will be denoted
\begin{equation}
	\Wline{x,y} \equiv \mathcal{P} \exp \left( -i\,g\int_x^y\ d\xi^\mu\ \Afield_\mu(\xi) \right)\ . 
	\label{eq-WLineStraight}
\end{equation}
Note that $(\Wline{x,y})^\dagger = \Wline{y,x}$. To keep notation concise, concatenations of Wilson lines along straight paths will be written as
\begin{equation}
	\Wline{x,y}\, \Wline{y,z} \equiv \Wline{x,y,z}
	\label{eq-WLineConcat}
\end{equation}
and accordingly for more than two line sections.

\section{Lightcone Coordinates}
\label{sec-lightconecoord}

We denote cartesian base vectors of Minkowski space $\hat n_0,\ldots,\hat n_3$. The non-vanishing components of the metric tensor $g_{\mu\nu} = \hat n_\mu \cdot \hat n_\nu$ are $g_{00} = - g_{11} = - g_{22} = - g_{33} = 1$. With base vectors ``on the lightcone'' given by
\begin{equation}
	\hat n_{\pm} \equiv \frac{1}{\sqrt{2}} \left( \hat n_0 \pm \hat n_3 \right)\ ,
\end{equation}
any four-vector $v$ has a decomposition
\begin{equation}
	v = v^+ \hat n_+ + v^- \hat n_- + v_\prp\ ,
\end{equation}
where $v_\prp$ is a spacelike vector in transverse direction, i.e. $v_\prp \equiv v^1 \hat n_1 + v^2 \hat n_2$, $v_\prp \cdot v_\prp \leq 0$.
We also introduce a corresponding Euclidean two-component vector $\vect{v}_\prp \equiv (v^1,v^2)^T$, $\vprp{v} \cdot \vprp{v} \geq 0$.
In light cone notation, the metric tensor is given by $g_{+-} = g_{-+} = - g_{11} = - g_{22} = 1$. 
The scalar product of two general four-vectors $v$ and $w$ thus reads
\begin{equation}
	v \cdot w = v^+ w^- + v^- w^+ + v_\prp \cdot w_\prp = v^+ w^- + v^- w^+ - \vprp{v} \cdot \vprp{w}\ .
\end{equation}
Projections onto the transverse directions can be accomplished with the tensor $g_\prp^{\mu\nu}$ whose only non-vanishing components are $- g_{\prp 11} = - g_{\prp 22} = 1$. The totally antisymmetric Levi-Civita symbol $\epsilon^{\mu\nu\rho\sigma}$ follows the convention $\epsilon^{0123}=\epsilon^{-+12}=1$. Its transverse projection $\epsilon_\prp^{\mu\nu} \equiv \epsilon^{-+\mu\nu}$ follows the convention $\epsilon_\prp^{12}=1$. In the two-dimensional, Euclidean transverse space we will denote it as a Levi-Civita symbol $\vect{\epsilon}_{\prp ij}$, with the transverse indices $i,j\in\{1,2\}$ and following the convention $\vect{\epsilon}_{\prp 1 2}=1$.

We choose a nucleon momentum $P$ on the light cone
\begin{equation}
	P = P^+ \hat n_+ + \frac{m_N^2}{2 P^+} \hat n_-\ .
	\end{equation}
In the frame we want to work in, the nucleon travels with large momentum in $\hat n_3$-direction, i.e.
$P^+ \gg m_N \gg P^-$.
The nucleon spin vector $S$ shall be parameterized as
\begin{equation}
	S = - \Lambda\, \frac{m_N}{2 P^+}\, \hat n_- + \Lambda\, \frac{P^+}{m_N}\, \hat n_+ + S_\prp\ .
	\label{eq-spinparam}
	\end{equation}
It fulfills $S \cdot P = 0$ and shall be normalized according to $-S^2 = \Lambda^2 + \vprp{S}^2 = 1$.

\section{Tensors in Minkowski Space}
\vspace{-2em}
\begin{align}
\epsilon^{0123} & = +1\,, &
\gamma^5 & = i \gamma^0 \gamma^1 \gamma^2 \gamma^3\,, & 
\sigma^{\mu\nu} & = \frac{i}{2}[\gamma^\mu, \gamma^\nu]\,, & 
\sigma^{\mu \nu}\gamma^5 & = \frac{i}{2} \epsilon^{\mu \nu \alpha \beta} \sigma_{\alpha \beta}\,. 
\label{eq-tensorsmink}
\end{align}

\section{Dirac Spinors of free particles in Minkowski Space}
\label{sec-diracspinorsmink}

For a nucleon of momentum $P$ and with a polarization vector $S$ fulfilling $S^2 = -1$, $S \cdot P = 0$, we define the Dirac spinor $U(P,S)$ such that
\begin{equation}
	U(P,S) \overline{U}(P,S) = (\slashed{P} + m_N)\ \frac{1}{2}( \Eins - \slashed{S} \gamma^5 ) \ .
	\label{eq-freespinor}
\end{equation}
Then 
\begin{equation}
	1 = \frac{1}{2 m_N}\, \overline{U}(P,S) U(P,S)\,, \quad
	S^\mu = \frac{1}{2 m_N}\, \overline{U}(P,S) \gamma^\mu \gamma^5 U(P,S)\ .
	\label{eq-spinorsubs}
	\end{equation}

\section{Gordon Identities in Minkowski Space}
\begin{minipage}{\textwidth}
\begin{align}
	\overline{U}(P,S')\ \gamma^\mu\ U(P,S) 
	&= \overline{U}(P,S')\ \frac{P^\mu}{m_N}\ U(P,S) \label{eq-gordonidfirst} \ ,\\
	\overline{U}(P,S')\ \gamma^\mu \gamma^5\ U(P,S) 
	&= \overline{U}(P,S')\ i \sigma^{\mu \nu} \gamma^5 \frac{P_\nu}{m_N}\ U(P,S) \ ,\\
	\overline{U}(P,S')\ \sigma^{\mu \nu}\ U(P,S) 
	&= \overline{U}(P,S')\ \epsilon^{\mu \nu \rho \sigma} \gamma_\rho \gamma^5 \frac{P_\sigma}{m_N}\ U(P,S) \ ,\\
	\overline{U}(P,S')\ \gamma^5\ U(P,S) 
	&= 0\ ,
	\label{eq-gordonidlast}
\end{align}
see, e.g., Ref.\,\cite{Diehl01}.
\end{minipage}

\section{Trace Projections in Minkowski Space}

The relation between a Dirac matrix $\Phi$ and its projection $\Phi^{[\GammaOp]}$ is defined as
\begin{equation}
	\Phi^{[\Gamma^{\mathrm{op}}]} \equiv \frac{1}{2}\, {\Tr}_\mathrm{D} (\ \Phi\ \Gamma^{\mathrm{op}}\ )\ .
	\label{eq-conv-phitrace}
\end{equation}
Then
\begin{equation}
	\Phi
	= \frac{1}{2}\, \Eins\, \Phi^{[\Eins]} 
	+ \frac{1}{2}\, \gamma_\mu\, \Phi^{[\gamma^\mu]} 
	+ \frac{1}{4}\, \sigma_{\mu \nu}\, \Phi^{[\sigma^{\mu \nu}]}
	- \frac{1}{2}\, \gamma_\mu \gamma^5\, \Phi^{[\gamma^\mu \gamma^5]}
	+ \frac{1}{2}\, \gamma^5\, \Phi^{[\gamma^5]}\ .
	\end{equation}

\section{Euclidean Space}
\label{sec-euklidean}

Euclidean four-vectors shall be denoted $x_\muE$ $(\muE=1..4)$. We stick to common conventions:
\begin{equation*}
	x_\Eu{4} \equiv i x^0 = i x_0\,,\hspace{10pt}
	x_\Eu{j} \equiv x^j = - x_j\,,\hspace{10pt}
	\epsilon_{\Eu{1}\Eu{2}\Eu{3}\Eu{4}}=+1\,.
\end{equation*}
\begin{equation}
	\gamma_\Eu{4} \equiv \gamma_0\,,\hspace{10pt}
	\gamma_\Eu{j} \equiv -i \gamma^j\,,\hspace{10pt}
	\gamma_\Eu{5} \equiv \gamma_\Eu{1} \gamma_\Eu{2} \gamma_\Eu{3} \gamma_\Eu{4} = -\gamma^5\,,\hspace{10pt}
	\sigma_{\muE\nuE} \equiv \frac{i}{2}[\gamma_\muE,\gamma_\nuE]\,,\hspace{10pt}
	\slashedE{p} \equiv \gamma_\muE p_\muE\,.
	\label{eq-euklidrules}
\end{equation}
where $j=1..3$. It follows that 
\begin{equation}
	\partial_\Eu{4} = -i \partial_0\ ,\quad
	\partial_\Eu{j} = \partial_j\ .
\end{equation}
We have the following rules for rewriting an expression in Euclidean notation (in order of precedence):
\begin{enumerate}
	\item If we need to raise or lower an index before we can apply one of the rules below, we multiply by $-1$.
	\item $\epsilon^{\mu \nu \alpha \beta} \rightarrow -i \epsilon_{\muE \nuE \alphaE \betaE}$\,.
	\item $\sigma^{\mu \nu} \rightarrow - \sigma_{\muE \nuE}$\,,\quad $\gamma^5 \rightarrow -\gamma_\Eu{5}$\,.
	\item $\gamma^\mu \rightarrow i \gamma_\muE$\,.
	\item $\partial_\mu \rightarrow \partial_\muE$\,.
	\item Upper indices $\mu$ become lower indices $\muE$.
\end{enumerate}
Useful identities:
\begin{equation}
	\{\gamma_\muE,\gamma_\nuE\} = 2 \delta_{\muE \nuE}\ ,	\quad
	x \tcdot x \equiv x_\mu x^\mu = - x_\muE x_\muE\ \equiv - x \cdotE x\ ,\quad
	\gamma^\mu \partial_\mu = i \gamma_\muE \partial_\muE\ ,\quad
	\slashed{p} = -i \slashedE{p}\ .
	\end{equation}
The Dirac equation in Euclidean space reads
\begin{equation}
	(\slashedE{\partial}+m)\psi = 0	\ .
	\end{equation}
The Dirac spinor of a free nucleon fulfills in Euclidean space
\begin{equation}
	U(P,S) \overline{U}(P,S) = (-i \slashedE{P} + m_N)\ \frac{1}{2}( \Eins - i \slashedE{S} \gamma_\Eu{5} )\ . 
	\label{eq-freespinorE}
\end{equation}
The trace projections keep their form:
\begin{equation}
	\Phi
	= \frac{1}{2}\, \Eins\, \Phi^{[\Eins]} 
	+ \frac{1}{2}\, \gamma_\muE\, \Phi^{[\gamma^\muE]} 
	+ \frac{1}{4}\, \sigma_{\muE \nuE}\, \Phi^{[\sigma^{\muE \nuE}]}
	- \frac{1}{2}\, \gamma_\muE \gamma_\Eu{5}\, \Phi^{[\gamma^\muE \gamma_5]}
	+ \frac{1}{2}\, \gamma_\Eu{5}\, \Phi^{[\gamma_5]}\ .
	\end{equation}
The Wilson line becomes
\begin{equation}
	\WlineC{\mathcal{C}} \equiv \mathcal{P} \exp \left( i\,g\int_{\mathcal{C}}\ d\xi_\muE\ \Afield_\muE(\xi) \right) = \mathcal{P} \exp \left( i\,g\int_0^1 d\lambda\ \mathcal{C}'_\muE(\lambda)\ \Afield_\muE(\mathcal{C}(\lambda)) \right)\ .
	\label{eq-WLineAlongPathLat}
\end{equation}
For a path on the lattice $\mathcal{C}^\lat = (x^{(n)},x^{(n-1)},\ldots,x^{(2)},x^{(1)},0)$ connecting the lattice sites at space-time locations $x^{(i)}$, we introduce gauge links as products of link variables according to
\begin{equation}
	\WlineClat{\mathcal{C}^\lat}\ \equiv\ U(x^{(n)},x^{(n-1)})\, \cdots\, U(x^{(3)},x^{(2)})\,U(x^{(2)},x^{(1)})\, U(x^{(1)},0)\ .
	\label{eq-lat-gaugelink}
\end{equation}

%
%

\chapter{Details}
\vspace{-2em}
\section{Parametrization of the Correlators}

\subsection{\texorpdfstring{$\MDM$}{M} structures}

\begin{align}
	\MDM_\Eins 
		&= 2\ A_1\ \ ,\nonumber \\
	\MDM_{\gamma^\mu} 
		&= \frac{2}{m_N}\ A_2\ P^\mu 
		 + \frac{2}{m_N}\ A_3\ k^\mu 
		 + \toddmark{\frac{2}{m_N^2}\ A_{12}\ \epsilon^{\mu \nu \alpha \beta}  P_\alpha k_\beta\ \gamma_\nu \gamma^5} \ ,\nonumber \\
	\MDM_{\sigma^{\mu \nu}} 
		&= \toddmark{\frac{2}{{m_N}^2}\ A_4\ \left( P^\mu k^\nu - P^\nu k^\mu \right)} 
		 + \frac{2}{m_N}\ A_9\ \epsilon^{\mu \nu \alpha \beta} P_\beta\ \gamma_\alpha \gamma^5 \nonumber \\
		&+ \frac{2}{m_N}\ A_{10}\ \epsilon^{\mu \nu \alpha \beta} k_\beta\ \gamma_\alpha \gamma^5 
		 + \frac{2}{m_N^3}\ A_{11}\ \epsilon^{\mu \nu \alpha \beta} k_\alpha P_\beta\ \slashed{k} \gamma^5 \ ,\nonumber \\
	\MDM_{\gamma^\mu \gamma^5} 
		&= -2\ A_6\ \gamma^\mu \gamma^5 
		 - \frac{2}{{m_N}^2}\ A_7\ P^\mu\ \slashed{k}\gamma^5 
		 - \frac{2}{{m_N}^2}\ A_8\ k^\mu\ \slashed{k} \gamma^5 \ ,\nonumber \\
	\MDM_{\gamma^5} 
		&= \toddmark{2i\ A_5\ \slashed{k} \gamma^5} \ .
	\label{eq-Mstructs}
	\end{align}

\begin{align}
	\tMDM_\Eins 
		&= 2\ \tAmp_1\ \nonumber \ ,\\
	\tMDM_{\gamma^\mu} 
		&= \frac{2}{m_N}\ \tAmp_2\ P^\mu 
		 + 2 i\,m_N\ \tAmp_3\ \elll^\mu 
		 + \left[ 2 i\ \tAmp_{12}\ \epsilon^{\mu \nu \alpha \beta}  P_\alpha \elll_\beta\ \gamma_\nu \gamma^5 \right] \ ,\nonumber \\
	\tMDM_{\sigma^{\mu \nu}} 
		&= \left[ 2 i\ \tAmp_4\ \left( P^\mu \elll^\nu - P^\nu \elll^\mu \right) \right]
		 + \frac{2}{m_N}\ \tAmp_9\ \epsilon^{\mu \nu \alpha \beta} P_\beta\ \gamma_\alpha \gamma^5 \nonumber \\
		&+ 2 i\, m_N\ \tAmp_{10}\ \epsilon^{\mu \nu \alpha \beta} \elll_\beta\ \gamma_\alpha \gamma^5 
		 - 2 \,m_N\ \tAmp_{11}\ \epsilon^{\mu \nu \alpha \beta} \elll_\alpha P_\beta\ \slashed{\elll} \gamma^5 \ ,\nonumber \\
	\tMDM_{\gamma^\mu \gamma^5} 
		&= -2\ \tAmp_6\ \gamma^\mu \gamma^5 
		 - 2 i\ \tAmp_7\ P^\mu\ \slashed{\elll}\gamma^5 
		 + 2\,{m_N}^2\ \tAmp_8\ \elll^\mu\ \slashed{\elll}\gamma^5 \ ,\nonumber \\
	\tMDM_{\gamma^5} 
		&= \left[ - 2\,m_N\ \tAmp_5\ \slashed{\elll} \gamma^5 \right]\ .
	\label{eq-Mtildestructs}
	\end{align}

An explicit expression for $\MDM_{\GammaOp}$ is
\begin{equation}
	\MDM_{\GammaOp}(k,P;\mathcal{C}_\elll) = \int \frac{d^4\elll}{(2\pi)^4}\ e^{-ik \cdot \elll}\ \bra{0} \bar \varphi^\dagger_N(\vect{P})\ \bar \quark(\elll)\, \GammaOp\ \WlineC{\mathcal{C}_\elll}\ \quark(0)\ \varphi^\dagger_N(\vect{P}) \ket{0}\ ,
\end{equation}
where $\varphi^\dagger_N(\vect{P}) \equiv \int d^3 x\ e^{-i\vect{P}\cdot\vect{x}}\, \psi_N^\dagger(x) |_{x^0=0}$ creates nucleons from the vacuum using the nucleon field $\psi_N(x)$. 

\subsection{Table of Ratios}
\begin{minipage}{\linewidth}
\begin{centering}
\renewcommand{\arraystretch}{2}
\begin{tabular}{|c|c|c|l|}
\hline
QDP code & $\GammaOp$ (Euclid.) & $\GammaOp$ (Mink.) & $\frac{1}{2} R^\ren_{\GammaOp}(\tau, \vect{P},\vect{\elll})$ \\ \hline
0 & $\Eins$ & $\Eins$ & $\displaystyle \frac{m_N}{E(P)}\,\tilde A_1$ \\
1 & $\gamma_\Eu{1}$ & $- i \gamma^1$ & 
	$\displaystyle - \frac{i}{E(P)}\,\tilde A_2\,\vect{P}_1 + \frac{m_N^2}{E(P)}\, \tilde A_3\, \vect{\elll}_1  - m_N\, \tilde A_{12}\, \vect{\elll}_2$ \\
2 & $\gamma_\Eu{2}$ & $- i \gamma^2$ & 
	$\displaystyle \frac{m_N^2}{E(P)}\,\tilde A_3\,\vect{\elll}_2 + m_N\,\tilde A_{12}\,\vect{\elll}_1$ \\
3 & $\frac{1}{2}[\gamma_\Eu{1},\gamma_\Eu{2}]$ & $i \sigma^{12}$ & 
	$\displaystyle -\frac{m_N}{E(P)}\,\tilde A_4\,\vect{\elll}_2 \vect{P}_1 + i\, \tilde A_9 + i m_N^2\, \tilde A_{11} \,(\vect{\elll}_3)^2$ \\
4 & $\gamma_\Eu{3}$ & $- i \gamma^3$ & 
	$\displaystyle \frac{m_N^2}{E(P)}\,\tilde A_3\,\vect{\elll}_3$ \\
5 & $\frac{1}{2}[\gamma_\Eu{1},\gamma_\Eu{3}]$ & $i \sigma^{13}$ & 
	$\displaystyle -\frac{m_N}{E(P)}\,\tilde A_4\,\vect{\elll}_3 \vect{P}_1 - i m_N^2\, \tilde A_{11} \,\vect{\elll}_2 \vect{\elll}_3$ \\
6 & $\frac{1}{2}[\gamma_\Eu{2},\gamma_\Eu{3}]$ & $i \sigma^{23}$ & 
	$\displaystyle i m_N^2\, \tilde A_{11} \,\vect{\elll}_1 \vect{\elll}_3$ \\
7 & $- \gamma_\Eu{4} \gamma_\Eu{5}$ & $\gamma^0 \gamma^5$ & 
	$\displaystyle i m_N\, \tilde A_7\, \vect{\elll}_3$ \\
8 & $\gamma_\Eu{4}$ & $\gamma^0$ & 
	$\displaystyle \tilde A_2 - \frac{i m_N}{E(P)}\,\tilde A_{12}\,\vect{\elll}_2 \vect{P}_1$ \\
9 & $\frac{1}{2}[\gamma_\Eu{1},\gamma_\Eu{4}]$ & $\sigma^{01}$ & 
	$\displaystyle i m_N\, \tilde A_4 \,\vect{\elll}_1  - \frac{i m_N^2}{E(P)}\,\tilde A_{10}\,\vect{\elll}_2$ \\
10 & $\frac{1}{2}[\gamma_\Eu{2},\gamma_\Eu{4}]$ & $\sigma^{02}$ & 
	$\displaystyle i m_N\, \tilde A_4 \,\vect{\elll}_2 + \frac{1}{E(P)}\,\tilde A_9\,\vect{P}_1 $ \\
   & & & $\displaystyle  + \frac{i m_N^2}{E(P)}\,\tilde A_{10}\,\vect{\elll}_1 + \frac{m_N^2}{E(P)}\,\tilde A_{11}\,(\vect{\elll}_3)^2 \vect{P}_1$ \\
11 & $\gamma_\Eu{3} \gamma_\Eu{5}$ & $i \gamma^3 \gamma^5$ & 
	$\displaystyle -\frac{i m_N}{E(P)}\,\tilde A_6 - \frac{i m_N^3}{E(P)}\,\tilde A_8\,(\vect{\elll}_3)^2$ \\
12 & $\frac{1}{2}[\gamma_\Eu{3},\gamma_\Eu{4}]$ & $\sigma^{03}$ & 
	$\displaystyle i m_N\, \tilde A_4 \,\vect{\elll}_3  - \frac{m_N^2}{E(P)}\,\tilde A_{11}\,\vect{\elll}_2 \vect{\elll}_3 \vect{P}_1$ \\
13 & $-\gamma_\Eu{2} \gamma_\Eu{5}$ & $-i \gamma^2 \gamma^5$ & 
	$\displaystyle \frac{i m_N^3}{E(P)}\,\tilde A_8\,\vect{\elll}_2 \vect{\elll}_3$ \\
14 & $\gamma_\Eu{1} \gamma_\Eu{5}$ & $i \gamma^1 \gamma^5$ & 
	$\displaystyle - \frac{i m_N^3}{E(P)}\,\tilde A_8\,\vect{\elll}_1 \vect{\elll}_3 - \frac{m_N}{E(P)}\,\tilde A_7\,\vect{\elll}_3 \vect{P}_1$ \\
15 & $\gamma_\Eu{5}$ & $-\gamma^5$ & $\displaystyle -\frac{m_N^2}{E(P)}\,\tilde A_5\,\vect{\elll}_3$ \\
\hline
\end{tabular} \par
\renewcommand{\arraystretch}{1}%
\end{centering}%
\vspace{5pt}
\tabcaption{
Plateau values of the ratios $R_{\GammaOp}^\ren(\tau, \vect{P},\vect{\elll})$ in terms of the amplitudes $\tilde A_i$. Here we employ the LHPC conventions for $\Gamma^\text{2pt}$ and $\Gamma^\text{3pt}$, i.e. the nucleons are spin-projected along the $z$-axis. We choose the nucleon momentum $\vect{P} = ( \vect{P}_1, 0, 0 )= ( P_\Eu{1}, 0, 0 )  $, and the quark separation is $\vect{\elll} = (\vect{\elll}_1, \vect{\elll}_2, \vect{\elll}_3) = (\elll_\Eu{1}, \elll_\Eu{2}, \elll_\Eu{3})$. Naturally, we have $\elll^0 = \elll_\Eu{4} = 0$ on the lattice. We scan through the different Dirac contractions $\GammaOp$, sticking to the conventions of the \texttt{QDP++} programming library \cite{QDPManual}. Note that $\tilde A_4$, $\tilde A_5$ and $\tilde A_{12}$ vanish for straight link paths.
}%
\label{tab-ratios}%
\end{minipage}

\subsection{General Fourier Transformation relating the \texorpdfstring{$\tAmp_i$}{amplitudes} to \TMDs}
\label{app-tmdpdfft}

From eq.\,(\ref{eq-phitransformpre}), we see that we need to rewrite Fourier transformations of the form
\begin{align}
Y_{j_1,j_2,\ldots,j_n} & \equiv \int \frac{d(\elll \cdot P)}{2\pi}\ e^{-i (\elll \cdot P) x} \int \frac{d^2 \vprp{\elll}}{(2\pi)^2}\ e^{i \vprp{\elll} \cdot \vprp{k}} \nonumber\\ & \times \vect{\elll}_{\prp j}\,\vect{\elll}_{\prp j_2}\cdots\vect{\elll}_{\prp j_n}\ (\elll^-)^m\ F(\elll^2,\elll \cdot P)\quad \Big \vert_{\elll^+ = 0}
\label{eq-transvft}
\end{align}
as an integration with respect to Lorentz-invariant quantities. Note that $\elll^2 \big \vert_{\elll^+=0} = -\vprp{\elll}^2$. We first concentrate on the integral with respect to $\vprp{\elll}$:
\begin{equation}
Y^\prp_{j_1,j_2,\ldots,j_n} \equiv \int \frac{d^2 \vprp{\elll}}{(2\pi)^2}\ e^{i \vprp{\elll} \cdot \vprp{k}} \ \vect{\elll}_{\prp j_1}\,\vect{\elll}_{\prp j_2}\cdots\vect{\elll}_{\prp j_n}\, F(-\vprp{\elll}^2,\elll \cdot P)\ . 
\end{equation}
We express the $\vect{\elll}_{\prp j}$ in terms of derivatives and integrate with respect to the angular degree of freedom:
\begin{multline}
Y^\prp_{j_1,j_2,\ldots,j_n} = \left( -i \frac{\partial}{\partial \vect{k}_{\prp j_1}} \right) \cdots \left( -i \frac{\partial}{\partial \vect{k}_{\prp j_n}}\right)\ \int \frac{d^2 \vprp{\elll}}{(2\pi)^2}\ e^{i \vprp{\elll} \cdot \vprp{k}} \  F(-\vprp{\elll}^2,\elll \cdot P) \\
= \left( -i \frac{\partial}{\partial \vect{k}_{\prp j_1}} \right) \cdots \left( -i \frac{\partial}{\partial \vect{k}_{\prp j_n}}\right)\ \int_0^\infty \frac{d(-\elll^2)}{2(2\pi)} \int_0^{2\pi} \frac{d\theta}{2\pi}\ e^{i\, \sqrt{-\elll^2}\, |\vprp{k}|\, \cos \theta}\  F(-\vprp{\elll}^2,\elll \cdot P) \\
= \left( -i \frac{\partial}{\partial \vect{k}_{\prp j_1}} \right) \cdots \left( -i \frac{\partial}{\partial \vect{k}_{\prp j_n}}\right)\ \int_0^\infty \frac{d(-\elll^2)}{2(2\pi)} \ J_0(\sqrt{-\elll^2} |\vprp{k}|)\  F(\elll^2,\elll \cdot P) \ .
\label{eq-transvftstep}
\end{multline}
Obviously the integral only depends on the absolute value of $\vprp{k}$, so we can replace
\begin{equation}
\frac{\partial}{\partial \vect{k}_{\prp j}} \rightarrow \frac{\partial |\vprp{k}|}{\partial \vect{k}_{\prp j}}\, \frac{\partial}{\partial |\vprp{k}|}
= \frac{\vect{k}_{\prp j}}{|\vprp{k}|} \sqrt{-\elll^2} \frac{\partial}{\partial \left( \sqrt{-\elll^2}\,|\vprp{k}|\right)}
= \frac{\vect{k}_{\prp j}}{m_N}\ \frac{1}{z}\, \frac{\partial}{\partial z}\ (-\elll^2) m_N^2 \ \frac{1}{m_N}\ .
\end{equation}
In the last step we have introduced the abbreviation $z \equiv \sqrt{-\elll^2}\,|\vprp{k}|$.
We now make use of a property of the Bessel function \cite{rade2000}
\begin{equation}
	\frac{1}{z} \frac{\partial}{\partial z} \left( \frac{J_p(z)}{z^p} \right)= - \frac{J_{p+1}(z)}{z^{p+1}}\ ,
\end{equation}
which holds for any non-negative integer $p$. Applying this iteratively in (\ref{eq-transvftstep}), we get
\begin{equation}
Y^\prp_{j_1,j_2,\ldots,j_n} = i^n \frac{\vect{k}_{\prp j_1} \cdots \vect{k}_{\prp j_n}}{m_N^n} \ \int_0^\infty \frac{d(-\elll^2)}{2(2\pi)} \ 
\frac{J_n(\sqrt{-\elll^2}\,|\vprp{k}|)}{\left(\sqrt{-\elll^2}\,|\vprp{k}|\right)^n} (-\elll^2 m_N^2)^n \left[ \frac{1}{m_N^n} F(\elll^2,\elll \cdot P) \right]\ .
\end{equation}
Regarding the integral with respect to $\elll \cdot P$, we only have to substitute $\elll^- = (\elll \cdot P)/P^+ \big \vert_{\elll^+=0}$ in eq.\,(\ref{eq-transvft}). Summarizing our results, we can write
\begin{equation}
Y_{j_1,j_2,\ldots,j_n} = \frac{\vect{k}_{\prp j_1} \cdots \vect{k}_{\prp j_n}}{m_N^n}\  
T_{m,n}(x,|\vprp{k}|)\left[ \frac{1}{(P^+)^{m}\ m_N^n}\,F(\elll^2,\elll \cdot P) \right]\ ,
\end{equation}
where we have defined
\begin{multline}
T_{m,n}(x,|\vprp{k}|)\left[ \langle\text{expression}\rangle \right] \equiv \\
i^n \int \frac{d(\elll \cdot P)}{2\pi}\ e^{-i (\elll \cdot P) x}\ (\elll\cdot P)^m\ \int_0^\infty \frac{d(-\elll^2)}{2(2\pi)} 
\frac{J_n(\sqrt{-\elll^2}\,|\vprp{k}|)}{\left(\sqrt{-\elll^2}\,|\vprp{k}|\right)^n} (-\elll^2 m_N^2)^n \ \left\langle \text{expression}\right\rangle\ .
\end{multline}
Note that the operation $T_{m,n}$ has the dimension $\text{mass}^{-2}$.

\section{Ingredients to the Perturbative Calculation}
\label{sec-pertingred}

The inverse gluon propagator for the MILC gluon action is given by \cite{Bistro}
\begin{align}
	\tilde D_{\muE\nuE}(k) & = 
	\hat{\kappa}_\muE \hat{\kappa}_\nuE + 
	g_{\muE\nuE} \left( 
		\left(1 - (c_2 + c_3) \hat{\kappa}^2\right) \sum_{\Eu{\rho}\neq\muE}\hat{\kappa}_\Eu{\rho}^2
		- \left(	\sum_{\Eu{\rho}} \hat{\kappa}_{\Eu{\rho}}^4 - \hat{\kappa}_\muE \hat{\kappa}_\muE \hat{\kappa}^2 \right) (c1 - c2 - c3) 
		\right) \nonumber \\
   & - \hat{\kappa}_\muE \hat{\kappa}_\nuE  (1 - g_{\muE\nuE})  \left(
			1 - c_1 (\hat{\kappa}_\muE^2 + \hat{\kappa}_\nuE^2) - (c_2 + c_3) \sum_{\Eu{\rho}\neq\muE,\nuE} \hat{\kappa}_\rho^2 
			\right) \ ,
	\label{eq-MILCinvprop}
\end{align}
where $\hat{\kappa}_\muE \equiv 2 \sin(a k_\muE / 2)$. Note that the expression $\tilde D_{\muE\nuE}(k)$ alone corresponds to the gauge with $\xi=1$ in eq.\,(\ref{eq-definversegluprop}). The parameters $c_i$ are determined according to
\begin{align}
	c_0 &= 5/3\,, &
	c_1 &= \frac{-1}{12 u_0^2} (1 + 0.4805 \alpha_s)\,, &
	c_2 &= 0\,, &
	c_3 &= \frac{-5}{3 u_0^2} 0.03325 \alpha_s\,.
\end{align}
where $\alpha_s = -4 \log(u_0)/3.0684$. 

For perturbative calculations with HYP blocking, the coefficients $\tilde h_{\muE\nuE}(k)$ are given by \cite{DeGrand:2002va,Bistro}
\begin{align}
	\tilde h_{\muE\nuE}(k) & = \delta_{\muE\nuE} \left( 1- \frac{\alpha_1}{6} \sum_{\Eu{\rho}} \hat{\kappa}_{\Eu{\rho}}^2\, \Omega_{\muE\Eu{\rho}}(k)\right) + \frac{\alpha_1}{6} \hat{\kappa}_\muE\, \hat{\kappa}_\nuE\, \Omega_{\muE\nuE}(k) \ ,\nonumber \\
	\Omega_{\muE\nuE}(k) & = 1+ \alpha_2(1+\alpha_3) - \frac{\alpha_2}{4}(1+2\alpha_3)(\hat{\kappa}^2 - \hat{\kappa}^2_\muE -\hat{\kappa}^2_\nuE) + \frac{\alpha_2 \alpha_3}{4} \prod_{\Eu{\eta} \neq \muE,\nuE} \hat{\kappa}^2_\Eu{\eta}\ .
\end{align}
where the coefficients $\alpha_i$ have been chosen as specified in section \ref{sec-HYPsmear}.

\section{Gluon-Exchange Corrections to the Static Potential}
\label{sec-glucorr}

\begin{figure}[tbp]
	\centering%
	\subfloat[][]{%
		\label{fig-glucorrunsm}%
		\includegraphics[clip=true]{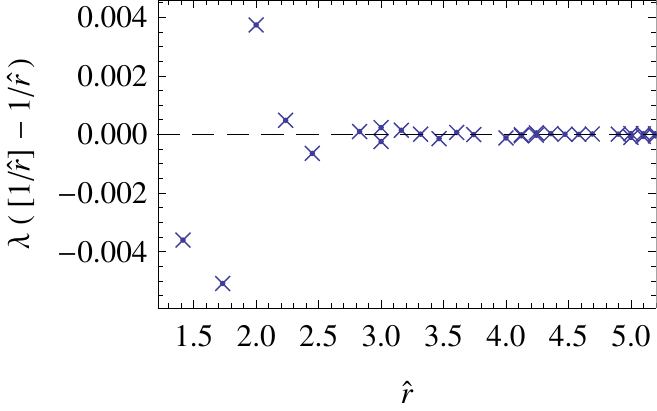}%
		}\hfill%
	\subfloat[][]{%
		\label{fig-glucorrsm}%
		\includegraphics[clip=true]{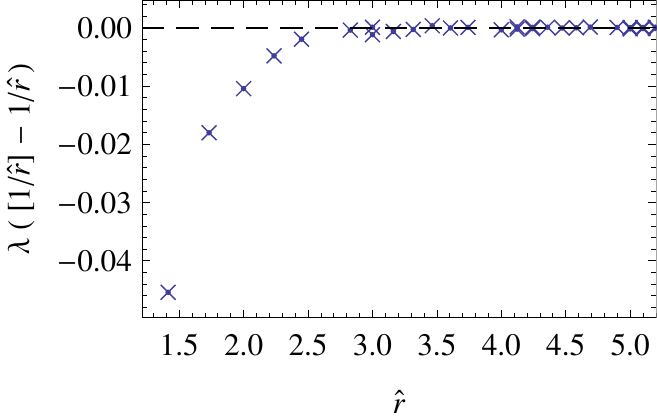}%
		}\par%
	\caption[gluon exchange corrections]{%
		Corrections to the static quark potential as obtained from lattice perturbation theory for the coarse-m020 ensemble, plotted versus $\hat r = |\hat{\vect{r}}|$ for various lattice vectors $\hat{\vect{r}}$. For $\lambda$, we have taken the central value from our fits to lattice data. \par
		\subref{fig-glucorrunsm} Unsmeared ensemble, $\lambda = 0.30(35)$, \par
		\subref{fig-glucorrsm} Smeared ensemble, $\lambda = 0.442(29)$.
		\label{fig-glucorr}	
		}
\end{figure}

In Fig.~\ref{fig-glucorr} we show the size and shape of the corrections to the static quark potential as we apply them in our fits in section \ref{sec-implstaticpot}. We remark that the large uncertainty in $\lambda$ for the unsmeared ensemble is a harmless symptom caused by the small number of configurations and a reduced set of lattice vectors $\hat{\vect{r}}$ evaluated for the unsmeared ensemble. 
The qualitative features of the corrections plotted in Fig.~\ref{fig-glucorrunsm} resemble very much those shown for a different action in Ref.~\cite{Bali:2000gf}, in particular we note that the sign of the corrections alternates quickly. In contrast, for the HYP smeared ensemble, the corrections are systematically negative, and much larger in size. Provided our implementation of the HYP smeared propagator is correct, these results are an indication that HYP smearing introduces sizeable lattice artefacts at distances smaller than about three lattice spacings. The displeasing feature of these artefacts is that they are not easy to recognize in the data because of their ``smooth'' dependence on $\hat r$. 

\section{Fourier Transform of the High-\texorpdfstring{$\vprp{k}$}{Momentum} Behavior}
\label{sec-highkt}

Suppose $\Phi^{[\gamma^+](1)}(\vprp{k};P,S) = f_1^{(1)}(\vprp{k}) \approx b/\vprp{k}$ for some real constant $b$ above some threshold $|\vprp{k}| \gtrsim k_\mmin$. The corresponding contribution to the amplitude $\tAmp_2(\elll^2,0)$ reads
\begin{multline}
	\int_{|\vprp{k}| > k_\mmin} d^2 \vprp{k}\ \exp(-i \vprp{k} \cdot \vprp{\elll})\ \frac{b}{\vprp{k}^2} 
	= b \int_{k_\mmin}^\infty d|\vprp{k}|\ \frac{2\pi J_0(|\vprp{k}|\,|\vprp{\elll}|)}{|\vprp{k}|} \\
	= 2 \pi b \int_{k_\mmin |\vprp{\elll}|}^\kappa du\ \frac{J_0(u)}{u} + 2 \pi b \int_\kappa^\infty du\ \frac{J_0(u)}{u} \ .
\end{multline}
We now make an approximation valid for $k_\mmin |\vprp{\elll}| < \kappa \ll 1$. Taylor expansion yields $J_0(u) = 1 + \mathcal{O}(u^2)$, so we get
\begin{multline}
	2\pi b\, \ln \left(\frac{\kappa}{k_\mmin |\vprp{\elll}|} \right) + 2\pi b \int_\kappa^\infty du\ \frac{J_0(u)}{u} + \mathcal{O}(\kappa^2) + \mathcal{O}(k_\mmin^2 |\vprp{\elll}|^2) = \\
	- 2\pi b\, \ln \left(\frac{|\vprp{\elll}|}{l_0} \right) + \mathcal{O}\left((k_\mmin |\vprp{\elll}|)^2\right)\ . 
\end{multline}
for an appropriate choice of $l_0$.

\section{Error Propagation for the Gaussian Parametrization}
\label{sec-widtherrprop}

Let us derive a simple estimate how systematic errors in $\delta m$ affect the width of our Gaussian fits.
Suppose we ``fit'' a Gaussian curve to the lattice data at just two quark separations $l_1$ and $l_2$:
\begin{align}
	c_j \exp( - l_1^2 / \sigma_j^2 ) & = \renZ^{-1}_{\psi,z}\ \exp\left( -\delta m\, l_1 \right)\  \tAmp^\text{unren}_j(-l_1^2,0) \nonumber \ ,\\
	c_j \exp( - l_2^2 / \sigma_j^2 ) & = \renZ^{-1}_{\psi,z}\ \exp\left( -\delta m\, l_2 \right)\  \tAmp^\text{unren}_j(-l_2^2,0)\ .
\end{align}
Solving this for $1/\sigma_j^2$ and differentiating with respect to $\delta m$, we obtain a formula for the error in the inverse width of the Gaussian:
\begin{equation}
	\frac{\partial}{\partial(\delta m)}\, \left[\frac{1}{\sigma_j^2}\right] = \frac{1}{l_1 + l_2} \quad \Rightarrow \quad
	\Delta \left[ \frac{4}{\sigma_j^2} \right] =  \frac{4}{a(l_1 + l_2)} \Delta[\delta \hat m]\ .
	\label{eq-widtherrprop}
\end{equation}
For $l_1$ and $l_2$, we simply take our fit range: $l_1 = 0.25 \units{fm}$, $l_2 = L/2$.

\section{\texorpdfstring{$\vprp{k}$}{k}-Moments}
\label{sec-kprpmomentsformalism}

In general, $\vprp{k}$-moments of the TMD quark-quark correlator are of the form
\begin{equation}
	\int d^2 \vprp{k}\ \vect{k}_{\prp j_1} \cdots \vect{k}_{\prp j_n}\ \Phi^{[\GammaOp](1)}(\vprp{k};P,S)\ .
\end{equation}
Assuming the integral above exists, and assuming that $\tilde \Phi^{[\GammaOp]}(\elll,P,S)$ vanishes quickly enough for large $\vprp{\elll}$, we may we may rewrite the expression above with the help of eq.\,(\ref{eq-deffirstmellin}) as
\begin{equation}
	\frac{1}{P^+} \int d^2 \vprp{k} \int \frac{d^2\vprp{\elll}}{(2\pi)^2}\ \tilde \Phi^{[\GammaOp]}(\elll,P,S)\ \frac{\partial}{\partial i \vect{\elll}_{\prp j_1}} \cdots \frac{\partial}{\partial i  \vect{\elll}_{\prp j_n}}\ e^{i \vprp{k}\cdot \vprp{\elll} } \big\vert_{\elll^+ = \elll^- = 0}\ .
\end{equation}
After $n$ partial integrations, and carrying out the Fourier transform, we obtain 
\begin{equation}
	\frac{i^n}{P^+} \frac{\partial}{\partial \vect{\elll}_{\prp j_1}} \cdots \frac{\partial}{\partial \vect{\elll}_{\prp j_n}}\ \tilde \Phi^{[\GammaOp]}(\elll,P,S) \big\vert_{\elll = 0}\ .
\end{equation}
In order to analyze a particular channel, we insert the parametrization of $\tilde \Phi^{[\GammaOp]}$ in terms of amplitudes $\tAmp_i$ according to eq.\,(\ref{eq-phitildetraces}). Here we can make use of the relation
\begin{align}
	\frac{\partial}{\partial \vect{\elll}_{\prp j}}\ \tAmp_i(\elll^2,0) & = 
		2\, \vect{\elll}_{\prp j}\ \frac{\partial}{\partial (-\elll^2)} \tAmp_i(\elll^2,0) \ .
\end{align}
Thus we find
\begin{align}
	\int d^2 \vprp{k}\ \Phi^{[\gamma^+](1)}(\vprp{k};P,S) & = 2 \tAmp_2(0,0) \ ,\nonumber \\
	\int d^2 \vprp{k}\ \vprp{k}\ \Phi^{[\gamma^+](1)}(\vprp{k};P,S) & = 2 i \vprp{\elll} \frac{\partial}{\partial(-\elll^2)}\,2\tAmp_2(\elll^2,0) \big\vert_{\elll=0} \ ,\nonumber \\
	\int d^2 \vprp{k}\ \vprp{k}^2\ \Phi^{[\gamma^+](1)}(\vprp{k};P,S) & = 
	-4 \left( \frac{\partial}{\partial(-\elll^2)} + (-\elll^2) \frac{\partial^2}{\partial(-\elll^2)^2} \right) 2 \tAmp_2(\elll^2,0)\big\vert_{\elll=0} \ , \nonumber \\
	\int d^2 \vprp{k}\ \Phi^{[\gamma^+\gamma^5](1)}(\vprp{k};P,S) \big\vert_{\vprp{S}=0} & = -  2 \Lambda\, \tAmp_6(0,0) \ ,\nonumber\\
	\int d^2 \vprp{k}\ \vprp{k}\ \Phi^{[\gamma^+\gamma^5](1)}(\vprp{k};P,S) \big\vert_{\Lambda=0} & = 
	- \left( \vprp{S} + 2\, \vprp{\elll} (\vprp{S} \tcdot \vprp{\elll}) \frac{\partial}{\partial (-\elll^2)} \right)\, 2 m_N\tAmp_7(\elll^2,0) \big\vert_{\elll=0} \ .
	\label{eq-ktmomentsfromamps}
\end{align}

\end{appendices}

\bibliography{Musch_thesis}
\bibliographystyle{mybibstyle2} 


\cleardoublepage
\newpage
\pagestyle{empty}
{\huge\bfseries Acknowledgments\par \vspace{0.4cm}}
\phantomsection 
\addcontentsline{toc}{chapter}{Acknowledgments}
This work has been initiated and supervised by Dr. Philipp H\"agler, who set up a lattice gauge theory research group at the TU M\"unchen in 2005, financed by a grant from the Emmy Noether program of the German Research Foundation (DFG). In the past three years Philipp has supported me and my project within his research group thoroughly in a sincere and professional manner. I profitted from his scientific experience and insight in many fruitful discussions. His contacts to experts and computing resources throughout the world proved to be essential for the success of this work. 

In equal measure, I am obliged to Prof. Wolfram Weise. As a member of his chair, I very much enjoyed the creative and uncomplicated atmosphere he cultivates in seminar talks and in passionate scientific discussions.
I am particularly thankful to him for supporting my travels to meet experts of the field.

This thesis would not have been possible without the data that has been made available to us by the LHPC and MILC/USQCD collaborations. It would take up too much space here to list all names of the people who contributed to the production of these data. 

Particular thanks are due to Prof. John Negele and the members of his chair for their hospitality during a four week visit at the Massachusetts Institute of Technology in Boston, USA. It has been an honor to meet a very active part of the LHPC collaboration, and I appreciated very much the critical and important questions that were raised concerning my project. The trip was funded to a large part by the German Academic Exchange Service (DAAD).

I am indebted to to Carleton DeTar, Doug Toussaint and Robert Suger for providing access to very fine lattice configurations of the USQCD collaboration, and to Alexei Bazavov, for giving me details on their  renormalization prescription.

Throughout the project, Philipp and I were kindly invited several times to the Universit\"at Regensburg by Prof. Andreas Sch\"afer for discussions with him and his colleagues Meinulf G\"ockeler, Prof. Vladimir Braun and Prof. Gunnar Bali. We are thankful for indispensable insights, valuable advice and crucial references to literature.

In my programs, I have taken over \toolkit{C++} code snippets developed by Dru Renner, now at DESY Zeuthen, Germany. Regarding the parametrization, I profited from notes of Piet Mulders (Vrije Universiteit Amsterdam, Netherlands). Thanks are also due to Markus Diehl at DESY Zeuthen for very useful comments regarding the behavior at high transverse momentum, and to Igor Cherednikov (JINR, Dubna, Russia) for clarifications regarding his publications.

I should not fail to mention my gratitude towards all members of the T39 group for the great team atmosphere and for stimulating discussions, in particular with Alexander Laschka, Thomas Hell, Simon R\"o{\ss}ner, Nino Bratovi\'c, Dr. Martin G\"urtler and Prof. Norbert Kaiser. Thanks also extend to Dr. Stefan Recksiegel, who aided me in setting up the local computer cluster. 

Last but not least, I would like to say thanks to all those who enabled me to enjoy this interesting time of my life so much, especially to my girlfriend Anna, to whom I have to apologize for devoting so much time to research. I remain most indebted to my family, for their continuing support.



\newpage
\pagestyle{empty}
{\Large\bfseries Correction History\par \vspace{0.4cm}}

The following corrections have been made with respect to the version published on the server of Technische Universit\"at M\"unchen on June 19, 2009:
\begin{itemize}
\item 2009-06-26: Corrected signs in eq.~(\ref{eq-Ylinedef}).
\item 2009-07-09: Added a factor 2 on the left hand side of eq.~(\ref{eq-multrencond}).
\end{itemize}

\end{document}